\documentclass[rmp,eqsecnum,showpacs,floats,superscriptaddress]{revtex4}

\usepackage[latin1]{inputenc}
\usepackage[english]{babel}
\usepackage{graphicx}
\usepackage{setspace}
\usepackage{psfrag}
\usepackage{amssymb}
\usepackage{amsmath}
\usepackage{amscd}
\usepackage{eucal}
\usepackage{color}
\usepackage{bm}
\input{epsf}
\newcommand{\nc}{\newcommand}
\nc{\ovl}{\overline}
\nc{\bea}{\begin{eqnarray}}
\nc{\eea}{\end{eqnarray}}
\nc{\be}{\begin{equation}}
\nc{\ee}{\end{equation}}

\newcommand{\parent}[1]{\left( #1\right)}

\nc{\intps}{\int  {\rm d} {\bf q}\int {\rm d }{\bf p}}
\nc{\intbps}{\int {\rm  d} {\bf \bar q}\int {\rm d }{\bf \bar p}}
\nc{\intbpsp}{\int {\rm  d} {\bf \bar q}'\int {\rm d }{\bf \bar p}'}

\begin{document}
\title{Decoherence, Entanglement 
and Irreversibility in Quantum Dynamical Systems with Few Degrees of Freedom}

\begin{singlespace}
\author{Ph.~Jacquod} 
\affiliation{Department of Physics, University of Arizona, 
1118 E. Fourth Street, Tucson, AZ 85721}
\author{C.~Petitjean}
\affiliation{Institut I -- Theoretische Physik,
Universit\"at Regensburg, Universit\"atsstrasse 31, 
D-93053 Regensburg, Germany}
\date{\today}

\begin{abstract}
This review summarizes and amplifies on recent investigations of
coupled quantum dynamical systems with few degrees of freedom
in the short wavelength, semiclassical limit. Focusing on the correspondence
between quantum and classical physics,
we mathematically formulate and attempt to answer three
fundamental questions: (i) How can one drive a small dynamical quantum
system to behave classically ? (ii) What determines the rate at which two
single-particle quantum--mechanical subsystems become entangled when they interact ? 
(iii) How does
irreversibility occur in quantum systems with few degrees of freedom ? 
These three questions are posed in the context of the quantum--classical
correspondence for dynamical systems with few degrees of freedom, 
and we accordingly rely on two short-wavelength approximations
to quantum mechanics to answer them -- 
the trajectory-based semiclassical approach on one hand, and
random matrix theory on the other hand. We construct novel investigative
procedures towards decoherence and the emergence of classicality out
of quantumness in dynamical systems coupled to external degrees of freedom. 
In particular we show how dynamical properties of chaotic classical 
systems, such as local exponential instability in phase-space, 
also affects their quantum counterpart. For instance, it is often the
case that the fidelity with which a quantum state is reconstructed after an imperfect
time-reversal operation decays with the Lyapunov exponent of the
corresponding classical dynamics. For not unrelated reasons, but perhaps more
surprisingly,
the rate at which two interacting quantum subsystems become entangled can also be governed
by the subsystem's Lyapunov exponents. 
Our method allows at each stage in our investigations
to differentiate quantum coherent effects -- those related to phase interferences -- from
classical ones -- those related to the necessarily extended envelope of quantal
wavefunctions. This makes it clear that all occurences of Lyapunov exponents we witness
have a classical origin, though they require rather strong decoherence effects to
be observed. We 
extensively rely on numerical experiments to illustrate our findings and
briefly comment on possible extensions to more complex problems involving 
environments with many interacting dynamical systems, going beyond the
uncoupled harmonic oscillators model of Caldeira and Leggett. 
\end{abstract}
\maketitle 

\end{singlespace}

\begin{spacing}{1.22}

\tableofcontents

\end{spacing}

\newpage

\section{Introduction}
\subsection{Preamble}

It is certainly not an exaggeration to say that quantum mechanics has revolutionized
the way we see and apprehend the world surrounding us. Daily experience tells us that 
material objects have well defined position, extension and velocity, and that the three can 
be measured simultaneously. Then why should microscopic objects instead be
represented by probability clouds whose evolution is governed by a wave equation ? 
Interacting quantum systems are even more intriguing: after some finite interaction time, the
subsystems lose at least part of their individuality in that they can no longer be
described by a set of coordinates of their own. This entanglement property of quantum systems
lies in strong contrast with classical interacting systems -- the moon is still the moon and its 
dynamics can be described by a finite set of coordinates, well separated from the coordinates
of the earth, despite
millions of years of orbital partnership. There is no classical counterpart to entanglement.
These and many other celebrated peculiarities
of quantum mechanics have left many a physicist suspicious
about the validity of quantum mechanics, or at least doubtful that it is a complete theory and 
often at a loss
to give it an understandable interpretation. Yet,
decades of experimental tests and theoretical developments have totally comforted us -- quantum
theory has been confirmed to a precision without precedent.
On the purely mathematical front, quantum mechanics does not require an interpretation, it is a well
defined algorithm that performs perfectly well without ever failing.
Still the relationship between quantum and classical physics has to be clarified. For once, 
a new scientific theory should not only be successful where the older one failed, one 
additionally expects
that it reproduces the theory it is supposed to supersede
in the latter's regime of validity -- this is the correspondence principle. 
How comes then that 
the world surrounding us, despite being made of quantum mechanical building blocks, behaves
classically most of the time ? How does this -- at least apparent -- classicality emerge out of
quantumness ? Over the years, more and more precise answers have been given to those questions
on the quantum-classical correspondence. The current consensus is
that, first, quantum systems can never be totally isolated from their environments, and that, second,
even tiny
couplings to many, fast moving external degrees of freedom are often
sufficient to erase quantum coherence and to drive a quantum system's 
time-evolution away from the
Schr\"odinger equation towards, say, a Liouvillian evolution. Simultaneously, information about the
exact state of the system gets lost in the entanglement generated between system and environment.
Entropy increase follows, and the lost information never returns.
It is not our purpose here to discuss this scenario
in all details, as it has been described in reviews and textbooks~\cite{Joo03,Zur03}. 
Yet, we revisit some related aspects, with a focus put on dynamical properties
of quantum systems with few degrees of freedom, systems who often exhibit complex behaviors
due to the chaotic dynamics of their classical counterpart. Because our focus is on dynamical
aspects, let us first briefly 
discuss, at a qualitative level, what are the respective trademarks of classical and 
quantal dynamical systems.

Classical dynamical systems are deterministic. For any given initial condition,
the state of the system at any later (or earlier) time is uniquely determined by the equations of
motion. Restricting ourselves to Hamiltonian systems, the phase-space dynamics
is unitary and in particular characterized by the Liouville conservation of phase-space volumes. 
Despite this unitarity, dynamical nonlinearities and 
chaos can emerge when there are not enough constants of
motion to restrict the dynamics to invariant tori. When this happens, the behavior of
the system becomes unpredictable beyond a certain time horizon. This is due
to local exponential instability, the trademark of classical chaotic behavior, where two almost
indistinguishable initial conditions -- two sets of position and momentum coordinates
differing only by a minute phase-space displacement -- eventually 
move away from one another at an exponential rate. The impossibility of determining
initial conditions with infinite precision effectively results in unpredictability and an apparently
random behavior of classical chaotic systems beyond a certain time horizon. 
Extending that horizon
is in principle possible, but requires an exponentially finer resolution of the initial 
condition. Chaotic behavior does not require
large numbers of degrees of freedom, but already occurs in two-dimensional
autonomous (i.e. energy-conserving) classical Hamiltonian systems. 
Yet, chaotic Hamiltonian systems do not lose
their deterministic nature~\cite{Lic92,Gut90,Cvi05}. 

The situation is both similar and quite different in quantum dynamical systems. The time-evolution
defined by Schr\"odinger's equation is equally deterministic and unitary as the Liouville
flow. For a given initial wavefunction, the corresponding 
future (or past) wavefunctions are uniquely determined
at any given time, and the Hilbert space norm of the wavefunction
is conserved. Statistical unpredictability notoriously arises due to the projective measurement of that
wavefunction, but mathematically speaking, that does not make the time-evolution of the
wavefunction any less deterministic.
Quantum systems however strongly differ from classical systems in that 
they are described by extended wavefunctions -- not phase-space points -- whose 
Schr\"odinger time-evolution
is unitary in either position or momentum space -- not in phase-space. The
symplectic nature of the Liouville evolution is not present in the quantum world, and this prohibits
the emergence of chaos in quantum mechanics in the sense of local exponential 
phase-space instability, at least for long enough times. There does not seem to be
anything such as quantum chaos from a dynamical point of view, or if it exists, it must be 
quite different from classical chaos. A comment is in order here, which we will restate
several times in this review.
The importance of time scales should not be underestimated,
and it has been realized that, in the spirit of the Ehrenfest theorem, the center
of mass motion of narrow wavepackets does 
exhibit local exponential instability at short times~\cite{Haa92}.
That behavior gets however lost at longer times, 
once the spreading of the wavepacket renders the definition of its
center of mass practically impossible or at least irrelevant. 
Assuming an exponential spread of the wavepacket
with the system's Lyapunov exponent, this defines an {\it Ehrenfest time} scale
$\tau_{\rm E}$~\cite{Lar68,Ber78,Ber79,Chi81,Chi88},
which is the time it takes for the
underlying classical chaotic dynamics to exponentially stretch an initial narrow wave packet to
the linear system size. The Ehrenfest time is a break time for the classical-quantal correspondence in
isolated systems. Once this threshold is crossed, quantum coherent effects set in that need to be
taken into account by the theory.
The rule of thumb is that quantum mechanical wavepackets of spatial extension $\nu$
(say, the minimal wavelength 
authorized by Heisenberg's uncertainty principle) follow 
classical dynamics at times shorter
than $\tau_{\rm E}$, qualitatively because until then the number of classical trajectories on which
they propagate is not sufficient to give rise to important interference effects. At
larger times, the dynamical quantum-classical correspondence breaks down as the proliferation
of classical trajectories exploring very different regions of phase-space
gives rise to multiple interferences between pairs of paths. In chaotic
systems, the crossover
between these two regimes is rather sharp, thanks 
to the exponential spreading of the
wavepacket extension, $\nu \rightarrow \nu \exp[\lambda t]$, with the Lyapunov exponent $\lambda$
of the corresponding classical dynamics. Once one reaches a spread comparable to, say,
the system size $L$, the notion of a center of mass of the wavepacket is no longer well
defined -- this occurs roughly at the Ehrenfest time $\tau_{\rm E} = \lambda^{-1} \ln L/\nu$.
The argument of the logarithm is a semiclassically 
large parameter defining the semiclassical
limit $L/\nu \rightarrow 0$, and as such it is often identified with an inverse effective 
Planck's constant, $\hbar_{\rm eff} \equiv \nu/L$.

Despite these discrepancies in the dynamical behaviors of quantum and classical systems,
there is still a one-to-one correspondence between classical integrals of motion and good
quantum numbers. One might thus
wonder if and how quantum systems with a complete set
of good quantum numbers differ from quantum systems lacking some of them.
Integrability is indeed equally well defined in classical and in quantum mechanics, however 
the theories
differ in how much dynamical freedom is gained once perturbations destroy good quantum
numbers or integrals of motion. The search for
signatures of chaos in quantized, classically chaotic systems defines
the field of quantum chaos~\cite{Haa01,Cas95,Cvi05,Gut90}. 

These discrepancies in the dynamical behavior of quantal and classical systems raise 
a number of issues, many of them related to the correspondence principle. How comes, for
instance, that macroscopic systems clearly exhibit chaotic dynamical behaviors, despite
their being made of quantum building blocks ? If there is no quantum chaos, how comes there
is classical chaos at all ? As fundamental is the question of the robustness of classical and
quantal systems with few degrees of freedom
when submitted to external perturbations. One qualitatively expects that
any perturbation, no matter how small, significantly alters the time-evolution of classical
chaotic systems. Perturbations first kick initial conditions some distance away from where
they were, then chaos does the rest. The perturbation effectively generates a certain amount 
of uncertainty in the initial condition which 
blows up exponentially with time. Classical chaotic systems seem therefore to be 
extremely sensitive to
perturbations -- one sometimes speak of hypersensitivity -- 
much more so than regular or integrable systems. 
Some care has to be taken in how the question of the sensitivity is asked, however,
and it should be stressed that chaotic systems taken as an ensemble are characterized by
some rather large degree of universality -- the individual behavior of a given system is not much
different from the average behavior of the ensemble. This universality is often called for, for instance
it is largely used in investigations of the classical fidelity~\cite{Pro02,Ben02,Eck03,Ben03c,Ben03b}. 
Regular or integrable systems, on the other
hand are characterized by large system-dependent deviations from average behaviors, and
special care has to be taken when discussing averages and fluctuations in this case.

How sensitive are quantal systems ? Here it might well be expected that quantal
systems also exhibit a strong sensitivity to perturbations, not because of the classical
dynamical scenario we just sketched, but because quantumness lives in
Hilbert spaces. Small perturbations generate pseudo-random relative phase shifts of the time-evolved
wavefunction components. In the semiclassical, short-wavelength limit, the number of these 
components becomes larger and larger. One thus expects that at large enough times, 
the scalar product between two wavefunctions, time-evolved from the same initial
wavefunction, but under the influence of two slightly different Hamiltonians, will be down
by a prefactor exponentially small in 
the variance of the phase shift distribution. 
In other words, this dephasing mechanism can generate orthogonality
between the actual (dephased) and the ideal (not dephased) wavefunction. 

These two
mechanisms for sensitivity to external perturbations are obviously very different. The former
is dynamically driven, and the perturbation is invoked only to generate a slight kick in the initial
condition, while the latter is entirely due to the perturbation, generating dephasing of the action
integrals accumulated on an otherwise unperturbed dynamics.
They are
specific to the classical or quantal character of the dynamical system under consideration.
The former mechanism originates from the decay of overlap of spatially extended
wavefunction envelopes -- this is analogous to the decay of overlap of Liouville distributions,
and in this sense this mechanism is classical in nature. The latter mechanism, on the other hand,
emerges from the accumulation of uncorrelated phase shifts in the wavefunction components -- it
is of purely quantal origin and has no classical counterpart. 
At short times both mechanisms can influence quantum systems and which mechanism
is relevant depends on the balance between the average stability of classical orbits and
the rate of dephasing.

These aspects of the quantum-classical correspondence have been thorougly investigated
over the past decades, and the search for quantum signatures of chaos has provided much insight
into how classical dynamics manifests itself
in quantum mechanics~\cite{Cas95,Haa01,Gut90}. The basic question is "can one
determine from a system's quantum properties whether the classical limit of
its dynamics is chaotic or regular ? And if yes, how ?". 
One very successful approach has been
to look at the spectral statistics, in particular the
distribution of level spacings~\cite{Boh84}. An altogether different, more recent
approach, advocated by Sarkar and Satchell~\cite{Sar88}, and Schack and Caves~\cite{Sch93}, 
has been to 
investigate the sensitivity of the quantum dynamics to perturbations
of the Hamiltonian -- the problem we have just outlined qualitatively. 
This approach goes back to the early work of 
Peres~\cite{Per84} and has attracted new interest recently in connection
with the study of decoherence, entanglement generation in coupled dynamical
system and quantum irreversibility. It is the purpose of this review to discuss recent progresses
made in this dynamical approach to quantum chaos. Our focus is on quantal systems
at large quantum numbers/short wavelength, in the so-called semiclassical limit.
We devote most of our attention to the mathematical formulation of and the (inevitably
incomplete) answer to three fundamental questions pertaining to the 
relationship between classical and quantum physics. The first one is\\[-1mm]

{\it How and when does a quantum mechanical system start to behave classically ?}\\[-1mm]

\noindent Decades of experimental investigations
have confirmed the validity of quantum theory to an unprecedented level, and a large
variety of fundamental
experimental tests have been passed with an A$^+$.
Double-slit experiments have been performed where quantum objects as
large as molecules have produced interference fringes~\cite{Jon74,Nai03} , 
the Aharonov-Bohm effect~\cite{Ehr49,Aha59} has been implemented in transport through
mesoscopic systems~\cite{Web85,Cha85,Osa86}, and quantum nonlocality, as predicted in the
EPR paradox~\cite{Ein35} has been illustrated via the experimental determination of
Bell inequalities~\cite{Asp81}. This list of quantum-mechanically driven 
phenomena is much longer, of course, and includes
phenomena such as superfluidity and superconductivity, Bose-Einstein condensation
or ferro- and antiferromagnetism, all of them cooperative phenomena that
occur at macroscopic scales, yet
cannot be explained without quantum mechanics.
Still, it is our daily experience that the world surrounding us, 
despite being made out of quantum mechanical building blocks, behaves 
classically most of the time. 
This suggests that, one way or another,
classical physics emerges out of quantum mechanics, at least for sufficiently
large systems. How and when does this happen ? 
The Copenhagen interpretation, that observations of the quantum world as we make them
are made
with macroscopic, therefore classical apparatuses,
while having been of great comfort to many a physicist, does not
answer the question satisfactorily. It merely pushes the problem a bit further, towards the question
"{\it what makes a measurement apparatus classical ? }" or in the words of 
Zurek~\cite{Zur93}
" {\it where is the border ? }" between classical and quantum mechanics ?
Instead, today's common understanding of this quantum--classical
correspondence is based on the realization that no quantum mechanical
system -- finite-sized almost by definition -- is ever fully
isolated, and it is unavoidable that its behavior is modified by its coupling to environmental
degrees of freedom. This requires to extend the theory to larger Hilbert spaces, including
external degrees of freedom modeling the environment, the rest of the universe or a heat bath
(all three denominations usually referring to the same concept). The latter degrees of freedom
are eventually integrated out
following a precise procedure -- the outcome depends on when this is done.
It is then hoped that a large regime of parameters exists where the 
coupling to the environment
destroys quantum interferences without modifying the system's classical 
dynamics. As a matter of fact, it is often argued that
such a coupling induces loss
of coherence on a time scale much shorter than it relaxes the 
system~\cite{Alt82,Joo85,Joo03,Zur93,Zur03,Bra01}. Decoherence originates from the coupling
to a large number of external degrees of freedom over which no control
can be imposed nor direct observation made. 
Once these degrees of freedom are integrated out of the problem, 
the reduced problem containing only the degrees of freedom of the system under observation
has (partially or totally) lost its quantum coherence. Quantal wavefunctions
no longer evolve according to Schr\"odinger's equation, instead, when decoherence is 
complete, they are fully represented by their squared amplitude only, the latter
evolving with Hamilton's equation. This is the broad picture.
Does it generically apply to specific systems, or are there some 
refinements to be implemented from case to case ? How big should the 
environment be for the quantum-classical crossover to occur ? These are
some of the related questions we are interested in below. Decoherence has been extensively
treated in a variety of contexts, it has been the subject of textbooks and rather large 
reviews~\cite{Joo03,Zur03}, and our purpose here is not to cover all or even a fraction of this
rather large literature. Instead we focus on dynamical systems with few degrees of freedom
in the semiclassical limit. In that limit, some approximations that are made for larger systems,
coupled to larger environments, are not necessarily legitimate and new behaviors occur. 
On the plus side, more generic
environments, and system-environment couplings can be considered under not too restrictive
assumptions, and we even expect that the approach we present below 
is scalable, in that it can be further developed
to treat larger systems coupled to complex environments with a large number of interacting
chaotic degrees of freedom. Often, our assumptions are legitimated by mathematically rigorous
results on classical dynamical systems, such as structural stability and shadowing 
theorems~\cite{Kat96}, which allow to find the dominant, stationary phase contributions to
our semiclassical expressions by pairing 
classical trajectories of slightly different Hamiltonians. As long as shadowing can be invoked,
the problem treated is that of pure dephasing, without momentum nor energy relaxation. There are
regimes where pure dephasing is sufficient to kill all coherent effects, and the resulting dynamics
is classical, given by the classical counterpart of the system's Hamiltonian -- in particular, the coupling
to the environment does not lead to renormalization/changes in the parameters of the Hamiltonian
or to the addition of new terms in it.

An alternative way of presenting decoherence is to say that, because of 
the coupling between them, system and environment become entangled. 
What does that mean ? The concept was already pretty much defined, at least
qualitatively, by Schr\"odinger in 1935~\cite{Sch35}. 
We quote him: 

{\it When two systems (\ldots) enter 
into temporary interaction
(\ldots), and when after a time of mutual influence the systems separate 
again, then they can no longer be described in the same way as before,
viz. by endowing each of them with a representative of its own.} 

At the quantum level, 
initially well separated subsystems lose at least part of their individuality when they interact,
and the global
quantum state describing the sum of the subsystems can no longer be represented into
a product of well-defined states of the subsystems taken individually. 
Quantumness
is not lost globally, of course, and the system as a whole -- the sum of the system under observation and 
of environmental degrees of freedom -- evolves coherently in the quantum sense of the
Schr\"odinger evolution. However, because of entanglement, 
the system loses its coherence once it is observed separately from its 
fast moving environment. The rate at which decoherence occurs is thus related to
the rate at which entanglement is generated between system and environment.
In the spirit of Schr\"odinger's above formulation, one is naturally led to
ask the second question of interest in this review \\[-1mm] 

{\it What determines the rate at which two interacting quantum systems
become entangled ?}\\[-1mm]

\noindent In particular, one might wonder if this rate is
solely determined by the interaction
between the two sub--systems or if it also depends on the
underlying classical dynamics, even perhaps on the states initially occupied
by the sub--systems. Of interest is also to determine the different regimes of interaction
and the corresponding rates of entanglement generation or its functional dependence
in time. Also, one might wonder how these rates scale with the dimension/number of degrees
of freedom of the environment. These are some aspects of this second question that we
discuss in this review.

The third, final question we ask is\\[-1mm]

{\it How irreversible are quantum mechanical systems with few degrees of
freedom compared to their classical counterpart ?}\\[1mm]

\noindent At first glance, this latter problem seems unrelated to the 
first two problems of decoherence and entanglement. The connection
emerges when, following the late Asher Peres, we observe that  
simple mechanisms of irreversibility exist in classical dynamical 
systems with few degrees of freedom, that 
cannot be exported to quantum mechanics~\cite{She83,Per84}. The chaos hierarchy
ensures that classical chaotic systems
exhibit mixing and exponential sensitivity to initial conditions
in phase space~\cite{Lic92,Gut90,Cvi05}. Irreversibility 
directly follows from these two 
ingredients, once they are combined with the unavoidable finite resolution
with which the exact state of the system can be determined. 
This finite
resolution blows up exponentially with time, so that a time-reversal operation
inevitably misses the initial state, if it is performed after a time logarithmic in
the resolution scale. In other words, to be successful, a time-reversal operation 
requires to determine the system's state with 
an accuracy exponential in the time at which it is performed.

Finite resolutions do not blow up under Schr\"odinger time-evolutions, 
moreover, they are better tolerated by quantum mechanical systems which are 
discrete by nature. The classical mechanism for irreversibility just underlined
is therefore invalidated by quantum mechanics. Instead, Peres
argued that quantum irreversibility originates from 
unavoidable uncertainties in the system's Hamiltonian. Once again,
uncontrolled external degrees of freedom are invoked, this time
to justify the finite resolution with which one can determine the 
Hamiltonian governing the system's dynamics -- and not the state the system
occupies. The coupling to external degrees of freedom generates entanglement
between the system and the environment, and information about 
the exact state of the system gets lost, never to return. Irreversibility sets in, and one hopes
that it can effectively be
quantified by the fidelity (unless explicitly stated otherwise
we set $\hbar \equiv 1$ throughout this article)
\begin{equation}\label{eq:def_LE2}
{\cal M}_{\rm L}(t)=\left\vert\left \langle \psi_0 \left\vert 
\exp[i H t] \exp[-i H_0 t]
\right \vert  \psi_0 \right\rangle\right \vert^2,
\end{equation} 
with which an initial quantum state $\psi_0$ is reconstructed after its
time evolution is imperfectly reversed at time $t$.  Below, 
${\cal M}_{\rm L}$ is called indifferently fidelity or Loschmidt echo -- the latter
denomination has been introduced by Jalabert and Pastawski~\cite{Jal01}, 
to stress its connection to
the gedanken time-reversal experiment proposed by Loschmidt in his
argument against Boltzman's H-theorem~\cite{Los76} --
and unless stated otherwise, it refers to an average taken over an ensemble of comparable
initial states $\psi_0$.
The difference $\Sigma \equiv H-H_0$ between the Hamiltonians governing 
forward and time--reversed propagations originates from the imperfect
knowledge one has over the microscopic ingredients governing the system's dynamics.
It turns out that in some instances, the problem of decoherence and
entanglement generation can be mapped onto the problem of irreversibility as formulated in 
Eq.~(\ref{eq:def_LE2}).
We now proceed to illustrate this statement and express in more quantitative terms
the connection between the three central questions we asked above.
We do this with a simple example.

Consider a quantum two-level system in the form of a spin-$1/2$.
Initially, we prepare that spin in a normalized, 
coherent superposition,
\begin{equation}
| \psi_0 \rangle_{\rm sys} = \alpha |\uparrow \, \rangle + \beta |\downarrow \,\rangle,
\;\;\;\;\;\;\;\; |\alpha|^2 + |\beta|^2 = 1,
\end{equation}
and let it evolve with time. A pure quantum-mechanical time-evolution
is unitary, and will therefore not alter the quantumness
of this state, in the sense that the product $\alpha \beta^*$ 
of the off-diagonal matrix elements of the density matrix oscillates
in time with constant (i.e. non-decaying) amplitude. Unavoidably,
however, the system is coupled to external degrees of freedom, and we
therefore extend the description of the initial state to
\begin{equation}\label{eq:initial_product}
|\Psi_0 \rangle = | \psi_0 \rangle_{\rm sys} \otimes 
|\phi_0 \rangle_{\rm env},
\end{equation}
where subscripts have been introduced to differentiate the degrees of
freedom of the two-level system (sys), on which our (i.e. the observer's)
interest focuses, from the external, environmental degrees of freedom (env),
on which no measurement is directly performed.
The dynamics of $\Psi $ is equally
quantum-mechanical as the dynamics that $\psi$ would follow if the system
were perfectly isolated. 
At this stage, however, we must depart from a pure quantum-mechanical
treatment of the problem, essentially 
because we -- i.e. the observer -- are focusing our
interest on the system's degrees of freedom only. This measurement process
projects the problem onto a basis with less degrees of freedom. In other words,
to provide for a description of the observed dynamics of the system,
the environment has to be removed from the problem. To achieve this, 
the standard procedure is to consider the time-evolution of the density
matrix
\begin{subequations}
\begin{eqnarray}
\rho_0 &=& |\Psi_0 \rangle \langle \Psi_0|, \\
\rho (t) &=& \exp[-i {\cal H} t] \; \rho_0 \; \exp[i {\cal H} t], 
\end{eqnarray}
\end{subequations}
and to {\it reduce} it to a local (system) density matrix by
integrating out the degrees of freedom of the 
environment~\cite{Joo03,Zur03},
\begin{eqnarray}\label{eq:tracing}
\rho_{\rm red} (t) &=& {\rm Tr}_{\rm env} \Big[ 
\exp[-i {\cal H} t] \; \rho_0 \; \exp[i {\cal H} t] \Big] .
\end{eqnarray}
The procedure one has to follow is to trace the environmental degrees
of freedom out of the time-evolved
density matrix. No decoherence is obtained, quite trivially,
if, for instance, one traces over the initial pure density matrix, then time-evolve the result.
The amplitude of the 
off-diagonal matrix elements of $\rho_{\rm red} (t)$ is now
decaying with time. 
The trace in Eq.~(\ref{eq:tracing})
can be exactly performed in specific situations only. For instance, 
the problem is significantly simplified if one freezes the intrinsic
dynamics of the two-level system and takes a system-environment interaction
with von Neumann form,
\begin{eqnarray}
{\cal H} &=& {\mathbb I}_{\rm sys} \otimes H_{\rm env}  + |\uparrow \rangle \langle \uparrow | \otimes H_{\uparrow} 
+ |\downarrow \rangle \langle \downarrow | \otimes H_{\downarrow}.
\end{eqnarray}
In this case, the diagonal
matrix elements $\rho_{\rm red}^{\sigma,\sigma}$, $\sigma=\uparrow,\downarrow$
are time-independent,
\begin{subequations}
\begin{eqnarray}
\rho_{\rm red}^{\uparrow,\uparrow} = |\alpha|^2,\;\;\;\;\;\;\;\;\;
\rho_{\rm red}^{\downarrow,\downarrow} =|\beta|^2,
\end{eqnarray}
\end{subequations}
on the other hand, the off diagonal elements 
$\rho_{\rm red}^{\uparrow,\downarrow}$
of the reduced density matrix are found to
evolve as
\begin{eqnarray}\label{eq:odiag_spin}
\rho_{\rm red}^{\uparrow,\downarrow} (t) &=& \alpha \beta^* \;
\langle \phi_0|\exp[i (H_{\rm env} + H_\downarrow) t]
\exp[-i (H_{\rm env} + H_\uparrow) t] |\phi_0 \rangle.
\end{eqnarray}
The described procedure
is probability-conserving,
${\rm Tr}[\rho_{\rm red}] \equiv 1$, moreover, it preserves the Hermiticity of
the reduced density matrix, $\rho_{\rm red}^{\downarrow,\uparrow} = 
(\rho_{\rm red}^{\uparrow,\downarrow})^*$.  
Quantum coherent effects are however carried by
the off-diagonal matrix elements which now become time-dependent. For instance
a measurement of the $x-$component of the spin gives
\begin{equation}
{\rm Tr}[ \hat{\sigma}_x \; \rho_{\rm red} (t) ] = 2 \, {\rm Re} \; \rho_{\rm red}^{\uparrow,\downarrow} (t),
\end{equation}
where $\hat{\sigma}_x$ is the corresponding Pauli matrix. The
time-dependence of this measurement is thus determined by 
\begin{equation}\label{eq:fidelity_amplitude}
f(t)
= \langle \phi_0|\exp[i (H_{\rm env} + H_\downarrow) t]
\exp[-i (H_{\rm env} + H_\uparrow) t] |\phi_0 \rangle,
\end{equation}
a quantity which is often referred to as the {\it fidelity amplitude}.
It is straightforward to see that
if $H_\uparrow \ne H_\downarrow$, $|f(t)|$ decays with time.
The decay of the off-diagonal matrix elements of $\rho_{\rm red}$
is commonly associated with the phenomenon of {\it decoherence},
which, as in this simple example, often affects only marginally, if at all, the behavior
of the diagonal matrix elements of $\rho_{\rm red}$. 
Decoherence occurs because system and environmental degrees of
freedom become entangled in the sense that the state of the global system
can no longer be represented by a product state, even
as an approximation, once the
coupling between system and environment has been given enough time to act. 
Therefore, the time-evolution of the reduced density matrix containing only 
the system degrees of freedom is no longer
governed by a Schr\"odinger/von Neumann equation. Whether the diagonal of
$\rho_{\rm red}$ is affected or not is of course basis-dependent, and 
decoherence, or the generation
of entanglement between the two subsystems 
can be quantified by the basis-independent purity 
\begin{equation}\label{eq:purity}
{\cal P}(t) \equiv {\rm Tr}[ \rho^2_{\rm red}(t) ]
\end{equation}
of the reduced density matrix, which is equal to one only in absence
of entanglement. The connection to decoherence is clear -- $ {\cal P}(t) $ is a basis-independent
measure of the relative weight carried by the off-diagonal matrix elements of $\rho_{\rm red}(t)$, those
containing information on interferences between different wave-components. It simultaneously
turns out that the vanishing of these matrix elements, the decay of $ {\cal P}(t) $, is indicative
of whether $\rho(t)$ can be factorized as the product of two density matrices pertaining
to each subsystems, with $ {\cal P}(t)=1$ corresponding to full factorizability.
For globally pure states (i.e. pure states of the system+environment) ${\cal P}(t)$ is an appropriate 
measure of entanglement, physically equivalent to the von Neumann entropy 
${\cal S}(t)=- {\rm Tr}[ \rho_{\rm red}(t) \ln \rho_{\rm red}(t)]$~\cite{Ved98,Mil99},
and both measures are monotonously related to the nonseparability of the pure total density matrix
$\rho$. The advantage of working with ${\cal P}(t)$ is that it is mathematically much easier to handle.
It is also important to note that both ${\cal P}(t)$ and ${\cal lS}(t)$ 
are symmetric and remain the same if one exchanges the roles
of the two subsystems.

For the above example of a spin-$1/2$ the purity reads
\begin{equation}\label{eq:purity_spin}
{\cal P}(t) = |\alpha|^4 + |\beta|^4 + 2 |\alpha|^2 |\beta|^2 \; |f(t)|^2.
\end{equation}
Eqs.~(\ref{eq:fidelity_amplitude}) and (\ref{eq:purity_spin}) give a somehow unifying picture of how
the a priori unrelated concepts of decoherence, entanglement generation 
and quantum reversibility are connected. 
In our simple example, the decay of the off-diagonal matrix elements of $\rho_{\rm red}$
is given by the fidelity ${\cal M}_{\rm L}(t) = |f(t)|^2$ with which $\phi_0$ (the initial state of the
 {\it environment}) is reconstructed after an imperfect time-reversal operation is performed at
  time $t$. Simultaneously, 
this short discussion illustrates that, strictly speaking, a direct connection between 
${\cal M}_{\rm L}$ and decoherence exists only under specific 
assumptions on the Hamiltonians governing
the coupled dynamics of system and environment. 
After this presentation of the main questions around which discussions to come
will orbit, we present a still
general and qualitative discussion of the behavior of quantum systems coupled to
external degrees of freedom, which leads us to introduce other mathematical quantities, besides
${\cal P}(t)$ and ${\cal M}_{\rm L}$, on which our
interest will focus.

\subsection{Echo experiments -- going beyond Loschmidt}

Obviously, the Loschmidt echo of Eq.~(\ref{eq:def_LE2}) is only a phenomenological measure
of quantum reversibility,
where the coupling to (not necessarily identified)
external degrees of freedom is modeled by the perturbation $\Sigma$,
acting on the system's degree of freedom only. 
A true 
microscopic approach to reversibility instead requires to start with a global system,
including an environment with a dynamics of its own, 
which one eventually integrates out. The extra degrees of freedom are intended to model
the unavoidable coupling of the system under interest to the rest of the universe. These degrees of freedom are, in principle,
so numerous that they can absorb any amount of information that the system has. 
The lost information never returns back to the system -- 
at least not within physical times, say up to the age of the universe -- and irreversibility sets in. 
It is then highly desirable
to figure out the conditions under which ${\cal M}_{\rm L}$ 
is obtained from this procedure. Let us be more specific and consider
an initial product state as, e.g. in Eq.~(\ref{eq:initial_product}),
which evolves during a time $t$ with the Hamiltonian
\begin{equation}
{\cal H}_{\rm f} = H_{\rm sys} \otimes  {\mathbb I}_{\rm env} +  {\mathbb I}_{\rm sys}
\otimes H_{\rm env} + {\cal H}_{\rm c}. 
\end{equation}
One then performs a time-reversal operation on the system degrees of
freedom only, and let the state evolve during an additional time $t$
under the influence of the partially time-reversed Hamiltonian
\begin{equation}
{\cal H}_{\rm b} = H_{\rm sys}' \otimes  {\mathbb I}_{\rm env} + {\mathbb I}_{\rm sys}
\otimes H_{\rm env}' + {\cal H}_{\rm c}'. 
\end{equation}
Perfect control over the degrees of freedom of the system can be assumed,
$H_{\rm sys}' = - H_{\rm sys}$, however there is no reason to believe that
a perfect time-reversal operation can be performed on 
environmental degrees of freedom -- by definition one has no control over them. 
Hence, $H_{\rm env}'$ and ${\cal H}_{\rm c}'$
are in general different from $-H_{\rm env}$ and $-{\cal H}_{\rm c}$. 
Reversibility is quantified by the probability that
after $2 t$ the central system returns to its initial state,
regardless of the environment. The quantity of interest thus reads
\begin{equation}\label{irrevtest}
{\cal M}_{\rm B} (t) = \Big \langle \big\langle \psi_0  \big|    
{\rm Tr}_{\rm env} \Big[
\exp[-{\it i }{\cal H}_{\rm b} t ] \exp[-{\it i}{\cal H}_{\rm f} t ] |\Psi_0\rangle\langle\Psi_0|
\exp[ {\it i }{\cal H}_{\rm f} t ] \exp[ {\it i}{\cal H}_{\rm b} t ]
\Big] \big| \psi_0 \big\rangle \Big \rangle_{\phi_{\rm env}} \, ,
\end{equation}
where the initial state is given in Eq.~(\ref{eq:initial_product}).
Because one has no control over the environment,
its fast evolving degrees of freedom are traced out. Moreover, one
averages over its initial state $\phi_{\rm env}$, as it cannot be prepared.
This is indicated by 
the outermost brackets in Eq.~(\ref{irrevtest}).
In Ref.~\cite{Pet06a}, we introduced ${\cal M}_{\rm B}(t)$ 
and dubbed it the {\it Boltzmann echo} to stress its 
connection to Boltzmann's
counterargument to Loschmidt that time cannot be  
inverted for all components of a system with many degrees of 
freedom. 
We will see below that, 
in the weak coupling limit when ${\cal H}_{\rm c}$ has a weaker effect
than the imperfection $H_{\rm sys}+H_{\rm sys}'$ in the inversion of the arrow of time
for the system, the decay of $ {\cal M}_{\rm B} (t)$
is indeed the same as that of ${\cal M}_{\rm L} (t)$ with $H_0=H_{\rm sys}$
and $H=-H_{\rm sys}'$. This justifies {\it a posteriori} the introduction
of ${\cal M}_{\rm L} (t)$ as a measure of quantum reversibility. 
However, there is a crossover to an interaction--governed decay as ${\cal H}_{\rm c}$
increases against $H_{\rm sys}' + H_{\rm sys}$.
In that regime, reversibility is governed by ${\cal H}_{\rm c}$,
regardless of the precision with which the time-reversal operation is
performed. 

The properties of the Boltzmann echo are discussed
in more details below in Chapter~\ref{section:boltzmannecho}. 
In the weak coupling regime it is reasonable to expect that integrating out 
the external degrees of freedom leaves us with
an effective time-dependent perturbation $H-H_0 = \Sigma_{\rm eff}(t)$
acting solely on the system's degrees of freedom. The explicit time-dependence of $\Sigma_{\rm eff}$
emerges from the environment's intrinsic dynamics, and often it is sufficient to only specify
how the environment's dynamics affects the
correlation function $\langle \Sigma_{\rm eff}({\bf x}+\delta {\bf x}, t+\delta t) \, 
\Sigma_{\rm eff}({\bf x}, t) \rangle_{{\bf x},t} \propto f(|\delta {\bf x}|/\xi_0) \, g(\delta t/\tau_0)$,
and how $f$ and $g$ decay. 
The fidelity under an imperfect time-reversal with
a time-dependent perturbation is investigated in
Ref.~\cite{Cuc06a} where, not surprisingly, 
earlier results on the decay of the Loschmidt echo
are reproduced. The decay rates in this case are
given either by the correlation time $\tau_0$ or the correlation length $\xi_0$ of $\Sigma_{\rm eff}(t)$.
Our analysis of the Boltzmann echo shows that investigating reversibility in quantum dynamical
systems with the time-dependent Loschmidt echo is justified only when the coupling between system and environment dominates the imperfection in the time-reversal operation [the perturbation in
Eq.~(\ref{eq:def_LE2})].

Investigations of the
Loschmidt echo are to some extent experimentally motivated. Echo experiments
abound in 
nuclear magnetic resonance~\cite{Hah50,Rhi70,Zha92,Lev04,Pas00,Pas95}, 
optics~\cite{Kur64}, 
cavity quantum electrodynamics~\cite{And03,And04,And06},
atom interferometry~\cite{Su06,Wu07a,Wu08},
cold atomic gases~\cite{Buc00,Cuc06b},
microwave cavities~\cite{Scha05,Schaf05,Hoe07}
and superconducting circuits~\cite{Nak02} among others.
Except for the microwave experiments, 
all these investigations are based on the same principle of a sequence
of electromagnetic pulses whose purpose it is to reverse the sign of hopefully
dominant terms in the Hamiltonian
by means of effective changes of coordinate axes. 
Imperfections in the pulse sequence result instead in
$H_0 \rightarrow -H_0-\Sigma$, and one therefore 
expects the Loschmidt echo to capture the physics of these experiments. 
As already mentioned, this line of reasoning 
deliberately neglects the fact that the time-reversal operation affects at best
only part of the system, for instance because the system is composed of 
so many degrees of freedom, that the time arrow can be inverted only for
a fraction of them. 
Another related issue is that subdominant terms in the Hamiltonian
are in principle not time-reversed -- these include for instance
the nonsecular terms in the Nuclear Magnetic Resonance (NMR) 
Hamiltonian for spin echoes~\cite{Sli92}  -- and affect echo experiments
in a way that is not necessarily correlated with how well the time reversal
operation seems to be performed. Both these aspects have to be kept in mind
when discussing echo experiments, and both motivate the investigations of
the Boltzmann echo of Eq.~(\ref{irrevtest}).

Equally important, most experimental
set-ups measure the return
probability of only a small 
part of the system's degrees of freedom. For instance,
the NMR spin echo experiments -- which provided the original
motivation for Jalabert and Pastawski's work on ${\cal M}_{\rm L}$~\cite{Jal01} --
measure the polarization echo~\cite{Pas95,Pas00,Lev04}
\begin{equation}\label{pecho}
{\cal M}_{\rm PE}(t) = 2 \langle\Psi_0 | \exp[i {\cal H}_{\rm f} t]
\exp[i {\cal H}_{\rm b} t] \, \hat{I}_0^y \, \exp[-i {\cal H}_{\rm b} t]
\exp[-i {\cal H}_{\rm f} t] |\Psi_0 \rangle,
\end{equation}
on a given site labeled ``0'' of a large lattice, starting with an initial
random many-body state $\Psi_0$, with prepared polarization on the 0$^{\rm th}$ 
site only. 
The polarization echo essentially differs from a many-body Loschmidt echo by 
the
presence of the local spin operator $\hat{I}_0^y$ instead of 
$|\Psi_0 \rangle \langle \Psi_0 |$. Perhaps the main puzzle posed by these experiments is
the existence of a perturbation-independent decay~\cite{Pas00} -- how can it be that the decay
of $ {\cal M}_{\rm PE}(t) $ remains the same when the perturbation (perhaps the nonsecular 
terms in the NMR Hamiltonian ${\cal H}_{\rm f}$ that cannot be time-reversed) 
is effectively made weaker and weaker ?
It is this theoretical search that indirectly led to the discovery of the perturbation-independent
Lyapunov decay of ${\cal M}_{\rm L}(t) \propto \exp[-\lambda t]$~\cite{Jal01},
which, of course,
is unable to explain
the experimental data. First, it is unclear (at least to the authors of the present manuscript)
what is the Lyapunov exponent of a NMR Hamiltonian of spins on a lattice; second, the 
experimentally observed saturated decay is Gaussian, whereas the
Lyapunov decay is exponential; and most importantly third, 
the Lyapunov decay is observed as the perturbation is cranked up, whereas the experiments
observe a saturation upon weakening of the perturbation.

A similar sandwiching as in Eq.~(\ref{pecho}) 
also occurs when one investigates the workability of a quantum
computer. The necessary switching on and off of spin-spin couplings in these machines inevitably
generates errors in the evolution of entangled many-body states, and Georgeot and Shepelyansky's
work on the many-body counterpart of the fidelity ${\cal M}_{\rm L}(t)$ suggested that error generation
would proceed at a much faster rate in a many-body quantum chaotic computer than in a regular
computer~\cite{Geo00}. By extrapolation, they concluded that many-body quantum chaos,
in the sense of interaction-induced mixing of noninteracting many-body states~\cite{Abe90,Jac97b}
would inevitably render any quantum computer inoperative. The authors of Ref.~\cite{Geo00}
did not consider the
fact that the computer can still work properly as long as errors can 
be corrected. Can we put that in mathematical form ? The answer is yes. To see how this can be
done, we suppose that, for the task at hand, $M$ qubits would be sufficient if the
computation were performed ideally, i.e. without errors.
To allow for error corrections, quantum computers instead work on an extended Hilbert space
of $N > M$ qubits, such that different output states of the computation
can be unambiguously differentiated, despite error generation. This can be achieved, as  long as 
error generation
does not lead two different initial states in the code space to a sizable scalar product.
The code space is then a $2^M$-dimensional Hilbert
space, embedded in the full $2^N$-dimensional Hilbert space of the $N$-qubit Hamiltonian,
such that any two states it contains cannot be confused, even after a number $K$ of  errors
has corrupted each of them.
This number is in principle a function of $M$ and $N$.
Representing the qubits by 
spin-$1/2$ and error operators with Pauli matrices, the condition
for the code space is that for any two $N$-qubit states in it, one has
\begin{equation}
\langle\psi_{0}|\hat{\sigma}^{\alpha_{1}}_{n_{1}}\hat{\sigma}^{\alpha_{2}}_{n_{2}}
\ldots\hat{\sigma}^{\alpha_{k}}_{n_{k}}|\psi'_{0}\rangle=0,
\end{equation}
with $ 1\leq k\leq 2K$. This condition must hold 
for any two $N$-spin states $\psi_0$ and $\psi_0'$
in the code space (not in the total
Hilbert space)
and any sequence of Pauli matrices $\hat{\sigma}_{n_i}^{\alpha_i}$ acting on
different spins labeled $n_i$. The application of one Pauli matrix 
to any of the $N$ spins counts as one error. The code space is the ensemble of $N$-body states,
initially encoding $M$-qubits code states,
which, even after $K$ errors remain orthogonal to one another -- to each 
such $N-$body state $\psi_0$
belongs an error
space, spanned by applying a sequence of up to $K$ Pauli matrices on $\psi_0$, and the
so defined $2^M$ error spaces are orthogonal to one another. 
Error correction after a time $t$ is successful if
$\psi(t)$ lies in the error space of the ideal state to which the initial state $\psi_0$ should have
evolved. Silvestrov, Schomerus and Beenakker quantified
the probability of successful error correction with~\cite{Sil01}
\begin{eqnarray}
F(t)&=&|{\cal D}\psi(t)|^{2} = \langle\psi_{0}|e^{iH_0t} e^{i H t}{\cal
D}e^{-iHt} e^{-iH_0t}|\psi_{0}\rangle\label{Ftdef0}, \\
{\cal D}&=&\sum_{p=0}^K \, \sum_{\{n,\alpha\}}\frac{1}{p!}\sigma_{n_{1}}^{\alpha_{1}}
\ldots\sigma_{n_{p}}^{\alpha_{p}}|\psi_{0}\rangle
\langle\psi_{0}|\sigma_{n_{1}}^{\alpha_{1}} \ldots\sigma_{n_{p}}^{\alpha_{p}} \, ,
\end{eqnarray}
where $H_0$ is the ideal Hamiltonian modeling the computation sequence and $H$ the
perturbed ones, generating computational errors.
The projector ${\cal D}$ plays a role similar to the one
played by the polarization operator $\hat{I}_0^y$ in the polarization echo,
Eq.~(\ref{pecho}).
For a model of randomly interacting Heisenberg spins on a $d$-dimensional 
lattice, subjected (or not) to a magnetic
field, Silvestrov, Schomerus and Beenakker 
obtained a lower bound for $F(t)$ which  
is independent of the integrability of the Hamiltonian, and, perhaps more importantly,
of the number $M$ of encoded qubits. 
The appearance of the Pauli matrices in $F(t)$
has the important consequence that
the operative time of a quantum computer is increased by a parametrically large prefactor, and that 
whether many-body quantum chaos is at work or not is irrelevant.
These two facts contradict
Ref.~\cite{Geo00}. The disagreement is of course 
due to the inclusion (or not) of error correction in the theory, and
Ref.~\cite{Sil01} included them indirectly, by invoking rigorous
results from nonconstructive
theorems -- one knows that
error correction codes exist with certain scaling properties, 
for instance relating
$M$ (how many qubits one needs to perform the task at hand) 
and $K$ (how many errors one estimates must be corrected at most) 
to $N$ (how many qubits one effectively needs in total). 
In real-life situations, under the 
assumption that quantum computers exist, it is not at all given that optimal error correction codes
are easily implemented. Therefore, the truth lies somewhere between the conclusions of
Ref.~\cite{Geo00} (assuming no error correction) and Ref.~\cite{Sil01} (assuming 
mathematically optimal error correction).

Treating such quantities as the many-body polarization echo of Eq.~(\ref{pecho}),
or the fidelity in quantum computers with error correction goes beyond the scope of this
review, and we do not discuss these concepts 
further. Still one might wonder how much of the
original, many-body NMR problem is still included in the single-particle fidelity. In the absence of
theory for many-body echoes the answer is hard to guess. One line of logic, due to Pastawski, 
somehow follows the celebrated 
spherical cow paradigm so dear to the heart of many a physicist.
It maps an original complex, many-body problem to
a much simpler single-particle problem. First
the full three-dimensional NMR Hamiltonian is reduced to a one-dimensional
quantum spin chain. 
The latter is then transformed into a model of noninteracting fermions by means
of a Jordan-Wigner transformation, and if one restricts oneself to single-spin excitations,
the problem becomes identical to a single-particle Loschmidt echo problem. 
Obviously, each of 
these steps is generally far from being exact. The philosophy is instead to
investigate simpler toy
models that are known to be solvable, while trying to retain the subtleties of the original
problem -- in the case just discussed, one considers a XY model which, compared to an Ising model,
still contains quantum kinetic terms. It thus seems that, for some specific situations at least,
single-particle echoes are still indicative of the behavior of many-body echoes. This conclusion
is in agreement with numerically observed similarities (RMT behavior most notably) between complex
many-body systems and chaotic systems with few degrees of 
freedom~\cite{Bro81,Abe90,Jac97b,Geo97,Geo00,Flam96a,Flam96b}. 

There are many instances in physics where one is interested in time-dependent correlation
functions of the form
\begin{eqnarray}\label{eq:t_correl}
Y ({\bf P},t)=\Big\langle  \exp[-i{\bf P}\cdot {\bf \hat{r}}]
\exp[ i H_0 t] \exp[i{\bf P}\cdot {\bf \hat{r}}] 
\exp[-i H_0 t] \Big\rangle .
\end{eqnarray}
Examples include spectroscopies such as neutron scattering, M\"ossbauer $\gamma$-ray,  and certain electronic transitions in molecules and 
solids~\cite{Hel87,Lax74,Lov84,vanH54}. More generally, any
measurement of momentum or position time correlators -- or combinations of 
the two -- can be viewed as a fidelity experiment under certain 
phase space displacements. 
In these  spectroscopies, momentum boosts or position 
shifts take place with little or no change in the potential, thus only one Hamiltonian appears
in $Y({\bf P},t)$.
In Eq.~(\ref{eq:t_correl}), the brackets represent an ensemble average, 
which can be a thermal average, or an average over
a given set of initial states,  
$\bf{\hat{r}}$ is the position 
operator of the nuclei and $H_0$ is the typical Hamiltonian of the 
target system. 
The thermal ensemble average of the correlation function can be 
written~\cite{Hel87}
\begin{eqnarray}
Y({\bf P},t)& \approx &\frac{1}{Q}\int \frac{d^{2N}\psi}{\pi^N} \; 
\Phi(\psi) \langle \psi| \exp[i H_{\bf P}t] \exp[-iH_0t] |\psi\rangle , \label{auto}
\end{eqnarray}
where $Q={\rm Tr}  \left[ \exp[-\beta H_0] \right]$,
$|\psi \rangle$ are coherent states with $N$ degrees of freedom, and the 
thermal weight $\Phi (\psi) \rightarrow 
\exp[-\beta H_{cl}(\psi)]$ at high temperatures. 
The notation $\exp[i H_{\bf P}t]=\exp[-i{\bf P}\cdot \hat{\bf r}] 
\exp[i H_0 t] \exp[i {\bf P}\cdot \hat{\bf r}]$ suggests that we
identify the kernel of the integral 
\begin{equation}
f(t)=\langle \psi |\exp[-i {\bf P}\cdot \hat{\bf r}]
\exp[i H_0t] 
\exp[i {\bf P}\cdot \hat{\bf r}]
\exp[-i H_0 t] |\psi\rangle
\end{equation} 
with a fidelity amplitude, i.e. the kernel of a Loschmidt echo problem.
This motivated the investigations of Ref.~\cite{Pet07a}, where
the momentum {\it displacement echo} was introduced,
\begin{equation}\label{decho}
{\cal M}_{\rm D}(t)=\vert f(t)\vert^2 = \big|\langle \psi|
\exp[i H_{\bf P}t] \exp[-i H_0 t] |\psi \rangle \big|^2.
\end{equation}
This quantity is discussed below in Section~\ref{section:displacement}.
The fidelity approach to the calculation of quantum correlation function
has also been used and further developed by Vani\v{c}ek~\cite{Van04}.
Other quantities such as the reduced and purity fidelity, which are more or less closely associated 
with the Loschmidt and Boltzmann echoes and the
purity of reduced density matrices, are discussed in Ref.~{\cite{Gor06}.

Below we deal with many, but not all, of the quantities just introduced. 
In the context of reversibility in quantum mechanics, 
Section~\ref{section:fidelity} and~\ref{section:wigner}, 
our attention focuses
on the Loschmidt echo~(\ref{eq:def_LE2}) as well as on the displacement 
echo~(\ref{decho}). Our discussion on
entanglement and decoherence follows in Section~\ref{section:entanglement}, where it
is centered on the purity ${\cal P}(t)$ of the reduced density
matrix, Eq.~(\ref{eq:purity}). The Boltzmann echo of Eq.~(\ref{irrevtest}) is the focus
of Section~\ref{section:boltzmannecho}.
In Table~\ref{table:table1}, we give a list
of the quantities of central interest in this review, with a mention of
where they are defined and discussed. The Loschmidt echo 
with prepared initial state, ${\cal M}_T(t)$, will be introduced momentarily.

\begin{table}
\begin{tabular}{|l||c|c|c|c|c} 
\hline
 & Name & Mathematical definition & Where ? \\
\hline
\hline
${\cal M}_{\rm L}(t) $& Loschmidt echo & Eq.~(\ref{eq:def_LE2}) & Chapters~\ref{section:fidelity} and~\ref{section:wigner}\\
\hline
${\cal M}_T (t) $& Loschmidt echo with prepared initial state & 
Eq.~(\ref{prepare_echo}) & Chapter~\ref{section:subplanck} \\
\hline
${\cal M}_{\rm D}(t) $& Displacement echo & Eq.~(\ref{decho}) & Chapter~\ref{section:displacement} \\
\hline
${\cal M}_{\rm B}(t) $& Boltzmann echo & Eq.~(\ref{irrevtest}) &
Chapter~\ref{section:boltzmannecho} \\
\hline
${\cal P}(t) $& Purity & Eq.~(\ref{eq:purity}) & Chapter~\ref{section:entanglement} \\
\hline
\end{tabular}
\caption{\label{table:table1} Quantities of central interest in this review and where we discuss them.}
\end{table}

\subsection{Scope and goals of this review, and what it is not about}

A low-energy quantum particle occupying the ground-state and perhaps few 
low-lying excited states of a confined quantum system has no choice but to be 
spatially extended over most of the available 
volume. This is independent of whether the confinement potential is chaotic or regular. 
It is hard to imagine how 
external sources of noise would affect the dynamics in such a way that
it reproduces the classical dynamics of a confined classical 
point-like particle. A direct quantum--classical correspondence
obviously presupposes that the considered system is
semiclassical in nature, in the sense that relevant quantum--mechanical
length scales such as de Broglie or Fermi wavelengths are small enough 
compared to classical length scales. Only then is the comparison of the 
quantum system to its classical counterpart meaningful. 
Stated otherwise, quantum-classical comparison requires that one 
considers the limit of large quantum numbers. This regime is sometimes referred
to as the $\hbar \rightarrow 0$ limit, in the sense, for instance, that higher-order terms
in an expansion in $\hbar$ are neglected -- a well-known example being 
the WKB approximation~\cite{Cvi05}.
Particularly
useful and appealing approaches in that limit are
semiclassical methods, which are based on expansions of quantum mechanical 
quantities in the ratio $\nu/L \ll 1$ of
a quantum--mechanical length scale $\nu$
(which below is the system's de Broglie wavelength)
with a classical length scale $L$ (which in the following is
the linear system size). The quantum--classical comparison of
course goes both ways, and a defining aspect of the field of quantum chaos
has always been to try and identify clear manifestations of the classical
phase-space dynamics in quantum systems. In that sense, the finding of Jalabert and
Pastawski~\cite{Jal01}, that the Loschmidt echo sometimes exhibits a 
time-dependent
decay governed by the system's Lyapunov exponent is certainly another
strong motivation for using semiclassical methods. 
In this review article we heavily rely on these
methods.

A powerful statistical alternative to semiclassics, also valid in the short wavelength limit,
is provided by Random Matrix Theory (RMT)~\cite{Meh91,Haa01}. 
RMT was born in nuclear physics in the late 50's and developed into a powerful
mathematical theory in the 60's, most notably by Wigner, Dyson, Gaudin and Mehta~\cite{Meh91}.
Nuclear physicists of the time were trying to understand the spectra of heavy nuclei.
Instead of attempting to describe the system microscopically  --
a procedure that is anyway doomed to fail in a strongly interacting system of two hundred
particles or more, where, additionally, the interaction potential is not well known -- Wigner proposed
to rely on a statistical description of the problem, where the nucleus' Hamiltonian is
replaced by a random Hermitian matrix. Only general symmetry requirements are enforced, 
depending on whether
the system is time-reversal symmetric, spin rotational symmetric, or not. Up to these
constraints, one assumes that all entries in the Hamiltonian matrix are randomly distributed.
This defines three ensembles of random matrices. RMT has been very successful in providing for
a statistical description of spectra of heavy nuclei and complex systems in 
general~\cite{Bro81,Meh91,Guh98}. 

The equivalence between the RMT and semiclassical approaches 
in confined quantum chaotic systems
seem to hold for two-point correlation functions~\cite{Sie01,Sie02,Heu07,Mul04}
(this theory neglect diffraction effects, which might or might not be legitimate). 
This equivalence is put to use numerous times in the present article, and we show below that 
there is a one-to-one correspondence between the RMT and
semiclassical decays of the purity
of Eq.~(\ref{eq:purity}) and of the fidelity of Eq.~(\ref{eq:def_LE2}),
under the assumption that RMT corresponds to systems with an infinite
classical Lyapunov exponent $\lambda$. This is 
qualitatively motivated by the absence of finite classical time scales in 
RMT, and by the condition for equivalence expressed 
in Ref.~\cite{Sie01,Sie02,Heu07,Mul04} that the underlying
classical system has local exponential divergence with $\lambda > 0$.
Together, these two conditions formally require $\lambda \rightarrow \infty$ for
a full RMT-semiclassical equivalence in the time--domain. 
In the context of the Loschmidt echo or the purity of reduced density matrices, this correspondence
is visible already  at a purely mathematical/technical level: RMT averages require pairings
of wavefunction components, which are in a one-to-one correspondence with
pairings of classical trajectories required by semiclassically motivated stationary phase
approximations. This point is further discussed below.

Throughout this article, our approach is statistical in essence, and we
concentrate on calculating quantities averaged over an ensemble of different
initial conditions $\psi_0$ or perturbations $\Sigma$. For this average to be
meaningful, one requires that all chosen $\psi_0$ lie in the same connected 
region of phase space, and have a similar character. Below we 
consider ensembles
of initial Gaussian wavepackets, pure and mixed superpositions of
Gaussian wavepackets, as well as pure initial random states. 
Averaging over an ensemble of initial Gaussian wavepackets justifies the
stationary phase conditions from which all semiclassical results derive. 
We argue that
these averages are meaningful in chaotic systems, which exhibit small
fluctuations. The situation is more contrasted in regular systems,
where averages and individual realizations can exhibit strongly different
behaviors. While Loschmidt echoes often exhibit a high degree of
universality -- the latter is summarized in Table~\ref{table:table2} below -- it is
worth mentioning that echoes under local perturbations exhibit 
interesting specificities that are not present in the echoes under
global (or at least strongly non-local) perturbations we consider in 
this review article. Echoes under 
phase-space displacement are also very special for qualitatively similar reasons. 
While we will discuss displacement echoes below in 
Chapter~\ref{section:displacement}, 
we refer the reader to Refs.~\cite{Gou07,Hoe07,Gou08} for theories and 
experiments
on echoes under local perturbations.

One of the first idea that comes to mind when facing the task of 
calculating ${\cal P}(t)$ or 
${\cal M}_{\rm L}(t)$ 
is to Taylor expand the complex exponentials in these expressions as
$\exp[\pm i H t] = 1 \pm i H t - H^2 t^2/2+...$,
and keep only the terms of
lowest nontrivial order -- they turn out to be quadratic in time. This linear response 
approach, with various refinements, has been reviewed in Ref.~\cite{Gor06},
and we will not discuss it much here. Let us just mention that, 
in a way similar
to the semiclassical approach, linear response delivers time-dependent decays 
given by classical correlators. 

In this review article, we restrict our discussion to 
quantum ballistic systems, by opposition to quantum disordered,
diffusive systems. Considered in Ref.~\cite{Ada03}, the latter can be treated using
the impurity Green's function technique instead of semiclassics or RMT.
Perhaps worth mentioning is the prediction that the Loschmidt echo in
quantum disordered systems may exhibit decays with different rates at different
times, even after the initial time-transient. In quantum chaotic systems, this can only
happen for the echo ${\cal M}_{T}$ of an initial state prepared by time-evolving
a Gaussian wavepacket for a time $T$ with the forward propagating Hamiltonian
$H_0$ [see Eq.~(\ref{prepare_echo}) and Section~\ref{section:subplanck} below],
or in intermediate, crossover regimes of perturbation~\cite{Cer02}. 
For more details on echoes in quantum disordered systems we refer
the reader to Ref.~\cite{Ada03}.

Kottos and co-authors considered a somehow modified version of the
Loschmidt Echo of Eq.~(\ref{eq:def_LE2}), 
\begin{equation}\label{eq:def_LE3}
{\cal M}_{\rm K}(t_0,t_1)=\left\vert\left \langle \psi_0 \left\vert 
\exp[i H t_1] \exp[-i H_0 t_0]
\right \vert  \psi_0 \right\rangle\right \vert^2,
\end{equation} 
with not necessarily equal propagation times $t_0$ and $t_1$. They found in particular that,
somewhat surprisingly, the value of $t_1$ which maximizes ${\cal M}_{\rm K}(t_0,t_1)$
is very often different from $t_0$. We will not discuss these works any
further here, and refer the reader to Refs.~\cite{Hil04,Hil06,Kot03} for details.

Recently, the fidelity under time-reversal of many-body systems has attracted some attention
in the context of interacting fermions~\cite{Man06,Piz07b,Man09} 
and cold atomic gases or Bose-Einstein condensates~\cite{Bod07,Man08,Cuc06b}, with some focus on
quantum criticality~\cite{Ng06,Qua06,Zan07,Alv07}.
While generally very interesting, we do not discuss these works any further here, as they 
certainly will soon deserve a review of their own. For the same reason, we do
not discuss entanglement in many-body 
systems~\cite{San03,San04,San05,San05a,Brow07,Vio07a,Ami08}, though we
will comment on possible routes leading there following
our analytical approaches.

\begin{figure}
\includegraphics[width=14cm,angle=0]{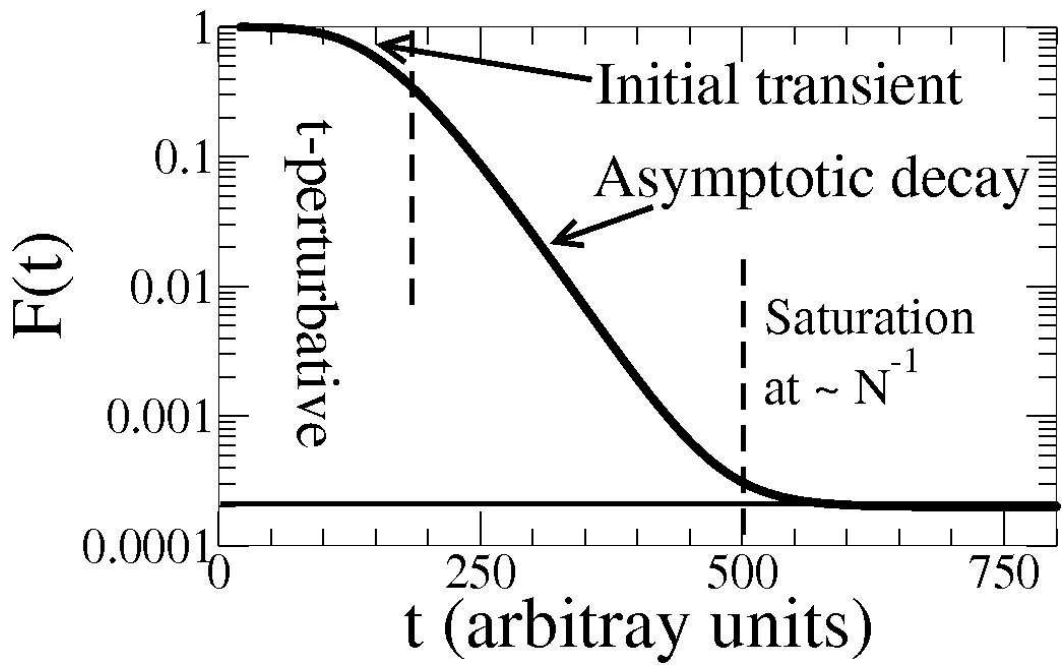}
\caption{\label{fig:fig_sketch_LE} Sketch of the successive decay regimes
 of $F(t) = {\cal P}(t) $,
${\cal M}_{\rm L}(t )$, ${\cal M}_{\rm B}(t) $, ${\cal M}_{\rm D}(t) $ or ${\cal M}_T(t)$
as a function of time. 
There is a short-time transient regime, well captured by first-order
perturbation theory in time, followed by an asymptotic decay, and eventually
by saturation at a value given by the inverse $N^{-1}$ of the size  of the considered 
Hilbert space / the effective Planck constant. The asymptotic decay is typically
exponential or Gaussian in chaotic systems. In coupling regimes
which are intermediate between first-order perturbation theory and 
golden rule regime, the decay can even be first exponential, then 
Gaussian~\cite{Cer02}.
In regular systems, the asymptotic decay is typically algebraic, and $F(t)$ usually
saturates above $N^{-1}$. This sketch corresponds to an average taken, e.g., over
an ensemble of initial states. For a given initial states, fluctuations are observed, in particular,
there is no long-time saturation, and instead one observes quantum revivals.}
\end{figure}

\subsection{Short survey of obtained results}

The behavior of $F(t) = {\cal P}(t) $,
${\cal M}_{\rm L}(t )$, ${\cal M}_{\rm B}(t) $, ${\cal M}_{\rm D}(t) $ or ${\cal M}_T(t)$ 
averaged over initial states is qualitatively sketched in 
Fig.~\ref{fig:fig_sketch_LE}. A short-time transient is followed by an 
asymptotic decay and finally by saturation. The level of saturation is
easily determined by ergodicity as $N^{-1}$, in terms of the Hilbert space
size $N$, i.e. the number of states in a complete orthogonal basis of the system.
In a cubic system in $d$ dimensions, one has $N=(L/\nu)^d$, with $\nu$ the
particle wavelength. The short-time 
transient is generically parabolic, as is easily obtained from
short-time perturbation theory. Our interest in this review focuses
on the intermediate,
asymptotic decay regime, which lies between these two, somewhat trivial,
regimes. For the sake of completeness, we nevertheless mention and sometimes
briefly discuss the other two regimes whenever needed.
In the semiclassical limit, it turns out that the behavior of
${\cal M}_{\rm L}(t )$, ${\cal M}_{\rm B}(t) $ or ${\cal M}_{\rm D}(t) $ are closely related, and we therefore
first focus this short survey of existing results on 
the Loschmidt echo. These results are summarized 
in Table~\ref{table:table2}. We next briefly comment on the similarities
and discrepancies between the Loschmidt echo and the 
purity, taken either as a measure of entanglement
between two sub-systems of a bipartite systems, or as a measure of
decoherence. At this point, we
warn the reader that this survey by no means claims to be exhaustive. Our purpose here
is to present a comprehensive table summarizing generic echo 
and purity behaviors. Accordingly, we 
deliberately omit exotic -- but certainly interesting -- 
behaviors occurring in specific situations,
such as the fidelity freeze occurring for perturbations lacking
first order contribution~\cite{Pro05a}, or for specific choices
of initial states~\cite{Wein03}, as well as 
parametric changes with time in the decay of ${\cal M}_{\rm L}$ in 
systems with diffractive impurities~\cite{Ada03},
or in the crossover between two parametric regimes of 
perturbation~\cite{Cer02}.

What determines the asymptotic decay of ${\cal M}_{\rm L}$ ?
Quite obviously, it should depend on the strength of $\Sigma$,
and it was shown in Ref.~\cite{Jac01} that the relevant measure for this strength is provided
by the average  golden rule spreading $\Gamma= 2 \pi \overline{|\langle
\alpha^{(0)}|\Sigma|\beta^{(0)} \rangle|^2}/\delta$
of eigenstates $\alpha^{(0)}$  of $H_0$ over the eigenbasis
$\{ \alpha \}$
of $H$ induced by the difference $\Sigma = H-H_0$. Different decay regimes are obtained
depending on the balance of $\Gamma$ with two additional 
energy scales~\cite{Jac01}:
the energy bandwidth $B$ of the unperturbed Hamiltonian $H_0$,
and the level spacing $\delta = B \hbar_{\rm eff}$, with 
the effective Planck's constant $\hbar_{\rm eff}=\nu^d/\Omega$, 
given by the ratio of the wavelength volume to the
system's volume. Parametrically, these three regimes are

(I) the weak perturbation regime, $\Gamma < \delta$,

(II) the golden rule regime, $\delta \lesssim \Gamma \ll B$, and

(III) the strong perturbation regime, $\Gamma > B$.

\noindent These three regimes are differentiated by the behavior of 
the local spectral density of perturbed states over the unperturbed ones --
we come back to this below.
In Ref.~\cite{Jac01}, the golden rule regime was first defined by
bounds on the strength of the perturbation $\Sigma$ for which 
the local density of eigenstates of $H_0$
over the eigenstates of $H$ acquires a Lorentzian shape.
Accordingly, the Lyapunov decay 
${\cal M}_{\rm L}(t)\propto \exp[-\lambda t]$ to be
discussed below occurs in the golden rule regime for the local spectral density of states.

\begin{figure}
\includegraphics[width=12cm,angle=0]{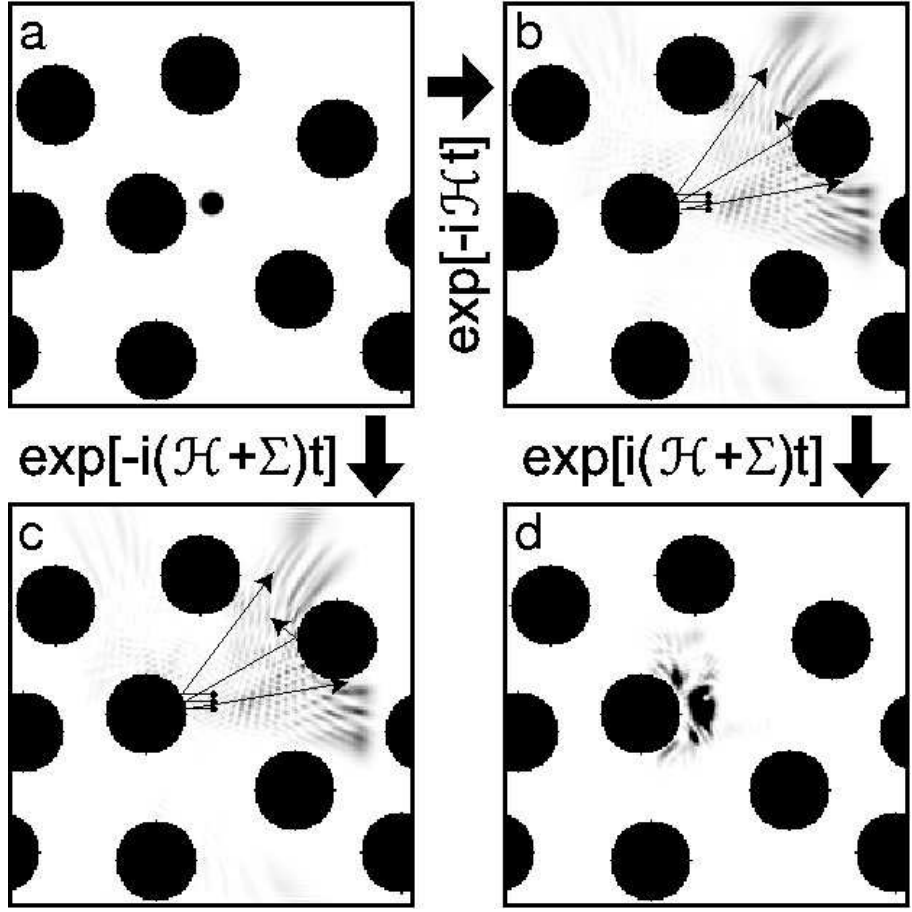}
\caption{\label{fig:fig_cucchietti} Wavepacket evolution in a Lorentz gas.
The initial wavepacket $|\psi_0\rangle$  is represented in the top left 
panel by a small dark spot. The large disks are fixed hard-wall scatterers.
The top right and bottom left panel show 
$|\psi_{\rm F} \rangle = \exp[-i H_0 t] | \psi_0 \rangle$ and $|\psi_{\rm R} \rangle = \exp[-i H t] | \psi_0 \rangle$ respectively. From the
point of view of their spatial distribution, $|\psi_{\rm F} \rangle$ and 
$|\psi_{\rm R} \rangle$ look very similar, and one would naively expect
$1-{\cal M}_{\rm L}(t) \ll 1$. This is not the case however, as the components
of $|\psi_{\rm F} \rangle$ are pseudo-randomly out of phase
with respect to those of $|\psi_{\rm R} \rangle$. This results in a 
strong discrepancy 
between initial (top left) and final (bottom right panel) wavepackets whose scalar product
gives $f(t)$.
(Figure taken from Ref.~\cite{Cuc02a},
with permission. Copyright (2002) by the American Physical Society. 
http://link.aps.org/doi/10.1103/PhysRevB.70.035311)}
\end{figure}

To understand the decays prevailing in these three regimes, 
we start by making the
trivial, though somehow enlightening statement that the
decay of ${\cal M}_{\rm L}$ is governed by the scalar product
$\langle \psi_{\rm R} | \psi_{\rm F} \rangle $ of two normalized 
wavefunctions
$|\psi_{\rm F} \rangle = \exp[-i H_0 t] | \psi_0 \rangle$ and 
$|\psi_{\rm R} \rangle = \exp[-i H t] | \psi_0 \rangle$. The
magnitude of this scalar product
is determined by 
(i) the spatial overlap of the two wavefunctions -- a classical quantity, not much different
from the overlap of two Liouville distributions -- and
(ii) phase interferences between the two wavefunctions -- 
a purely quantum mechanical effect. A decay of ${\cal M}_{\rm L}$ due to
smaller and smaller spatial overlaps is easy to understand at the classical
level already. Because $\Sigma = H-H_0 \ne 0$, both wavepackets visit
different regions of space, and the overlap between these two regions 
decreases with time. This mechanism however sets in for a classically
sizable perturbation $\Sigma$, in a sense that will be defined shortly.
Weak perturbations do not sensibly reduce the spatial overlap of
$|\psi_{\rm F} \rangle$ and $|\psi_{\rm R} \rangle$, even on time scales 
where a significant decay of ${\cal M}_{\rm L}$ is observed. Instead, ${\cal M}_{\rm L}$ 
decays due to mechanism (ii) above, i.e. the fact that
different components of $|\psi_{\rm F} \rangle$ and $|\psi_{\rm R} \rangle$
acquire uncorrelated phase differences generated by $\Sigma$.
This mechanism is illustrated in Fig.~\ref{fig:fig_cucchietti}. The spatial distribution of the 
initial state $|\psi_0\rangle$ is depicted in the top left panel, and the top right and bottom left panel 
show its time-evolution under $H_0$ and $H_0+\Sigma$, respectively.
Even though the spatial
probability distributions $|\langle {\bf r} |\psi_{\rm F} \rangle|^2$ and
$|\langle {\bf r} |\psi_{\rm R} \rangle|^2$ look almost identical -- 
compare the wave patterns on the
top right and bottom left
panels --  ${\cal M}_{\rm L}$ is significantly smaller than one because of phase
randomization. This can be inferred from the very different
initial and final probability cloud, $|\langle {\bf r} |\psi_0 \rangle|^2$ (top left) and
$|\langle {\bf r}| \exp[i H t] \exp[-i H_0 t] | \psi_0 \rangle|^2$ (bottom right).

Strong perturbations
on the other hand ergodize $\exp[i Ht]\exp[-iH_0t] |\psi_0\rangle$
very fast, so that overlaps are not relevant either, in the sense that ${\cal M}_{\rm L}$ decays
with time scales associated with the longitudinal flow, much
shorter than the typical time scale $\lambda$ of overlap decays.
It turns out that overlaps of wavepackets only rarely determine the
asymptotic decay of the Loschmidt echo in quantum chaotic systems. It is in fact the rule rather than the
exception that ${\cal M}_{\rm L}$ decays because of 
$\Sigma-$induced dephasing of $|\psi_{\rm F} \rangle$
against $|\psi_{\rm R} \rangle$ -- we are obviously discussing relative dephasing due to 
the absence of $\Sigma$ in the forward time-evolution.  
Additionally, wavefunction overlaps are relevant
only for specific choices of classically meaningful 
$\psi_0$, such as narrow Gaussian wavepackets, or position 
states~\cite{Jal01,Van03a,Jac02,Iom04b}.
When it is relevant, the overlap decay is very sensitive to the 
dynamics generated by the unperturbed Hamiltonian, but is mostly insensitive
to $\Sigma$.

Let us discuss this more quantitatively.
The condition $\Gamma < \delta$ for the weak perturbation regime (I) legitimates the use
of first-order perturbation theory in $\Sigma$, in which case the relative dephasing
between $|\psi_{\rm F}\rangle$ and $|\psi_{\rm R}\rangle $
is weak and leads to 
a Gaussian decay ${\cal M}_{\rm L}(t) \simeq \exp(-\sigma_1^2 t^2)$. The decay rate is
given by 
$\sigma_1^2 \equiv \overline{\langle \alpha^{(0)}|\Sigma^2|\alpha^{(0)}\rangle}
-\overline{\langle\alpha^{(0)}|\Sigma|\alpha^{(0)}\rangle^2} $, averaged over the ensemble
$\{\alpha^{(0)}\}$ of eigenstates of $H_0$~\cite{Per84}. 
The dephasing is of course strongest in the strong perturbation regime (III)
where it generically leads to another asymptotic Gaussian decay
${\cal M}_{\rm L}(t) \simeq \exp(-B^2 t^2)$~\cite{Jac01} (perhaps excepting specific systems with pathological density of states). The intermediate
golden rule regime (II) is of much interest, in that it witnesses 
the competition
between overlap decay and dephasing decay. For classically chaotic 
systems, the decay of 
${\cal M}_{\rm L}$ is exponential, ${\cal M}_{\rm L}(t)
\simeq \exp[-{\rm min}(\Gamma,\lambda) t]$, with a rate set by the smallest
of $\Gamma$ -- characterizing dephasing -- and the system's Lyapunov exponent 
$\lambda > 0$ -- characterizing the decay of spatial overlaps~\cite{Jac01}. 
The physics behind this quantum--classical competition is that 
both overlap and dephasing mechanisms are simultaneously 
at work here and they both originate from explicitly separable contributions
to ${\cal M}_{\rm L}$. They are therefore additive.
Because they both lead to
exponential decays, the decay of ${\cal M}_{\rm L}$ is therefore governed by
the slowest of the two. The situation is different in 
regular systems, where slightly perturbed wavepackets move away from
unperturbed ones at an algebraic rather than exponential rate. Accordingly,
one expects a power-law decay of ${\cal M}_{\rm L}$~\cite{Jac03} (see also Ref.~\cite{Eme02}).
These results are summarized in Table~\ref{table:table2}.

\begin{table}
\begin{tabular}{|l||c|c|c|c|c} 
\hline

$ {\cal M}_{\rm L}(t) $ & {\bf Regime of validity} & 
{\bf First method of derivation} & $\psi_0$ & $H_0$ \\
\hline
\hline
$1-\sigma_0^2 t^2$ & $ t \ll \sigma_0^{-1} $ & First order PT in $t$ & Any  & Any \\
\hline
$\exp[-\sigma_1^2 t^2]$ & $ \sigma_1 \ll \delta $ & First order PT in $\Sigma$ &Any & Any \\
\hline
$\exp[-\Gamma t]$ & $\delta \lesssim \Gamma \ll B $ & RMT, semiclassics  & Any  & Any  \\
& $\lambda > \Gamma $& & &
\\
\hline
$\exp[-\lambda t]$ & $\delta \lesssim \Gamma \ll B $ & Semiclassics & Classically & Chaotic \\
& $\lambda < \Gamma $& & meaningful  & \\
\hline
$(t_0+t)^{-\alpha}$ & $\delta \lesssim \Gamma \ll B $ & Semiclassics & 
Classically & Regular \\
& & & meaningful  & \\
\hline
$\exp[-B^2 t^2] $ & $\Gamma > B $& RMT & Any  & Any \\
\hline
$N^{-1}$ & $ t \rightarrow \infty$ & RMT & Any  & Any \\
\hline
\end{tabular}
\caption{\label{table:table2}Summary of the different decays and decay regimes for 
the average Loschmidt echo $ {\cal M}_{\rm L}(t)  $. The treatment 
of regular systems assumes that no selection rule
exists for transitions induced by $\Sigma$. This might be hard to 
achieve in regular systems. Accordingly, 
the power-law decay in the table's 
fifth row is to be taken with a grain
of salt. The asymptotic saturation $
{\cal M}_{\rm L}(\infty)
= N^{-1}$ at the inverse Hilbert space size is also based on the same 
assumption. If selection rules exist, $
{\cal M}_{\rm L} $ saturates at a
larger value. Exotic behaviors occurring in specific situations
such as fidelity freeze (for phase-space displacements or perturbation without
first-order contribution) have been deliberately omitted from this table.
In this table, as in the rest of the 
article, actions are expressed in units of $\hbar$, which we accordingly set equal to one.}
\end{table}

The rough classification presented here
is based on the scheme of Ref.~\cite{Jac01} which relates the 
behavior of ${\cal M}_{\rm L}$ in quantum dynamical systems with smooth
potentials to the Fourier transform of the local spectral density of 
eigenstates of $H_0$ over the eigenbasis of $H$~\cite{Jac01,Wis02b}.
Accordingly, regime (II) corresponds to the range
of validity of Fermi's golden rule, 
where the local spectral density has a Lorentzian
shape~\cite{Wig57,Jac95,Fra95,Fyo95,Jac01,Wis02b}. 
A similar correspondence has been emphasized between the local spectral density of
states and the return probability~\cite{Coh00}. It should be stressed however that 
the Fourier transform of ${\cal M}_L (t)$ would be equal to 
the local spectral density of states, in exactly the same way as the return probability,
only if the initial state $\psi_0$ were an eigenstate of $H_{0}$ (or of $H$). 
The choice of $\psi_0$ is largely irrelevant in
the golden rule regime, but it is essential that $\psi_0$ is classically meaningful 
(a narrow wavepacket or a position state) for a decay rate
given by the Lyapunov exponent.

Other investigations beyond this qualitative picture 
have focused on deviations from the behavior 
$\propto \exp[-{\rm min}(\Gamma,\lambda)t]$ in regime (II) due to action 
correlations in weakly chaotic systems~\cite{Wan04,Wan08a,Wan08b}. 
Quantum disordered systems 
with diffractive impurities (not with smooth potentials)
have been predicted to exhibit golden rule decay $\propto
\exp[-\Gamma t]$ and Lyapunov decay $\propto \exp[-\lambda t]$ 
in different time intervals for otherwise fixed 
parameters~\cite{Ada03}, while another crossover has been shown to occur
between an exponential decay at short times and a Gaussian decay
at long times in the crossover
regime between (I) and (II)~\cite{Cer02}. Let us finally mention 
Ref.~\cite{Gar03} which showed that, for open systems,
the Lyapunov decay of ${\cal M}_{\rm L}(t) - {\cal M}_{\rm L}(\infty)$ 
is followed at times larger than the Ehrenfest time (to be defined below)
by a decay governed by Ruelle-Pollicot 
resonances~\cite{Rue86a,Rue86b,Cvi05}. While certainly interesting from a 
mathematical point of view, this decay is barely noticeable in practice and 
we will not discuss it any further. 

Investigations of the dependence of ${\cal M}_{\rm L}$ on the choice
of the initial state considered the Loschmidt echo for a prepared
initial state $\psi_T = \exp[-i H_0 T] \psi_0$ obtained by
evolving
a Gaussian wavepacket $\psi_0$ during a time $T$~\cite{Kar02}. One is then interested
in the quantity
\begin{eqnarray}\label{prepare_echo}
{\cal M}_{T}(t) & = & |\langle \psi_T | \exp[i H t] \exp[-i H_0 t] |\psi_T \rangle|^2 \nonumber \\
&=& |\langle \psi_0 | \exp[i H_0 T] \exp[i H t] \exp[-i H_0 t] \exp[-i H_0 T]
|\psi_0 \rangle |^2.
\end{eqnarray}
The preparation time obviously does not lead to additional dephasing, and
therefore the perturbation-dependent decay $\exp[-\Gamma t]$ does not depend
on $T$. However, the wavepacket spreads during the preparation, and therefore,
the overlap of the two wavefunctions
$|\psi_{\rm F} \rangle = \exp[-i H_0 t] \exp[-i H_0 T] | \psi_0 \rangle$ 
and $|\psi_{\rm R} \rangle = \exp[-i H t] \exp[-i H_0 T]| \psi_0 \rangle$
picks up an additional dependence $\propto \exp[-\lambda T]$, which turns the Lyapunov decay into
$\propto \exp[-\lambda (t+T)]$. These results are discussed in 
Section~\ref{section:displacement}. They were first 
obtained in Ref.~\cite{Jac02}.

The displacement echo ${\cal M}_{\rm D}(t) $ introduced in Eq.~(\ref{decho})
is remarkable in that the perturbation does not lead to dephasing
between otherwise unperturbed trajectories. In the regime
$\delta \lesssim \Gamma \ll B$, ${\cal M}_{\rm D}(t) \propto \exp[-\lambda t]$
only decays because the momentum displacement leads to the decrease of 
wavefunction overlaps. This is not the full story, however, as for small displacements,
this overlap cannot decay to its minimal, ergodic value. In this case,
the short-time (but still asymptotic) exponential decay with the Lyapunov
exponent is followed by a quantum freeze at a displacement-dependent value
which can exceed the ergodic value $N^{-1}$ by orders of magnitude if the displacement is small.
It seems that the easiest way to observe direct manifestations of the classical Lyapunov exponent
in quantum mechanics is the displacement echo.
It is remarkable that, according to 
both trajectory--based semiclassics and RMT,
the purity ${\cal P}(t)$ of the reduced density matrix
in bipartite interacting dynamical systems exhibits the same phenomenology as
${\cal M}_{\rm L}$, up to short-time
discrepancies~\cite{Jac04a,Pet06b,Tan02,Fuj03},  provided one replaces
$\delta$,  $B$ and $\Gamma$ with two-particle level spacing and bandwidth 
$\delta_{2}$ and $B_2$ and the interaction-induced golden rule broadening $\Gamma_2$
of two-particle states. For ${\cal P}(t)$, the Lyapunov decay goes into the sum of two
exponentials with both particle's Lyapunov exponent,  
$\exp[-\lambda t] \rightarrow \exp[-\lambda_1  t] + \exp[-\lambda_2 t]$.
Mathematically speaking, the parallel behaviors of ${\cal M}_{\rm L}$ and
${\cal P}(t)$ come from the fact that both semiclassics and RMT 
rely on pairing -- of either classical trajectories
(motivated by a stationary phase approximation), 
or of wavefunction components (originating from the assumed RMT 
invariance of the distribution of eigenfunction components against basis 
transformation~\cite{Meh91}). This effectively leads to a decay
of ${\cal P}(t)$ given by either dephasing generated by the coupling between
particles, or the decay of overlaps of two initially identical wavefunctions 
evolving under two Hamiltonians differing by their coupling to a second 
particle with different initial conditions. After RMT pairing of
wavefunction components 
or semiclassical pairing of classical paths, the mathematics
of ${\cal P}(t)$ is mostly the same as that of ${\cal M}_{\rm L}(t)$.

We now know that the purity ${\cal P}(t)$  and the Loschmidt echo ${\cal M}_{\rm L}(t)$
have essentially similar behaviors, in that the decays they exhibit are in a one-to-one 
correspondence. What about the Boltzmann echo ? Its definition, Eq.~(\ref{irrevtest}),  
puts it somehow in between 
${\cal P}(t)$  and ${\cal M}_{\rm L}(t)$, one thus
expects that it exhibits the same variety of decays. This is indeed 
the case, up to the important caveat that the rate of 
all perturbation dependent decays is given by the sum of a
term depending on the accuracy with which the system is time-reversed
and a term depending on the coupling between the two subsystems.
Also, there is no dependence on the dynamics of the uncontrolled
subsystem since the corresponding degrees of freedom are integrated out of the problem, and the symmetry exhibited by ${\cal P}(t)$ 
between the two subsystems is lost.

With this we end this voluntarily short and nonexhaustive survey of previously obtained results.
Before going into details of the derivation of these results, we give the outline of this review.

\subsection{Outline}

In Section~\ref{section:fidelity}, we discuss reversibility in 
quantum dynamical systems with few degrees of freedom. We
focus on the Loschmidt echo, Eq.~(\ref{eq:def_LE2}), and describe both
the semiclassical and the RMT approaches in some details. This lays the 
foundation for the use of these analytical methods in later sections. 
Using these two methods of choice in this review, we calculate both the
average fidelity and its mesoscopic fluctuations, computed over ensembles of
spatially distinct, but structurally similar initial states.
In the last two Chapters of 
Section~\ref{section:fidelity}, the discussion digresses somehow from
${\cal M}_{\rm L}$ towards the more specific, but experimentally
relevant displacement echoes, for which we stress the connections
and the differences with the standard Loschmidt echo. 
In Section~\ref{section:wigner}
we revisit several aspects of the Loschmidt echo, this time following a phase-space 
approach. The approach is partially motivated by recent discussions on 
sub-Planck scale structures in the Wigner functions. Their existence
is well established and certainly not put in doubt, however, we
comment on whether they are relevant for understanding quantum
reversibility and decoherence. While there is no observed 
behavior of the Loschmidt echo that cannot be explained by analytical
real-space methods, our phase-space approach
based on Wigner functions 
is very instructive in emphasizing the quantum-classical competition
between the two sources of decay of ${\cal M}_{\rm L}$ -- dephasing
due to imperfect time-reversal and decay of overlap of initially
identical wavepackets evolving with two different dynamics. It complements, and does not
invalidate arguments relating dephasing to sub-Planck scale structures~\cite{Zur01,Kar02}.
In Section~\ref{section:entanglement}
we address the problem of how entanglement between two dynamical 
sub-systems is generated once they start to interact. Here, in some similarity
with Section~\ref{section:fidelity}, we witness a 
quantum-classical competition between coupling-induced and dynamically-induced
generation of entanglement. 
In Section~\ref{section:boltzmannecho} we discuss realistic reversibility
experiments in presence of coupled uncontrolled degrees of freedom -- the problem of the
Boltzmann echo.
There, the fidelity decay rate is bounded from below by the unavoidable generation of
entanglement with the uncontrolled degrees of freedom. We argue that 
this might well have been observed
experimentally in Ref.~\cite{Pas00}.
Conclusions and final discussions are presented in 
Section~\ref{section:conclusion}.

\section{Irreversibility in Quantum Mechanics - the Loschmidt echo}\label{section:fidelity}

Our aim in this chapter is to investigate quantum
irreversibility in dynamical systems with few degrees of freedom by means of the
fidelity of Eq.~(\ref{eq:def_LE2}). We
stress right away that, despite frequent claims to the contrary,
our investigations have little -- if anything -- to do with the second law
of thermodynamics, Boltzmann's $H$-theorem, and the emergence of irreversibility in large 
systems with macroscopic numbers of interacting
degrees of freedom. A probabilistic solution to the irreversibility
paradox and the Boltzmann-Loschmidt controversy~\cite{Los76} was already 
given in the late nineteenth century~\cite{Bol96} and, with certain 
refinements, still holds to this day~\cite{Leb99}. The argument can 
straightforwardly be extended to quantum mechanics --
both quantum and classical macroscopic systems become irreversible
in essentially the same way~\cite{Leb99}. The situation is however different for 
microscopic systems with few degrees of freedom. 
Simple mechanisms of irreversibility already exist at the microscopic level 
in chaotic classical systems with few degrees of freedom, where the 
properties of ergodicity and mixing
ensure that, after a sufficiently long evolution,
two initially well separated phase-space distributions evenly fill
phase-space cells on an arbitrarily small scale (of course smaller scales require
longer evolutions). Since phase-space points can
never be located with infinite precision -- one might 
think of unavoidable round-off errors in numerical simulations, external 
sources of noise or
finite measurement resolution --  irreversibility sets in
after mixing has occurred on a scale smaller than the typical
phase-space resolution scale. This mechanism cannot be carried
over to quantum systems, however, mostly 
because the Schr\"odinger time-evolution is unitary, in either real- or
momentum-space, and that a phase-space resolution on a scale comparable to
Planck's constant is sufficient (see however Section~\ref{section:wigner} for a discussion
of sub-Planck scales in phase-space representations of quantum mechanics).
The coarse-graining of phase-space that is effectively brought by 
unavoidable finite resolutions of the state of the system, and which is one of the two key
ingredients of the just described scenario for classical irreversibility, is obviously less efficient in 
quantum systems -- they are discrete by nature.
Microscopic quantum systems are generically stable under
time-reversal, even when their classical counterpart is irreversible~\cite{She83}. 

This picture is however incomplete.
Peres, pointing out that quantum systems can never 
be considered isolated,
suggested accordingly to investigate quantum irreversibility 
at the microscopic level through the fidelity [we rewrite
Eq.~(\ref{eq:def_LE2})]
\begin{equation}\label{eq:def_LE}
{\cal M}_{\rm L}(t)=\left\vert\left \langle \psi_0 \left\vert 
\exp[i H t] \exp[-i H_0 t]
\right \vert  \psi_0 \right\rangle\right \vert^2,
\end{equation}    
with which a quantum state $ \psi_0$ can be reconstructed by 
inverting the dynamics after a time $t$ with a perturbed Hamiltonian 
$H =H_0 +\Sigma $~\cite{Per84}. Because of its connection
with the gedanken time-reversal experiment proposed by Loschmidt in his
argument against Boltzman's H-theorem \cite{Los76},
${\cal M}_{\rm L}$ has been dubbed the {\it Loschmidt echo} by
Jalabert and Pastawski~\cite{Jal01}, hence the subscript ``L'' in
Eq.~\eqref{eq:def_LE}. The present section is concerned with the calculation of ${\cal M}_{\rm L}$
as a measure of reversibility for small quantum dynamical systems. We first present
a semiclassical calculation, which we then compare to a RMT calculation. Our
analytical predictions are next confirmed by numerical experiments. We finally 
investigate an offspring of the Loschmidt echo, the displacement echo defined 
above in Eq.~(\ref{decho}).

\subsection{Semiclassical approach to the Loschmidt echo}\label{section:semicl}

Semiclassical approaches have been successfully applied in various forms
to the calculation of the fidelity~\cite{Jal01,Cer02,Jac03,Cuc04a,Van03a,Van04,Wan05,Com05a,Pet05,Wan05b,Wan07,Com07a}.
It is probably fair to
say that, while these works certainly amplified on Ref.~\cite{Jal01} and improved it, 
they mostly only confirmed the 
most important result obtained there, that under certain circumstances, the quantum mechanical fidelity
in chaotic dynamical systems decays at a rate determined by the classical Lyapunov exponent. 
The search for classical Lyapunov exponents in quantum mechanics is a
celebrated problem in quantum chaos \cite{Haa92,Per93,Haa01}, and a significant
part of the importance of
Ref.~\cite{Jal01} was to analytically predict that the decay of ${\cal M}_{\rm L}(t)$ can sometimes
be governed by Lyapunov exponents. This is not the first occurrence, however, of a Lyapunov
exponent in the time-evolution of 
a quantum system. Zurek and Paz~\cite{Paz94} predicted that the rate of increase
of the von Neumann entropy $S$ of an inverted Harmonic oscillator weakly coupled to a sufficiently
warm heat bath would increase linearly with the rate $\lambda_{\rm iho}$
at which two neighboring trajectories
move exponentially away from one another, $\partial_t S = \lambda_{\rm iho}$. 
They called $\lambda_{\rm iho}$ the system's 
"Lyapunov exponent", though strictly speaking,
the inverted harmonic oscillator is integrable with a vanishing Lyapunov exponent -- its inverse
parabolic potential generates sensitivity to initial conditions but no folding. They
nevertheless made the leap of faith that, under the same conditions,
chaotic dynamical systems with true positive Lyapunov exponent $\lambda > 0$ have
$\partial_t S = \lambda$. This is certainly not a trivial
step, as truly chaotic systems not only exhibit local exponential instability, but stretching,
contracting and folding of phase-space distributions, which certainly have an effect on wavepacket
dynamics. Yet, Zurek and Paz's prediction was later confirmed by Miller and Sarkar in their
numerical analysis of the kicked rotator -- a model that can be tuned to be truly chaotic -- coupled to a bath of noninteracting harmonic 
oscillators~\cite{Mil99b}. The intuition gained in the study of the inverted oscillator seems to be valid,
at least up to some extent.
The question is still whether such occurrences of classical Lyapunov exponents 
in quantum mechanics require large heat bath, and if they are restricted to 
the high
temperature regime, as suggested by subsequent refinements of the theory of
Paz and Zurek. 
Perhaps more importantly, can one {\it analytically} investigate quantal systems
with a well-defined, truly chaotic limit -- in the mathematically rigorous definition discussed above -- 
and predict the emergence of a Lyapunov-driven behavior of some of its properties in a well-defined
regime of parameters ?

Jalabert and Pastawski gave a positive answer to the first part of that question, but
only specified that their approach is valid for a {\it quantum mechanically large, but classically weak perturbation}. They made no comment on
what this quantitatively means. That second, equally important part of the question was answered by
Jacquod, Silvestrov and Beenakker~\cite{Jac01} 
who obtained precise parametric bounds and quantitative estimates for the validity
of the theory of Ref.~\cite{Jal01}
from a comparison of semiclassics with 
RMT. To make a long story short, Ref.~\cite{Jac01} argued that,
first, in a regime to be determined, the decay of ${\cal M}_{\rm L}$ is given by the sum of the
two semiclassical decays $\propto \exp[-\Gamma t] + \exp[-\lambda t]$, both terms
being multiplied by prefactors of order one. This
implicitly follows from the calculation of Ref.~\cite{Jal01}, but was not explicitly stated there.
Second, by analogy with RMT, which relates the decay of ${\cal M}_{\rm L}$ with the Fourier
transform of the local density of states -- the energy-resolved 
projection of eigenstates of $H_0$ over the basis of eigenstates of $H$ --
the decay term $\propto \exp[-\Gamma t]$ was predicted to occur 
whenever the local spectral density of states is Lorentzian. 
Ref.~\cite{Jac01} concluded that 
the regime of validity of Ref.~\cite{Jal01} is defined by the regime of perturbation 
leading to a Lorentzian local density of states. A RMT approach identified this regime as  
$\delta \lesssim  \Gamma \ll B$ in Ref.~\cite{Jac95,Fra95,Fyo95}, based on rather general grounds.
It thus appears that 
{\it quantum mechanically large} means
that the perturbation broadens eigenstates to an energy width larger than the level spacing, 
thereby making the spectrum effectively continuous, while
{\it classically weak} means that this broadening must be much smaller than the system's bandwidth.
When these two conditions are met, the above argument predicts 
${\cal M}_{\rm L} \propto \exp[-{\rm min} (\Gamma,\lambda) t]$. Looking back, these statements 
and this line of reasoning sound
almost trivial. It is therefore 
important to recall that the range of applicability of the semiclassical
theory of ${\cal M}_{\rm L}$ was not known before Ref.~\cite{Jac01}.

Semiclassical methods apply to the case of classically relevant initial states $\psi_0$, such as the
narrow Gaussian phase-space wavepackets considered in this chapter. Real-space semiclassics
also relies on stationary phase approximations, which implicitly assumes that enough action phase
has been accumulated on the considered classical trajectories, and for times at least shorter
than the time it takes to resolve the discreteness of the quantum spectrum -- beyond that,
purely quantum effects set in which are not captured by semiclassics.
These points, which we rephrase more quantitatively below, have to be
kept in mind -- the method presented in this section applies to the regime of asymptotic decay
and of saturation of ${\cal M}_{\rm L}$, but not to the short-time initial transient regime.
Additionally, as just mentioned, the perturbation $\Sigma$ has to be
quantum-mechanically large -- semiclassics as presented in this chapter
does not apply to the first-order perturbation regime -- but classically small. 
The semiclassical 
results to be presented in this chapter thus are not valid outside
the regime defined by $ \delta \lesssim \Gamma \ll B$~\cite{Jac01}.
These gaps in the theory will be filled in the next section on RMT.

Here we extend the work of Jalabert and Pastawski~\cite{Jal01}
beyond the special case of an extended impurity perturbation potential. It was indeed pointed out
in Ref.~\cite{Jac03} (but probably known to the authors of Ref.~\cite{Jal01}) 
that only bounds on the decay in time of classical correlators matter in
the semiclassical calculation of ${\cal M}_{\rm L}$ -- at least  up to a
phenomenological constant which eventually can be related to the golden rule spreading
$\Gamma$~\cite{Jac01,Cer02}.
The semiclassical calculation of the average Loschmidt echo has already
been described in great details in several publications, therefore we only repeat the steps
that are required to make this section self-consistent.

\subsubsection{Ensemble average}\label{subsub_average}

The semiclassical approach to the Loschmidt echo requires that  the initial state
is classically meaningful, that it is either a position state or a narrow wavepacket in
phase-space. Semiclassics can also be extended to coherent or
incoherent superpositions of such states, provided one can neglect the mutual overlap of the
different states in the superposition. In this chapter, we
consider an initial narrow
Gaussian wavepacket $\psi_0({\bf r}_0') = 
(\pi \nu^2)^{-d/4} \exp[i {\bf p}_0 \cdot ({\bf r}_0'-{\bf r}_0)-
|{\bf r}_0'-{\bf r}_0|^2/2 \nu^2]$ in $d$ dimensions. To time-evolve it, we 
use the semiclassical approximation for the time-evolution kernel discussed in Appendix \ref{appendix:semiclassics}~\cite{Cvi05,Gut90,Haa01}
\begin{subequations}\label{propwp}
\begin{eqnarray}
\langle {\bf r}|
\exp(-i H_0 t) |\psi_0\rangle  =  \int d{\bf r}_0'
\sum_s K_s^{H_0}({\bf r},{\bf r}_0';t) \psi_0({\bf r}_0'), \\
\label{eq:propscl}
K_s^{H_0}({\bf r},{\bf r}_0';t)  =  \frac{C_s^{1/2}}{(2 \pi i)^{d/2}} 
\exp[i S_s^{H_0}({\bf r},{\bf r}_0';t)-i \pi \mu_s/2].
\end{eqnarray}
\end{subequations}
\noindent The semiclassical propagator $K_s^{H_0}({\bf r},{\bf r}_0';t) $ is expressed
as a sum over classical trajectories (labeled $s$)
connecting ${\bf r}$ and ${\bf r}_0'$ in the time $t$.
For each $s$, the partial propagator contains
the action integral $S_s^{H_0}({\bf r},{\bf r}_0';t)$ along $s$,
a Maslov index $\mu_s$, and
the determinant $C_s$ of the stability matrix. Once this time-evolution is inserted into expressions
involving more than one time-evolution operator, quantum coherent effects 
can be captured via
nontrivial phase interferences involving two or more classical trajectories with different action
phases. In the cases investigated in this review, such nontrivial effects already occur
at the level of diagonal pairing, setting classical trajectories pairwise 
equal to one another, where one element of the pair feels the effect of the perturbation
(corresponds to $H$) while the other one does not (as it corresponds to $H_0$). 

What is the range of validity of the semiclassical approach ? 
To answer this question, one needs to discuss 
the hierarchy of important time scales we consider in this review in some more details.
The semiclassical propagator, Eq.~(\ref{eq:propscl}), is
derived from the Feynman-Kac path integral expression for the quantum time-evolution
operator, once a stationary phase condition is enforced. The latter requires that 
the action phase accumulated on almost all paths in the path integral 
(not only the classical ones) is much larger than 1 (in units  of $\hbar$). This requires a minimal time
which can be estimated as $\tau_{\rm min} = 1/E$, with the energy $E$ of the system.
For larger times, the Hamiltonian flow generates enough action phase to justify
a stationary phase condition. The approach also breaks down at longer times,
and certainly loses its validity once the discreteness of the spectrum is resolved, i.e.
for times longer than the Heisenberg time $\tau_{\rm H} = \hbar/\delta$ with the level
spacing $\delta$ of the system considered. (This time is determined by a standard uncertainty
relation with the level spacing, hence its name.)
Earlier breakdowns can occur due to the
proliferation of conjugate points, and it has been numerically observed
that the semiclassical 
approach permits to calculate the time
evolution of smooth, initially localized wavepackets with a reasonable accuracy 
up to algebraically long times in the effective Planck's constant
$\propto {\cal O}(\hbar_{\rm eff}^{-a})$ (with $a>0$)~\cite{Tom91,Hel93}.
The semiclassical methods employed in this review are not applicable outside
the time interval $[\tau_{\rm min},\tau_{\rm H}]$. This interval is parametrically very large
in $\hbar_{\rm eff}$ in the semiclassical limit.

Beside these two quantum time scales, classical time scales limit the applicability of our
approach in that they give bounds for some statistical assumptions we have to make as we
go along. 
Closed chaotic systems are characterized first by
the time of flight $\tau_{\rm f} = L/v$, i.e. the time it takes to cross the system once. This time
is however so small -- the velocity of the particle becoming larger and large in the
semiclassical limit -- that we neglect it altogether and set it equal to zero. 
A second important classical
time scale is the Lyapunov time $\lambda^{-1}$,
roughly giving the time it takes for local exponential instability to set in.
Next one has
the ergodic time $\tau_{\rm erg}$ measuring the time it takes for an initial condition
to have visited most of its available phase-space. This time scale is also assumed
to be very short. Still one has to keep in mind that the sum rules we employ are often
justified by ergodicity assumptions which break down at times shorter than
$\tau_{\rm erg}$.
There remains one important quantal time scale to discuss, and it is 
the Ehrenfest time $\tau_{\rm E}=\lambda^{-1} \ln L/\nu$. This time is logarithmic 
in $\hbar_{\rm eff}=\nu/L$, is always
much shorter than $\tau_{\rm H}$ in the semiclassical limit, and gives bound for the
onset of coherence effects in semiclassics.
In summary, our semiclassical approach is valid in a parametrically large regime
of time, bounded from below by either a classical ergodic time or the time it takes
to accumulate enough action phase to justify stationary phase approximations, and bounded
from above by a time algebraically large in $\hbar_{\rm eff}$, which is smaller or
equal to 
the Heisenberg time it takes to resolve the discreteness of the underlying
quantum spectrum.

Within the semiclassical approximation the fidelity reads, not so elegantly,
\begin{eqnarray}\label{moft}
{\cal M}_{\rm L} (t) & = & \int d{\bf r} \int d{\bf r}_{01} \int d{\bf r}_{02} \;
\psi_0({\bf r}_{01})\psi_0^*({\bf r}_{02}) \;
\sum_{s_1,s_2} K_{s_1}^{H_0}({\bf r},{\bf r}_{01};t) \;
 [K_{s_2}^H({\bf r},{\bf r}_{02};t)]^*  \nonumber \\
 & \times & \int d{\bf r}' \int d{\bf r}_{01}' \int d{\bf r}_{02}' \;
\psi_0^*({\bf r}_{01}')\psi_0({\bf r}_{02}') \;
\sum_{s_3,s_4} [K_{s_3}^{H_0}({\bf r}',{\bf r}_{01}';t)]^*  \;
K_{s_4}^H({\bf r}',{\bf r}_{02}';t) \, .
\end{eqnarray}
This expression is easily obtained by inserting four semiclassical time-evolution kernels, 
Eq.~(\ref{propwp}), into Eq~(\ref{eq:def_LE}).
The fidelity is given by
a six-fold integral over initial and intermediate (at time-reversal) positions, with additionally
a four-fold sum over classical trajectories. To reduce this to a useful, tractable  expression, one
first notices that because $\psi_0$ is a narrow Gaussian wavepacket centered on
${\bf r}_0$, one can linearize the integrand in ${\bf r}_{0i}={\bf r}_{0}+\delta {\bf r}_{0i}$ and 
${\bf r}_{0i}'={\bf r}_{0}+\delta {\bf r}_{0i}'$, then perform the resulting (Gaussian) integrals
over initial positions.
The second step is to enforce semiclassically motivated stationary phase conditions
that reduce the
four-fold sum over classical paths to three dominant terms, two involving a two-fold sum, one
involving a single sum over classical paths. 
The structure of the semiclassical approximation to the average fidelity at this stage
is sketched on the right-hand side of Fig.~\ref{fig:fig_LEavg}.
Classical trajectories are represented
by a full line if they correspond to $H_0$
and a dashed line for $H$, with an arrow indicating the direction
of propagation. 
For a given initial condition ${\bf r}_0$, each contribution consists in four classical paths
connecting ${\bf r}_0$ to two final evolution points ${\bf r}$ and ${\bf r}'$. 

\begin{figure}
\includegraphics[height=4.5cm]{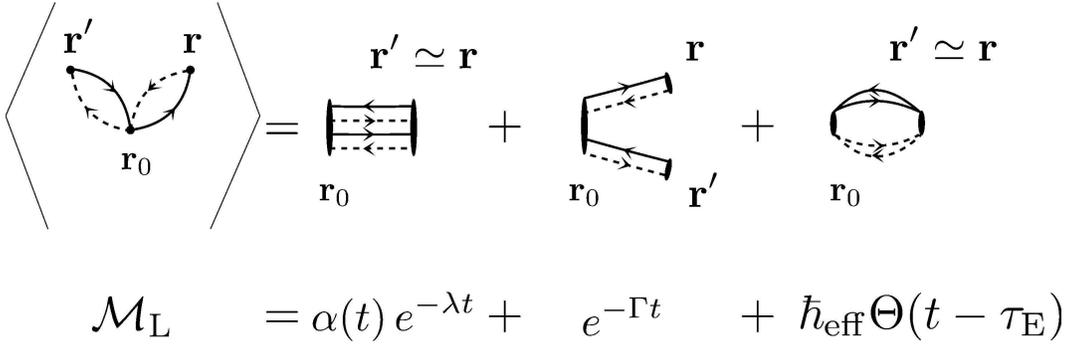}
\caption{\label{fig:fig_LEavg} Diagrammatic representation of the 
average fidelity ${\cal M}_{\rm L}$  and the trajectory pairings leading to the Lyapunov decay
$\propto \exp[-\lambda t]$ (for chaotic systems -- for regular ones this contribution
gives an algebraic decay), the golden rule
decay $\propto \exp[-\Gamma t] $ and the
long-time saturation of ${\cal M}_{\rm L}$. The semiclassical fidelity is expressed as a four-fold sum
over unperturbed (solid lines) and perturbed (dashed lines) 
classical trajectories (left-hand side). This sum is reduced to single and double sums after
semiclassically motivated stationary phase conditions are enforced (right-hand side). The exponential decay with the Lyapunov exponent (first term on the right-hand side) goes into an algebraic decay
for regular systems (see Table~\ref{table:table2}).}
\end{figure}

The motivation for enforcing a stationary phase condition on
the action phase differences $S_{s_1}({\bf r},{\bf r}_0;t)-
S_{s_2}({\bf r},{\bf r}_0;t)$ 
and $S_{s_3}({\bf r},{\bf r}_0;t)-
S_{s_4}({\bf r},{\bf r}_0;t)$ 
appearing in Eq.~(\ref{moft}) -- the phases are contained in the
partial semicassical propagators $K_s$, see Eq.~(\ref{propwp}) -- 
is that we calculate the fidelity averaged over an ensemble of
initial Gaussian wavepackets $\psi_0$. As the center of mass
${\bf r}_0$ of these initial states is moved, the phase differences
fluctuate, so that
the only contributions that survive the average are those which minimize
these fluctuations. The dominant such contribution is obtained from 
the diagonal approximation $s_1=s_2$, $s_3=s_4$.
But how can this be justified, given that $s_{1,3}$ are classical trajectories
generated by $H_0$, while $s_{2,4}$ correspond to a different
Hamiltonian $H=H_0+\Sigma$ ? The answer is that 
setting $s_1 = s_2$ for two trajectories generated by two
different chaotic Hamiltonians $H = H_0 + \Sigma$ is 
justified by shadowing and the structural stability of hyperbolic systems
for not too large $\Sigma$~\cite{Kat96}. In the context of the
fidelity, this point was first mentioned in Refs.~\cite{Cer02,Van03a},
and we discuss it further below in Chapter~\ref{section:displacement}.
Here we only mention what this qualitatively means. Structural stability/shadowing theorems
state that almost all trajectories of 
slightly perturbed classical hyperbolic systems come in a one-to-one correspondence 
with the trajectories of the corresponding unperturbed hyperbolic system, in that 
for each unperturbed trajectory, there exists a perturbed trajectory which 
remains
in its immediate vicinity. The two trajectories do not share endpoints in general, but it does not matter
for our purpose here, all we need being that they stay a distance less than the quantum
mechanical resolution apart. Setting $s_1=s_2$ thus means that we chose $s_2$ to be the
shadow of $s_1$.

While strictly speaking shadowing and structural stability theorems apply to 
uniformly hyperbolic systems only, numerical investigations have shown that
generic chaotic systems also display structural stability and shadowing
of trajectories upon not too strong perturbations~\cite{Gre90}. Therefore, setting
$s_1=s_2$ is justified for chaotic systems.
There is no such principle that justifies setting the diagonal pairing of
trajectories in regular or integrable systems, however, and, despite
several convincing numerical confirmations, the
results on regular systems to be presented below must be 
considered cautiously. Sort of ironically, the calculation to be presented is
more reliable for chaotic than regular systems. But this makes a lot of sense,
given that universality applies to chaotic systems, not to regular ones, which 
typically exhibit largely fluctuating system-dependent behaviors.

After the linearization around ${\bf r}_0$ and 
this first stationary phase approximation, ${\cal M}_{\rm L}(t)$ is reduced
to a double sum over classical paths $s$ and $s'$ 
and a double integration over coordinates ${\bf r}$ and
${\bf r}'$,
\begin{eqnarray}\label{mtot}
{\cal M}_{\rm L}(t) & = & (\nu^2/\pi)^d \int d{\bf r} \int d{\bf r}'
\sum_{s,s'} C_s C_{s'} \exp[i \delta S_s({\bf r},{\bf r}_0;t) - 
i \delta S_{s'}({\bf r}',{\bf r}_0;t) ] \nonumber \\
& & \times \exp(-\nu^2 |{\bf p}_s-{\bf p}_0|^2-\nu^2 
|{\bf p}_{s'}-{\bf p}_0|^2),
\end{eqnarray}
\noindent with $ \delta S_s({\bf r},{\bf r}_0;t) =
S_s^H({\bf r},{\bf r}_0;t)-S_s^{H_0}({\bf r},{\bf r}_0;t)$.
Our strategy next is to differentiate between contributions in Eq~(\ref{mtot}) where 
the trajectories $s$ and $s'$ are correlated ($s\simeq s'$, within a spatial resolution $\nu$) 
from those where they are not. We call the
correlated contribution the {\it diagonal contribution}, and
the uncorrelated one the {\it nondiagonal contribution} by
some abuse of language, even though both contributions already emerge from
the diagonal approximation $s_1 \approx s_2$ we made to go from Eq.~(\ref{moft}) to 
Eq.~(\ref{mtot}). These two sets of contributions are quite different in essence, and they
lead to fundamentally different decays. 
We argue in Appendix \ref{appendix:semiclassics}
that the decay of the diagonal contribution is
governed by the decay of the overlap of 
$|\psi_{\rm F} \rangle = \exp[-i H_0 t] | \psi_0 \rangle$ with
$|\psi_{\rm R} \rangle = \exp[-i H t] | \psi_0 \rangle$, while the
behavior of the nondiagonal contribution is determined by the
$\Sigma$-induced dephasing between the wavepacket propagating along 
$s$ and the one propagating along $s'$. 
Consequently, the diagonal contribution have a classical decay determined by the
disappearance of wavefunction overlap. The latter occurs exponentially fast in chaotic
systems, due to their characteristic exponential instability of neighboring
orbits, but is much slower, it is in fact algebraic, in regular systems. Simultaneously, the 
decay of the nondiagonal contributions  is governed by perturbation-generated dephasing
between the forward and backward propagation along $s$ and $s'$. Because we neglected
the effect of the perturbation on the classical trajectories -- an approximation that was
justified by invoking structural stability and shadowing --
this dephasing is of purely
quantal nature. It typically leads to an exponential decay in chaotic systems, and to a Gaussian
decay in regular systems which often have surviving correlations. The Gaussian 
decay is however often
masked because in regular systems, the diagonal contribution generates 
a much slower algebraic decay.
We also note that
the diagonal contribution sensitively depends
on whether $H_0$ is regular or chaotic, while the nondiagonal contribution
is generically insensitive to the nature
of the classical dynamics set by $H_0$, provided that
the perturbation Hamiltonian $\Sigma$ induces enough mixing of eigenstates 
of $H_0$, and in particular that it has no common integral of motion
with $H_0$.

In addition, there is a third contribution depicted in Fig.~\ref{fig:fig_LEavg} which
corresponds to the long-time saturation of ${\cal M}_{\rm L}(t)$.
We seem to be the first to notice that 
the latter can also be calculated semiclassically. To see this, one has to go
back one step before the diagonal approximation
leading to Eq.~(\ref{mtot}). After one performs the linearization around ${\bf r}_0$ on Eq~(\ref{moft}), 
one has
\begin{eqnarray}\label{full_semicl}
{\cal M}_{\rm L}(t) & = & (\nu^2/\pi)^d \int d{\bf r} \; \int d{\bf r}' 
\sum_{s_1,s_2,s_3,s_4} K_{s_1}^{H_0}({\bf r},{\bf r}_0;t)
[K_{s_2}^{H}({\bf r},{\bf r}_0;t)]^*  [K_{s_3}^{H_0}({\bf r}',{\bf r}_0;t)]^*
K_{s_4}^{H}({\bf r}',{\bf r}_0;t) \nonumber \\
&& \times \exp\big(-\nu^2 \big[|{\bf p}_{s_1}-{\bf p}_0|^2
+|{\bf p}_{s_2}-{\bf p}_0|^2 + |{\bf p}_{s_3}-{\bf p}_0|^2 
+|{\bf p}_{s_4}-{\bf p}_0|^2\big]/2 \big).
\end{eqnarray}
Pairing the trajectories as $s_1=s_3$ and $s_2=s_4$ cancels exactly all action phases. On the negative side, this pairing
simultaneously requires ${\bf r} \simeq {\bf r}'$ within the
wavelength resolution $\nu$, a restriction that results in a reduction of its contribution by 
a prefactor $\hbar_{\rm eff}$. 
The calculation of this term is described in Appendix \ref{appendix:semiclassics},
and we do not repeat it here. One gets a time-independent 
contribution
\begin{equation}\label{longtimesaturationLE}
{\cal M}_{\rm L}(\infty) = \hbar_{\rm eff} \; \Theta(t > \tau_{\rm E}),
\end{equation}
corresponding to the long-time saturation of ${\cal M}_{\rm L}$. This term 
requires that uncorrelated paths exist between ${\bf r}_0$ and ${\bf r}
\simeq {\bf r}'$ (see the rightmost contribution sketched in 
Fig.~\ref{fig:fig_LEavg}) and therefore does not exist for times
shorter than the Ehrenfest time
$\tau_{\rm E} \equiv \lambda^{-1} |\ln[\hbar_{\rm eff}]|$. 

With all this we write
\begin{equation}
{\cal M}_{\rm L}(t)={\cal M}_{\rm L}^{\rm (d)}(t)+{\cal M}_{\rm L}^{\rm (nd)}(t)+{\cal M}_{\rm L}(\infty).
\end{equation}
The trajectory pairings that lead to these three terms are summarized in Fig.~\ref{fig:fig_LEavg}.
The above splitting of ${\cal M}_{\rm L}(t)$ into three terms is not only mathematically convenient,
it is physically meaningful. The first term is phase-independent and we are momentarily going to
argue that it decays with the decay of the overlap of $\psi_0$ evolving under $H_0$ with
itself, when it is evolved under $H=H_0+\Sigma$. It is of purely classical origin -- this is generic
of semiclassically computed terms of maximal diagonal pairing (another 
example is
the Drude conductance in the semiclassical theory of transport~\cite{Bar91}). The second term is
perturbation-dependent and within the semiclassical approach, it decays with the variance of
the phase difference accumulated along paired trajectories due to the presence of the
perturbation. Strictly speaking, it is also a diagonal contribution. Its quantumness, however, 
originates in that the perturbation $\Sigma$ affects the action phase accumulated along only one
of the trajectories. In other words, dephasing due to $\Sigma$ is taken into account. The third
contribution finally corresponds to the unbreakable, time-independent, 
ergodic core of ${\cal M}_{\rm L}(t)$, i.e.
that part which correspond to minimal overlap of two random, ergodic wavefunctions. On average
they are not orthogonal (this would require a degree of correlation which gets lost in the
long time evolution under two different Hamiltonians) and this is why ${\cal M}_{\rm L}(t)$
saturates at a finite value inversely proportional to the Hilbert space volume. Let us have a quick
look at these decays and their origin in some more details.

The calculation of all these contributions is presented in Appendix ~\ref{appendix:semiclassics}.
The decay of the nondiagonal contribution is governed by
the action phases accumulated on the uncorrelated paths $s \ne s'$. It is thus legitimate to
perform the phase averaging separately for $s$ and $s'$ with 
\begin{equation}\label{eq:phase}
\langle \exp[i \delta S_s] \rangle = 
\exp(-{\textstyle \frac{1}{2}} \langle \delta S_s^2 \rangle ) = \mbox{}
\exp\left(-{\textstyle \frac{1}{2}} \int_0^t d{\tilde t} \int_0^t d{\tilde t}' 
\langle \Sigma[{\bf q}(\tilde{t})]
\Sigma[{\bf q}(\tilde{t}')] \rangle \right).
\end{equation}
Here ${\bf q}(\tilde{t})$ lies on path $s$ with 
${\bf q}(0)={\bf r}_0$ and ${\bf q}(t)={\bf r}$. 
In Ref.~\cite{Jal01}, a specific assumption was made about the perturbation potential, that 
it corresponds to a random distribution of extended impurities. This gives an exponential decay
of the correlator in Eq.~(\ref{eq:phase}). 
Here we go beyond this
approach, noting that, for chaotic systems, one generically observes fast decays of
correlations. Under the assumption that
$\Sigma$ and $H_0$ have no common integral of motion, so that
$\delta S_s$ fluctuates fast and randomly enough, 
the correlator of $\Sigma$ decays fast with time, which
gives the golden rule decay 
\begin{equation}\label{eq:nondiagdecay}
{\cal M}_{\rm L}^{({\rm nd})}(t) \propto \exp(-\Gamma t), \;\;\;\; \;{\rm with} \;\;\;\;
\Gamma t \equiv  \frac{1}{2} \int_0^t d{\tilde t} \int_0^t d{\tilde t}' 
\langle \Sigma[{\bf q}(\tilde{t})]\,
\Sigma[{\bf q}(\tilde{t}')] \rangle,
\end{equation}
regardless of whether $H_0$ is chaotic or regular. This conclusion, that the golden rule decay 
holds whether $H_0$ is regular or chaotic, 
can also be obtained via a fully quantum mechanical 
approach based on random-matrix theory assumptions for $\Sigma$, in which case
the invariance under unitary transformations of the distribution
of $\Sigma$ is sufficient to obtain the exponential decay 
${\cal M}_{\rm L}^{\rm (nd)}(t) \propto \exp(-\Gamma t)$, irrespective of the
distribution of $H_0$. However it has to be noted that the whole argument
relies on the assumption
that the perturbation correlator in Eq.~(\ref{eq:nondiagdecay}) decays 
faster than $t^{-1}$, also in regular systems. While perturbations can
be tailored to meet this assumption, there are certainly cases where
the correlator oscillates in time or even saturates at a finite,
nonzero value at large times. Several instances have been recorded where the decay of
${\cal M}_{\rm L}(t)$ in regular systems is
Gaussian rather than exponential~\cite{Pro02,Pro03a,Gor06}. This might reflect
the nondecaying behavior of the correlator of $\Sigma$, but 
also indicates
that the diagonal contributions do not exist, presumably because of lack of shadowing
of unperturbed classical orbits by classical ones, i.e. the double diagonal
approximation is not justified there.

The calculation of the diagonal contribution is detailed in Appendix~\ref{appendix:semiclassics_avg}.
With Eqs.~(\ref{eq:expansion_phase}), (\ref{eq:divergence}),
(\ref{clt}), and (\ref{correl}), 
Eq.~(\ref{mtot}) gives for the diagonal contribution to 
the Loschmidt echo
\begin{eqnarray}\label{echo_diag}
{\cal M}_{\rm L}^{\rm (d)}(t) & = & (\nu^2/\pi)^d \int d{\bf r}_+ \int d{\bf r}_- \sum_s C_s^2
\exp\bigl(-\frac{1}{2} U \tau \; {\bf r}_-^2\bigr) 
\exp(-2 \nu^2 |{\bf p}_s-{\bf p}_0|^2),
\end{eqnarray}
with $\tau=t/6$ for regular systems and 
$\tau=\lambda^{-1}(1-\exp[-\lambda t]) \simeq \lambda^{-1}$ for chaotic
systems, and $U$ is defined in Eq.~(\ref{correl}) from the correlator of 
derivative of $\Sigma$,
\begin{equation}
\langle \partial_i \Sigma[{\bf q}(\tilde{t})] \partial_j 
\Sigma[{\bf q}(\tilde{t}')] \rangle = U \delta_{ij} 
\delta({\tilde t}-{\tilde t}'). 
\end{equation}
 The rest of the calculation is straightforward.
The Gaussian integration over ${\bf r}_- \equiv 
{\bf r}-{\bf r}'$ ensures that ${\bf r} \approx {\bf r}'$, 
and hence ${\bf r}_+ \equiv ({\bf r}+{\bf r}')/2 \approx {\bf r}$. One further uses
one $C_s$ (which is the determinant of a Jacobian)  
to perform a change of variables from ${\bf r}_+$
to ${\bf p}_s$. For the remaining $C_s$ we take into account
the algebraic stability of regular systems with 
$C_s \propto t^{-d}$ (regularized at short times with $t_0$) 
to be contrasted with the
exponential instability of chaotic systems with $C_s \propto \exp[-\lambda 
t]$. One finally arrives at
\begin{eqnarray}\label{diagdecay}
{\cal M}_{\rm L}^{\rm (d)}(t) & \propto & \left \{
\begin{array}{cc}
t^{-d}, & {\rm regular \; systems \; with \;} U \tau < \nu^{-2}, \\
t^{-3d/2}, & {\rm regular \; systems \; with \;} U \tau > \nu^{-2}, \\
\exp[-\lambda t], &{\rm chaotic \; systems}.
\end{array} \right.
\end{eqnarray} 
These decays are rather insensitive to the choice (\ref{correl}) 
of a $\delta$-function force correlator. 
Even a power-law decaying correlator 
$\propto |{\tilde t}-{\tilde t}'|^{-a}$ 
reproduces Eqs.\ (\ref{diagdecay}) at large enough times, provided $a \ge 1$.
These diagonal decays make a lot of sense, they actually agree rather well with our intuition,
based on decays of overlaps of classical phase-space distributions. Translating the perturbation --
which is assumed to be classically small -- into slight phase-space displacements, one expects
that the local exponential instability of chaotic systems leads to an exponential decay of these
overlaps -- this is confirmed by works on the classical fidelity, at least in 
some regime~\cite{Pro02,Ben02,Eck03,Ben03c,Ben03b}.

Our semiclassical approach thus predicts that, up to the long-time
saturation at the effective Planck's constant, the
Loschmidt echo is given by the sum of the diagonal and nondiagonal terms,
\begin{eqnarray}\label{eq:final_decay}
{\cal M}_{\rm L} (t) = {\cal M}_{\rm L}^{\rm (d)}(t)
+ {\cal M}_{\rm L}^{\rm (nd)}(t) \propto \left\{
\begin{array}{cc}
t^{-d}, & {\rm regular \; systems \; with \;}  U \tau < \nu^{-2}, \\
t^{-3d/2}, & {\rm regular \; systems \; with \;}  U \tau > \nu^{-2}, \\
\alpha e^{-\lambda t} + e^{-\Gamma t}, & {\rm chaotic \; systems}.
\end{array} \right.
\end{eqnarray} 
These results are valid in the asymptotic regime, past  
the initial parabolic transient (see Fig.~\ref{fig:fig_sketch_LE}), and
as such they lose their validity at short times -- Eqs.~(\ref{eq:final_decay})
does not predict a singularity at $t=0$ for regular systems, nor
${\cal M}_{\rm L}(t=0) = 1+\alpha > 1$ for chaotic systems ! 
The predicted decays are 
parametric in essence, and are smoothly connected to the initial,
short-time transient decay via 
weakly time-dependent prefactors of order one.
This is confirmed by numerical works.
It has to be kept in mind that the
results given in Eq.(\ref{eq:final_decay}) are averages over an ensemble 
of initial Gaussian wavepackets
$\psi_0$. This is required to justify the semiclassical
stationary phase approximations from which these results derive. 

Still the dominant, diagonal contribution to the fidelity for regular systems has been
derived under the assumption that correlations decay fast enough, Eq.~(\ref{correl}). This 
is not always
satisfied in regular systems, where it is actually the rule, rather than the exception,
that correlators such as the one in Eq.~(\ref{correl}) decay more slowly than $t^{-1}$, i.e. 
$a< 1$. Assuming a constant correlator
\begin{equation}
\langle \partial_i \Sigma[{\bf q}(\tilde{t})] \partial_j 
\Sigma[{\bf q}(\tilde{t}')] \rangle = U' \delta_{ij} 
\end{equation}
results in $\tau = t^2/8$ in Eq.~(\ref{echo_diag}), which leads for $U' t^2/8 > \nu^{-2}$
to an accelerated, but still power-law decay of the diagonal 
contribution to the fidelity, ${\cal M}_{\rm L}(t) \propto t^{-2d}$, in regular systems. 
We believe that the decay of the {\it average} fidelity in regular systems is generically algebraic,
however, the exponent with which ${\cal M}_{\rm L}(t)$ decays can vary from case to case.
It is well possible that the fidelity calculated for individual $\psi_0$ exhibits different behaviors,
as the Gaussian ones reported in Refs.~\cite{Pro02,Pro03a,Gor06}, 
since it is an average over an ensemble of
$\psi_0$ that effectively leads to the integral over ${\bf r}_-$ in Eq.~(\ref{echo_diaga}). Without
that integration, one has a Gaussian time-dependence.

In both regular and chaotic systems, the decay of
${\cal M}_{\rm L}^{\rm (nd)}(t)$ reflects the stability of nearby orbits,
$C_s \propto \exp[-\lambda t]$ for chaotic, $C_s \propto t^{-d}$ for regular
systems. This is not the full story in regular systems, however,
where the correlator (\ref{correl}) contributes another $t^{-d/2}$ for $U t/6 > \nu^{-2}$, or
$t^{-d}$ for $U' t^2/8 > \nu^{-2}$.
Compared to the ``classical fidelity'',
i.e. the overlap of
classical phase-space distributions~\cite{Pro02,Ben02,Eck03,Ben03c,Ben03b},
the quantum fidelity decays faster in regular systems, because 
dephasing does not totally decouple from overlap. The same 
effect also occurs in chaotic systems where, however, it gives
a subdominant, algebraic correction to the exponential Lyapunov decay of overlaps. 
This is hardly noticeable.

Given the respective specificities of classical and quantum mechanical dynamics, and the
structure of their equations of motion, it is at first glance quite surprising to observe
Lyapunov exponents in the dynamics of quantal systems as directly as in the decay of 
the Loschmidt echo. The semiclassical approach presented above 
is however transparent enough that one can
trace back the origin of the Lyapunov decay to the stability
of chaotic classical trajectories. One concludes that the fidelity decays exponentially at the 
Lyapunov rate in precisely the same way as the overlap of
two classical phase-space distributions, initially identical, but evolving under the influence of
two slightly different Hamiltonians. Wisdom comes with experience and once this
mathematical observation is done, having a Lyapunov decay does not come as a surprise after all. 
But this is only one side of the story, which in particular neglects the second contribution
to the Loschmidt echo, the one we called nondiagonal and which 
carries quantum coherence. In chaotic systems, we have seen that diagonal and
nondiagonal contributions decay exponentially with time, so that their sum decays effectively
with the weakest of the two decay rates, 
${\cal M}_{\rm L}(t) \propto \exp[-{\rm min}(\Gamma,\lambda) \, t]$. This a very simple formula, which
initially looked too simple to be true, even to two of the authors of 
Ref.~\cite{Jac01} where it first appeared.
It actually contains a lot more 
physics than meets the eyes at first glance. Most of all, it states that quantum mechanically strong, but 
classically weak perturbations can generate a decay of ${\cal M}_{\rm L}(t)$ which is totally governed by
dephasing. There is thus a parametrically large regime where external perturbations have 
no observable classical dynamical effect, yet lead to the decay of the
Loschmidt echo. In that regime, the handwaving argument, that ${\cal M}_{\rm L}(t)$
decays because the perturbation first leads to a small phase-space displacement which is
subsequently exponentially amplified by the underlying classical dynamics is incorrect. It neglects
the fact that pairing of trajectories is still possible between two slightly different Hamiltonian. This
is the second surprise, and this time it remains a surprise even retroactively!
It is concepts so deeply rooted into classical dynamics as shadowing and
structural stability that allow dephasing to occur so fast that the perturbation at its origin has
effectively no dynamical effect. In that regime, dephasing or decoherence cannot be apprehended
by paradigms based on phase-space displacements.

\subsubsection{Mesoscopic fluctuations}

Fluctuations of a physical quantity often contain more information 
than its average. For example, quantum signatures of classical
chaos are absent of the average density of states, but strongly 
affect spectral fluctuations~\cite{Haa01}. Here, we investigate the fluctuations
of the Loschmidt echo as the initial state is modified. We will see that Lyapunov exponents
can be extracted from the fluctuations of ${\cal M}_{\rm L}$ over a larger range
of parameters than from the average of ${\cal M}_{\rm L}$. However no fundamentally
new physics emerges from fluctuations.

Ref.~\cite{Sil03} presents the first
investigation of the properties of ${\cal M}_{\rm L}$ beyond its average. It shows
that, for classically large perturbations, $\Gamma \gg B$, 
${\cal M}_{\rm L}$ is dominated
by very few exceptional events, so that the
fidelity for a typical initial state is better described by 
$\exp[\overline{\ln({\cal M}_{\rm L})}]$, and that ${\cal M}_{\rm L}$ 
does not
fluctuate for times longer than the Ehrenfest time. Ref.~\cite{Pet05}
showed however that
these conclusions do not apply to the regime of classically
weak but quantum-mechanically strong perturbation, instead they are valid when
the perturbation is
classically large. 
In that regime, ${\cal M}_{\rm L}(t)$ measures what can still be successfully
recovered after a hopelessly imperfect time-reversal operation is performed. Accordingly, ref.~\cite{Sil03} states that 
some recovery is possible if the time-reversal operation is performed soon enough that the perturbation
has no time to propagate ergodically. Some numerical data 
for the distribution of ${\cal M}_{\rm L}$ in the
weak perturbation regime were presented in Ref.~\cite{Gor04a}. Here, we
focus on chaotic
systems -- we discuss only very briefly fluctuations of ${\cal M}_{\rm L}$ in regular or
integrable systems at the end of this section -- and investigate the behavior of the variance $\sigma^2({\cal M}_{\rm L})$ of the
fidelity in the golden rule regime from a semiclassical point of view.

\begin{figure}
\begin{psfrags}
\psfrag{ro}{${\bf r}_0$}
\psfrag{r1}{${\bf r}_1$}
\psfrag{r2}{${\bf r}_2$} 
\psfrag{r3}{${\bf r}_3$}
\psfrag{r4}{${\bf r}_4$}
\psfrag{s1}{$ s_1 $}
\psfrag{s2}{$ s_2 $} 
\psfrag{s3}{$ s_3 $}
\psfrag{s4}{$ s_4 $}
\psfrag{s5}{$ s_5 $}
\psfrag{s6}{$ s_6 $} 
\psfrag{s7}{$ s_7 $}
\psfrag{s8}{$ s_8 $}
\includegraphics[height=5cm]{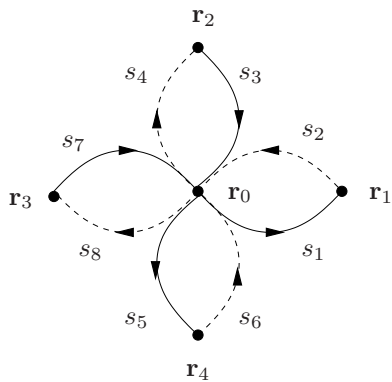}
\end{psfrags}
\caption{\label{fig1_echoflucs} Diagrammatic representation of the 
squared fidelity 
${\cal M}_{\rm L}^2(t)$. (Figure taken from Ref.~\cite{Pet05}. Copyright (2005) by the American Physical Society.)}
\end{figure}

We want to calculate ${\cal M}_{\rm L}^2$. Squaring Eq.~(\ref{moft}), we see
that it is given by
eight sums over classical paths and twelve spatial integrations. 
We use the same tricks as for the average fidelity, first that spatial integrations
over initial wavefunction coordinates can be brought to Gaussian form after
the corresponding integrands have been linearized around ${\bf r}_0$ --
eight of the twelve spatial integrals can be calculated in this way -- and second
to identify stationary phase solutions justified by our averaging over ${\bf r}_0$.
The starting point is
\begin{equation}\label{eq:m2}
{\cal M}_{\rm L}^2(t) = \int \prod_{j=1}^4 d{\bf r}_j
\sum_{s_i;i=1}^8 \;  \exp[i (\Phi^{H_0}-\Phi^{H}-\pi \Xi /2)] 
\; \prod_i C_{s_i}^{1/2} \left(\frac{\nu^2}{\pi}\right)^{d/4}
\exp(-\nu^2\delta{\bf p}_{s_i}^2/2).
\end{equation}
Here we introduced the staggered 
sum $\Xi=\sum_{i=0}^{3} (-1)^i (\mu_{s_{2i+1}} -
\mu_{s_{2i+2}} )$ of Maslov indices and the momentum
difference $\delta{\bf p}_{s_i}={\bf p}_{s_i}-{\bf p}_0$.
The right-hand side of Eq.~(\ref{eq:m2}) 
is schematically described in Fig.~\ref{fig1_echoflucs}, where, as before,
classical trajectories are represented
by a full line if they correspond to $H_0$
and a dashed line for $H$, with an arrow indicating the direction
of propagation. 
In the semiclassical limit $S_s \gg 1$ (we recall that actions are expressed
in units of $\hbar$), and upon average over $\psi_0$,
Eq.~(\ref{eq:m2}) is dominated by terms which satisfy
a stationary phase
condition, i.e. where the variation of the two differences of action phases
\begin{subequations}
\begin{eqnarray}
\Phi^{H_0}&=&S^{H_0}_{s_1}({\bf r}_1,{\bf r}_0;t)-
S^{H_0}_{s_3}({\bf r}_2,{\bf r}_0;t) +
S^{H_0}_{s_5}({\bf r}_4,{\bf r}_0;t)-
S^{H_0}_{s_7}({\bf r}_3,{\bf r}_0;t), \\
\Phi^{H}&=&
S^{H}_{s_2}({\bf r}_1,{\bf r}_0;t)\;-
S^{H}_{s_4}({\bf r}_2,{\bf r}_0;t)\; +
S^{H}_{s_6}({\bf r}_4,{\bf r}_0;t)\;-
S^{H}_{s_8}({\bf r}_3,{\bf r}_0;t),
\end{eqnarray}
\end{subequations}
has to be minimized. These stationary phase
terms are easily identified from the diagrammatic
representation as those where two classical trajectories $s$ and $s'$ 
of opposite direction of propagation
are {\em contracted}, i.e. $s \simeq s'$, up to a quantum resolution given
by the wavelength $\nu$. As mentioned above, contracting $s$ (generated by $H_0$)
with $s'$ (generated by $H=H_0 + \Sigma$) is
justified by the structural stability of hyperbolic systems
for not too large $\Sigma$~\cite{Kat96}.
Paths contractions are represented in
Fig.~\ref{fig2_echoflucs} by bringing two lines together in parallel.
Contracting either two dashed or two full lines allows for
an almost exact cancellation of the actions, hence an almost 
perturbation-independent contribution, up to a contribution arising
from the finite resolution $\nu$ with which the two paths overlap. 
However when a full line is
contracted with a dashed line, the resulting contribution still
depends on the action $\delta S_s = -\int_s \Sigma ({\bf q}(t),t)$
accumulated by the perturbation along the classical path $s$,
spatially parametrized as ${\bf q}(t)$.
Since we are interested in the variance $\sigma^2({\cal M}_{\rm L}) = 
\langle {\cal M}_{\rm L}^2 \rangle-\langle {\cal M}_{\rm L} \rangle^2$ (this is indicated by brackets
in Fig.~\ref{fig2_echoflucs}) we must subtract the nonconnected 
terms contained in $\langle {\cal M}_{\rm L}^2 \rangle$, i.e. those 
corresponding to independent contractions in each of the two
subsets $(s_1,s_2,s_3,s_4)$ and $(s_5,s_6,s_7,s_8)$. The result is that
all contributions to $\sigma^2({\cal M}_{\rm L})$ require pairing of spatial
coordinates, $|{\bf r}_i-{\bf r}_j| \le \nu$, for at least one
pair of indices $i,j=1,2,3,4$ -- in particular, this has the consequence that there is
no $ \exp[-2 \Gamma t]$-term.

\begin{figure*}
\begin{center}\nonumber 
\begin{psfrags}
\psfrag{ro}{${\bf r}_0$}
\psfrag{r1}{${\bf r}_1$}
\psfrag{r2}{${\bf r}_2$} 
\psfrag{r3}{${\bf r}_3$}
\psfrag{r4}{${\bf r}_4$} 
\psfrag{r13}{${\bf r}_1\! \simeq \!{\bf r}_3$}
\psfrag{r12}{${\bf r}_1\! \simeq\!{\bf r}_2$}
\psfrag{r14}{${\bf r}_1\! \simeq\!{\bf r}_4$}  
\psfrag{r24}{${\bf r}_2\! \simeq \!{\bf r}_4$}
\psfrag{SIGMA}{$\sigma^2({\cal M}_{\rm L})$}
\psfrag{LYAPDOUB}{$ { \alpha^2 e^{-2\lambda t} }$}
\psfrag{LYAPSINGLET}{${ 2 \alpha e^{-\lambda t}e^{-\Gamma t } }$}
\psfrag{SATSINGLET}{${ 2\hbar_{\rm eff}e^{-\Gamma t }\Theta(t-\tau_{\rm E}) }$}
\psfrag{SATSAT}{${ \hbar_{\rm eff}^2\Theta(t-\tau_{\rm E}) }$}
\includegraphics[height=3.cm]{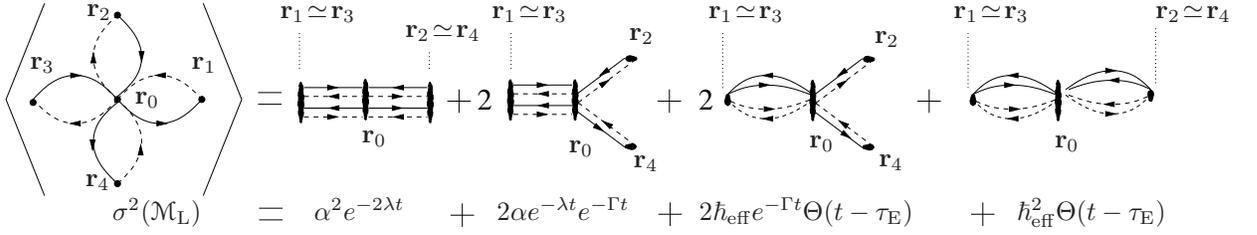}
\end{psfrags}
\end{center}
\vspace{-0.4cm}
\caption{\label{fig2_echoflucs} Diagrammatic representation of the 
averaged fidelity variance
$\sigma^2({\cal M}_{\rm L})$ and the three time-dependent contributions that 
dominate semiclassically, together with the contribution giving the
long-time saturation of $\sigma^2({\cal M}_{\rm L})$. There is no $\exp[-2 \Gamma t]$-term.
(Figure taken from Ref.~\cite{Pet05}. Copyright (2005) by the American Physical Society.)}
\end{figure*}

With these considerations, the four dominant contributions to 
the fidelity variance are depicted on the right-hand side of Fig.~\ref{fig2_echoflucs}. We
calculate them one by one in Appendix~\ref{appendix:semiclassics_flucs}.

The first one corresponds to $s_1=s_2 \simeq
s_7=s_8$ and $s_3=s_4 \simeq s_5=s_6$, which
requires ${\bf r}_1 \simeq {\bf r}_3$, ${\bf r}_2 \simeq {\bf r}_4$, it gives a contribution
\begin{subequations}\label{sigma1lambda}
\begin{eqnarray}
\sigma_1^2&=& \alpha^2 \exp[-2\lambda t], 
\end{eqnarray}
\end{subequations}
where $\alpha$ is the same as in Eq.~(\ref{eq:final_decay}).

The second dominant term is obtained from  
$s_1=s_2 \simeq s_7=s_8$, $s_3=s_4$ and $s_5=s_6$, with
${\bf r}_1 \simeq {\bf r}_3$, or equivalently $s_1=s_2$, $s_7=s_8$ and  
$s_3=s_4 \simeq s_5=s_6$ with ${\bf r}_2 \simeq {\bf r}_4$. This 
term comes therefore with a multiplicity of two, and one obtains
\begin{equation}\label{sigma2lg}
\sigma_2^2 \simeq 2 \alpha \exp[-\lambda t] \exp[-\Gamma t].
\end{equation}

The third and last dominant time-dependent term arises
from either $s_1=s_7$, $s_2=s_8$, $s_3=s_4$, $s_5=s_6$ and 
${\bf r}_1\simeq{\bf r}_3$, or $s_1=s_2$, $s_3=s_5$, $s_4=s_6$, $s_7=s_8$ and 
${\bf r}_2\simeq{\bf r}_4$. It thus also has a multiplicity of two. One gets,
\begin{eqnarray}\label{sigma3g}
\sigma^2_3 \simeq 2 \hbar_{\rm eff} \exp[-\Gamma t]
\Theta(t-\tau_{\rm E}).
\end{eqnarray}
This term exists only for 
times larger than the Ehrenfest time. For shorter times,
$t<\tau_{\rm E}$, the third diagram on the right-hand side 
of Fig.~\ref{fig2_echoflucs} 
goes into the second one, and the corresponding contributions is included
in $\sigma_2^2$. It emerges at larger times and renders hopeless to 
witness the Lyapunov decay after $\tau_{\rm E}$.

Subdominant terms are obtained by
higher-order contractions, for instance setting ${\bf r}_2 \simeq {\bf r}_4$
in the second and third graphs on the right hand-side
of Fig.\ref{fig2_echoflucs}. They either decay faster, or are of higher
order in $\hbar_{\rm eff}$, or both. We only discuss the
term which gives the dominant long-time saturation at the ergodic value
$\sigma^2({\cal M}_{\rm L}) \simeq \hbar_{\rm eff}^{2}$, and refer the
reader to Ref.~\cite{Pet07b} for a detailed calculation of subdominant
terms. For 
$t>\tau_{\rm E}$, there is a phase-free, hence time-independent
contribution with four different paths, resulting
from the contraction $s_1=s_7$, $s_2=s_8$, $s_3=s_5$, $s_4=s_6$,
and ${\bf r}_1 \simeq {\bf r}_3$, ${\bf r}_2 \simeq {\bf r}_4$.
Its contribution is sketched as the fourth diagram on the
right-hand side of Fig.~\ref{fig2_echoflucs}. It gives
\begin{eqnarray}\label{eq:satsat} 
\sigma^2_4 &= &  \left(\frac{\nu^2}{\pi}\right)^{2d} 
\Big\langle\int d{\bf r}_1d{\bf r}_3 \sum C_{s_1}C_{s_2} \;
\exp[-\nu^2 (\delta{\bf p}_{s_1}^2  +\delta{\bf p}_{s_2}^2 )]  
\Theta(\nu-|{\bf r}_1-{\bf r}_3|)\Big\rangle^2. 
\end{eqnarray}
From the sum rule of Eq.~(\ref{sumrule1}), and again invoking the
long-time ergodicity of the semiclassical dynamics, Eq.~(\ref{saturation}), 
one obtains the long-time saturation of $\sigma^2({\cal M}_{\rm L})$,
\begin{equation}\label{eq:saturation}
\sigma^2_4 = \hbar_{\rm eff}^2 \Theta(t-\tau_{\rm E}).
\end{equation}
Note that for $t<\tau_{\rm E}$, this contribution does not exist by itself and is 
included in $\sigma_1^2$, Eq.~(\ref{sigma1lambda}).

According to our semiclassical approach,
the fidelity has a variance given to leading order by the sum of the
four terms of 
Eqs.~(\ref{sigma1lambda}), (\ref{sigma2lg}), (\ref{sigma3g}) and
(\ref{eq:saturation})
\begin{equation}\label{sigmascl}
\sigma^2_{\rm sc}({\cal M}_{\rm L}) = \alpha^2 \exp[-2 \lambda t] + 
2 \alpha \exp[-(\lambda + \Gamma) t] + 2 \hbar_{\rm eff} \exp[-\Gamma t] \Theta(t-\tau_{\rm E})
+\hbar_{\rm eff}^2 \Theta(t-\tau_{\rm E}). 
\end{equation}
We see that 
for short enough times -- before ergodicity sets in and the saturation
of ${\cal M}_{\rm L}(t) \simeq \hbar_{\rm eff}$ and $\sigma^2({\cal M}_{\rm L}) \simeq \hbar_{\rm eff}^{2}$
is reached --  the first term on the right-hand
side of (\ref{sigmascl}) dominates as long as $\lambda < \Gamma$.
For $\lambda > \Gamma$ on the other hand, $\sigma^2({\cal M}_{\rm L})$ exhibits
a behavior $\propto \exp[-(\lambda+\Gamma) t]$ for $t<\tau_{\rm E}$, turning
into $\propto \hbar_{\rm eff} \exp[-\Gamma t]$ for $t>\tau_{\rm E}$. 
Thus, in contrast to the average Loschmidt echo, its variance allows to
extract the Lyapunov exponent from the second term on the right-hand
side of Eq.~(\ref{sigmascl}) even when $\lambda > \Gamma$. Also
one sees that, unlike the strong perturbation regime $\Gamma \gg B$
\cite{Sil03}, ${\cal M}_{\rm L}$ continues
to fluctuate above the residual variance $\simeq \hbar_{\rm eff}^2$ up to
a time $ \simeq \Gamma^{-1} |\ln \hbar_{\rm eff}|$
in the semiclassical regime $B>\Gamma>\Delta$. For $\Gamma \ll \lambda$,
$\Gamma^{-1} |\ln \hbar_{\rm eff}| \gg \tau_{\rm E}$ and ${\cal M}_{\rm L}$ fluctuates 
beyond $\tau_{\rm E}$.

The above semiclassical approach breaks down at short
times for which not enough phase is accumulated to motivate
a stationary phase approximation. This time is very short, 
of the order of the inverse
energy of the particle, i.e. $O(\hbar_{\rm eff}^a)$, where $a \ge 0$ depends
on the system dimension and the energy-momentum relation. For 
$E \propto p^2$ and in two dimensions, one has $a=1$.
The short-time behavior of $\sigma^2({\cal M}_{\rm L})$ can instead be 
calculated using
a RMT-based perturbative approach, which we present in the next chapter.

In principle, the fluctuations of the Loschmidt echo in 
regular systems can also be calculated semiclassically. However, 
compared to the average echo, fluctuations contain higher order correlations,
and the already daring assumptions we made when calculating the average echo
for regular systems
become even
much riskier for the fluctuations. Therefore we here only mention that 
blindly applying the approach presented in Chapter~\ref{subsub_average}
replaces Eq.~(\ref{sigma1lambda}) with $\sigma_1^2 \propto t^{-a}$, 
$a=2d$ or $3d$, depending on the relation between the correlator
(\ref{correl}) and $\nu^2$ [this relation evolves in time from $a=2d$ at short times
to $a=3d$ at longer times, see the discussion below Eq.(\ref{diagdecay})].
This term then dominates the total fluctuations.
While it is quite realistic to expect the survival of larger
fluctuations for longer times in regular systems, this result should obviously be taken with a 
rather big grain of salt. It is actually expectable
that in regular systems, fluctuations are dominated by exceptional
events that are hard to capture with our statistical approach. 

What have we learned from this calculation of the mesoscopic fluctuations of the
Loschmidt echo ? Saddly enough, not much. It seems that, except at short times,
the behavior of the fluctuations are merely reproducing the behavior of the
average -- there is no novel regime that does not exist for the average, no new
physics emerging from fluctuations that is not present in the average. On the positive side,
we see that fluctuations are not large in chaotic systems, thus the average echo is representative
of individual events in chaotic systems.


\subsubsection{Afterthoughts on the semiclassical approach}
\label{afterthoughts}

The two semiclassical time-dependent contributions to the Loschmidt echo, 
Eqs.~(\ref{eq:nondiagdecay}) and (\ref{diagdecay}) 
are diagonal contributions -- they both
follow from setting $s_1 \simeq s_2$ in Eq.(\ref{moft}). 
The ``superdiagonal'' contribution, giving the Lyapunov decay, 
is clearly classical in nature -- its origin can be traced back to the
asymptotic behavior $C_s \propto \exp[-\lambda t]$ of the determinant of the stability
matrix, whose elements are given by derivatives of classical actions along a single
trajectory as a function of the initial and final position of that trajectory. There is no
quantumness in that object. Yet, the authors of Ref.~\cite{Cuc02a} present numerically obtained
exponential decays of ${\cal M}_{\rm L}(t)$ with the Lyapunov exponent which, they claim, goes on beyond
the Ehrenfest time. If this were truly the case, that would invalidate our argument about
the classicality of the Lyapunov decay. It is however important to identify
the relevant classical length which goes into the definition
of the Ehrenfest time.
An example is given by the spectrum of Andreev billiards -- ballistic billiards in partial contact with
a superconductor -- where the relevant Ehrenfest time scale $\tau_{\rm E} = - \lambda^{-1}
\ln[ \hbar_{\rm eff} \tau_{\rm D}^2 ]$ differs from the standard definition by a logarithmic
correction in the average return time $\tau_{\rm D}$ of a quasiparticle to the 
superconductor~\cite{Vav03,Sch05}. The point is however that different definitions of $\tau_{\rm E}$
differ only by a classical quantity. In the case of the Lorentz gas investigated
in Ref.~\cite{Cuc02a}, there are two
different $\tau_{\rm E}$ that can be defined, depending on whether one compares the
wavelength of the particle with the size $\zeta$ of the scatterers or the system size. The Lyapunov decay
observed in Ref.~\cite{Cuc02a} stops at the Ehrenfest time defined with the system size. The fact that the Lyapunov decay extends a bit 
beyond $\lambda^{-1} \ln \zeta/\nu$ is of marginal importance and does not invalidate our conclusion that the Lyapunov decay is classical in nature.

In recent years, semiclassics
has achieved a degree of sophistication which allows to calculate contributions beyond
the diagonal approximation~\cite{Mul04,Heu06,Jac06,Whi06,Rah05,Rah06,Ric02,Sie01,Sie02,Pet08,Whi08}, and one 
might wonder if these weak-localization corrections
would sensitively affect the decay of ${\cal M}_{\rm L}$. A direct calculation of these corrections
in the context of the Loschmidt echo has not been performed to this day, however we will
argue below, in the context of RMT, that these corrections are subdominant, in that they give
${\cal O}(N^{-1})$ corrections at $t=0$ and decay exponentially with time at a rate given
by $\Gamma$. Still, it would be interesting to find out if a weak localization to ${\cal M}_{\rm L}$ exists
with a Lyapunov dependence.

To close this chapter on the semiclassical approach to the Loschmidt echo, let us briefly
discuss Vani\v{c}ek's elegant {\it dephasing representation} approach~\cite{Van04,Van04a}. It rewrites
the fidelity amplitude $f(t)$ (with ${\cal M}_{\rm L}(t) = |f(t)|^2$) as
\begin{equation}
f(t) = \int {\rm d}{\bf q} {\rm d}{\bf p} \, W_{\psi_0} ({\bf q},{\bf p}) \, \exp[i \Delta S({\bf q},{\bf p};t)],
\end{equation}
in terms of the Wigner function $W_{\psi_0} ({\bf q},{\bf p})$ of the initial state,
and the action difference $\Delta S({\bf q},{\bf p};t)$ due solely to the perturbation acting
on the classical trajectory of duration $t$ starting at $({\bf q},{\bf p})$. Compared to the theory
we just presented, this treatment is perhaps 
more elegant in that it treats in a unified
way golden rule and Lyapunov decays, without the need to split the calculation into
diagonal and nondiagonal contributions. This new approach, so far, has only confirmed
what was already known from earlier semiclassical theories as presented above. Its application
to specific problems in, e.g., chemical reactions looks very promising, however.

\subsection{Random matrix theory of the Loschmidt echo}\label{section:ML_RMT}

We next calculate ${\cal M}_{\rm L}$ under the assumption that both $H_0$ and $H$
are quantum chaotic Hamiltonians that display RMT eigenvector
component statistics. To be more specific, we assume that the complex coefficients
of the expansion of $\psi_0$ over the eigenbasis of $H_0$ and $H$, 
\begin{eqnarray}
| \psi_0 \rangle = \sum_{\alpha=1}^N \langle \alpha^{(0)} | \psi_0 \rangle \; |\alpha^{(0)} \rangle , \;\;\;\;\;\;\;\;\;
| \psi_0 \rangle = \sum_{\alpha=1}^N \langle \alpha | \psi_0 \rangle \; |\alpha \rangle,
\end{eqnarray}
satisfy, to leading order in $N^{-1}$, the inverse number of basis 
states~\cite{Ber77b,Pri94,Pri95,Guh98,Mir00}
\begin{subequations}\label{contractions}
\begin{eqnarray}
\overline{\langle  \alpha^{(0)} | \psi_0 \rangle} &=&
\overline{\langle  \alpha | \psi_0 \rangle} = 0, \\
\overline{\langle  \alpha^{(0)} | \psi_0 \rangle \langle \psi_0 | \beta^{(0)} \rangle}  &=& 
\overline{\langle  \alpha | \psi_0 \rangle \langle \psi_0 | \beta \rangle}  = N^{-1} \; \delta_{\alpha,\beta}, \\
\label{4contractions} \overline{\langle  \alpha^{(0)} | \psi_0 \rangle \langle \psi_0 | \beta^{(0)} \rangle
\langle  \gamma^{(0)} | \psi_0 \rangle \langle \psi_0 | \delta^{(0)} \rangle}  &=& 
\overline{\langle  \alpha | \psi_0 \rangle \langle \psi_0 | \beta \rangle
\langle  \gamma| \psi_0 \rangle \langle \psi_0 | \delta \rangle} \\
&=&  N^{-2} 
[\delta_{\alpha,\beta} \delta_{\gamma,\delta} + \delta_{\alpha,\delta} \delta_{\beta,\gamma}
] \nonumber .
\end{eqnarray}
\end{subequations}
The bars denote averages taken over an ensemble of random Hamiltonians (up to constraints
of hermiticity and time-reversal or spin rotational symmetry~\cite{Meh91}) and the above equations
hold for generic choices of $\psi_0$, in particular excluding cases where $\psi_0$ is an
eigenstate of the Hamiltonian under consideration. 
Note that in Eq.~(\ref{4contractions}), 
we neglected the contraction $\delta_{\alpha,\gamma} \delta_{\beta,\delta}$
which exists only in time-reversal symmetric systems and leads to a subdominant weak localization
correction $\propto \exp[-\Gamma t]/N$.
The RMT approach to the Loschmidt echo was first mentioned, but not described 
in Refs.~\cite{Jac01}. More details were given later on in 
Refs.~\cite{Gor04a,Cuc02b,Hil06,Cer03}. Refs.~\cite{Sto04,Sto05} calculated 
the average fidelity amplitude
using supersymmetric methods~\cite{Efe97,Haa01}, which proved to agree 
extremely well with 
numerics on random matrices remarkably
accurately. Most spectacularly, a partial recovery of the fidelity amplitude
at the Heisenberg time $\tau_{\rm H}$
was emphasized. It is unclear how much of these findings affect the fidelity itself.
Here we sketch the so far unpublished approach that led to the 
results
presented in Refs.~\cite{Jac01}.

\subsubsection{Ensemble average -- leading order}

Our strategy in the RMT calculation of the Loschmidt echo is to insert the
resolutions of the identity
\begin{eqnarray}
I & = & \sum_{\alpha=1}^N |\alpha^{(0)} \rangle \langle \alpha^{(0)} |
= \sum_{\alpha=1}^N |\alpha \rangle \langle \alpha |
\end{eqnarray}
into Eq.~(\ref{eq:def_LE}).
With Eqs.~(\ref{contractions}), 
the average Loschmidt echo (and its variance, see below) then
depend on the projections
of the eigenstates of $H_0$ over the eigenbasis of $H$. The dominant term 
is
\begin{eqnarray}\label{eq:echo_estates}
{\cal M}_{\rm L}(t) &=& \left[\frac{1}{N}
\sum_{\alpha,\beta} \overline{|\langle \alpha | \beta^{(0)} \rangle|^2} \, \exp[ i(E_\alpha - E_\beta^{(0)}) t] \right]^2,
\end{eqnarray}
with the eigenvalues $E_\beta^{(0)}$ and $E_\alpha$ of $H_0$ and $H$
respectively.
It is seen that RMT relates 
the fidelity to the local spectral density of 
states, 
\begin{equation}
\rho_{\rm ldos}(E) = \left\langle \sum_\alpha |\langle \alpha | \beta^{(0)} \rangle|^2 \delta(E+E_\beta^{(0)}-E_\alpha) \right\rangle_{\beta^{(0)}},
\end{equation}
a relationship which, it seems, cannot capture the
Lyapunov decay~\cite{Jac01,Coh02}.
Three regimes of perturbation are
differentiated with the level spacing $\delta$, the golden rule spreading $\Gamma= 2 \pi \overline{|\langle \alpha^{(0)}|\Sigma|\beta^{(0)} \rangle|^2}/\delta$ and the bandwidth $B$~\cite{Wig57,Jac95,Fra95,Fyo95,Jac01,Wis02b}. They are
\bea\label{eq:1part_spread}
\overline{|\langle \alpha | \beta^{(0)} \rangle|^2}
& = & \left\{
\begin{array}{cc}
\delta_{\alpha,\beta}, & \Gamma < \delta,  \\
(\Gamma \delta / 2 \pi) \big/[(E_\alpha-E_\beta^{(0)})^2+\Gamma^2/4], & \;\; \delta \lesssim \Gamma \ll B, \\
N^{-1}, & \Gamma > B,
\end{array}
\right.
\eea
\noindent From these expression and Eq.~(\ref{eq:echo_estates}) one obtains
the three asymptotic decays of the average Loschmidt echo, to leading order
\begin{eqnarray}\label{eq:echo_leading}
{\cal M}_{\rm L}(t) = \left\{
\begin{array}{cc}
\exp[-\sigma_1 t^2], & \Gamma < \delta, {\rm regime \,(I)}, \\
\exp[-\Gamma t], & \;\;\;\;\; \delta \lesssim \Gamma \ll B, {\rm regime \, (II)}, \\
\exp[-B^2 t^2], & \;\;\; \Gamma > B, {\rm regime \, (III)},
\end{array}
\right.
\end{eqnarray}
with the RMT result $\sigma_1^2 \equiv {\rm Tr} \Sigma^2/N$. The contractions
in Eq.~(\ref{contractions}) also give us the long-time saturation 
\begin{equation}\label{eq:saturationrmt}
{\cal M}_{\rm L}(\infty) = N^{-1}.
\end{equation}
The equivalence between semiclassics and RMT in the golden rule regime is 
achieved
assuming that RMT corresponds to a chaotic system with infinite Lyapunov 
exponent, and thus vanishingly small Ehrenfest time.

Let us also note that the short-time parabolic 
decay ${\cal M}_{\rm L}(t) = 1-\sigma_0^2 t^2$, with the RMT average
$\sigma_0^2=\sigma_1^2$, is equally easily obtained after the 
time-evolution exponentials
are Taylor expanded to second order, $\exp[\pm iH_{0}t]=1 \pm i H_{0}t - H_{0}^2 t^2 /2
+ O(H_{0}^3 t^3)$. Finally, assuming a normalized perturbation operators $\Sigma$
 with a spectrum of eigenvalues in the interval $[-\epsilon,\epsilon]$ -- this requires a
 scaling of its matrix elements as $\Sigma_{ij} \sim \sqrt{\hbar_{\rm eff}}$ -- RMT
 gives the parametric estimates 
 \begin{eqnarray}
 \Gamma \sim \epsilon^2 \big/ B \hbar_{\rm eff}^{2} \, , \\
 \sigma_{0,1}^2 \sim \epsilon^2 \big/ \hbar_{\rm eff}^{2} \, ,
 \end{eqnarray}
 and accordingly, the condition for the golden rule regime, $\delta \lesssim \Gamma \ll B$
translates into $B \hbar_{\rm eff}^{1/2} \le 
 \epsilon \big/ \hbar_{\rm eff} \ll B$. This range is parametrically large in the semiclassical
 parameter $\hbar_{\rm eff} \ll 1$. Still it requires a vanishing $\Gamma \ll B \hbar_{\rm eff}$
 which legitimates to invoke shadowing when constructing a semiclassical theory.
  
It is interesting to note that the RMT
contractions leading to the dominant decay terms, Eqs.~(\ref{eq:echo_leading}),
is in direct correspondence with the first diagonal approximation $s_1=s_2$
done in the semiclassical approximation to obtain Eq.~(\ref{semicl}).
What do we mean by that ? Semiclassically, one writes the fidelity amplitude
as
\bea\label{eq:2.31}
\langle \psi_0 \vert 
e^{i H t} e^{-i H_0 t}
\vert  \psi_0 \rangle &=& \int{\rm d}{\bf r}
{\rm d}{\bf r}_0' {\rm d}{\bf r}_0''
\sum_{s_1,s_2}  K_{s_1}^{H_0}({\bf r},{\bf r}_0';t) 
[K_{s_2}^{H}({\bf r},{\bf r}_0'';t)]^*
\langle {\bf r}_0' | \psi_0 \rangle \, \langle \psi_0 | {\bf r}_0''\rangle .
\eea
Invoking next the narrowness of the initial state $\psi_0$ and enforcing a
stationary phase condition leads to ${\bf r}_0'={\bf r}_0''$ and
$s_1=s_2$. RMT on the other hand expresses the fidelity amplitude as
\bea\label{eq:2.32}
\langle \psi_0 \vert 
e^{i H t} e^{-i H_0 t}
\vert  \psi_0 \rangle &=& \sum_{\alpha,\beta,\gamma}
\langle \beta^{(0)}| e^{i H t} |\gamma \rangle \langle \gamma | e^{-i H_0 t}
| \alpha^{(0)} \rangle \,
\langle \alpha^{(0)} | \psi_0 \rangle \, \langle \psi_0 | \beta^{(0)} \rangle .
\eea
Similarly to setting ${\bf r}_0'={\bf r}_0''$ and pairing the trajectories in Eq.~(\ref{eq:2.31}),
Eq.~(\ref{eq:2.32}) requires to set $\alpha^{(0)} = \beta^{(0)}$.
No further pairing of trajectories, nor contractions are required to
obtain the golden rule decay.
Similarly, the long-time saturation term is 
obtained within RMT by contractions similar to the
trajectory pairing giving Eq.~(\ref{longtimesaturationLEa}).

\subsubsection{A quick and incomplete remark on weak localization}

Eq.~(\ref{4contractions}) generates subdominant terms which 
exist only in presence of time-reversal symmetry. These are usually called weak localization 
corrections, in analogy with coherent corrections 
to electronic transport~\cite{Akk07,Imr02}. 
The calculation of these terms proceeds along the same
lines as for the leading order contribution to ${\cal M}_{\rm L}$, and it is seen that they lead to
initially ($t=0$) subdominant contributions of order ${\cal O}(N^{-1})$, furthermore having 
an exponential (golden rule regime) or Gaussian (strong perturbation regime)
time-dependent decay. These corrections are only marginally relevant at best and it is
doubtful that they can be observed numerically, mostly because
the prefactor in front of the golden rule
decay ${\cal M}_{\rm L} \propto \exp[-\Gamma t]$ is determined by the
initial transient and is therefore system-dependent. In our opinion, there
is unfortunately
no way one can unambiguously observe these weak localization corrections.

Weak localization corrections have yet
to be calculated using semiclassics, and it is
therefore unclear at this time whether they exhibit a $\lambda$-dependence 
or not in regime (II). Strictly speaking, there is no weak localization 
correction 
in the perturbative regime (I), in the sense that no additional
term exists in presence of time-reversal symmetry that disappears when 
this symmetry
is broken. Note, however, that 
$\sigma_1^2$ itself depends on the eigenfunctions
of $H_0$.
We finally note that 
there is no weak localization correction for the initial parabolic transient
either, as the average decay rate $\sigma_0$ does not directly depend on $H_0$.

\subsubsection{Mesoscopic fluctuations}

The variance $\sigma^2({\cal M}_{\rm L})$ can also be calculated using the RMT approach
just used for the average Loschmidt echo. 
In the golden rule regime
(II), the semiclassical result of Eq.~(\ref{sigmascl}) is replaced by
\begin{equation}\label{sigmarmt}
\sigma^2_{\rm RMT} =  \frac{2}{N} e^{-\Gamma t} + \frac{1}{N^2}.
\end{equation}
These two terms correspond to the two $\lambda$-independent terms in the semiclassical variance
of Eq.~(\ref{sigmascl}), once again illustrating the one-to-one correspondence between semiclassics
at infinite Lyapunov exponent and RMT in the golden rule regime. RMT allows to explore the
short-time regime preceding the semiclassically reachable regime, and
to get the short-time behavior of
$\sigma^2({\cal M}_{\rm L})$, we
Taylor expand the time-evolution exponentials 
$\exp[\pm iH_{(0)}t]=1 \pm i H_{(0)}t - H_{(0)}^2 t^2 /2 + ...
+ O(H_{(0)}^5 t^5)$. The resulting expression 
for $\sigma^2({\cal M}_{\rm L})$ contains matrix elements such
as $\langle \psi_0 | H_{0}^a | \psi_0 \rangle$, $a=1,2,3,4$, whose mesoscopic average
are evaluated using Eqs.(\ref{contractions}) and their generalization
up to the product of eight coefficients $\langle \psi_0 | \alpha^{(0)} \rangle$~\cite{Mir00}.
Keeping non-vanishing terms of lowest order in $t$,
one has a quartic onset 
$\sigma^2({\cal M}_{\rm L}) \simeq (\overline{\sigma_0^4}-\overline{\sigma_0^2}^2)t^4$ for
$t \ll \sigma_0^{-1}$, with $\sigma_0 \equiv 
[(\langle\psi_0|\Sigma^2|\psi_0\rangle
-\langle\psi_0|\Sigma|\psi_0\rangle^2)]^{1/2} $. RMT gives
$(\overline{\sigma_0^4}-\overline{\sigma_0^2}^2) \propto (\Gamma B)^2$,
with a prefactor of order one.
From this and Eq.~(\ref{sigmascl}) one concludes that
$\sigma^2({\cal M}_{\rm L})$ has a nonmonotonous behavior, i.e. it first rises
at short times, until it decays after a time $t_c$ which one can evaluate
by solving $\sigma^2_{\rm sc}(t_c) = (\Gamma B)^2 t_c^4$.
In the regime $B>\Gamma >\lambda$ one gets
\begin{subequations}\label{tc}
\begin{eqnarray}
t_c&=&\left(\frac{\alpha_0}{\Gamma B}\right)^{1/2+d}
\Bigg[1-\lambda \left(\frac{\alpha_0}{\Gamma B}\right)^{1/2+d}\frac{1}{2+d}
+O\left(\lambda^2
\left\{\frac{\alpha_0}{\Gamma B}\right\}^{2/2+d} \right) \Bigg], \\
\label{sigmatc}
\sigma^2(t_c) &\simeq& (\Gamma B)^2
\left(\frac{\alpha_0}{\Gamma B}\right)^{4/2+d} \Bigg[
1-\frac{4 \lambda}{2+d}  
\left(\frac{\alpha_0}{\Gamma B}\right)^{1/2+d} +O\left(\lambda^2
\left\{\frac{\alpha_0}{\Gamma B}\right\}^{2/2+d} \right)  \Bigg].
\end{eqnarray}
\end{subequations}
Here, we explicitly took the $t$-dependence  
$\alpha(t) = \alpha_0 t^{-d}$ into account [see Eq.~(\ref{sigma1lambda})]. We further estimate
$\alpha_0 \propto (\Gamma \lambda)^{-d/2}$
by setting the Lyapunov time
equal to few times the time of flight through a correlation
length of the perturbation potential. This is generically the case for
simple dynamical systems such as billiards or maps. We then obtain 
$\sigma^2(t_c) \propto (B/\lambda)^{2d/2+d} \gg 1$. Because
$0 \le {\cal M}_{\rm L}(t) \le 1$, this value is
however bounded by ${\cal M}_{\rm L}^2(t_c)$. Since in the other regime
$\Gamma \ll \lambda$, one has $\sigma^2(t_c) \simeq
2 \hbar_{\rm eff} [1-(2 \hbar_{\rm eff})^{1/4} \sqrt{\Gamma/B}]$
we predict that $\sigma^2(t_c)$ grows during the crossover from
$\Gamma \ll \lambda$ to $\Gamma > \lambda$, until it saturates at a
non-self-averaging value, $\sigma(t_c)/{\cal M}_{\rm L}(t_c) \approx 1$,
independent of $\hbar_{\rm eff}$ and $B$, with possibly a weak
dependence on $\Gamma$ and $\lambda$. 

These considerations conclude our analytical
calculation of the Loschmidt echo, its average and fluctuations. Our findings extend the
standard universality connecting RMT and semiclassics in chaotic systems. This relation is
somehow less trivial in the time domain considered here, where the Lyapunov exponent
enters the game. It is largely absent of spectral correlations, where the equivalence of the
two approaches only requires to have chaos, i.e a positive Lyapunov exponent, independently
of its precise values -- important
time scales include the period of the shortest periodic orbit of the Heisenberg time which are not
related to the Lyapunov time in any way. Here, we have seen that the equivalence between RMT
and semiclassics is only complete when $\lambda \rightarrow \infty$. When this is not the case,
still with $\lambda > 0$, details of the spatial dynamics that are absent of RMT sometimes influence
the decay of the fidelity, leading in particular to its Lyapunov decay. 

\subsection{Lyapunov exponent, what Lyapunov exponent ?}
\label{section:lyapunov?}

Our calculation show that the Lyapunov exponent in the time-evolution 
of the fidelity emerges from  the determinant $C_s$ of the stability matrix, which has
the asymptotic form $C_s \propto \exp[-\lambda t]$. 
Physically, this stability can be related to the decaying
overlaps of slightly displaced wavepackets.
The Lyapunov exponent 
is, rigorously speaking, defined as a long-time limit of the local exponential 
stretching due to the chaotic dynamics~\cite{Lic92}, and the above asymptotic form is
valid only at large times. The numerical
experiments we are about to present, on the other hand, show a Lyapunov decay of 
the Loschmidt echo for rather short times. 
One might thus wonder what really is the observed Lyapunov exponent, and whether it
really is connected with the system's true, mathematically defined Lyapunov exponent.

Classically, the answer would be to invoke the ergodicity
of chaotic systems in order to replace the long-time average one
takes when numerically determining the Lyapunov exponent
(see Ref.~\cite{Ben76}) with a spatial average over a set of homogeneously
distributed phase-space initial conditions. This is actually what we do in
our numerical investigations of the Loschmidt echo -- the average 
${\cal M}_{\rm L}$ is calculated over an ensemble of initial states
$\psi_0$. For initial Gaussian wavepackets, this ensemble corresponds
classically to taking different initial conditions in phase-space.
From this line of reasoning, one concludes that, in the appropriate
regime, ${\cal M}_{\rm L}(t) \propto \exp[-\lambda t]$ with the true
classical Lyapunov exponent. 

This is not the full story, however, since averaging over  
different $\psi_0$ averages $\langle C_s \rangle \propto \langle
\exp[-\lambda t] \rangle \ne \exp[-\langle \lambda \rangle t]$~\cite{Sil03}, 
so that the observed
Lyapunov decay is sensitive to spatial and/or time variations of 
the ``finite-time'' Lyapunov exponent~\cite{Sch02,Sil06}. We show below in
several instances that ${\cal M}_{\rm L}$ often decays with a rate
smaller than the true classical Lyapunov exponent,
${\cal M}_{\rm L}(t) \propto \exp[-\lambda_0 t]$,
$\lambda_0 < \lambda$. But then how do we know that we are truly
witnessing the predicted Lyapunov decay ? 
First, because the decay is exponential
and is perturbation-independent --
cranking up the strength of the perturbation leaves the decay slope
unchanged. Second, because, as the chaoticity of the problem changes, so
does the slope of the decay -- changing the true Lyapunov exponent also
changes the decay rate $\lambda_0$ of the Loschmidt echo in such a way
that $d \lambda_0/d \lambda > 0$ and there is a one-to-one monotonous correspondence
between $\lambda$ and $\lambda_0$. Third, because the decay
disappears if one considers classically meaningless initial states --
such as random states -- and that if one takes coherent superpositions of
$M$ Gaussian wavepackets as initial states, the decay becomes $M^{-1} \exp[-\lambda_0 t]$.
We believe that these are three minimal conditions to be satisfied before
one concludes that the Lyapunov decay of the Loschmidt echo has been
observed.
These three behaviors are checked at one point or another in the numerical
simulations we are about to present.

\subsection{Numerics -- The Loschmidt echo in quantum maps}

We present numerical checks of our theories, obtained 
from two different dynamical systems, the kicked top, which we
use to check our results on the average Loschmidt echo, and the kicked 
rotator, with which
we investigate the properties of $\sigma^2({\cal M}_{\rm L})$. 
Most of the data 
to be presented are extracted 
from Refs.~\cite{Jac01,Jac03,Pet05}. Several other dynamical systems 
have been numerically experimented in the literature, among them 
billiards~\cite{Wis02a,Wis03}
and Lorentz gases~\cite{Cuc02a}, and it has been found that 
${\cal M}_{\rm L}$
exhibits the same behavior as for the maps discussed here. Maps however present the advantages
of being easily tunable from regular to fully chaotic -- this is impossible for billiards, nor for the
Lorentz gas -- while allowing for large Hilbert spaces, thus
small effective Planck's constant, and rather short computation times. 

\subsubsection{Ensemble-averaged fidelity}\label{kicked_top}

In this paragraph we present numerical confirmation of our semiclassical and random matrix
theories for the fidelity ${\cal M}_{\rm L}(t)$ averaged over ensembles of initial states $\psi_0$. 
To this end, we use the kicked top model, and
the numerical procedure is succintly described in Appendix~\ref{appendix:ktop}. For more
details on the kicked top, we refer the reader to Refs.~\cite{Haa87,Haa01}

We first numerically extracted the dependence of the Lyapunov exponent
$\lambda$ on $K$ using the method of Benettin et al.~\cite{Ben76}. We do this because,
first, we want to know whether we are in the chaotic regime or not -- for the kicked top, there
is a crossover between regular and chaotic behavior driven by the kicking parameter $K$ -- 
and second, because we need to know $\lambda$ with enough accuracy
if we want to give full numerical
confirmation to the predicted decay ${\cal M}_{\rm L}(t) \propto \exp[-\lambda t]$.
Our results 
are plotted in the inset to the top left panel of Fig.~\ref{fig:fig1_echo1}.
The error bars reflect the spread in $\lambda$ in different regions of phase
space. In particular the presence of islands of stability at low values of $K$ for which
the dynamics is mixed results in much larger fluctuations of $\lambda$,
i.e. much larger error bars.
For $K\gtrsim 9$ the error bars vanish, which reflects the fact that the system becomes
fully chaotic.

\begin{figure}
\includegraphics[width=7.5cm,angle=0]{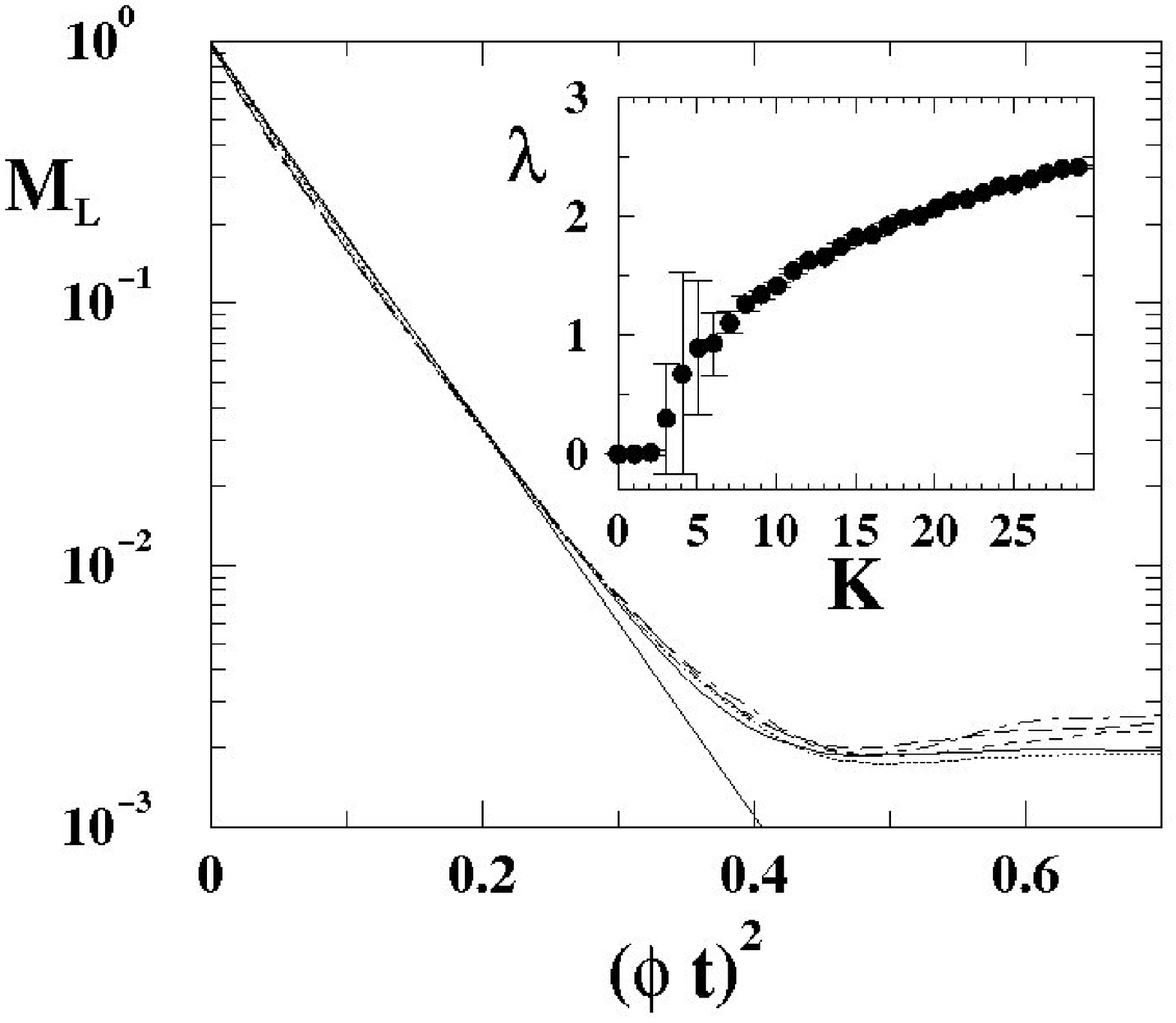}
\includegraphics[width=7.5cm,angle=0]{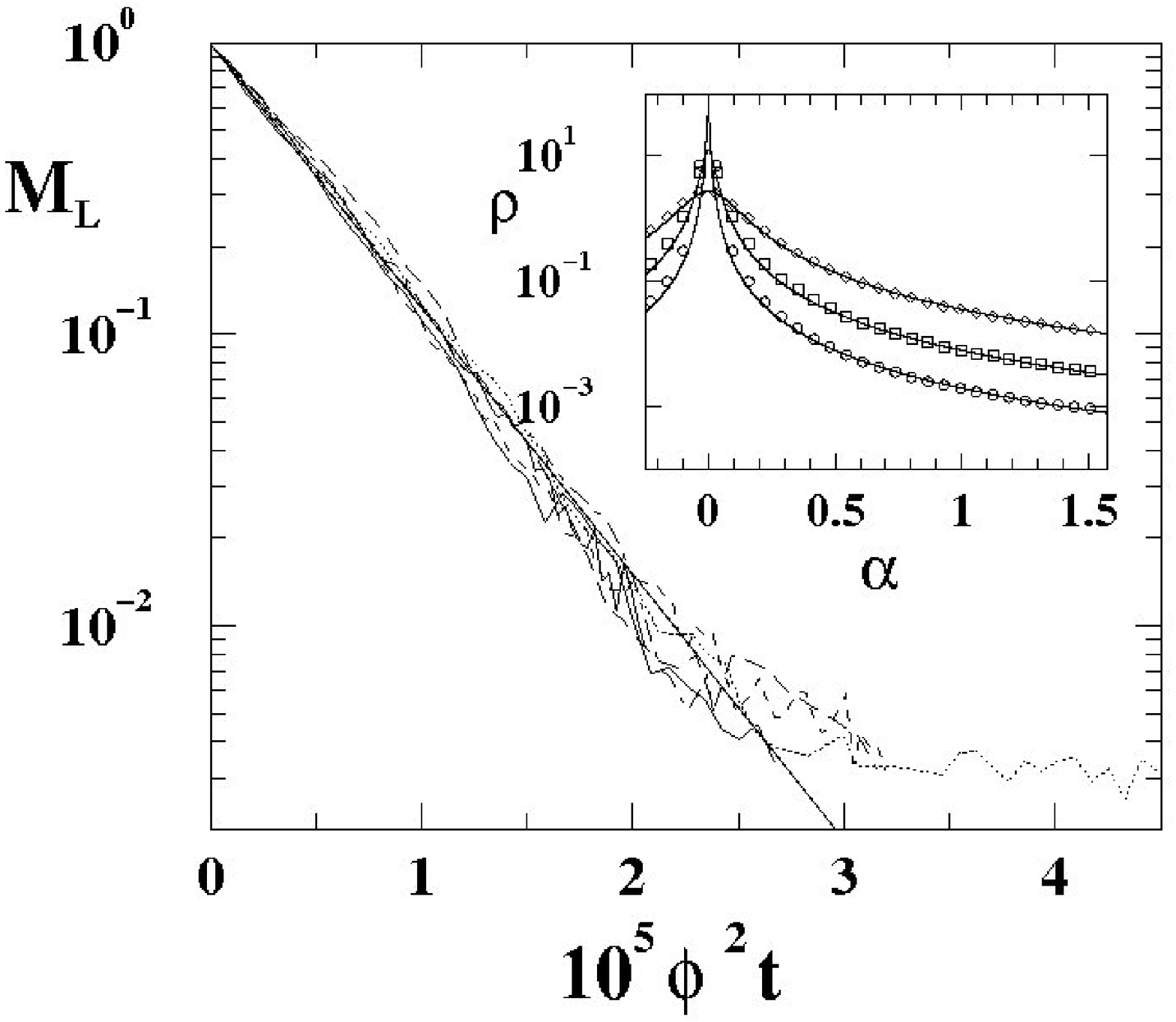} \\
\includegraphics[width=7.5cm,angle=0]{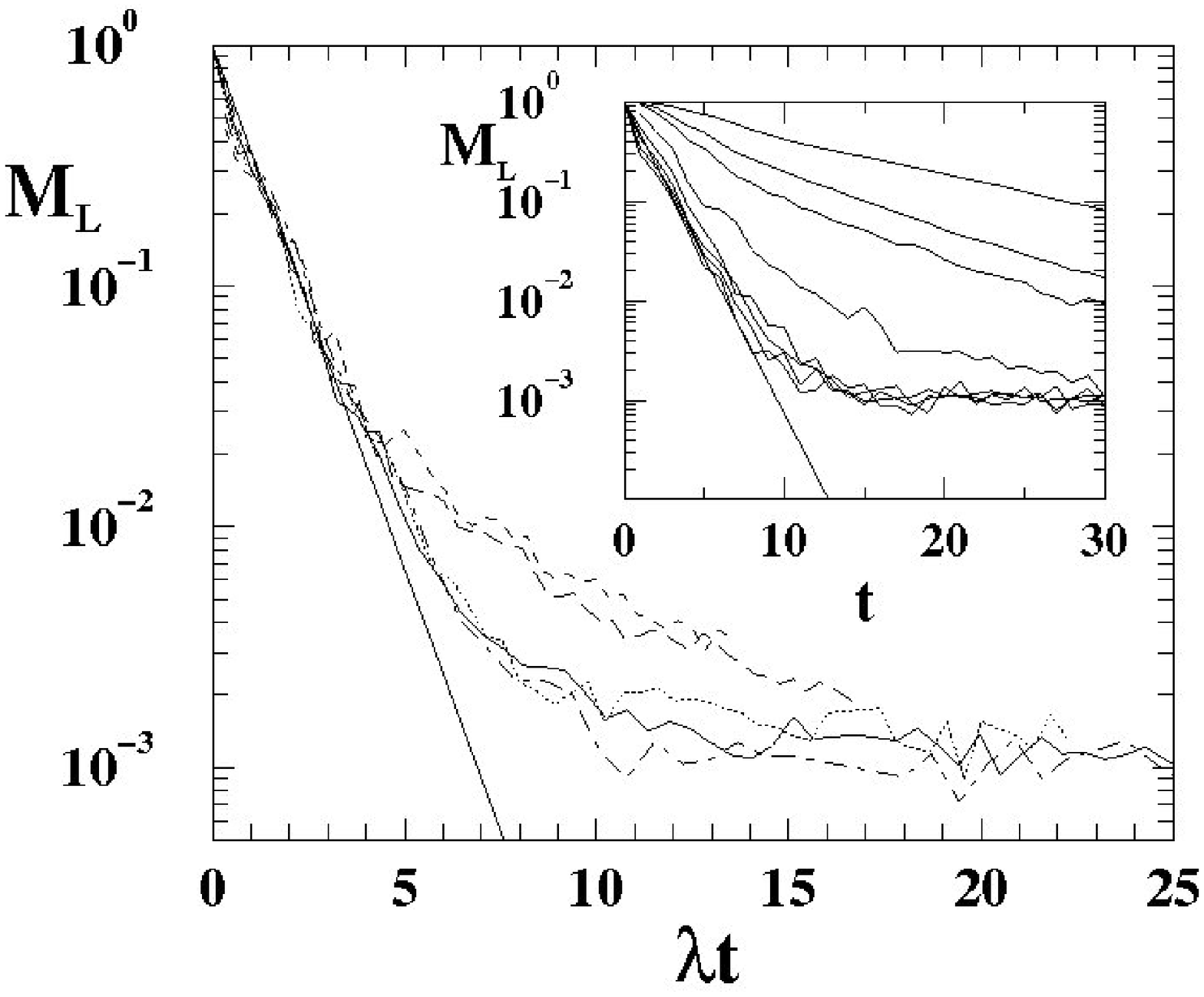}
\includegraphics[width=7.5cm,angle=0]{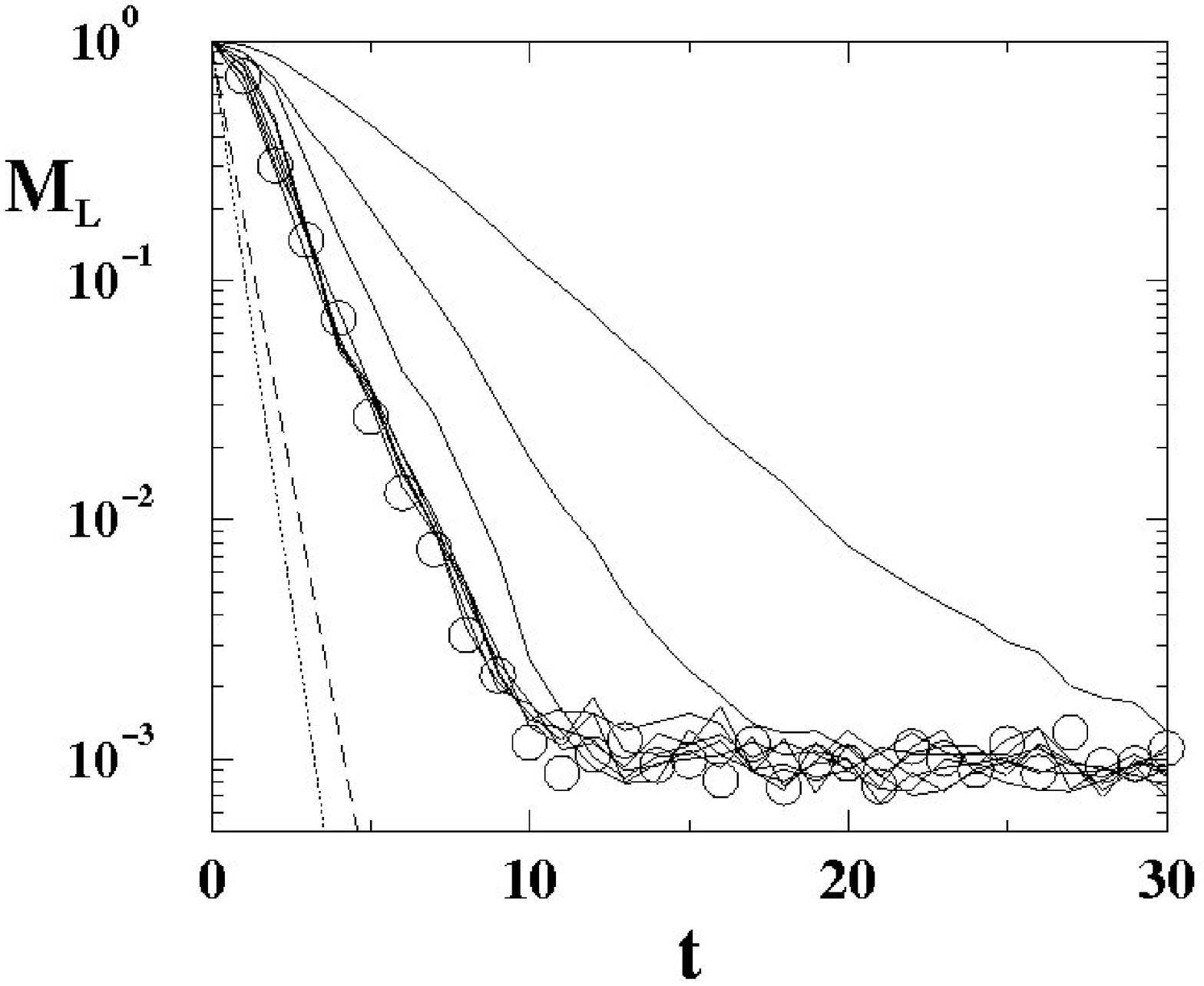}
\caption{\label{fig:fig1_echo1}
Various decays of the average fidelity ${\cal M}_{\rm L}$ for the quantum kicked top 
defined in Eqs.~(\ref{Fdef}) and (\ref{H1def}) with $S=500$.
{\bf Top left}: ${\cal M}_{\rm L}(t)$ in the weak perturbation regime with $\phi\in[10^{-7},10^{-6}]$ and
$K=13.1$, as a function of the squared
rescaled time $(\phi t)^2$. 
The straight line corresponds to the Gaussian decay (\ref{perturbation}). 
Inset : Numerically
computed Lyapunov exponent for the classical kicked top as a function of the 
kicking strength $K$ (see Ref. \protect\cite{Ben76}). The dots correspond to averages taken over
$10^4$ initial conditions.
The error bars reflect the distribution of different exponents
obtained with different initial conditions. 
{\bf Top right}: ${\cal M}_{\rm L}(t)$ in the golden rule regime with $\phi \in [10^{-4},10^{-3}]$, and
$K=13.1$, 17.1, 21.1,
as a function of the rescaled time $\phi^{2} t$. 
Inset: Local spectral density of states 
for $K=13.1$ and perturbation
strengths $\phi=2.5 \cdot 10^{-4}, 5 \cdot 10^{-4}$, $10^{-3}$. 
The solid curves are Lorentzian fits, from which the
decay rate $\Gamma \approx 0.84\, \phi^2 S^2$ is extracted.
The solid line in the main plot gives the decay ${\cal M}_{\rm L} \propto
\exp(-\Gamma t)$ with this value of $\Gamma$. There is no free parameter.
{\bf Bottom left}: ${\cal M}_{\rm L}(t)$ in the Lyapunov regime,
for $\phi=2.1 \cdot 10^{-3}$, $K=2.7$, 3.3, 3.6, 3.9, 4.2.
The time is rescaled with the Lyapunov exponent
$\lambda \in [0.22, 0.72]$. The straight solid line indicates the 
decay ${\cal M}_{\rm L} \propto \exp(-\lambda t)$. 
Inset: ${\cal M}_{\rm L}$ for $K=4.2$ and different
$\phi=j \cdot 10^{-4}$, $j=$1, 2, 3, 4, 5, 9, 17, 25. 
The decay slope saturates at the value $\phi \approx 1.7 \cdot 10^{-3}$
for which $\Gamma \approx \lambda$, even though $\Gamma$ 
keeps on increasing. This demonstrates the decay law ${\cal M}_{\rm L} \propto
\exp[-{\rm min}(\Gamma,\lambda) t]$.
{\bf Bottom right}:  ${\cal M}_{\rm L}(t)$ in the strong perturbation regime,
$\phi = j \cdot 10^{-3}$, ($j$=1, 1.5, 2, \ldots5) (solid curves)
and $K=21.1$, $\phi=3 \cdot 10^{-3}$ (circles). Dashed
and dotted lines show exponential decays with Lyapunov
exponents $\lambda = 1.65$ and 2.12, corresponding to 
$K=13.1$ and 21.1, respectively. The decay slope saturates at 
$\phi \approx 2.5 \cdot 10^{-3}$, when $\Gamma$ reaches the bandwidth.
(Figures taken from Ref.~\cite{Jac01}. Copyright (2001) by the American Physical Society.)}
\end{figure}

We choose the initial wave
packets $\psi_0$ as coherent states of the spin SU(2) group~\cite{Per86}, i.e.\ states
which minimize the Heisenberg uncertainty in phase space. For the kicked top, this
is the
sphere of radius $S$, on which the Heisenberg resolution is determined by the
effective Planck constant
$\hbar_{\rm eff} \sim S^{-1}$. The corresponding Ehrenfest time
is $\tau_{\rm E}=\lambda^{-1}\ln S$. The time-evolution is discrete and proceeds by consecutive
applications of the Floquet operators $F_0$ (for the unperturbed evolution) and $F$ (for the
perturbed evolution). These operators are defined in Eqs.~(\ref{Fdef}) and (\ref{H1def}).
We take $S=500$ and average
${\cal M}_{\rm L}(t=n)=|\langle\psi_0|(F^{\dagger})^{n}F_{0}^{n}|\psi_0\rangle|^{2}$ over 100
initial coherent states $\psi_0$.

We first show results in the fully chaotic regime $K>9$, 
where we choose the initial
states randomly over the entire phase space. The local spectral density 
$\rho(\alpha)$ of the
eigenstates of $F$ in the basis of the eigenstates of $F_{0}$ with 
eigenphases $\alpha$ is plotted for three different perturbation strengths $\phi$
in the inset to the top right panel of Fig. 2. The curves can be fitted
by Lorentzians from which we extract the spreading width $\Gamma$. We find that it is
given up to numerical coefficients by $\Gamma\simeq U^{2}/\delta$, $U\simeq\phi
\sqrt{S}$, $\delta\simeq 1/S$. The golden rule regime $\Gamma\gtrsim\delta$ is
entered at $\phi_{c}\approx 1.7\cdot 10^{-4}$. For $\phi\ll\phi_{c}$ we are in
the perturbative regime, where eigenstates of $F$ do not appreciably
differ from
those of $F_0$ and eigenphase differences can be calculated in first
order perturbation theory.
We then expect the Gaussian decay
\begin{equation}
{\cal M}_{\rm L} \propto\exp(-\sigma_1^2 t^{2}) \; \Rightarrow \; \ln {\cal M}_{\rm L} \propto (\phi t)^{2}.
\label{perturbation}
\end{equation}
This decay is evident in the top left panel of Fig.~\ref{fig:fig1_echo1}, which shows 
${\cal M}_{\rm L}$ as a function
of $(\phi t)^{2}$ on a semilogarithmic scale for $\phi \le 10^{-6}$.
The decay (\ref{perturbation}) stops when ${\cal M}_{\rm L}$ approaches the inverse 
$1/2S$ of the dimension of the
Hilbert space in agreement with our predictions.

For $\phi > \phi_c$ one enters
the golden rule regime, where the Lorentzian 
spreading of eigenstates of $F$ over those of $F_0$ results in
the exponential decay 
\begin{equation}
{\cal M}_{\rm L} \; \propto \; \exp(-\Gamma t) \; \Rightarrow \; \ln {\cal M}_{\rm L} \propto \phi^{2} t.
\label{goldenrule}
\end{equation}
The data presented in the top right panel of Fig.~\ref{fig:fig1_echo1} clearly confirm the validity 
of the scaling (\ref{goldenrule}). There is no dependence 
of ${\cal M}_{\rm L}$ 
on $K$ in this regime of moderate (but non-perturbative) values of $\phi$,
i.e. no dependence on the Lyapunov exponent, which
varies by a factor of 1.4 for the different values of $K$ used to generate the data
in the top right panel of Fig.~\ref{fig:fig1_echo1}.

For the kicked top model, it is hard to satisfy $\lambda<\Gamma$ in the fully chaotic regime, 
because values of $K>9$ already corresponds to $\lambda \gtrsim 1$
(see the inset to the top left panel of 
Fig.~\ref{fig:fig1_echo1}), while the band width $B$, the upper limit for
$\Gamma$, is $B=\pi/2$ (in units of $1/\tau$).
For this reason, when the perturbation strength $\phi$ is further
increased, the decay rate saturates at the band width ---
before reaching the Lyapunov exponent. This is shown
in the bottom right panel of
Fig.~\ref{fig:fig1_echo1}. There is no trace of a Lyapunov decay in this fully chaotic
regime.

To observe the Lyapunov decay ${\cal M}_{\rm L} \propto \exp[-\lambda t]$, 
we therefore reduce $K$ to values in the range $2.7 \le K \le 4.2$, 
which allows us to vary the Lyapunov exponent over a wider range 
between 0.22 and 0.72. In this range the 
classical phase space is mixed and we have coexisting regular and 
chaotic trajectories. We choose 
the initial coherent states in the chaotic region, which was numerically identified
through the participation ratio of the initial state. Because
the chaotic region still occupies more than 80\% of the
phase space for the smallest value of $K$ considered, we expect
nonuniversal effects (e.g. nonzero overlap of
our initial wavepackets with regular eigenfunctions of $F_0$ or $F$) to
be small if not negligible. Our theory predicts a
crossover from the golden rule decay (\ref{goldenrule}) to the
Lyapunov decay \cite{Jal01}
\begin{equation}
{\cal M}_{\rm L} \; \simeq \; \exp(-\lambda t) \; \Rightarrow \; \ln {\cal M}_{\rm L} \propto \lambda t,
\label{Lyapunov}
\end{equation}
once $\Gamma$ exceeds $\lambda$. This expectation is borne out by our
numerical simulations, see the bottom left panel of Fig.~\ref{fig:fig1_echo1}. Note that these
early numerics are unable to resolve the observed, effective Lyapunov exponent from the
system's true Lyapunov exponent (see the discussion above in Section~\ref{section:lyapunov?}).

We next operate the kicked top in the regular regime with $K=1.1$
to check the prediction given in the first line of Eq.~(\ref{eq:final_decay}).
In the left panel Fig.~\ref{fig:fig5_echo1} we show the decay of ${\cal M}_{\rm L}$ 
for $S=1000$ and different perturbation strengths $\phi$. For
weak perturbations, the decay of  ${\cal M}_{\rm L}$ is exponential, and not 
Gaussian as one would expect from first order perturbation 
theory. The reason why we do not witness
a Gaussian decay in that regime is that the perturbation operator gives 
no first order correction for low $K$. Indeed, for $K=1.1$,
eigenfunctions of $F_0$ are still almost identical to eigenfunctions of $S_y$,
so that diagonal matrix elements of $S_x$ vanish in this basis. 
We numerically obtained an
exponential decay $\propto \exp(-\gamma t)$ of the fidelity with
$\gamma \propto \phi^{1.5}$, which is to be contrasted with
the golden
rule decay $\propto \exp(-\Gamma t)$ with $\Gamma \propto \phi^2$.

\begin{figure}
\includegraphics[width=7.9cm,angle=0]{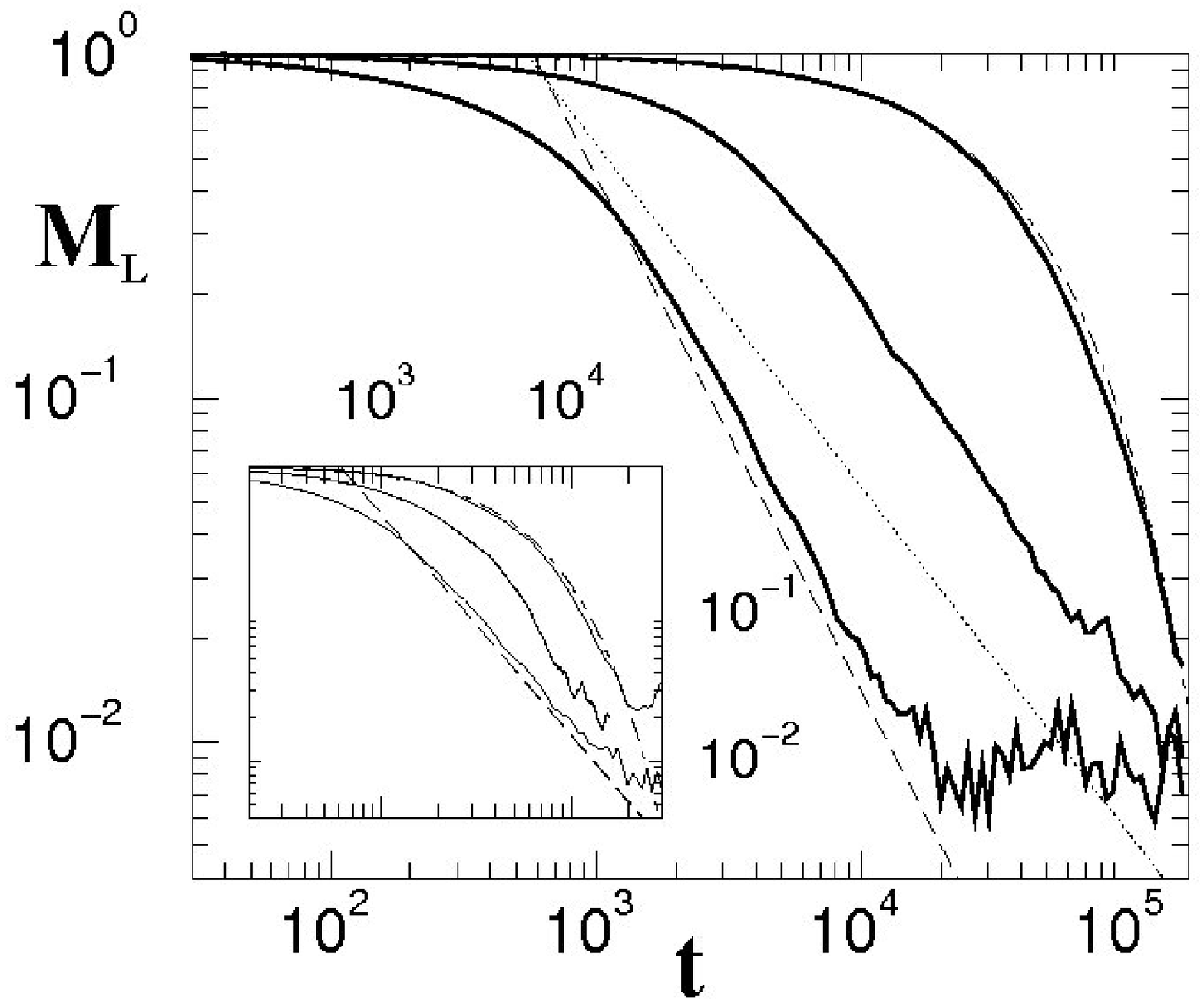}
\includegraphics[width=8.3cm,angle=0]{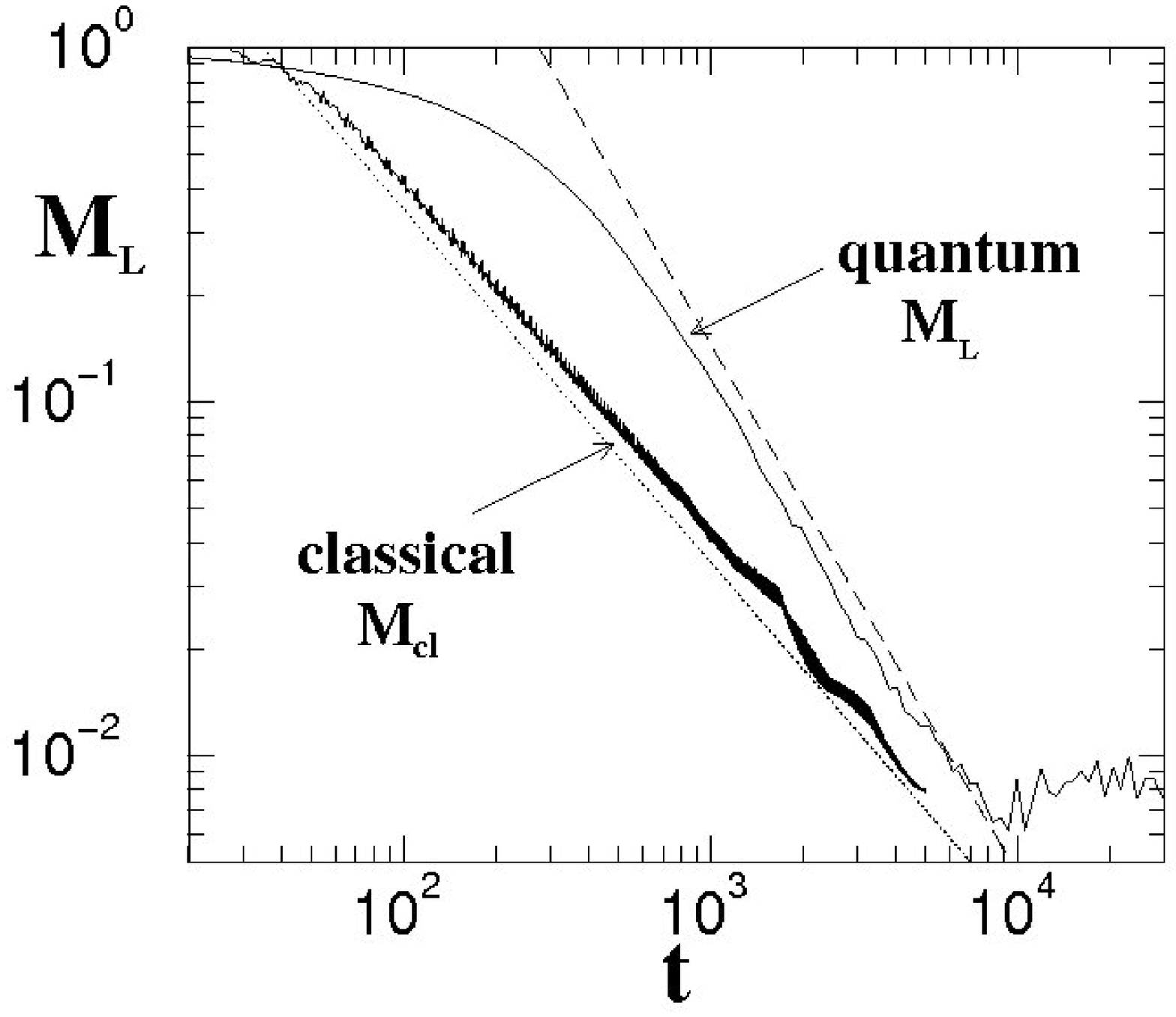}
\caption{\label{fig:fig5_echo1}
{\bf Left panel}:
Decay of ${\cal M}_{\rm L}$ for $S=1000$, 
$K=1.1$, and $10^{5} \,\phi=1.5$, 4.5, and 10 (thick
solid lines from right to left). The crossover from exponential
to power-law decay is illustrated by the dotted-dashed line 
$\propto \exp[-2.56 \cdot 10^{-5} \,t]$ and the dashed 
line $ \propto t^{-3/2}$. The dotted line gives the classical
decay $\propto t^{-1}$.
Inset: Decay of ${\cal M}_{\rm L}$ for $K=1.1$,
$\phi=10^{-4}$, and $S=250$, $500$, and $1000$ (solid lines from right 
to left). The dashed and dotted-dashed lines indicate
the power law $\propto t^{-3/2}$ and exponential 
$\propto \exp[-2 \cdot 10^{-4} \,t]$ decay, respectively.
These plots show that the $t^{-3/2}$ decay is reached either by
increasing the perturbation strength $\phi$ at fixed spin
magnitude $S$, or by increasing $S$ at fixed $\phi$.
{\bf Right panel}: Decay of the quantum fidelity ${\cal M}_{\rm L}$ for $S=1000$,
compared to the decay of the average overlap ${\cal M}_{\rm cl}$
of classical phase space distributions, both for the kicked top with $K=1.1$
and $\phi=1.7 \cdot 10^{-4}$. The initial classical
distribution extends over a volume $\sigma=10^{-3}$ of phase space,
corresponding to one Planck cell for
$S=1000$. The dotted and dashed lines
give the classical and quantum 
power law decays $\propto t^{-1}$ and $ \propto t^{-3/2}$,
respectively. (Figure taken from Ref.~\cite{Jac03}.)}
\end{figure}

As $\phi$ increases, and looking back at the left panel of Fig.~\ref{fig:fig5_echo1},
the decay of ${\cal M}_{\rm L}$ turns into the predicted  
power law $\propto t^{-3/2}$, which prevails as soon as one
enters the golden rule regime, i.e. for 
$\Gamma/\Delta \approx \phi^2 S^3 \ge 1$~\cite{Jac01}. 
One therefore expects the power law
decay to appear as $S$ is increased at fixed $\phi$,
which is indeed observed in the inset to the left panel of Fig.~\ref{fig:fig5_echo1}.

We also checked that these results are not sensitive to our choice of
Hamiltonian, by replacing $S_x$ in Eq.~(\ref{H1def}) with 
$S_z^2$, as used in Refs.~\cite{Pro03a,Gor06}) and also by
studying a kicked rotator as an alternative model to the kicked top.
These numerical results all give
confirmation of the power law decay predicted in 
Eq.~(\ref{eq:final_decay}) for regular systems.

It is instructive to contrast these results for the decay of the squared scalar product
of quantum wavefunctions with the decay of the overlap of classical 
phase space distributions, a ``classical fidelity'' problem that was
investigated in Refs.~\cite{Ben02,Pro03a,Eck03,Ben03c,Ben03b}. We assume that
the two phase space distributions $\rho_0$ and $\rho$ are initially
identical and evolve according to the Liouville equation of
motion corresponding to the classical limit map of the kicked
top~\cite{Haa87,Haa01}
\begin{eqnarray}\label{clmap}
\left\{
\begin{array}{cc}
x_{n+1} = &z_n \cos(K x_n) + y_n \sin(K x_n) \\
y_{n+1} = &-z_n \sin(K x_n) + y_n \cos(K x_n) \\
z_{n+1} = &-x_n, 
\end{array}
\right.
\end{eqnarray}
for two different Hamiltonians $H_0$ and $H$. We consider regular
dynamics and ask for the decay of the normalized phase space 
overlap 
\begin{equation}\label{classic}
{\cal M}_{\rm cl}(t)=\int d{\bf x}\int d{\bf p} \; \rho_0({\bf x},{\bf p};t) 
\; \rho({\bf x},{\bf p};t)/{\cal N}_{\rho},
\end{equation}
where ${\cal N}_{\rho}=(\int d{\bf x}\int d{\bf p} \; \rho_0)^{1/2}
(\int d{\bf x}\int d{\bf p} \; \rho)^{1/2}$.

We have found above that a factor $\propto t^{-d/2}$ in the decay
of the quantum fidelity ${\cal M}_{\rm L}(t) \propto t^{-3d/2}$ originates from the action 
phase difference and is thus of purely quantum origin.
One therefore expects a slower 
classical decay ${\cal M}_{\rm cl}(t) \propto C_s \propto t^{-d}$.
In the right panel to Fig.~\ref{fig:fig5_echo1} we show the decay of the 
averaged ${\cal M}_{\rm cl}$ taken over 
$10^4$ initial points within a narrow volume of phase space
$\sigma \equiv \sin \theta \delta \theta \; \delta \varphi$, for $K=1.1$ and
$\phi=1.7 \cdot 10^{-4}$. The decay is
${\cal M}_{\rm cl} \propto t^{-1}$, and clearly differs 
from the quantum decay $\propto t^{-3/2}$. 

The power law decay prevails
for classically weak perturbations, for which the center of mass 
of $\rho$ and $\rho_0$ stay close together. This condition is required 
by the diagonal approximation $s_1 = s_2$ leading to Eq.~(\ref{semicl}).
Keeping the de Broglie wavelength $\nu$
fixed, and increasing the perturbation strength $\phi$, the invariant
tori of $H_0$ start to differ significantly from those of $H$ 
on the resolution scale $\nu$, giving a threshold $\phi_{\rm cl} \approx \nu$.
Above $\phi_{\rm cl}$, the distance 
between the center of mass of $\rho_0$ and $\rho$ increases 
with time $\propto t$ and one
expects a much faster decay ${\cal M}_{\rm cl}(t) \propto \exp[-{\rm const}\times t^2]$ for 
classical Gaussian phase space distributions~\cite{Eck03}. 
In the quantum kicked top, $\nu=1/S$ and the threshold translates into
$\phi_{\rm cl} \sim 1/S$. It is quite remarkable that this coincides with the upper boundary of the 
golden rule regime. As long as one stays in that regime, the perturbation
will affect the phase in Eq.~(\ref{clt}), and result in the anomalous power law
decay $\propto t^{-3d/2}$. 

\subsubsection{Mesoscopic fluctuations of the Loschmidt echo}\label{section:k_rot}

In our investigations of the mesoscopic fluctuations of the Loschmidt echo, we rely on the
one-particle kicked rotator~\cite{Izr90}. The model presents the advantage that  it is the product of
two unitary time-evolution operator, each of them diagonal in either position or  momentum.
Time-evolutions can be calculated very efficiently via recursive calls to fast Fourier transforms.
The resulting speed increase in the algorithm, compared to the kicked top, allows to reach
systems size of $N=2^{18}$, several orders of magnitude larger than for the kicked top.
This is particularly advantageous for detecting Lyapunov decays. The one-particle kicked rotator
is briefly discussed in Appendix~\ref{appendix:1krot}. For further details we refer the reader
to Ref.~\cite{Izr90}.

\begin{figure}
\includegraphics[width=7.8cm]{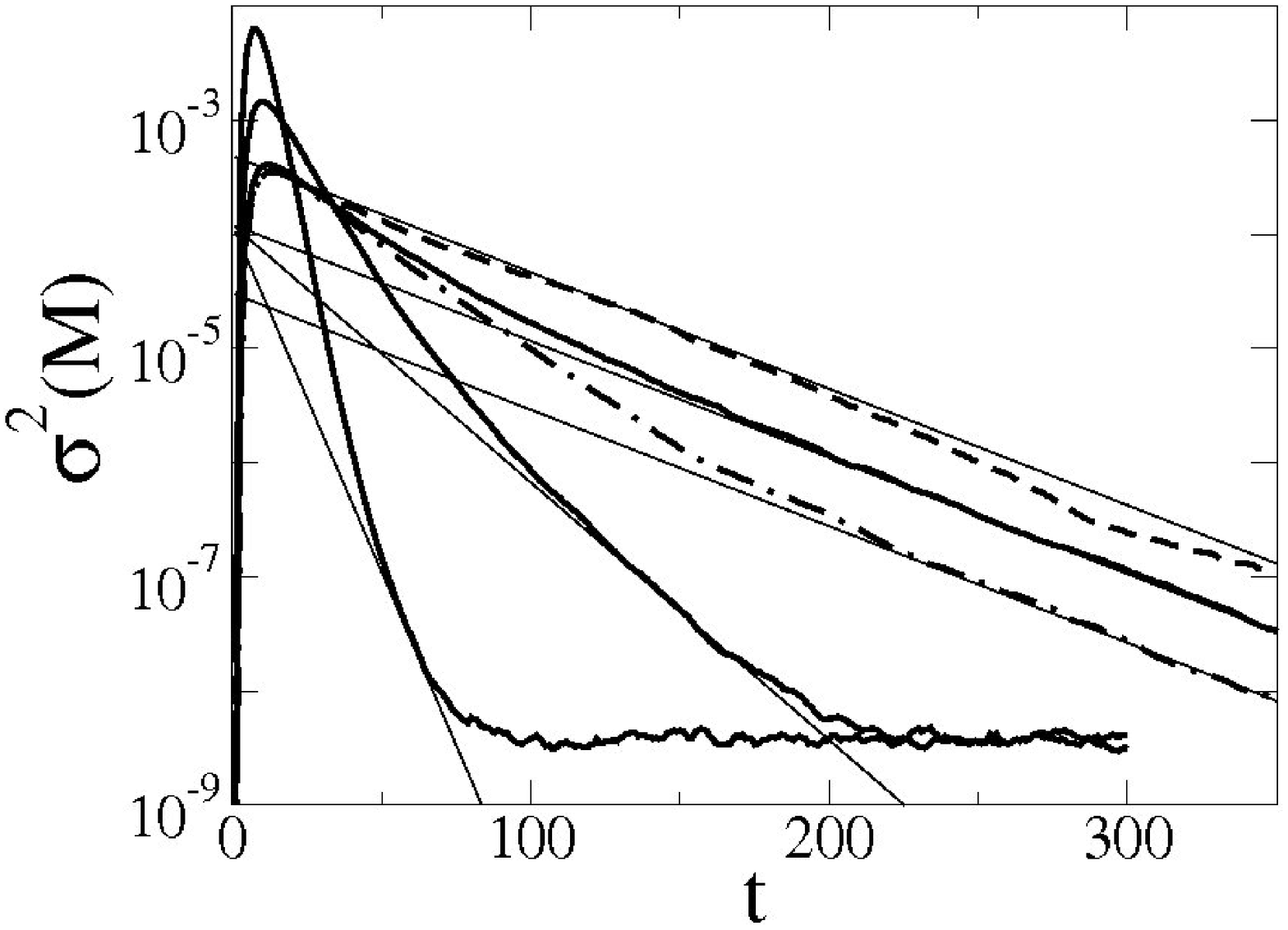} \;\;\;
\includegraphics[width=8cm]{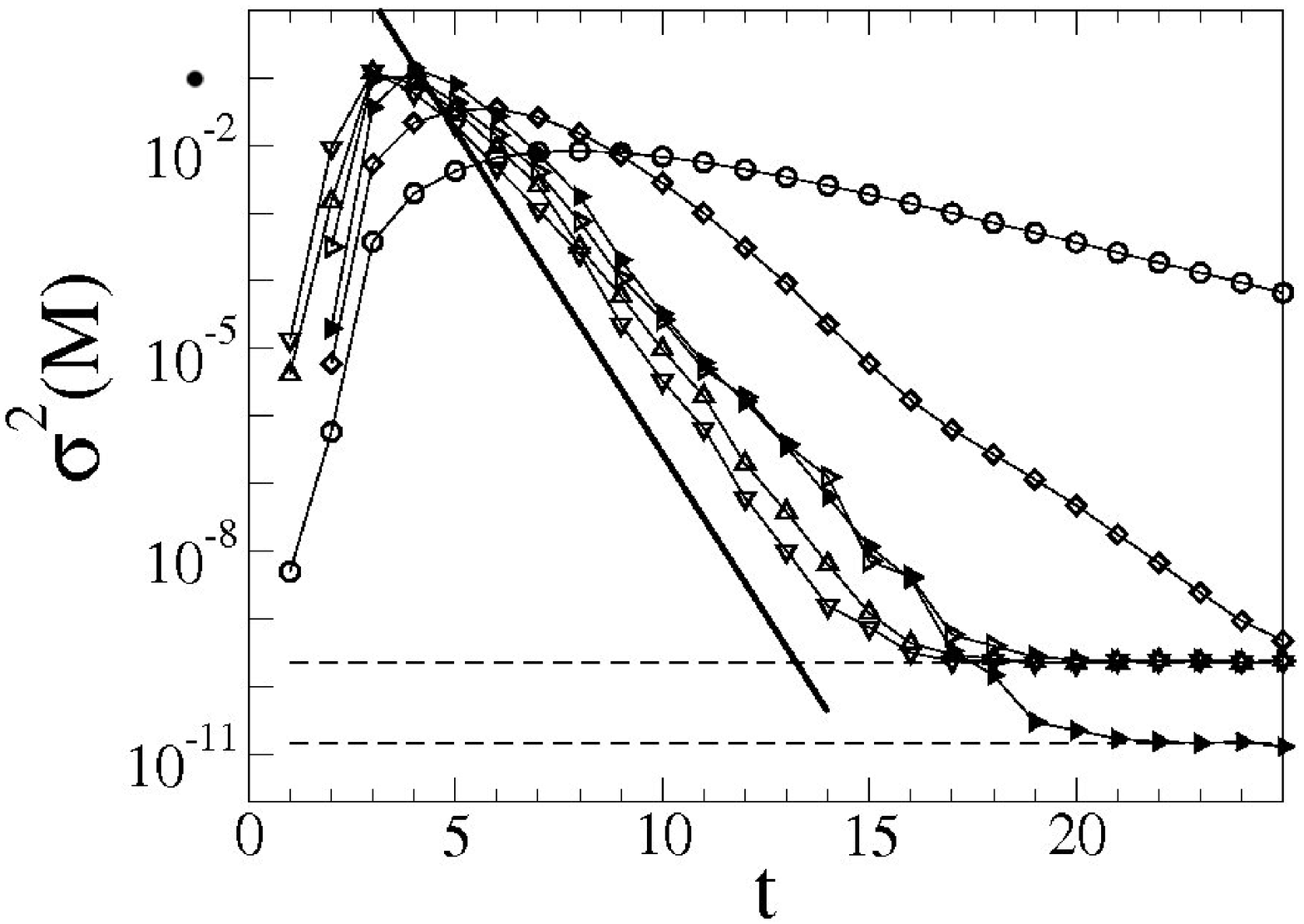}
\caption{\label{fig:fig4_echoflucs} 
{\bf Left panel}: Variance
$\sigma^2({\cal M}_{\rm L})$ of the fidelity vs. $t$ for weak $\Gamma \ll \lambda$,
$N=16384$ and $10^{5}\cdot  \delta K = 5.9$, $8.9$ and $14.7$ (thick
solid lines), $N=4096$ and  $\delta K = 2.4 \cdot 10^{-4}$ (dashed line)
and $N=65536$ and $\delta K = 1.48 \cdot 10^{-5}$ (dotted-dashed line).
All data have $K_0=9.95$. The thin solid lines indicate the
decays $= 2 \hbar_{\rm eff} \exp[-\Gamma t]$, with $\Gamma = 0.024 
(\delta K \cdot N)^2$; there is no adjustable free parameter.
The variance has been calculated from $10^3$ different
initial states $\psi_0$. 
{\bf Right panel}: Variance $\sigma^2({\cal M}_{\rm L})$ of the fidelity
vs. $t$ in the golden rule regime with $\Gamma \gtrsim \lambda$ for 
$N=65536$, $K_0=9.95$ and
$\delta K \in [3.9 \cdot 10^{-5},1.1 \cdot 10^{-3}]$ (open symbols),
and  $N=262144$, $K_0=9.95$, $\delta K=5.9 \cdot 10^{-5} $ (full triangles). 
The solid line is 
$\propto \exp[-2 \lambda_0 t]$, with an exponent $\lambda_0 = 1.1$, 
smaller than the Lyapunov exponent $\lambda=1.6$, because the fidelity
averages $\langle \exp[-\lambda t] \rangle$ (see text).
The two dashed lines give $\hbar_{\rm eff}^2=N^{-2}$.
In all cases, the variance has been calculated from $10^3$ different
initial states $\psi_0$.
(Figure taken from Ref.~\cite{Pet05}. Copyright (2005) by the American Physical Society.)}
\end{figure}

We numerically illustrate the validity of 
our analytical theory for the 
variance $\sigma^2({\cal M}_{\rm L})$ of the Loschmidt echo.
We determine the dependence of $\Gamma$ on the system's parameter by 
investigating the local spectral density of eigenstates of $F$
over those of $F_0$. We found that it has a Lorentzian shape with a 
width $\Gamma \simeq 0.024 (\delta K  \cdot N)^2 \propto (\delta
K/\hbar_{\rm eff})^2$, with a very weak dependence of $\Gamma$ in $K_0$, 
in the range $B = 2 \pi \gg  \Gamma \gtrsim \delta = 2 \pi/N$. 
We focus on $\sigma^2$ in the golden rule regime with 
$\Gamma \ll \lambda$. Data are shown in the left panel of Fig.~\ref{fig:fig4_echoflucs}.
One sees that $\sigma^2({\cal M}_{\rm L})$ first rises, up to a time
$t_c$, after which it decays. The maximal value 
$\sigma^2(t_c)$ in that regime increases with increasing perturbation,
i.e. increasing $\Gamma$. Beyond $t_c$, the decay of $\sigma^2$
is very well captured by Eq.~(\ref{sigma3g}), once enough
time has elapsed. This is due to the increase of $\sigma^2(t_c)$
above the self-averaging value $\propto \hbar_{\rm eff}$ as $\Gamma$ increases.
Once the influence of the peak
disappears, the decay of $\sigma^2({\cal M}_{\rm L})$ is very well captured by $\sigma^2_3$
given in Eq.~(\ref{sigma3g}), 
without any adjustable free parameter. Finally, at large times,
$\sigma^2({\cal M}_{\rm L})$ saturates at the value $\hbar_{\rm eff}^2 = 
N^{-2}$, as
predicted by Eqs.~(\ref{sigmascl}) and (\ref{sigmarmt}).

As $\delta K$ increases, so does $\Gamma$ and $\sigma^2({\cal M}_{\rm L})$ decays faster
and faster to its saturation value until 
$\Gamma \gtrsim \lambda$. Once $\Gamma$ 
starts to exceed $\lambda$, the decay saturates at $\exp(-2 \lambda t)$. 
This is shown in the right panel of Fig.~\ref{fig:fig4_echoflucs}, which corroborates the Lyapunov decay
of $\sigma^2({\cal M})$ predicted by Eqs.~(\ref{sigma1lambda}). 
In agreement with our discussion in Chapter~\ref{section:lyapunov?}, we see that 
the decay exponent slightly differs from the Lyapunov exponent 
$\lambda=\ln[K/2]$. This is due to the fact 
that the fidelity averages $\langle C_s \rangle \propto \langle 
\exp[-\lambda t] \rangle \ne \exp[-\langle \lambda \rangle t]$ over
finite-time fluctuations of the Lyapunov exponent 
\cite{Sil03}. At long times, $\sigma^2({\cal M}_{\rm L},t \rightarrow \infty) = 
\hbar_{\rm eff}^2$ saturates at 
the ergodic value,
as predicted. Finally, it is seen in both panels of Fig.~\ref{fig:fig4_echoflucs} 
that $t_c$ decreases as the perturbation is
cranked up. Moreover, there is no $N$-dependence of $\sigma^2(t_c)$ at fixed
$\Gamma$. These two facts are in qualitative and quantitative
agreement with Eq.~(\ref{tc}). 

The numerics on true dynamical systems presented in this
section qualitatively and quantitatively confirm the results of 
both the semiclassical theory and RMT in their respective regime of validity. 

\subsection{Displacement echoes: classical decay and quantum freeze}\label{section:displacement}

So far we have discussed quantum reversibility from the rather general point of
view of Eq.~(\ref{eq:def_LE}). Our approach
has been statistical in nature and applies to generic perturbations,
in the minimal sense that they do not commute with the unperturbed Hamiltonian.
The point has been 
made above that specific families of echoes naturally
occur when the problem at hand requires to investigate correlation 
functions such as the one in Eq.~(\ref{eq:t_correl}),
\begin{eqnarray}\label{eq:t_correl_repeat}
Y ({\bf P},t)=\Big\langle  \exp[-i{\bf P}\cdot {\bf \hat{r}}]
\exp[ i H_0t] \exp[i{\bf P}\cdot {\bf \hat{r}}] 
\exp[-i H_0t] \Big\rangle .
\end{eqnarray}
This quantity is of interest, for instance, 
in spectroscopies such as neutron scattering, 
M\"ossbauer $\gamma$-ray,  and certain electronic transitions in molecules and 
solids~\cite{Hel87,Lax74,Lov84,vanH54}, and more generally whenever the
problem at hand requires some knowledge of
momentum or position time correlators -- or combinations of 
the two. For instance, the differential cross section  for incoherent neutron scattering and M\"ossbauer emission/absorption can be written as~\cite{Lov84}
 \begin{equation}\label{diffcross}
 \frac{d^2 \sigma}{d { \Omega}  d E}  = 
\frac{|{\bf P}_{\rm out}|}{|{\bf P}_{\rm in}|} \, \frac{\ell^2_{\rm i}}{4 \pi} \, {\cal S}_{\rm i} ({\bf P} , \omega),
 \end{equation}
in terms of the solid scattering angle ${ \Omega} $,  
the total incoherent scattering length $\ell_{\rm i}$, the initial and final neutron momenta 
${\bf P}_{\rm in}$ and ${\bf P}_{\rm out}$ and the momentum transfer ${\bf P}={\bf P}_{\rm out}-
{\bf P}_{\rm in}$. It turns out that the incoherent scattering response function 
$ {\cal S}_{\rm i} ({\bf P} , \omega)$
can be expressed  in terms of  the Fourier transform  
 \begin{equation}\label{respfunc}
{\cal S}_{i} ({\bf P} , \omega)  = 
\frac{1}{2 \pi   {\cal N} }  \int {\rm d} t \, e^{-{\it i} \omega t}  \sum_{j} Y\parent{\bf{P},t}
 \end{equation}
of the correlation function $Y\parent{\bf{P},t}$ given in Eq.~(\ref{eq:t_correl_repeat}).
This establishes the physical relevance of our investigations of displacement echoes
for experiments on incoherent scattering. 

The operator inside the bracket of Eq.~(\ref{eq:t_correl_repeat}) is similar to the kernel of the
Loschmidt echo -- it is given by a forward and a backward time-evolution.
In this case, however, both are governed by the same Hamiltonian
$H_0$, but the backward propagation is sandwiched between two
momentum boost operators. Writing
\begin{equation}
\exp[i H_{\bf P} t] = \exp[-i{\bf P}\cdot {\bf \hat{r}}]
\exp[ iH_0t] \exp[i{\bf P}\cdot {\bf \hat{r}}],
\end{equation}
the kernel of Eq.~(\ref{eq:t_correl_repeat}) 
goes into a true Loschmidt echo kernel, and one would expect 
all the results presented earlier in this chapter to apply to
the {\it displacement echo}
\begin{equation}\label{decho_repeat}
{\cal M}_{\rm D}(t)= \big|\langle \psi_0|
\exp[i H_{\bf P}t] \exp[-i H_0 t] |\psi_0\rangle \big|^2.
\end{equation}
This line of reasoning is not quite correct, as we show below. The
displacement operator is very special in that, speaking semiclassical
language, it does not lead to phase accumulations along an otherwise
unperturbed trajectory. It is therefore unable to generate a golden
rule decay $\propto \exp[-\Gamma t]$. One consequence of this is that
in the golden rule regime $\delta \lesssim \Gamma \ll B$, ${\cal M}_{\rm D}$
exhibits only the Lyapunov
decay $\propto \exp[-\lambda t]$. This is however not the full story, as
the displacement generated by $\exp[\pm i\bf{P}\cdot \bf{\hat{r}}]$
leads to a reduction of the overlap of 
$|\psi_{\rm F} \rangle = \exp[-i H_0 t] | \psi_0 \rangle$ with
$|\psi_{\rm R} \rangle = \exp[-i H_{\rm P} t]| \psi_0 \rangle$,
which, for small displacements, depends on $t$ only for short times. The large time asymptotic --
the saturation ${\cal M}_{\rm D}(\infty)$ -- depends
on the distance over which the wavepacket is translated.
For not too large displacements, one has a {\it quantum freeze} of
the displacement echo, at values which can be orders of magnitude bigger than 
the minimal saturation value $N^{-1}$ of the Loschmidt echo.
This behavior is illustrated in Fig.~\ref{fig:fig1_heller}.
It obviously derives from some spatially resolved
dynamics, which cannot be captured by RMT. We therefore exclusively
rely on the semiclassical approach in this section.

What does the quantum freeze correspond  to physically ?  
It is the elastic component in any of the mentioned spectroscopies: 
M\"ossbauer, neutron, and molecular electronic, and was first identified by 
van Hove in connection with neutron scattering~\cite{Hov54}.  
To make a long story short, there is a finite probability, above 
the $N^{-1}$ statistical limit, of not having a quantum transition
to a new state, in spite of being ``hit''. This is the source, for example, 
of  the recoilless peak in M\"ossbauer spectroscopy.

Recent experimental efforts in atom interferometry motivate the
investigation of the real-space displacement echo,
\begin{subequations}\label{eq:spatial_displacement}
\bea
{\cal M}_{\rm D}(t)= \big|\langle \psi_0|
\exp[i H_{\bf X}t] \exp[-i H_0 t] |\psi_0\rangle \big|^2, \\
\hat{H}_{\bf X } = \exp[-{\it i} {\bf  X \cdot \hat{p} }] 
\exp[-{\it i} H_0   t ] \exp[{\it i} {\bf X \cdot \hat{p}} ],
\eea
\end{subequations}
instead of the momentum displacement 
echo (\ref{decho_repeat})~\cite{Su06,Wu07a,Wu08,Wu08b}.
These are so-called Talbot-Lau experiments that probe interferences of 
guided atomic waves through periodic potentials
in the form of optically formed gratings.
It is not our task here to describe these experiments and the
effects on which they are based in detail (for a very recent review on
atom interferometry, see Ref.~\cite{Cro07}), we nevertheless briefly
discuss why they are connected to Eq.~(\ref{eq:spatial_displacement}).
The discussion is kept at a qualitative level.

In Talbot-Lau experiments, a 
plane-wave incident on a transverse periodic potential -- a grating -- 
is split into
partial waves. The distance between the center of masses of these
partial waves increases linearly in time,
and behind the grating they interfere in such a way that they
produce a self-image of the grating structure at the Talbot distance 
$L_{\rm T}=2 d^2/\nu$. Here, $d$ gives the periodicity of the grating
and $\nu$ the de Broglie wavelength of the matter wave. 
This is the Talbot effect. Applying a second grating induces a back
effect and, possibly, the recombination of the partial waves. In the experiments,
an optical pulse was included between the two gratings
a distance ${\bf X}$ away from the first one~\cite{Su06,Wu07a}. This pulse is
devised to
generate a global momentum change $\exp[i{\bf X} \cdot {\bf \hat{p}}]$,
and the experiment thus  probes the real-space displacement echo of
Eq.~(\ref{eq:spatial_displacement}).
In the following paragraphs we discuss both spatial and momentum 
displacement echoes, illustrate their specificities and show how, not
surprisingly, they essentially behave in the same way in chaotic systems.

\begin{figure}
\includegraphics[width=12cm,angle=0]{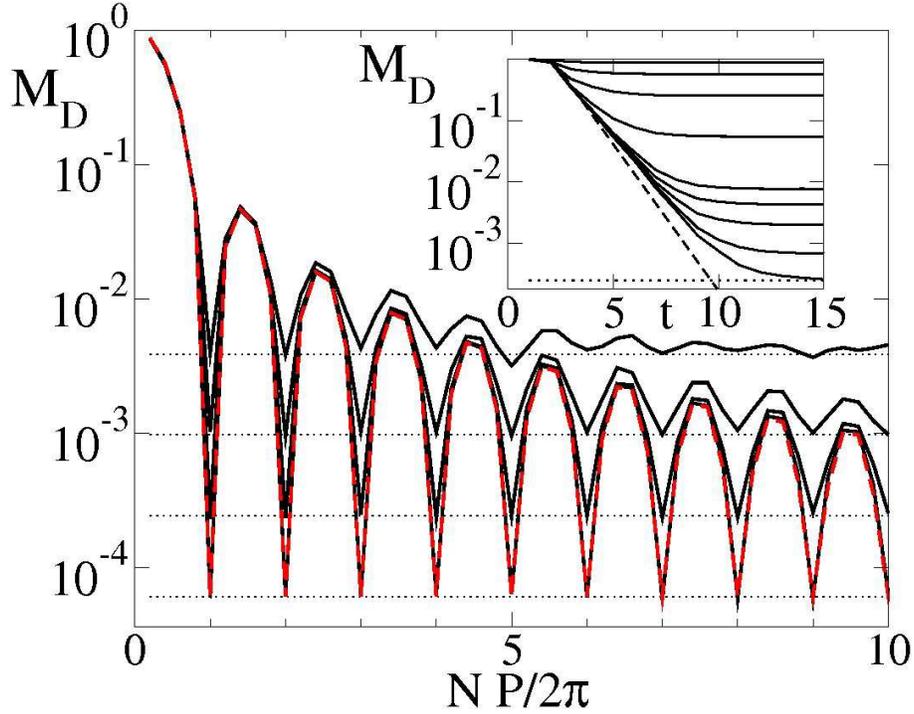}
\caption{\label{fig:fig1_heller} Main plot: Saturation value ${\cal M}_{\rm D}(\infty)$ of the displacement echo
as a function of the rescaled displacement 
$N P/2 \pi$ for the kicked rotator 
model with $N = 256$,  $ 1024$,  $ 4096$, $16384$ (full lines, from 
top to bottom). Data are obtained from 1000 different initial coherent states.
The dotted lines give the saturation at $N^{-1}.\,$ The red dashed line 
gives the theoretical prediction 
${\cal M}_{\rm D}(\infty) = 
{\rm Max} (4 \exp[-(\sigma P)^2/2] \sin^2(P L/2)\Big/(PL)^2 , N^{-1})$ for $N=16384$.
Inset: Quantum freeze of the displacement echo for kicking strength 
$K=10.09$, $N=4096$, and $P\in[0,2 \pi/N]$.
The dashed line gives the decay with the reduced Lyapunov exponent
$\lambda_0 = 1.1$. (Figure taken from Ref.~\cite{Pet07a}. Copyright (2007) by the American Physical Society.)}
\end{figure}

\subsubsection{Momentum displacement -- semiclassical theory}

We first discuss the validity of the
diagonal approximation used in Appendix~\ref{appendix:semiclassics}
[before Eq.~(\ref{semicl})]
for the semiclassical approach
to the average Loschmidt echo and show
why this approximation is even better for the displacement echo.
This diagonal approximation equates each 
classical trajectory
$s_1$ generated by an unperturbed Hamiltonian $H_0$ with a classical trajectory
$s_2$ generated by a perturbed Hamiltonian $H=H_0+\Sigma$.
It has already been mentioned that 
this procedure does not seem to be justified at first glance
in chaotic systems with local exponential instability. Instead one
would expect that
an infinitesimally small perturbation generates trajectories
diverging exponentially fast away from their unperturbed counterpart. 
Why then are we allowed to set $s_1 \simeq s_2$ ? Because of the shadowing and structural 
stability properties of hyperbolic systems~\cite{Kat96,Van04}.
Roughly speaking one can show that,
given a uniformly hyperbolic Hamiltonian system $H_0$, and a generic 
perturbation $\Sigma$,
each classical trajectory $s_2$ generated by the still 
hyperbolic but
perturbed Hamiltonian $H_0+\Sigma$ remains almost always arbitrarily close to 
one, and only one unperturbed trajectory $s_1$. 
In general the two trajectories do not share common endpoints,
however these endpoints are close enough that they
are not resolved quantum-mechanically.
This is illustrated in the left panel of Fig.~\ref{fig:fig2_heller}.
The semiclassical expression for the kernel of the Loschmidt echo involves a 
double sum
over the perturbed and the unperturbed classical trajectories,
so that both $s_2$ and $s_1$ are included. After a stationary 
phase condition, this double sum is reduced to a single sum where
$s_2$ and $s_1$ are equated -- this is done in Appendix~\ref{appendix:semiclassics}, just above Eq.~(\ref{semicl}). 
In other words, a semiclassical particle in a Loschmidt echo experiment
follows $s_1$ in the forward direction, and 
$s_2$ in the backward direction because this is the best way to minimize 
the action for weak enough perturbations. 
The action difference is simply given by the integral of the 
perturbation along the backward
trajectory. It is in general time-dependent and leads to a finite 
action phase difference $\delta S_{s_1,s_2}=S_{s_1}-S_{s_2}$, which 
dephases the two trajectories, and eventually generates the golden rule decay.
Strictly speaking, proofs of structural stability exist only for
uniformly hyperbolic systems. However, numerical investigations have shown that
generic chaotic systems such as the kicked rotator
also display structural stability and shadowing
of trajectories upon not too strong perturbations~\cite{Gre90}. Because the
threshold for the Golden rule regime puts a semiclassically small
parametric bound $\delta K \ll B \hbar_{\rm eff}$ on the strength of the perturbation, 
shadowing can be invoked in that regime in the semiclassical limit, 
where the perturbation becomes smaller and smaller.

\begin{figure}
\includegraphics[width=10.cm,angle=0]{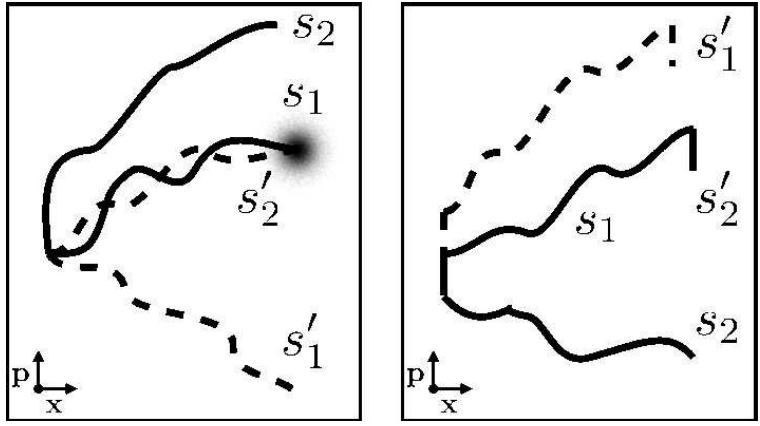}
\caption{\label{fig:fig2_heller} Illustrative view of structural stability. Left panel: 
generic perturbation, where $s_1$ and $s_2$ are two orbits of the 
unperturbed Hamiltonian, $s_1'$ is the orbit of the perturbed Hamiltonian 
with the same initial condition as $s_1$, while $s_2'$ is the orbit of 
the perturbed Hamiltonian with the same initial condition as $s_2$. 
The endpoints of $s_1$  and $s_2'$ are separated by less
than a quantum-mechanical
resolution scale (red shaded area). Right panel: phase space displacement. 
Labels are the same as in the left panel.
Note that $s_2'$ and $s_1$ lie on top of each other, up to the initial and final displacements.
(Figure adapted from Ref.~\cite{Pet07a}.)}
\end{figure}

In the case of a uniform phase-space displacement,
the diagonal approximation is even more straightforwardly justified.
This is so because any classical 
trajectory of the unperturbed 
Hamiltonian is also a trajectory of the perturbed Hamiltonian,
up to displacements at the trajectory's ends.
This is illustrated in the right panel of Fig.~\ref{fig:fig2_heller}.
The fact that the action
difference is time-independent here has the important consequence that 
the golden rule decay is replaced by 
a time-independent saturation term. The Lyapunov decay term is left almost
unaffected, as it 
depends on the classical measure of nearby trajectories with perturbed 
initial conditions
and does not depend on quantum action phases. We also note that for 
displacement echoes
there is no Gaussian perturbative decay, since phase space displacements 
do not change the spectrum of the system aside from some possible but 
irrelevant global shift.

Having discussed the justification of the diagonal approximation to the 
displacement echo, we present details of a semiclassical calculation of the displacement
echo in Appendix~\ref{appendix:semiclassics_displ}. Here also, one differentiate between
diagonal and nondiagonal contributions, depending on whether classical paths are
correlated or not. This results in two separate additive contributions to 
${\cal M}_{\rm D} (t )$,
\begin{subequations}
\begin{eqnarray}\label{corr2}
{\cal M}_{\rm D}^{\rm (d)}(t)&=& 
\alpha \, \exp[-({\bf P} \nu)^2/2] \, \exp[-\lambda t] \, ,\\
{\cal M}_{\rm D}^{\rm (nd)}(t) &=& 
\exp[-({\bf P} \nu)^2/2] \; g(|{\bf P}|L)\Big/(|{\bf P}| L)^2,
\end{eqnarray}
\end{subequations}
in terms of
an oscillatory function $g(|{\bf P}|L)=4 \sin^2(|{\bf P}|L/2)$ for $d=1$
and $g(|{\bf P}|L)=4 J_1^2(|{\bf P}|L)$ for $d=2$. For $d=3$, 
$g$ is given by Bessel and Struve functions.
This gives the total displacement echo 
\begin{eqnarray}\label{corrtot}
{\cal M}_{\rm D}(t) =
\exp[-({\bf P} \nu)^2/2] \, \left [\alpha \, \exp[-\lambda t] \, +
\frac{g(|{\bf P}|L)}{(|{\bf P}| L)^2} \right].
\end{eqnarray}

Eq.~(\ref{corrtot}) states that ${\cal M}_{\rm D}(t)$ is
the sum of a time-dependent decaying term of classical origin and a
time-independent term of quantum origin. For larger displacements, the latter can also be obtained 
within RMT. The prefactor
$\exp[-({\bf P} \nu)^2/2] \rightarrow 1$ in the semiclassical limit of constant displacement 
with $\nu \rightarrow 0$. It
is thus of little importance for us here. We see that generically, 
${\cal M}_{\rm D}(t)$ follows a classical exponential decay, 
possibly interrupted by a quantum freeze as long as the displacement
is not too large and $g(|{\bf P}|L)\Big / (|{\bf P}| L)^2 > N^{-1}$.
This fidelity freeze differs from the one found
by Prosen and \v{Z}nidari\v{c} in Ref.~\cite{Pro05a}. In our case,
the spectrum is left exactly unchanged 
by phase-space displacements, i.e. to all orders in perturbation theory.
This is why the freeze of ${\cal M}_{\rm D}(t)$ found here persists up to $t \rightarrow \infty$.
In Ref.~\cite{Pro05a}, only low-order corrections to the spectrum vanish, so that the freeze
is limited in time. We note that in the semiclassical limit,
${\cal M}_{\rm D}(t\rightarrow 0) \rightarrow 1$, 
because of the saturation of 
$\alpha(t \rightarrow 0) \rightarrow 1$
and the disappearance of 
uncorrelated contributions at short times.
Most importantly, there is no displacement- and time-dependent decay, i.e. 
no counterpart to the golden rule decay nor to the perturbative Gaussian decay 
for ${\cal M}_{\rm D}(t)$, because phase-space displacements 
leave the spectrum unchanged, up to a possible irrelevant homogeneous shift.

\begin{figure}
\includegraphics[width=12cm,angle=0]{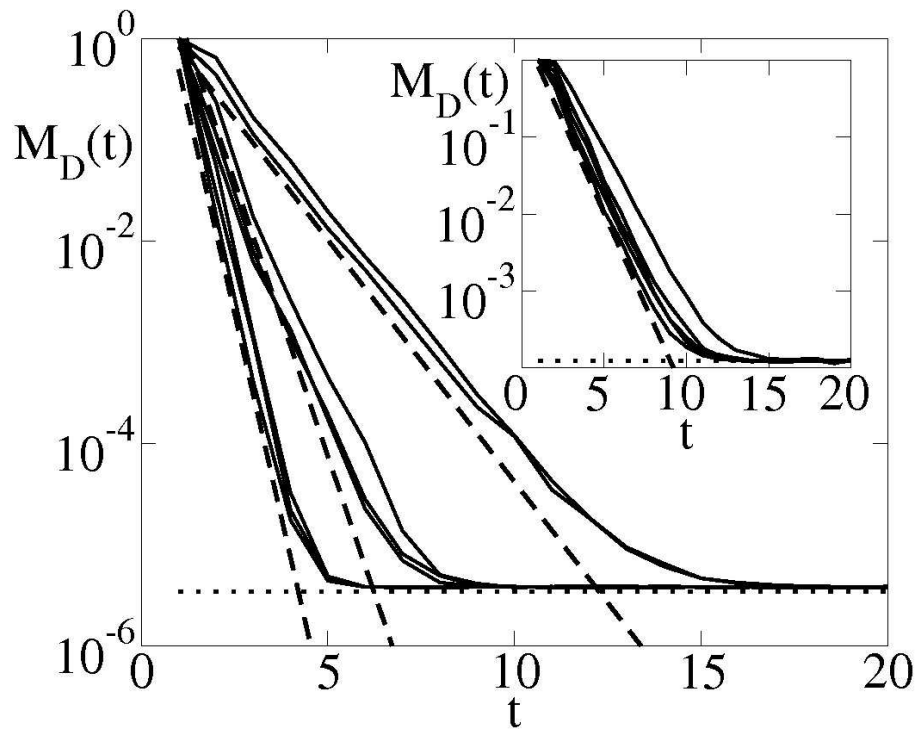}
\caption{\label{fig:fig3_heller} Main plot : Displacement echo ${\cal M}_{\rm D}(t)$ for the kicked rotator 
model with $N=262144$,
and displacements $P= m \times 2 \pi/N$, $m=10,\, 20, \,30$.
Averages have been performed over 10000 different initial coherent
states $\psi_0$. The full lines correspond to kicking strengths $K= 10$, 50 
and $200$ (from right to left). The dashed lines have been slightly shifted for clarity;
they give the predicted exponential decay $\exp[-\lambda_0 t]$ with 
$\lambda_0 = 1.1,\, 2.5,\, 3.7$. 
The dotted line gives the saturation at $N^{-1}.\,$
Inset : Displacement echo for $N=8192$, $K= 10.09 $, and 
displacements $P=2 \pi/N,\, 4 \pi/N, \ldots 10 \pi/N$.
Data are obtained from 1000 different initial coherent states.
The dashed line gives the predicted exponential decay given with
$\lambda_0 = 1.1$.
The dotted line gives the minimal saturation value at $N^{-1}.\,$
(Figure taken from Ref.~\cite{Pet07a}. Copyright (2007) by the American Physical Society.)}
\end{figure}

Displacement echoes are thus seen to be a very special subclass of Loschmidt 
echoes, where the quantum--classical competition between golden rule and 
Lyapunov decays does not take place. As a matter of fact, quantum coherence 
is of little importance for ${\cal M}_{\rm D}$ in the sense that the 
perturbation does not bring interfering paths out of phase. Quantumness
only affects ${\cal M}_{\rm D}$ in that it determines its long-time 
saturation, while the time dependence of ${\cal M}_{\rm D}$
is solely determined by the underlying classical dynamics. Accordingly,
displacement echoes are generically given by the sum of 
a classical decay and a quantum freeze term (\ref{corrtot}). 
Because phase-space displacements do not generate time-dependent
action differences, and because they vanish in
perturbation theory, there is no other time-dependent
decay. This is in strong contrast to the Loschmidt 
echo investigated in earlier chapters. 

\subsubsection{Momentum displacement -- numerical experiments}

We summarize the numerical results of Ref.~\cite{Pet07a} on the kicked rotator
model of Eq.~(\ref{kickrot}). We follow the numerical
procedure described in chapter~\ref{section:k_rot} but this time calculate
${\cal M}_{\rm D}(t)$ as in Eq.~(\ref{decho_repeat}).
We first focus in Fig.~\ref{fig:fig1_heller} on small displacements
$P \le  2 \pi / N$. The inset demonstrates that the
behavior of ${\cal M}_{\rm D}(t)$ clearly
follows Eq.~(\ref{corrtot}), with a quantum freeze at a displacement-dependent
value following a decay with a slope given by the Lyapunov exponent. We 
next show in
the main panel the $P$-dependence 
of the saturation value ${\cal M}_{\rm D}(\infty)$. 
The data fully confirm 
the algebraically damped oscillations
predicted in Eq.~(\ref{corrtot}) and shown as a red dashed line in Fig.~\ref{fig:fig1_heller} for
the case $N=16384$.

Next, we show in Fig.~\ref{fig:fig3_heller}
the behavior of the echo for displacements in the
range $P \gg 2 \pi/N$. In that regime, the uncorrelated
contribution ${\cal M}_{\rm D}^{\rm (nd)}(t) \ll N^{-1}$, it
thus plays no role. It is seen that the decay rate 
of the displacement echo strongly depends on the kicking strength $K$, but
is largely independent of the displacement $P$. We quantitatively found
that in that regime, ${\cal M}_{\rm D}(t) \approx \exp[-\lambda_0 t]$, 
in terms of a reduced Lyapunov exponent $\lambda_0$ 
[see the discussion in Chapter~\ref{section:lyapunov?}]. Most importantly, the absence of
other time-dependent decay allows to observe the Lyapunov decay with values
$\lambda_0$ significantly exceeding the bandwidth $B$. The displacement echo is the best place
in quantum mechanics to date where the Lyapunov exponent of the classical dynamics can
be observed. The inset shows
moreover, that lowering the displacement to the regime $P=m 2\pi/N$ 
with $m \le 5$ does not affect the decay rate of $M_{\rm D}(t)$. This 
confirms that there is no golden rule decay for the displacement echo.

\subsubsection{Spatial displacement -- semiclassical theory}\label{section:spatial_displacement}

Our standard semiclassical approach can be applied to 
Eqs.~(\ref{eq:spatial_displacement}). Compared to the momentum
displacement echo, the only difference is that it is now ore convenient to
use resolutions of identity in momentum space instead of real space, 
accordingly the semiclassical propagators are expressed in terms of
classical trajectories with well-defined initial momentum instead of position.
Eqs.~(\ref{corr2}) and (\ref{uncorrelated}) now become
\begin{eqnarray}\label{corr1_spatial}
{\cal M}_{\rm D}^{\rm (d)}(t)&=& 
\left(\frac{1} {\pi \nu^2}\right)^{d} 
\int {\rm d}{\bf p} {\rm d}{\bf p}^{\prime} \, 
\Theta(\nu^{-1}-|{\bf r}-{\bf r}'|)
\Big\langle \sum_{s} \,
 \tilde{C}_{s}^2\, e^{-\left[ (\mathbf{r}_s-\mathbf{r}_0)^2+(\mathbf{r}_s-\mathbf{r}_0-\mathbf{X})^2 \right]/\nu^2}\Big\rangle, \\
&=& 
\tilde{\alpha} \, \exp[-({\bf X}/\nu)^2/2] \, \exp[-\lambda t],
\end{eqnarray}
where the new determinant $\tilde{C}_s$ now measures the stability of the
spatial endpoint of $s$ upon a change of the initial momentum (instead
of the stability of the final momentum of $s$ as the starting point
is slightly displaced). This gives another prefactor $\tilde{\alpha}$ 
multiplying the Lyapunov decay, which, as $\alpha$ in Eq.~(\ref{corr2}),
is of order one and weakly time-dependent.  

Simultaneously, the uncorrelated contribution to ${\cal M}_{\rm D}$, 
Eqs.(\ref{uncorrelated}), now becomes
\begin{subequations}\label{uncorrelated_spatial}
\begin{eqnarray}
{\cal M}_{\rm D}^{\rm (nd)} (t ) 
&=& f({\bf X}) \; \tilde{\cal M}_{\rm D}^{\rm (nd)} (t) , \\
f({\bf X}) &=& g(|{\bf X}| p_0)\big /(|{\bf X}| p_0)^2, \\
\tilde{\cal M}_{\rm D}^{\rm (nd)}(t) &=& \left( \frac{1} {\pi\nu^2}\right)^{d} 
\left( \int {\rm d}{\bf p}
\sum_{s} \,
 \tilde{C}_{s}\,  
\exp -\frac{1}{2\nu^2}\Big[ (\mathbf{r}_s-\mathbf{r}_0)^2+ 
(\mathbf{r}_s-\mathbf{r}_0-{\bf X})\Big] 
\right)^2, \mbox{}
\end{eqnarray}
\end{subequations}
where $g(x) $ is the same as for the momentum displacement echo.

Summing the correlated and the uncorrelated contributions to ${\cal M}_{\rm D}(t)$ one finally obtains
\begin{eqnarray}\label{corrtot_spatial}
\langle{\cal M}_{\rm D} (t )\rangle &=&  \exp[ -{\bf X }^2/2\nu^2]  \; \; 
\left[ \tilde{\alpha}  \exp[ -\lambda t]  +  \frac {g(\vert {\bf X} \vert p_0 ) }{\left( \vert {\bf X} \vert p_0 \right)^2} \right],
\eea
which is the phase-space symmetric of Eq.~(\ref{corrtot}). This was
expected from the phase-space ergodicity of chaotic systems.

\subsubsection{Displacement echoes -- restoring the golden rule decay with 
external noise}

The absence of any golden rule decay in displacement echoes has to be taken 
with a grain of salt. In any realistic experiment, time-dependent
external sources of noise will affect the time-evolution. Taking them into
account requires to substitute
\begin{equation}
\exp[\pm i H t] \rightarrow {\cal T} \exp[\pm i \{H t \int_0^t dt' 
\Sigma(t')\}],
\end{equation}
in Eqs.(\ref{decho_repeat}) and (\ref{eq:spatial_displacement}).
Accordingly, random action phases are accumulated in the forward 
and backward time-evolutions, which do not cancel each other.
Under the same assumptions as in Eqs.~(\ref{eq:phase}) and (\ref{eq:nondiagdecay})
of a fast decay of phase correlations, one recovers a golden rule decay,
$\propto \exp[-\Gamma t]$ replacing the second term in brackets 
in Eqs.~(\ref{corrtot}) and (\ref{corrtot_spatial}), with $\Gamma$ defined
as in Eq.~(\ref{eq:nondiagdecay}).
If, on the other hand, the external sources of
noise are efficiently screened, this decay becomes slower and Gaussian. In both 
instances, the random phases have to compete with the Lyapunov decay -- this
is the only instance we know of where the alternative to the
exponential Lyapunov decay of the Loschmidt echo is Gaussian and not exponential.  
This is perhaps worth future investigations. 

This brings an end to this section. In the next section we use a phase-space 
representation of quantum mechanics to revisit some of the issues we just 
discussed.

\section{Irreversibility in Phase-Space Quantum Mechanics}\label{section:wigner}

The study of quantum mechanics in phase-space goes back to Weyl~\cite{Wey27,Wey31} and later
Wigner who introduced the phase-space representation
of the density matrix $\rho({\bf x},{\bf y})$~\cite{Wig32}
\begin{equation}\label{Wigner_f}
W_\rho({\bf q},{\bf p};t) = \frac{1}{\pi^d}  \int {\rm d}{\bf x}
\exp[2 i {\bf p} \cdot {\bf x}] \rho({\bf q}-{\bf x},{\bf q}+{\bf x};t).
\end{equation}
Since then, $W_\rho$ has been dubbed the Wigner function~\cite{Hil84}.
It is easily checked that $W_\rho$ is a real function. Because it is nonlocal, 
$W_\rho$ is not necessarily positive, and it is instructive to 
write it as the sum of a positive envelope -- having the meaning of
a probability distribution -- and an oscillating
part, $W_\rho = W_\rho^{\rm cl} + W_\rho^{\rm qm}$, with subscripts
obviously referring to {\it classical} and {\it quantum} parts.
Quantum mechanics can be rephrased using the Wigner function
representation, and following Ref.~\cite{Zur01} 
various investigations have analyzed the Loschmidt echo using 
$W_\rho$~\cite{Ada02,Cuc04a,Kar02}.
Expressed in terms of Wigner functions
$W_\rho^{H_0}$ (propagating with $H_0$) and $W_\rho^H$ (propagating with $H$) 
the Loschmidt echo reads
\begin{equation}\label{loschmidt_wigner}
{\cal M}_{\rm L}(t) = (2 \pi)^d \int {\rm d {\bf q}} \int {\rm d {\bf p}} \;
W_\rho^{H_0}({\bf q},{\bf p};t) W_\rho^H({\bf q},{\bf p};t).
\end{equation}
This latter equation is a special application of the trace product rule,
that the trace of two density matrices is equal to the 
phase space integral of the product of the two
corresponding Wigner functions,
 \begin{equation}\label{tracerule}
 {\rm Tr}\left[ \rho_{a} \rho_{b} \right] = 
(2 \pi )^{d}  \int {\rm d {\bf q}} \int {\rm d {\bf p}}
\; W_{\rho_a}({\bf q},{\bf p}) 
W_{\rho_b}({\bf q},{\bf p}).
\end{equation}
Using the semiclassical propagator for $W_\rho$~\cite{Rio02,Ada02,Dit06},
and splitting the Wigner function into a classical and a quantum part,
it is possible to
identify the classical and quantum coherent contributions to ${\cal M}_{\rm L}$,
and connect them to classical processes in phase-space.
More pedestrian uses of the Wigner representation have also been made in the context of
quantum reversibility and decoherence~\cite{Kar02}. 
It is our purpose in this chapter to review and discuss these phase-space 
investigations of quantum reversibility, and to find out if anything
new can be learned or new predictions made following this approach.

Besides being real-valued, $W_\rho$ is normalized,
\begin{equation}\label{wigner_norm}
\int {\rm d {\bf q}} \int {\rm d {\bf p}} \;
W_\rho({\bf q},{\bf p};t) = 1
\end{equation}
which expresses the conservation of probabilities. 
Moreover, 
if $\rho = |\psi \rangle \langle \psi|$ is pure, one has
\begin{equation}\label{wigner_pure}
(2 \pi)^d \int {\rm d {\bf q}} \int {\rm d {\bf p}}\; 
W^2_\rho({\bf q},{\bf p};t) = 1.
\end{equation}
This latter property is preserved under the Schr\"odinger / von Neumann
time--evolution, however as time goes by, it relies more and more on 
the quantum part $W_\rho^{\rm qm}$ of the Wigner function.
Noting that the off-diagonal elements of $\rho$
appear only in Eq.~(\ref{wigner_pure}) [and not in Eq.~(\ref{wigner_norm})], we can characterize decoherence in systems coupled to an external environment with the decay of
$(2 \pi)^d \int {\rm d {\bf q}} \int {\rm d {\bf p}} \; 
W_{\rho_{\rm red}}^2$, 
with the reduced density matrix $\rho_{\rm red}$ from which the external degrees
of freedom have been removed.
The trace product rule tells us that this quantity is actually nothing else 
but the purity ${\cal P}(t)$ of $\rho_{\rm red}$.

\subsection{Do sub-Planck scale structures matter ?}\label{section:subplanck}

\subsubsection{Why care about sub-Planck scale structures ?}

For pure quantum states, 
the Wigner function differs from the classical Liouville distribution in
that it can exhibit strong oscillations and even become negative. It has
been a known fact for quite some time that 
these oscillations occur on smaller scales, the
larger the total volume occupied by the corresponding wavefunction.
For instance, Ref.~\cite{Ami81} gives the Wigner function for 
a quantum superposition of two distant 
Gaussian wavepackets in one dimension as (we use the notation of
Ref.~\cite{Ami81},
where $2 q_0^2 = \nu^2$)
\begin{subequations}
\begin{eqnarray}
\psi(r) & = & (2 \pi q_0^2)^{-1/4}
\left[\exp(-|r-r_0|^2/4 q_0^2)
+  \exp(-|r+r_0|^2/4 q_0^2) \right], \\
\label{eq:3.6b}
W_\psi(q,p) & = & \exp[-2 \, (p \, q_0)^2] \left[
\exp(-(q-r_0)^2/2 q_0^2) 
+ \exp(-(q+r_0)^2/2 q_0^2) \right. \nonumber \\
&& \left. \;\;\;\;\;\;\;\;\;\;\;\;\;\;\;\;\;\;\;\;\;\; + 2 \cos(p \, r_0) \exp(-q^2/2 q_0^2) \right] .
\end{eqnarray}
\end{subequations}
The first two terms inside the square brackets in 
Eq.~(\ref{eq:3.6b}) are easy to interpret, and would still be there even if
we had considered an incoherent superposition. The third term, however,
finds its origin in the coherence of the superposition. The fact that it 
oscillates is not surprising {\it per se} -- quantum coherence is due to 
phase interferences -- however it is seen that the period of these oscillations
is inversely proportional to the distance $r_0$ between the two wavepackets.
Increasing $r_0$ thus gives more and more 
oscillation strips below a Gaussian envelope of Heisenberg resolution --
one gets structures in the Wigner function on arbitrarily small scales.

This is a very simple observation, which most likely was made before Ref.~\cite{Ami81}.
Yet, it looks like it was not easy to
accept that structures on scales smaller than Planck's constant can
develop from an initially smooth, quantum-mechanically time-evolved
wavefunction. In the words of Berry and Balasz~\cite{Ber79}:

{\it It seems obvious that Wigner's function $W(q, p, t )$ 
cannot follow the increasing complication
of ${\cal C}$} [the corresponding classical distribution of orbits]
{\it as $t \rightarrow \infty$. 
The reason is that quantum functions on phase space can surely
have no detail on areas smaller than $O(h)$, 
whereas ${\cal C}$ develops structure down to
arbitrarily fine scales.}

Even accepting that such structures exist,
one interpretation of the Heisenberg uncertainty principle (perhaps the most natural
one) 
is that phase-space structures on scales smaller than Planck's constant
have no observable consequence. The common wisdom would be then to disregard
sub-Planck phase-space structures as artifact of the Wigner representation,
with no physical content whatsoever. The assertion of 
Zurek~\cite{Zur01} 
that sub-Planck scale structures in the Wigner function enhance the 
sensitivity of a quantum state to an external perturbation,
therefore came out as particularly intriguing~\cite{Alb01}
and even controversial~\cite{Jor01}.
His argument can be summarized as follows. The 
overlap (squared amplitude of the scalar product) of
two pure quantum states $\psi$ and $\psi'$
is given by the phase-space integral of the product of their Wigner functions,
(from now on, we use $W_\psi$ for pure states, and $W_\rho$ for 
mixtures/reduced density matrices)
\begin{equation}
I_{\psi,\psi'} \equiv  |\langle \psi | \psi' \rangle |^2 = 
(2 \pi)^d \int  d{\bf q} \, d{\bf p} \; W_\psi \, W_{\psi'}.
\end{equation}
For an extended quantum state covering a large volume  
$A \gg 1$ of $2d$-dimensional phase space, 
the Wigner function $W_\psi$ exhibits oscillations from quantum interferences
on a scale 
corresponding to an action $\delta S \simeq 1/A^{1/d} \ll 1$ (remember that
we set $\hbar \equiv 1$, so that $A \gg 1$ stands for
$A \gg \hbar^d$).
These sub-Planck scale oscillations are brought out of phase by a
shift $\delta p$, $\delta x$ with $\delta p \cdot \delta x \simeq \delta S 
\ll 1$. Thus a $\psi'$ that is obtained from $\psi$ after 
even a modest phase-space shift
is then nearly orthogonal to $\psi$. Zurek concludes that
sub-Planck structures substantially enhance the sensitivity of a 
quantum state to an external perturbation. That $I_{\psi,\psi'}$ is sensitive to the phase-space
shift after which $\psi'$ is obtained from $\psi$ 
is easily seen without the need to
invoke Wigner functions. Let us, for the sake of the argument's simplicity consider a momentum shift,
$|\psi' \rangle = \exp[i {\bf p} \cdot \hat{\bf r}] |\psi \rangle$, then
one has
\begin{equation}
I_{\psi,\psi'} = \left| \int {\rm d}{\bf r} |\langle \psi | {\bf r} \rangle |^2 \exp[i {\bf p} \cdot {\bf r}] \right|^2,
\end{equation}
i.e. $I_{\psi,\psi'} $ is the Fourier transform of the real-space probability distribution of the wavefunction.
From well-known properties of the Fourier transform, it is straightforward to conclude that,
quite generically, 
the larger the spatial extension of $\psi$, the smaller
the maximal value of ${\bf p}$ for which $I_{\psi,\psi'} $ remains sizeable. The connection between
this real-space picture and Zurek's phase-space picture is not that easy to make, and therefore
we believe that his appealing 
argument deserves to be checked in more details. This is what we do in
this chapter. We follow a three-pronged approach. First we investigate the
Loschmidt echo for dynamically prepared initial states, Eq.~(\ref{prepare_echo}), which was
proposed by Karkuszewski, Jarzynski and Zurek as a direct measure of the importance of 
sub-Planck phase-space structures on irreversibility and 
decoherence~\cite{Kar02}. Second, we study  compass
states similar to those introduced by Zurek in Ref.~\cite{Zur01}, and focus on how fast their
fidelity decay as a function of the distance between the Gaussians forming the compass -- it is
this distance which determines the scale of oscillations in the Wigner function. Third, we
introduce incoherent compass mixtures, which do not contain the rapid phase-space
oscillations in their Wigner function that Zurek's coherent compasses have,
and investigate their (properly normalized) fidelity. If the sub-Planck scale argument holds,
the fidelity of the compass mixtures 
should decay more slowly than that of the coherent compasses. Let us
first present these three quantities and discuss our results before we derive them
rigorously in 
Chapter~\ref{section:preparation} and \ref{pure_vs_mixed}.

\begin{figure}
\includegraphics[width=12cm,angle=0]{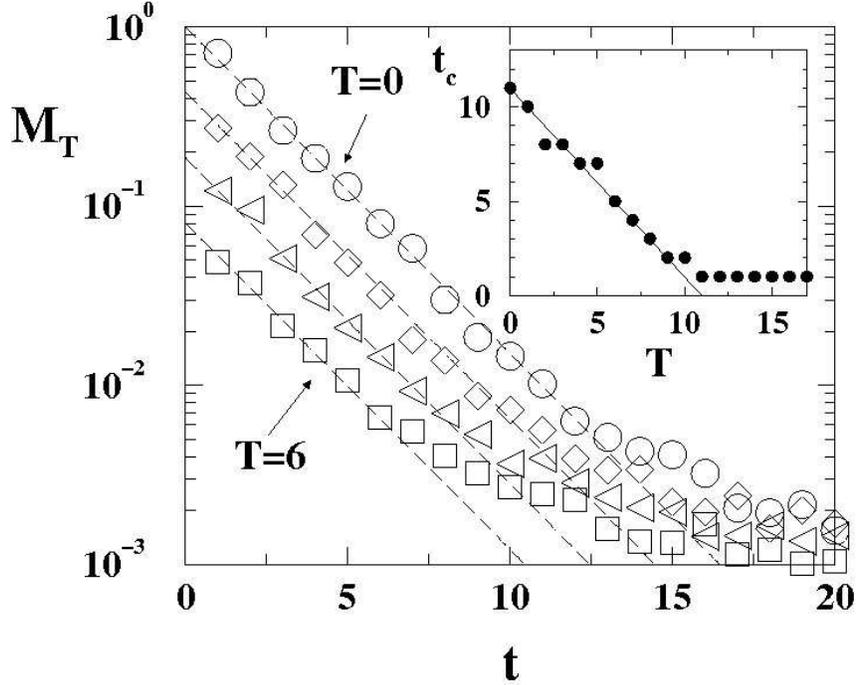}
\caption{\label{fig:fig1_echo2}
Decay of the average fidelity ${\cal M}_{T}$ for the kicked
top with parameters $\phi=1.2 \times 10^{-3}$,
$K=3.9$ and for preparation times
$T=0$ (circles), 2 (diamonds), 4 (triangles), and 6 (squares). In each case,
the dashed lines give the analytical decay 
${\cal M}_T = \exp[-\lambda(t+T)]$,
in the Lyapunov regime with $\lambda=0.42$.
Inset: threshold time $t_c$ at which ${\cal M}_T(t_c) = 10^{-2}$. 
The solid line gives the
analytical behavior $t_c = -\lambda^{-1} \ln {\cal M}_{Tc}-T$.
(Figure taken from Ref.~\cite{Jac02}. Copyright (2002) by the American Physical Society.)}
\end{figure}

\subsubsection{Brief outline of obtained results}

Ref.~\cite{Kar02} proposed to use the Loschmidt echo to investigate the sensitivity to external 
perturbation that sub-Planck scale structures bring about. The size of the structures is tuned by considering
prepared quantum states
$|\psi_T \rangle = \exp[-i H_0 T] |\psi_0 \rangle$, i.e.
initially narrow Gaussian
wavepackets $|\psi_0 \rangle$ which one evolves during a preparation 
time $T$ under the influence of a chaotic Hamiltonian $H_0$. As $T$ grows,
the wavepacket spreads, and for a chaotic $H_0$, $|\psi_T\rangle$
eventually covers the entire available phase-space. This is ensured by ergodicity.
When this happens, oscillations in $W_{\psi_T}$ occur on the
smallest possible scale. Zurek's argument suggests that as $T$ increases,
the fidelity 
\begin{eqnarray}\label{prepare_echo_repeat}
{\cal M}_{T}(t) & = & |\langle \psi_0 | \exp[i H_0 T] \exp[i H t] \exp[-i H_0 t] \exp[-i H_0 T]
|\psi_0 \rangle |^2,
\end{eqnarray}
for dynamically prepared initial states $|\psi_T \rangle$ should decay with $T$ and eventually
reach its minimum faster with $t$ at larger $T$.
More generally, we could prepare 
$|\psi \rangle =\exp(-i H_T T) |\psi_0 \rangle$ 
with a chaotic Hamiltonian $H_T$
that is different from $H_0$ and $H$. We assume $H_T=H_0$
for ease of notation, but the results we are about to present remain the same, regardless of this choice,
up to a possibly different Lyapunov exponent $\lambda_T$ for the preparation Hamiltonian
$H_T$.

We investigated ${\cal M}_{T}(t) $ in Ref.~\cite{Jac02} and, 
as a matter of fact, we found that the decay of ${\cal M}_T(t)$ can be accelerated with longer
preparation times $T$ of the initial state. This
is illustrated on Fig.~\ref{fig:fig1_echo2}, where numerical data 
for ${\cal M}_{T}(t)$ are shown with four different preparation times $T$, all
other parameters being kept fixed. One clearly sees that data for larger $T$ lie below those
with shorter $T$. 
This could be interpreted as
a confirmation of the above sub-Planck scale argument. The situation is more
complicated, however. Ref.~\cite{Jac02} found that one can accelerate the decay of the fidelity with 
the preparation time only when, for $T=0$, one has a Lyapunov decay of
${\cal M}_{\rm L}$. This is analytically shown below in Chapter~\ref{section:preparation}, see
Eq.~(\ref{prepdecay}).
The preparation leads to the disappearance of the
Lyapunov decay, in other words, it suppresses the classical contribution
to the Loschmidt echo, but has no effect on the quantum coherent
golden rule decay -- the latter is insensitive to the choice of initial
state (prepared or Gaussian wavepacket), as is shown by a semiclassical analysis, which we
corroborate by both RMT and numerics. 
We conclude that the accelerated decay with the preparation time $T$ is 
not due to the generation of sub-Planck scale structures.

It would however be premature to conclude that sub-Planck scale structures
have no effect on the decay of the Loschmidt echo.
Taking our inspiration from Ref.~\cite{Zur01} we therefore perform a second analysis 
on the compass states considered there. Compass states are coherent superpositions of 
four Gaussian wavepackets, 
\begin{eqnarray}\label{compass:pure}
\psi_{\rm c} ({\bf r}) 
& = & 
\frac{1}{2 \, (\pi \nu^2)^{d/4}}
\left\{
\exp[-|{\bf r}-{\bf r}_0|^2/2 \nu^2] +
\exp[-|{\bf r}+{\bf r}_0|^2/2 \nu^2] \right. \nonumber \\
&& \left. \;\;\;\;\;\;\;\;\;\;\;\;\;\; +
\exp[i {\bf p}_0 \cdot {\bf r}-|{\bf r}|^2/2 \nu^2]+
\exp[-i {\bf p}_0 \cdot {\bf r}-|{\bf r}|^2/2 \nu^2]
\right\}.
\end{eqnarray}
Here, we assumed that $|{\bf r}_0| \gg \nu$ so that the overlap between the Gaussians is negligible.
When this condition is not satisfied, the normalization prefactor in Eq.~(\ref{compass:pure})
has to be adapted.
With ${\bf p}_0 = {\bf r}_0/\nu^2$ (again with $\hbar \equiv 1$),
the four Gaussians form a compass rose on a two-dimensional phase-space
hyperplane (defined by ${\bf p}_0$ and ${\bf r}_0$) of phase-space. This 
is sketched in Fig.~\ref{fig:zurek}.

\begin{figure}
\includegraphics[width=12cm,angle=0]{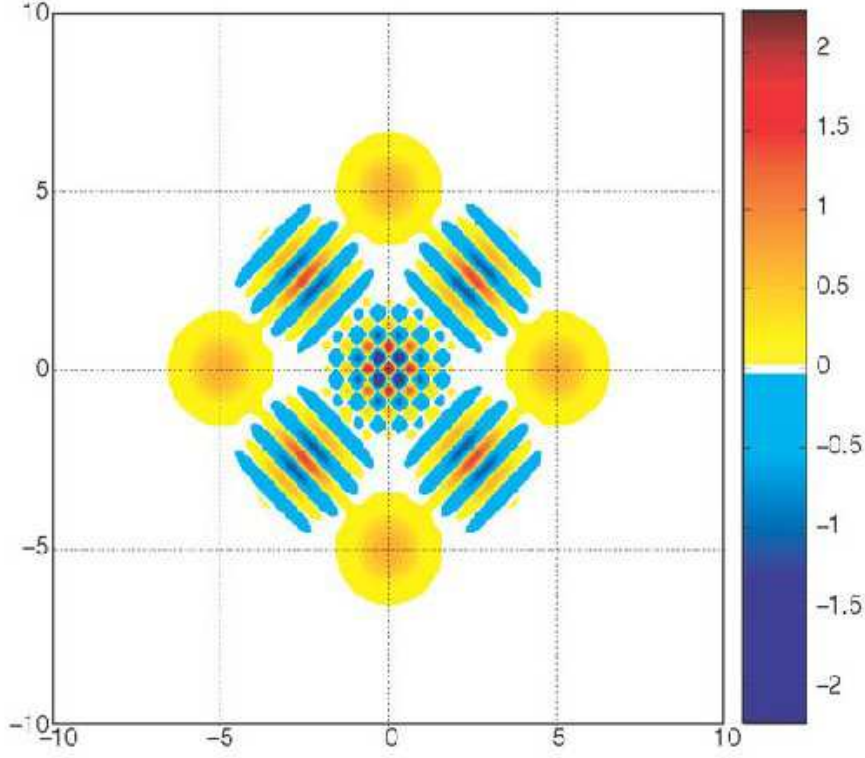}
\caption{\label{fig:zurek} Wigner representation of the
pure compass state of Eq.~(\ref{compass:pure}). The coherence of the
superposition is reflected in the oscillating patterns lying in-between 
the four Gaussian wavepackets (yellow circles). The checkerboard pattern
in the middle of the figure exhibits oscillations with smaller and smaller period as
the distance between the Gaussians increases. Eventually, the 
``squares'' of the central checkerboard cover an area smaller
than Planck's constant.
(Figure taken from Ref.~\cite{Zur01}, with permission.)}
\end{figure}

The Wigner function
for such compass states develops finer and finer structures as the distance 
between the Gaussians increases. Let us then investigate the fidelity ${\cal M}_{\rm L}(t)$
for the specific choice of compass states as initial states $\psi_0$, and look at how
${\cal M}_{\rm L}(t)$ decays as a function of the distance ${\bf r}_0$ (or equivalently ${\bf p}_0$) 
between the four Gaussians forming the compass.   
Again applying Zurek's argument, one
expects a faster decay of ${\cal M}_{\rm L}$ at larger ${\bf r}_0$.
This is confirmed in Fig.~\ref{fig:fig1_compass}, where averages of ${\cal M}_{\rm L}(t)$ are
performed for ensembles of compass states randomly distributed in phase-space with fixed
${\bf r}_0$. One sees that compasses with larger ${\bf r}_0$ decay faster, 
however, the slope of the asymptotic (in this case, golden rule) decay is the same, 
regardless of the distance between the Gaussians. These numerics show
that varying ${\bf r}_0$ 
affects only the initial short-time transient, which is slower when the Gaussian wavepackets
forming the compass state are closer, and faster when they are further apart. We warn the
reader not to be fooled
by the logarithmic vertical scale used in the main panel of Fig.~\ref{fig:fig1_compass} -- it
is erroneous to conclude
that increasing the phase-space extension of compass states 
has only a minute effect on the decay of ${\cal M}_{\rm L}(t)$.
The inset to Fig.~\ref{fig:fig1_compass} shows the same data as in the main panel, this time
on a normal scale, and it clearly demonstrates that varying ${\bf r}_0$ can have a critical 
impact -- it is, for instance, 
solely responsible for a decay over half of the total range of ${\cal M}_{\rm L}(t)$, 
from ${\cal M}_{\rm L}(t=2) = 1$ (violet curve with $d=\pi/1250$) to
${\cal M}_{\rm L}(t=2)=0.5$ (black curve with $d=\pi/2$).

\begin{figure}
\includegraphics[width=12cm,angle=0]{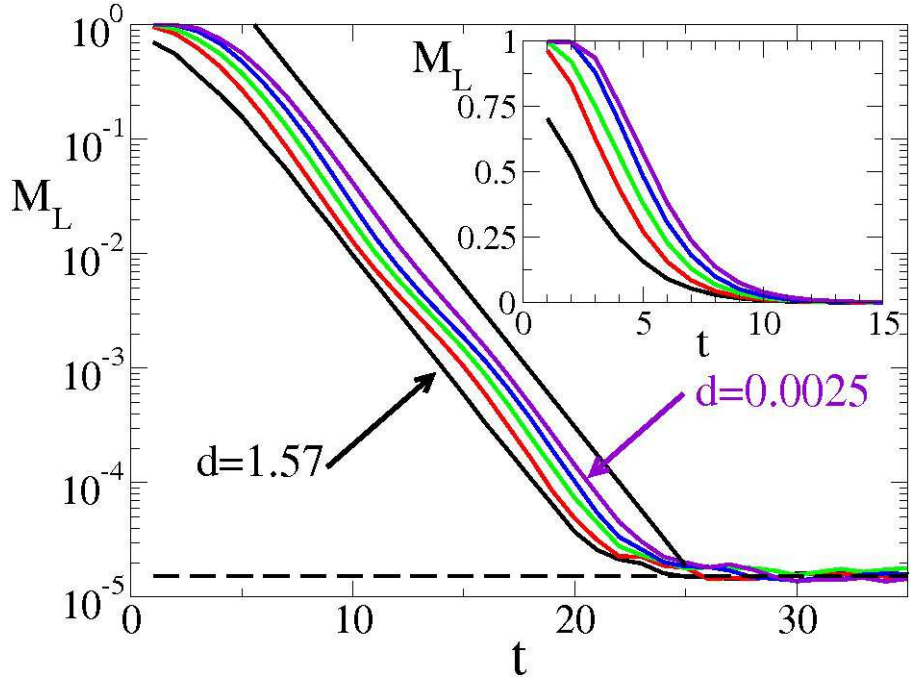}
\caption{\label{fig:fig1_compass} Decay of the Loschmidt echo
${\cal M}_{\rm L}$ for pure compass states $\psi_0$ 
separated by diagonal phase-space
distances $d=\pi/(2 \cdot 5^n)$, with $n=0$ (black), $n=1$
(red), $n=2$ (green), $n=3$ (blue), and
$n=4$ (violet). The model is the kicked rotator with
$K=9.95$, $\delta K = 7 \cdot 10^{-5}$, and $N=65536$. Data correspond to 
averages over 150 initial compass states, randomly positioned in phase-space,
but with fixed intergaussian distance $d$ [see Eq.~(\ref{compass:pure})]. 
The dashed line gives the
saturation at ${\cal M}_{\rm L}(\infty) = N^{-1}$ and the
solid line is a guide to the eye giving the decay $\exp[-\Gamma t]$, with
$\Gamma = 0.024 (\delta K N)^2 \simeq 0.56$. Only the initial transient depends on $d$, and one
sees that the asymptotic regime is entered earlier for larger $d$. Inset: the same data
as in the main panel with a normal instead of a logarithmic vertical axis.}
\end{figure}

So far, we looked at the decay of the fidelity as a function of the phase-space 
extension of initial pure states. All such initial states 
exhibit phase-space oscillations, because they are pure quantum superpositions.
To supplement these investigations we try and remove phase-space oscillations as much
as we can.
The fine structures in the Wigner function disappear if, instead of a coherent
superposition of four Gaussians we take a compass mixture
\begin{eqnarray}\label{compass:mixture}
\rho_{\rm cm} ({\bf r},{\bf r}') & = & \frac{1}{4 (\pi \nu^2)^{d/2}}\left\{
\exp[-(|{\bf r}-{\bf r}_0|^2+|{\bf r}'-{\bf r}_0|^2)/2 \nu^2] +
\exp[-(|{\bf r}+{\bf r}_0|^2+|{\bf r}'+{\bf r}_0|^2)/2 \nu^2] \right. \nonumber \\
&&  \;\;\;\;\;\;\;\;\;\;\;\;\;\; +
\exp[i {\bf p}_0 \cdot ({\bf r}-{\bf r}')-(|{\bf r}|^2+|{\bf r}'|^2)/2 \nu^2] 
\nonumber \\
&& \;\;\;\;\;\;\;\;\;\;\;\;\;\; +\left.
\exp[i {\bf p}_0 \cdot ({\bf r}'-{\bf r})-(|{\bf r}|^2+|{\bf r}'|^2)/2 \nu^2] \, \right\}.
\end{eqnarray}
Such a state differs from the compass states of Eq.~(\ref{compass:pure}) in that there is
no coherence between the separate Gaussians wavepackets forming
the state. In other words, pure compass states given by Eq.~(\ref{compass:pure})
read $\psi_c = \psi_N+\psi_S+\psi_E+\psi_W$ (with labels corresponding to the four cardinal
points) and the corresponding density matrix $\rho_{\rm cp} = |\psi_c \rangle \langle \psi_c|$ 
contains terms involving Gaussians at different cardinal points, i.e. $|\psi_N \rangle \langle \psi_S|$,
$|\psi_N \rangle \langle \psi_E|$
and so forth. This is not the case
for the compass mixture of Eq.~(\ref{compass:mixture}) which corresponds to 
$\rho_{\rm cm} = |\psi_N \rangle \langle \psi_N| +  |\psi_S \rangle \langle \psi_S| + 
 |\psi_E \rangle \langle \psi_E| +  |\psi_W \rangle \langle \psi_W|$. The phase-space picture 
for that mixture corresponds to the one in Fig.~\ref{fig:zurek} without the short-scale oscillations
between any two Gaussians (yellow circles).
 
Because the initial density matrix $\rho_{\rm cm}$  is a mixture we 
normalize the Loschmidt echo in
this case as
\be\label{eq:LE_mixture}
{\cal M}_{\rm L}(t) = 4 \, {\rm Tr}\big[\exp[-i H_0 t] \rho_{\rm c} \exp[i H_0t]
\, \exp[-i H t] \rho_{\rm c} \exp[i H t] \big],
\ee
to have ${\cal M}_{\rm L}(t=0)=1$ 
(this is fine as long as one can neglect the overlap between different 
Gaussians). This normalization does not affect the decay rate
of ${\cal M}_{\rm L}(t)$ but is introduced to facilitate direct comparison between the decays
of initial pure and mixed states. Eq.~(\ref{eq:LE_mixture}) gives a perfectly reasonable measure
of irreversibility for the specific mixtures defined in Eq.~(\ref{compass:mixture}).

The sub-Planck scale argument predicts that the
Loschmidt echo for pure compass states decays faster than it does
for the compass mixtures -- or that the latter are more easily reconstructed after an imperfect
time-reversal operation.
This is confirmed in Fig.~\ref{fig:fig2_compass}, where one clearly sees that, all other parameters
being kept constant, the Loschmidt echo for compass mixtures is always rather significantly 
larger than its counterpart for pure compass states. However, once again
the slope of
the asymptotic decay is the same for a pure compass state and a compass 
mixture. Only the short-time transient is affected by the presence or
absence of short-scale structures in the Wigner function. Below we
present analytical calculations corresponding to the numerical experiments
in Figs.~\ref{fig:fig1_echo2} and \ref{fig:fig2_compass}. These calculations
do not rely on phase-space considerations, yet, they 
perfectly agree with our numerical data. 
Sub-Planck scale arguments seem to be inspiring to many, however we feel more
comfortable with the calculations we are about to present. We are unaware of any numerical
nor analytical observation made in investigations of the Loschmidt echo and its offspring
that cannot be quantitatively captured by our semiclassical and RMT approaches.

\begin{figure}
\includegraphics[width=12cm,angle=0]{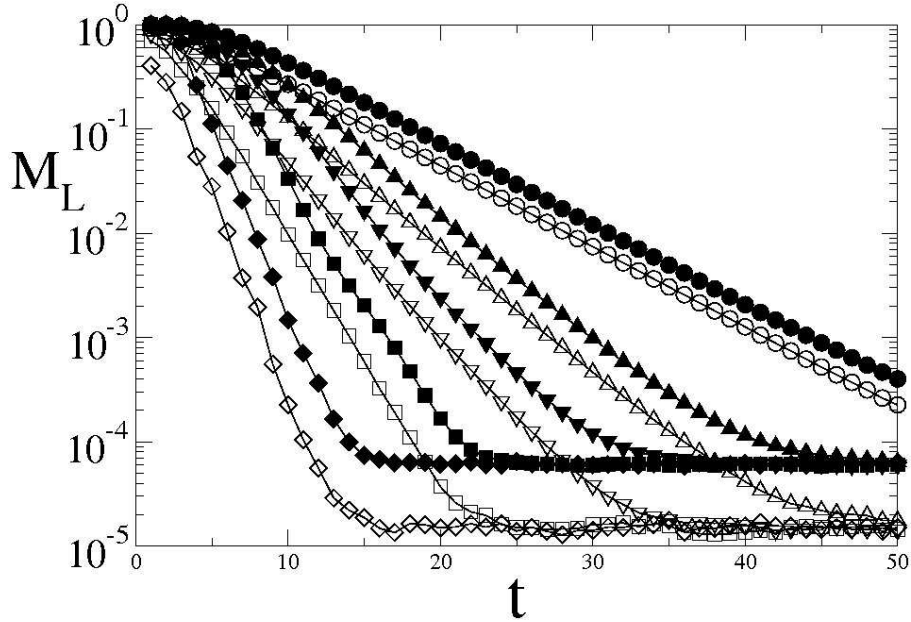}
\caption{\label{fig:fig2_compass} Decay of the Loschmidt echo
${\cal M}_{\rm L}$ for pure (open symbols) initial compass states
$\psi_0$ as given in Eq.~(\ref{compass:pure}) and for
mixed (full symbols) initial compass density matrix $\rho_0$
of Eq.~(ref{compass:mixture}). In both cases, the diagonal phase-space distance between 
the center of masses of the Gaussians forming the compass is $d=\pi/2$.
The model is the kicked rotator of Eq.~(\ref{kickrot}) with $N=65536$ and
$K=9.95$, $\delta K = 4 \cdot 10^{-5}$ (circles), 
$\delta K = 5 \cdot 10^{-5}$ (triangles up),  
$\delta K = 6 \cdot 10^{-5}$ (triangles down), 
$7 \cdot 10^{-5}$ (squares), and 
$2 \cdot 10^{-4}$ (diamonds).
Data correspond to 
averages over 250 initial states.}
\end{figure}

\subsubsection{The Loschmidt echo with chaotically prepared initial states}\label{section:preparation}

We start our analytical investigations of the impact that fine phase-space structures have on the
decay of the Loschmidt echo with a semiclassical calculation of the fidelity ${\cal M}_T(t)$
for dynamically prepared initial states. This treatment is later complemented with a RMT
calculation. 

The Lyapunov decay for ${\cal M}_{\rm L}(t)$ sensitively depends 
on the choice of
an initial narrow wavepacket $\psi_0$. For example, if $\psi_0$ is
a coherent superposition of $M$ nonoverlapping wavepackets, the diagonal
Lyapunov contribution to ${\cal M}_{\rm L}$ 
is reduced by a factor $1/M$, while the
golden rule contribution remains the same. Does the same phenomenon
occur for prepared initial states $\psi_T = \exp(-i H_{0} T) \psi_0$,
which for large $T$ can be seen as random superpositions of a large number
of overlapping Gaussians ? For an initial Gaussian wavepacket $\psi_0$,
the semiclassical approximation to Eq.~(\ref{prepare_echo_repeat}) gives
\begin{equation}\label{semiclt}
{\cal M}_T(t) = \Big |\int d{\bf r}
\sum_{s} [K_{s}^{H_\tau}({\bf r},{\bf r}_0;t+T)]^* 
K_{s}^{ H_0}({\bf r},{\bf r}_0;t+T) \exp[-\nu^2|{\bf p}_s-{\bf p}_0|^2 ]
\Big |^2 ,
\end{equation}
instead of Eq.~(\ref{semicl}). Here, one has a
time-dependent Hamiltonian $H_\tau=
H_0$ for $\tau < T$ and $H_\tau=H$ for $\tau>T$. 
We can apply the same analysis as above in chapter~\ref{section:semicl}
to the time-dependent Hamiltonian. Only the time interval $(T,t+T)$ of
length $t$ leads to a phase difference between $K_{s}^{H_\tau}$ 
and $K_{s}^{ H_0}$, because $H_\tau=
H_0$ for $\tau < T$. Hence the nondiagonal contribution 
${\cal M}^{\rm (nd)}_{T}(t)$ to 
${\cal M}_{T}(t)$, which is entirely due to this phase difference,
still decays $\propto \exp(-\Gamma t)$, independent of the 
preparation time $T$. This conclusion can also be reached with RMT,
according to which the averages given in Eqs.~(\ref{contractions}) 
do not depend on $\psi_0$.

The preparation does however have an effect on the 
diagonal contribution ${\cal M}_{T}^{({\rm d})}(t)$ to the fidelity. It
decays $\propto \exp[-\lambda (t+T)]$ instead of $\propto \exp(-\lambda t)$,
provided $t$, $T \gg \lambda^{-1}$. This is most easily
seen from the expression
\begin{eqnarray}\label{semicld}
{\cal M}^{({\rm d})}_T(t)&=&\int d{\bf r} \sum_{s} 
|K_{s}^{H_\tau}({\bf r},{\bf r}_0;t+T)|^2 \;
|K_{s}^{H_0}({\bf r},{\bf r}_0;t+T)|^2, 
\end{eqnarray}\noindent by following a path 
from its endpoint ${\bf r}$ to an intermediate point
${\bf r}_i$ reached after a time $t$. The time-evolution
from ${\bf r}$ to ${\bf r}_i$ leads to an exponential decrease
$\propto \exp(-\lambda t)$ as in Ref. \cite{Jal01}. Due to the classical
chaoticity of $H_0$, the subsequent evolution from
${\bf r}_i$ to ${\bf r}_0$ in a time $T$ brings in
an additional prefactor $\exp(-\lambda T)$. 
The combination of 
diagonal and nondiagonal contributions therefore results in the bi-exponential
asymptotic decay
\begin{equation}\label{prepdecay}
{\cal M}_T(t) \propto \exp(-\Gamma t) + \alpha \exp[-\lambda (t+ 
T)],
\end{equation}
\noindent with, as always, prefactors of order one multiplying each exponential
[see also the discussion following Eq.~(\ref{eq:final_decay}) above].
The Lyapunov decay prevails if $\Gamma > \lambda$ and $t>\lambda 
T/(\Gamma - \lambda)$, while the golden rule decay dominates if either
$\Gamma < \lambda$ or $t < \lambda T/(\Gamma -\lambda)$.
In both regimes the decay saturates when ${\cal M}_{T}$ has reached its 
minimal value $\hbar_{\rm eff}$. 
In the Lyapunov regime, this saturation
occurs at the
Ehrenfest time. When the preparation time $T \rightarrow \tau_{\rm E}$,
we have a complete decay within a time $\lambda^{-1}$ of the fidelity
down to its minimal value.

We give numerical confirmation to these analytical results.
We take the kicked top model defined in Eqs.~(\ref{H0def})
and (\ref{H1def}), and, as in chapter~\ref{kicked_top},
we choose $\psi_0$ as a coherent state of the spin SU(2) group.
The state is then prepared as $\psi_T = \exp(-i H_0 T) \psi_0$. 
We can reach the Lyapunov regime by 
selecting initial wavepackets centered 
in the chaotic region of the mixed phase
space for the Hamiltonian~(\ref{H0def}) with kicking 
strength $K=3.9$~\cite{Jac01}.
Fig.~\ref{fig:fig1_echo2}
gives a clear confirmation of the predicted decay 
$\propto \exp[-\lambda (t+T)]$ in the Lyapunov regime.
The additional decay 
induced by the preparation time $T$ can be quantified via 
the time $t_c$ it takes for ${\cal M}_T$ to reach a given
threshold ${\cal M}_{Tc}$. From the Lyapunov decay we expect
\begin{equation}\label{ttc}
t_c = -\lambda^{-1} \ln {\cal M}_{Tc}-T,
\end{equation}
provided ${\cal M}_{Tc} > \hbar_{\rm eff} = (2 S)^{-1} = 10^{-3}$ 
and $T < -\lambda^{-1} \ln {\cal M}_{Tc}$.
In the inset to Fig.~\ref{fig:fig1_echo2} we confirm this formula
for ${\cal M}_{Tc}=10^{-2}$. As expected, $t_c$ saturates at the first kick 
($t_c=1$) when $T \simeq -\lambda^{-1} \ln {\cal M}_{Tc} < \tau_{\rm E} = 
\lambda^{-1} \ln (2 S)$. Numerical 
results qualitatively 
similar to those shown in the inset to Fig.~\ref{fig:fig1_echo2}
were obtained in Ref. \cite{Kar02}.
This similarity is only qualitative, mainly because of the much
larger value ${\cal M}_{Tc}=0.9$ chosen in Ref.~\cite{Kar02}. For values of
${\cal M}_{Tc}$ close to 1, we expect that we can do perturbation theory
in $t$ which gives ${\cal M}_{T}(t) = 1-\exp(\lambda T) \sigma^2 t^2$, 
and hence $t_c = \sqrt{1-{\cal M}_{Tc}} \exp(-\lambda T/2)/\sigma$. Analyzing
the data presented in Fig. 2 of Ref. \cite{Kar02} gives the quite
realistic values
$\sigma \approx 0.042$ and $\lambda \approx 0.247$.

\begin{figure}
\includegraphics[width=12cm,angle=0]{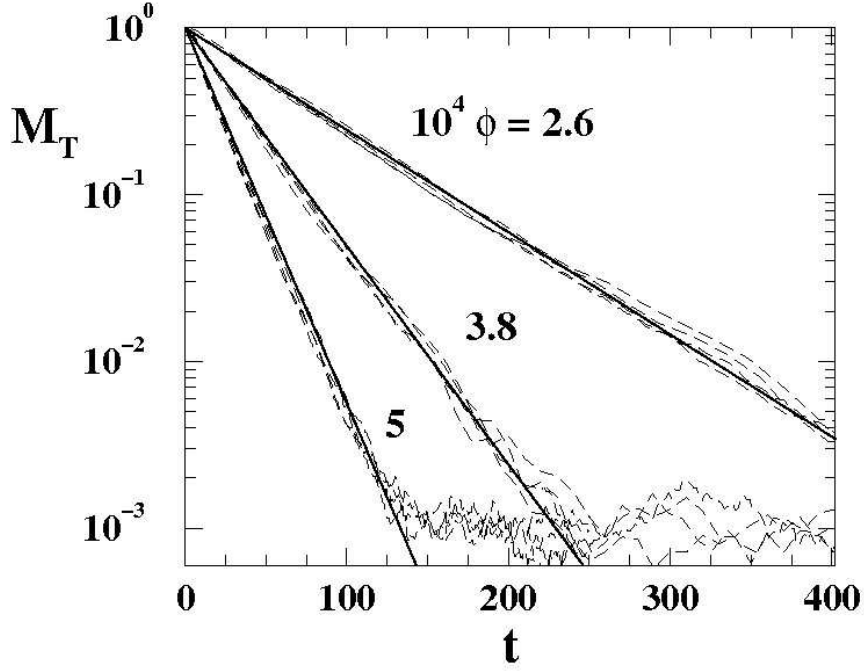}
\caption{ \label{fig:fig2_echo2}
Decay of ${\cal M}_T$ in the golden rule regime for 
$\phi=2.6 \times 10^{-4}$,
$3.8 \times 10^{-4}$, $5 \times 10^{-4}$, $K=13.1$, and for 
preparation times $T=0$, 5, 10, and 20 (nearly indistinguishable dashed
lines). The solid lines give the corresponding 
golden rule decay with $\Gamma = 0.84 \; \phi^2 S^2$ as obtained for
the kicked top in chapter~\ref{kicked_top}.
(Figure taken from Ref.~\cite{Jac02}. Copyright (2002) by the American Physical Society.)}
\end{figure}

We next illustrate the independence of ${\cal M}_{T}(t)$ 
on the preparation time $T$ in the golden rule regime, i.e. at larger
kicking strength $K$ when $\lambda > \Gamma$.
As shown in Fig.~\ref{fig:fig2_echo2}, the decay 
of ${\cal M}_{T}(t)$ is the same 
for the four different preparation times $T=0$, 5, 10, and 20. 
For these data, we estimate the Ehrenfest time as $\tau_{\rm E} \approx 7$,
so that increasing $T$ further does not increases the complexity of 
the initial state.

These numerical data give a clear confirmation of the 
semiclassical result (\ref{prepdecay}).
As summarized above in Table~\ref{table:table2}, there are five
different regimes for the decay of 
the Loschmidt echo in chaotic systems, and since
only two of them are captured by the semiclassical approach we used in this chapter, 
we finally argue that the chaotic preparation does not affect
the remaining three. The five regimes correspond to different
decays: \\
(i) Parabolic decay,
${\cal M}_{\rm L}(t) = 1-\sigma_0^2 t^2$, with
$\sigma_0^2 \equiv \langle\psi_0|\Sigma^2|\psi_0\rangle
-\langle\psi_0|\Sigma|\psi_0\rangle^2$, which
exists for any perturbation strength at short enough times. \\
(ii) Gaussian decay, ${\cal M}_{\rm L}(t) \propto 
\exp(-\sigma_1^2 t^2)$,
valid if $\sigma_1 \equiv \overline{\langle\alpha^{(0)}|\Sigma^2|\alpha^{(0)}\rangle}
-\overline{\langle\alpha^{(0)}|\Sigma|\alpha^{(0)}\rangle^2}$ is much smaller than the 
level spacing $\delta$. (As before, $\{\alpha^{(0)}\}$ is the set of eigenvectors of $H_0$.)\\
(iii) Golden rule decay, ${\cal M}_{\rm L}(t) \propto \exp(-\Gamma t)$, with
$\Gamma \simeq 2 \pi \overline{|\langle\alpha^{(0)}|\Sigma|\beta^{(0)}\rangle|^2}/\delta$, if $\delta \lesssim \Gamma \ll \lambda$.\\
(iv) Lyapunov decay, ${\cal M}_{\rm L}(t) \propto \exp(-\lambda t)$, if $\lambda < \Gamma$.\\
(v) Gaussian decay, ${\cal M}_{\rm L}(t) \propto \exp(-B^2 t^2)$, if $\Sigma$
is so large that $\Gamma$ is  larger than the energy bandwidth $B$ of $H$. 

We already saw that all these regimes, except regime (iv), can be dealt with 
quantum mechanically under the sole assumption that $H_0$ and $H$ are 
classically chaotic, using RMT. Using Eqs.~(\ref{contractions}), 
it is straightforward to show that the decay of the average fidelity in the 
three quantum regimes (ii), (iii), and (v) does not depend on the choice 
of the initial state, so that $\psi_0$ and $\exp[i H_0 T] \psi_0$
give the same average decay.

The faster decay of the Loschmidt echo with chaotic preparation of the initial
state was interpreted in Ref.~\cite{Kar02} as the accelerated decay resulting
from sub-Planck scale structures. The analysis presented 
in Ref.~\cite{Jac02}, and which we reproduce here suggests that in our numerics, 
we observe the same phenomenon. However, the fact that our numerical data is
described so well by Eq.~(\ref{ttc}) points to a classical rather than a
quantum origin of the decay acceleration. Indeed, Eq.~(\ref{ttc}) contains 
only the classical Lyapunov exponent as a system dependent parameter,
so that it cannot be sensitive to any fine structure in phase
space resulting from quantum interference.

\subsubsection{Pure compass states vs. compass mixtures}\label{pure_vs_mixed}

For a quantum superposition of $M$ nonoverlapping
Gaussian wavepackets
$\psi_0 = M^{-1/2} \sum_\alpha \phi_\alpha$, the Loschmidt echo reads
\be
{\cal M}_{\rm L,pure}(t) =  \big|M^{-1} \sum_{\alpha,\beta} 
\langle \phi_\alpha | \exp[i Ht] \exp[-i H_0 t] | \phi_\beta \rangle \big|^2.
\ee
This has to be contrasted with the normalized 
Loschmidt echo (\ref{eq:LE_mixture}) for mixed initial states
\be
{\cal M}_{\rm L,mixed}(t) = M^{-1} \sum_{\alpha,\beta}  \big|
\langle  \phi_\alpha | \exp[i Ht] \exp[-i H_0 t] |  \phi_\beta \rangle \big|^2.
\ee
The difference between these two quantities is best emphasized at short
times, where perturbation theory gives
\be
\delta {\cal M}(t) \equiv {\cal M}_{\rm L,mixed}(t)-{\cal M}_{\rm L,pure}(t) \simeq  M^{-1}
\sum_{\alpha \ne \beta} \langle  \phi_\alpha|
\Sigma^2 |  \phi_\beta \rangle t^2  \ge 0.
\ee
We see that the transient decay is slower, and therefore lasts longer
for the mixture. 
This agrees with Fig.~\ref{fig:fig2_compass}, where we have compass states
with $M=4$. We see that the asymptotic decay rate is the same, regardless
of whether one has a pure state or a mixture, however, the initial transient is
sensitive to that difference and is slower for mixtures (black symbols) than for
pure states (empty symbols).
Also in this figure, one sees that the asymptotic decay is the same,
regardless of whether the initial state is pure or mixed. We can estimate
$\delta {\cal M}(t)$ under our standard RMT assumptions that the wavepackets forming the
initial pure or mixed states obey similar relations as in Eq.~(\ref{contractions})
when they are projected onto the
eigenfunctions of the perturbation $\Sigma$. In terms of the variance $\epsilon^2$ of the
spectrum of $\Sigma$, one gets
\begin{equation}\label{dm}
\delta {\cal M}(t) \simeq \epsilon^2 (M-1) t^2.
\end{equation}
This is quite surprising -- at short times, the difference between the fidelities of a compass mixture and
of the corresponding pure state is proportional to the number $M-1$ 
of Gaussians the states are made of minus one. This has to be taken
with a grain of salt, of course, and since 
${\cal M}_{\rm L}(t) \in [0,1]$ is bounded, so is $\delta {\cal M}(t)$. In other words Eq.~(\ref{dm}) means
that the parametric border for the validity of the short-time perturbative regime depends on the
purity of the chosen initial compass state. The initial transient decays, in each situations, read
\begin{eqnarray} 
{\cal M}_{\rm L,pure}(t) & \simeq & 1- M^{-1} \sum_{\alpha,\beta} \langle  \phi_\alpha|
\Sigma^2 |  \phi_\beta \rangle t^2 \approx 1- M \epsilon^2 t^2 \, , \\
{\cal M}_{\rm L,mixed}(t) & \simeq & 1- M^{-1} \sum_\alpha \langle  \phi_\alpha|
\Sigma^2 |  \phi_\alpha \rangle t^2 \approx 1-\epsilon^2 t^2 \, .
\end{eqnarray}
These expressions, being the result of short-time perturbation theory, 
are only valid as long as the predicted ${\cal M}_{\rm L}(t)$ is still of order one. Thus one has two
different parametric borders for the breakdown of the initial parabolic 
transient. The latter prevails for $t < t_c$ with
\begin{subequations}\label{eq:thresholds}
\begin{eqnarray}
t_{c, {\rm pure}} &\propto& \epsilon^{-1} M^{-1/2}  \, , \;\;\;\;\; {\rm for} \; {\rm pure} \; {\rm states}\,  \\
t_{c, {\rm mixed}} &\propto& \epsilon^{-1}  \, , \;\;\;\;\;\;\;\;\;\;\;\; {\rm for} \; {\rm mixtures} \, .
\end{eqnarray}
\end{subequations}
Once $t_c$ is reached, one enters
the asymptotic decay and Eqs.~(\ref{eq:thresholds}) predict that 
the coherent superposition enters the asymptotic decay faster than the mixtures. 
When the asymptotic regime
is the exponential golden rule decay one predicts that the ratio of the
two fidelities is given by
\begin{equation}
\frac{{\cal M}_{\rm L,mixed}(t)}{{\cal M}_{\rm L,pure}(t)} \propto \exp[\Gamma (t_{c, {\rm mixed}}-t_{c, {\rm pure}} )]. 
\end{equation}
For the kicked rotator model we have been using, one has $\Gamma = 0.024 (\delta K \cdot N)^2$
with $\epsilon = \delta K$,
and this predicts 
${\cal M}_{\rm L,mixed}(t) \big/ {\cal M}_{\rm L,pure}(t) \approx \exp[a \cdot \delta K ]$ with some
constant $a$ depending on the Hilbert space size $N$ only. The data presented in 
Fig.~\ref{fig:fig2_compass} are consistant with this prediction with $a \approx 0.25$.

Both semiclassical
theory and RMT can be applied to the Loschmidt echo for 
pure (\ref{compass:pure}) or mixed (\ref{compass:mixture}) compass 
states in the asymptotic regime after the initial transient, and we now proceed to show that the
decay rates predicted by both methods 
does not depend on the purity of the initial compass state one choses. 
For a chaotic time-evolution one obtains
\bea\label{eq:decay_pure}
{\cal M}_{\rm L,pure} (t) &\propto&
\alpha  \exp[-\lambda t]\big/4 +  \exp[-\Gamma t] , \\
\label{eq:decay_mixed} {\cal M}_{\rm L,mixed} (t) &\propto&
\alpha' \exp[-\lambda t]\big/4 +  \exp[-\Gamma t].
\eea
The prefactors $\alpha$ and $\alpha'$ 
possibly has a weak time-dependence and the 
magnitude of the factors of order one multiplying both exponentials in Eqs.~(\ref{eq:decay_pure}) 
and (\ref{eq:decay_mixed})
is determined by the short-time decay of ${\cal M}_{\rm L}$ -- one has to smoothly connect the
initial transient to the asymptotic decay. 
This is the only place where the purity of the initial state matters, and as argued above, this prefactor
can be very sensitive to the purity of the chosen initial state.
There is no difference in decay rates, however.
We note that in both cases, 
the Lyapunov decay is reduced by a factor 1/4. As already mentioned
in Chapter~\ref{section:preparation}, this 
generalizes to $M^{-1}$ in the case of $M$ nonoverlapping Gaussians.

The RMT calculation giving the golden rule decay can be extended to stronger
perturbations, $\Gamma \gtrsim B$ and one gets ${\cal M}_{\rm L}(t) \propto 
\exp[-B^2 t^2]$, 
both for pure and mixed initial state. Finally, the long-time saturation value is
\bea
{\cal M}_{\rm L,pure} (\infty) &=& N^{-1}, \\
{\cal M}_{\rm L,mixed} (\infty) &=& 4 N^{-1},
\eea
with a discrepancy obviously arising from the normalization 
we introduced in Eq.~(\ref{eq:LE_mixture})
to ensure ${\cal M}_{\rm L,mixed}(0)=1$. This analysis quantitatively 
explains the dominant features of Fig.~\ref{fig:fig2_compass}.

In this chapter we have learned three things. First, the Lyapunov decay disappears for states
differing from classically meaningful states. For both coherent superpositions and mixture of
Gaussian wavepackets, the Lyapunov decay is multiplied by the inverse number of wavepackets
in the initial state. For prepared states, the preparation time leads to the stretching, squeezing and
folding of the wavepacket, all this leading to an average probability $\propto \exp[-\lambda t]$ to
stay close to an orbit for a time $t$, 
and thus to an additional prefactor $\sim \exp[-\lambda T]$ multiplying
the Lyapunov exponential in the decay of ${\cal M}_{\rm L}$ -- see Fig.~\ref{fig:fig1_echo2}. 
We believe this explains
the observed accelerated decay of ${\cal M}_{\rm L}$ for prepared states in Ref.~\cite{Kar02}.
Second, all other decays are largely insensitive to the form of the initial state, except
the initial time-perturbative transient, which is sensitive to whether one has a coherent
superposition or a mixture -- see Fig.~\ref{fig:fig2_compass}. The reason why the golden rule
decay, for instance, is largely insensitive to the chosen initial state is that it corresponds to 
pure dephasing, where phase-space shifts are totally absorbed by shadow orbits -- Zurek's
argument that minute phase-space shifts lead to fast total dephasing does not apply in that
regime because of this somehow couterintuitive, but mathematically rigorously proven
dynamical robustness of classical systems under perturbations. 
Third, for coherent superpositions
of Gaussian wavepackets,
the decay is faster the larger the distance between the Gaussians -- see Fig.~\ref{fig:fig1_compass}.
Here again, the decay acceleration comes solely from the initial transient. A better analytical
understanding of this latter behavior is certainly desirable.

\subsection{The Wigner function approach to the Loschmidt echo}\label{chapter:Wigner}

In this section we present a phase-space semiclassical calculation of the Loschmidt echo
based on Wigner functions. Because formulas are often discussed as they are derived,
we keep some technical details in the body of the paper. 

Eqs.~(\ref{wigner_norm}) and (\ref{wigner_pure}) are key constraints
when constructing a semiclassical theory for the time evolution of the
Wigner function. The main difficulty is that $W_\psi$ is bilinear in the wavefunction, 
which renders the propagator
for $W_\psi$ nonlocal. This obstacle in the construction of a semiclassical
propagator for $W_\psi$ was of course realized long 
ago~\cite{Hel76,Hel77,Ber77,Mar91}, however it was overcome only recently
via an elegant geometric construction~\cite{Rio02} (see also~\cite{Dit06}). 
Below we reformulate this approach and split $W_\psi$ into a sum of a
positive, smooth envelope $W_\psi^{\rm cl}$ whose propagator is
local, and an oscillating function $W_\psi^{\rm qm}$ which carries
quantum coherence and accordingly has a nonlocal time-evolution.
Eq.~(\ref{wigner_pure}) can be satisfied only when taking both
$W_\psi^{\rm cl}$ and $W_\psi^{\rm qm}$ into account.

For our choice of an initial narrow Gaussian wavepacket,
the Wigner function is a positive real function at $t=0$, and
the situation is optimally devised to investigate the 
emergence of the quantum coherent correction $W_\psi^{\rm qm}$.
Before we discuss the semiclassical approach, we briefly comment on earlier
approaches based on partial differential equations for the time-evolution of $W_\psi$.

\subsubsection{Time-evolution of the Wigner function: the Moyal product}\label{timeevolW}

The equation of motion for $W_\psi$ can be derived  
from the Von Neumann equation for the density matrix 
\be\label{vonneun} 
\frac{\partial \rho }{\partial t}  = -\frac{{\it i } }{\hbar}
\left[  H_{0} , \rho  \right], \;\;\;\;\; \rho(t=0) = |\psi \rangle \langle \psi|.
\ee
In this chapter, unlike in the rest of this review, we explicitly write
$\hbar$ where applicable to make some of our reasonings and discussions clearer.
Translating Eq.~(\ref{vonneun}) into the language of 
Wigner functions requires to
introduce the Moyal product~\cite{Moy47}, 
\be\label{Moyprod}
\left[{\cal A\cdot B}\right]({\bf q},{\bf p}) =  {\cal A}({\bf q},{\bf p}) \exp \left[-(i \hbar/2) \hat{\Lambda} \right] {\cal B}({\bf q},{\bf p}) 
={\cal B}({\bf q},{\bf p}) \exp \left[(i\hbar/2) \hat{ \Lambda} \right] {\cal A}({\bf q},{\bf p}), 
\ee
giving the phase-space representation (Weyl function~\cite{Wey31}))
of a product of operators in terms of their
Weyl functions ${\cal A}({\bf q},{\bf p})$ and 
${\cal B}({\bf q},{\bf p})$, and the operator
\be
\hat{ \Lambda}  =
\frac{\overleftarrow{\bf \partial}}{\bf  \partial p}
\frac{\overrightarrow{\bf \partial}}{\bf  \partial q}
-
\frac{\overleftarrow{\bf \partial}}{\bf  \partial q}
\frac{\overrightarrow{\bf \partial}}{\bf  \partial p}.
\ee
Applying Eq.~(\ref{Moyprod}) on Eq.~(\ref{vonneun}) yields the equation of
motion for the Wigner function, 
 \be\label{Moyalbrac}
  \frac{\partial W_\psi({\bf q},{\bf p}) }{\partial t  } =
  - \frac{2}{\hbar}  H_{0}({\bf q},{\bf p})  \sin\left[\frac{\hbar}{2 } \hat{ \Lambda} \right] W_\psi({\bf q},{\bf p}). 
\ee 
The right hand side of  Eq.~(\ref{Moyalbrac}) is called 
the Moyal bracket. When looking for a quantum-classical correspondence, it makes sense 
to expand the latter in powers of $\hbar$. This gives~\cite{Hil84} 
\be\label{Moyalbracexp}
  \frac{\partial W_\psi({\bf q},{\bf p}) }{\partial t  } = 
  \left\{ H_{0},  W_\psi \right\}
  + \sum_{n\geq1}
\frac{ (-1)^n  }{ (2n+1) !}  \left(\frac{\hbar}{2}\right)^{2n}
\frac{{\bf \partial }^{2n +1} H_0}{{\bf \partial q}^{2n +1} }
\frac{{\bf \partial }^{2n +1} W_\psi}{{\bf \partial p}^{2n +1} } \;,
\ee 
where we restricted ourselves to a Hamilltonian $H_0 = {\bf p}^2/2m + V({\bf q})$.
Eq.~(\ref{Moyalbracexp}) can be interpreted as a quantum Liouville equation,
where the time-evolution of $W$ is given by a classical, Poisson bracket term
to which quantum corrections are added. In the semiclassical
limit $\hbar \rightarrow 0$, 
naive dimensional analysis suggests to neglect the quantum correction
terms since they seem to depend on the square and higher powers of $\hbar$.
If the classical dynamics generated by $H_0$ is chaotic, this 
however misses the exponential growth of derivatives of the
Wigner function $\propto \exp[\lambda t]$ on the right-hand side of 
Eq.~(\ref{Moyalbracexp}) which follows from the squeezing, stretching and 
folding of the phase-space distribution. Then, for times longer than the 
Ehrenfest time $\tau_{\rm E}$, 
the second term on the right-hand side of Eq.~(\ref{Moyalbracexp}) is of the same order
of magnitude as the first term and quantum
corrections cannot be neglected. We now present an alternative semiclassical
approach which circumvents these difficulties and treats classical and quantum contributions to the
time-evolution of $W_\psi$ on an equal footing.
 
\subsubsection{The semiclassical propagator for the Wigner 
function}\label{semiclW}

We calculate the semiclassical time-evolution of the Wigner function for
an initial Gaussian wavepacket 
$\psi({\bf r}_0') = 
(\pi \nu^2)^{-d/4} \exp[i {\bf p}_0 \cdot ({\bf r}_0'-{\bf r}_0)-
|{\bf r}_0'-{\bf r}_0|^2/2 \nu^2]$. From here on, we
restore our convention that $\hbar \equiv 1$. 
At $t=0$, $W_\psi$ is Gaussian
\be\label{Initial-Wigner}
W_\psi({\bf q},{\bf p};t=0) = W_\psi^{\rm cl}({\bf q},{\bf p};t=0)
= \pi^{-d} \exp\left[-\vert{\bf q}-{\bf r}_0\vert^2/\nu^2\right]
\exp\left[-{ \nu^2}\vert{\bf p}-{\bf r}_0\vert^2\right].
\ee
It is in particular always positive, and can thus be interpreted as
a classical probability to measure
the system at $({\bf p},{\bf q})$ in phase-space. This property gets lost
with time as $W_\psi$ starts to develop oscillations, and is no longer
positive everywhere~\cite{Ber78,Ber79}.

\begin{figure}
\begin{center}
\resizebox{!}{4cm}{\includegraphics{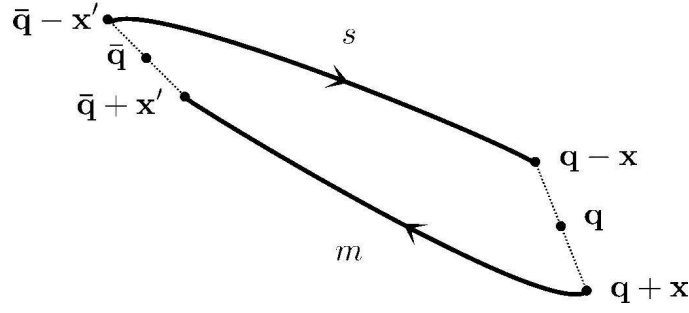}}
\end{center}
\caption{Geometric representation of the trajectory-based 
semiclassical propagator of Eq.~(\ref{semiclWp}) for the Wigner function.}
\label{Wigner_prop_fig}
\end{figure}

The semiclassical time-evolved Wigner function can be 
obtained by inserting the propagators of Eq.~(\ref{propwp})  
into Eq.~(\ref{Wigner_f}). One gets
\bea\label{Wignerprop}
W_\psi({\bf q},{\bf p};t) & = & \intbps \; {\cal K}({\bf q},{\bf p};
{\bar {\bf q}},{\bar {\bf p}};t) \; W_\psi({\bar {\bf q}},{\bar {\bf p}};0).
\eea
Because the Wigner function is bilinear in $\psi_0$, its propagator 
is expressed in terms of a double
sum over the product of two semiclassical wavefunction propagators,
\bea\label{semiclWp}
{\cal K}({\bf q},{\bf p};
{\bar {\bf q}},{\bar {\bf p}};t) & = & 2^{2d} \, 
\int {\rm d}{\bf x} {\rm d}{\bf x}' \;  
e^{2 i ({\bf p} \cdot {\bf x} - {\bar {\bf p}} \cdot {\bf x}')} \; 
\sum_{m,s} \,
K^{\ast}_{m}({\bf q}+{\bf x},{\bar {\bf q}}+{\bf x}';t) \;
K_{s}({\bf q}-{\bf x},{\bar {\bf q}}-{\bf x}';t) \nonumber \\
& = & (2/\pi)^{d} \,
\int {\rm d}{\bf x} {\rm d}{\bf x}'  \, \sum_{m,s}  \, ( C_m \; C_{s} )^{1/2} 
\exp[i \Phi_{m,s} +i \pi (\mu_m-\mu_s)/2].
\eea
where we define the action phase difference
\be
\Phi_{m,s} = 
2({\bf p} \cdot {\bf x} - {\bar {\bf p}} \cdot {\bf x}')
- S_{m}({\bf q}+{\bf x},{\bar {\bf q}}+{\bf x}';t)+ 
S_{s}({\bf q}-{\bf x},{\bar {\bf q}}-{\bf x}';t) .
\ee
A sketch of the paths involved in ${\cal K}$ is shown in 
Fig.~\ref{Wigner_prop_fig}. At this point, one readily realizes the main
difficulty in constructing ${\cal K}$: it is given by a double sum
over classical paths, which will therefore interfere. Our task now is to find
the leading stationary phase contributions in the semiclassical
limit of large actions $S_{m,s} \gg 1$.

\begin{figure}
\begin{center}
\resizebox{!}{4cm}{\includegraphics{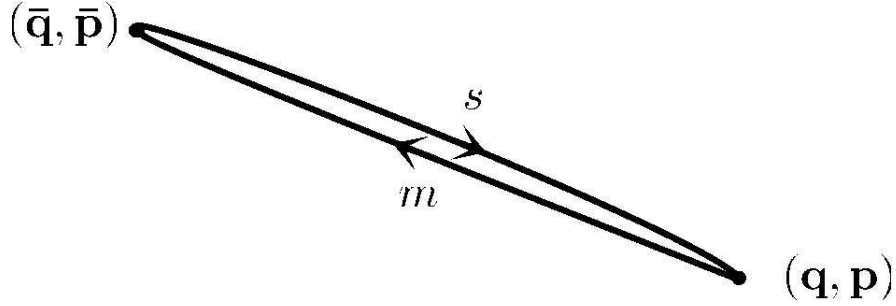}}
\end{center}
\vspace{-0.2cm}
\caption{Geometric interpretation of the local, Liouville contributions
to the Wigner function propagator given in  Eq.\ (\ref{liouvilleprop}). 
Those contributions correspond to classical paths connecting the initial
$({\bar {\bf q}},{\bar {\bf p}})$ and final $({\bf q},{\bf p})$ phase space points. }
\label{Wigner_liou}
\end{figure}

The first contribution is obtained by expanding $\Phi_{m,s}$ to first
order around ${\bf x} = {\bf x}'= 0 $. This leads to the pairing of
the trajectories $m \simeq s$ and 
correctly reproduces the Liouville flow (see Fig.~\ref{Wigner_liou})
\be\label{liouvilleprop}
{\cal K}^{\rm cl}({\bf q},{\bf p};
{\bar {\bf q}},{\bar {\bf p}};t) = \delta({\bar {\bf q}}(t)-{\bf q}) \;
\delta({\bar {\bf p}}(t)-{\bf p}).
\ee
This purely local propagator  ${\cal K}^{\rm cl}$ obviously fails to capture 
quantum contributions. We next enforce a stationary phase condition
on the global phase $\Phi_{m,s}$, i.e. search for solutions of 
\bea\label{stphasew}
\left \{
\begin{array}{ccc}
2 {\bf p}-\left( \left. \partial S_m \big/ \partial {\bf q} \,
\right|_{{\bf q}+{\bf x}} \left. 
+ \partial S_s \big/ \partial {\bf q} \, \right|_{{\bf q}-{\bf x}} \right)
& = & 0, \\
& & \\
2 {\bar {\bf p}}+\left( \left. 
\partial S_m \big/ \partial {\bar {\bf q}} \,
\right\vert_{{\bar {\bf q}}+{\bf x}'} \left. 
+\partial S_s \big/ \partial {\bar {\bf q}} \,
\right\vert_{{\bar {\bf q}}-{\bf x}'} \right) & = & 0.
\end{array}
\right. 
\eea 
We are led to define two chords with midpoints
$(  {\bf q} , {\bf p})$ and $(  {\bf \bar q} , {\bf \bar p})$  
respectively. This is shown in Fig.~\ref{Wigner_quan}. 
The stationary solutions defining the endpoints of these chords (and hence
the endpoints of the trajectories $s$ and $m$) are given by
$ ({\bf \bar q} \pm {\bf x}',{\bf \bar p} \pm {\bf \bar p}_c/2)$ 
and $({\bf  q} \pm {\bf x},{\bf p} \pm {\bf p}_c/2)$, where
$ {\bf  p}_c = {\bf p}_{s}^{\rm in} + {\bf p}_m^{\rm in}$ and 
$ {\bf {\bar p} }_c= {\bf p}_{s}^{\rm fin} + {\bf p}_m^{\rm fin}$ 
are given by the sum of initial and 
final momenta along $s$ and $m$. 
The coherent part ${\cal K}^{\rm qm}$ of ${\cal K}$ is 
obtained from those contribution with $s \ne m$ in Eq.~(\ref{semiclWp}),
with initial and final momenta on $s$ and $m$ as depicted in 
Fig.~\ref{Wigner_quan}. This contribution is thus strongly 
nonlocal. If we start with an initial Gaussian wavepacket
centered at $({\bf q}_0,{\bf p}_0)$, 
the wavepacket envelope forces ${\bf x}, {\bf p}_c \rightarrow 0$, and 
$({\bf q},{\bf p}) \rightarrow ({\bf q}_0,{\bf p}_0)$. The trajectories
$s$ and $m$ thus start from the same phase-space point, up to the
Heisenberg uncertainty. The existence of ${\cal K}^{\rm qm}$
begins as soon as the
classical dynamics generates well separated trajectories $s$ and $m$,
with nearby initial conditions inside
a unit phase space area (in units of $\hbar$) around
$({\bf q}_0,{\bf p}_0)$. In chaotic systems, the birth of ${\cal K}^{\rm qm}$
occurs at the Ehrenfest time $\tau_{\rm E}$. Beyond $\tau_{\rm E}$, coherence
and nonlocality develop and the phase-space evolution of a quantum
system deviates from the Liouvillian flow. The associated stationary phase 
difference $\Phi_{m,s}^{\rm spc}$ has a simple geometric meaning -- it is 
the symplectic area enclosed by $s$, $m$ and the chords~\cite{Rio02},
i.e. the shaded area in Fig.~\ref{Wigner_quan} -- note that this symplectic
area depends on the Hamiltonian considered.
The oscillations in the Wigner function thus become faster and
faster as this area increases, until eventually sub-Planck scale structures
are generated. In the next chapter, we 
discuss these points further and relate them to the
pure state condition $\intps W_\psi^2 = 1 $.
\begin{figure}
\begin{center}
\resizebox{!}{5cm}{\includegraphics{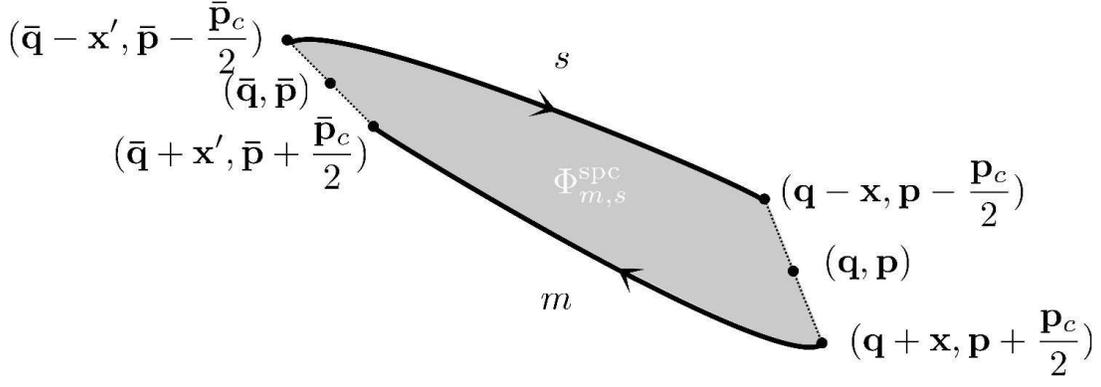}}
\end{center}
\caption{Geometric interpretation of the nonlocal, quantum contributions
to the Wigner function propagator. Those contributions correspond
to pairs of classical paths $(s, m) $ connecting pair of phase space points located symmetrically around the initial $({\bar {\bf q}},{\bar {\bf p}})$ and the final  $({\bf q},{\bf p})$ phase space points. The shaded area correspond to the reduced action $\Phi_{m, s}^{\rm spc}$ obtained from the stationary phase solution to Eq.~(\ref{semiclWp}).}
\label{Wigner_quan}
\end{figure}
 
\subsubsection{Reversibility, purity and the Wigner function}
In Eq.~(\ref{loschmidt_wigner}) we wrote
the Loschmidt echo in terms of Wigner functions.
In the particular case $H  = H_0$, 
${\cal M}_{\rm L}$ reduces to  the purity, which, since the time-evolution is unitary and
the initial state is pure, must satisfy
${\cal P}(t)=\intps \, W_{\psi}^2 =1$ at all times. One of our main tasks in our phase-space 
calculation of the Loschmidt echo is therefore
to ensure that the time-evolution is unitary at least at the level of
the integrated product of two Wigner functions.
Using the results of the previous chapter, we can write, perhaps not
too elegantly,
  \begin{eqnarray}\label{Wignerecho}
{\cal M}_{\rm L} (t )&=& 
(2 \pi)^{d}
 \int \! {\rm d} {\bf q} \,{\rm d} {\bf p} 
\int \! {\rm d} {\bf \bar q}_1 \, {\rm d} {\bf  \bar p}_1
\int \! {\rm d} {\bf \bar q}_2 \, {\rm d} {\bf \bar p}_2 \;
 {\mathbb K} ({\bf q}, {\bf p} ; {\bf \bar q}_1 ,  {\bf  \bar p}_1;{\bf \bar q}_2 ,  {\bf  \bar p}_2;t ) \nonumber \\
&& \;\;\;\;\;\;\;\;\;\;\;\;\;\;\;\;\;\;\;\;\times 
\, W_{\psi}  ({\bf \bar q}_{\rm 1} , {\bf \bar  p}_{\rm 1}; 0)\,  
\, W_{\psi}  ({\bf \bar q_2},{\bf  \bar p}_2;0), \,
 \end{eqnarray}
where we defined the -- even less elegant -- propagator for the Loschmidt echo
\bea\label{LE-Wigner}
{\mathbb K} ({\bf q}, {\bf p} ; {\bf \bar q}_1 ,  {\bf  \bar p}_1;{\bf \bar q}_2 ,  {\bf  \bar p}_2;t ) = &&{\cal K}({\bf q}, {\bf p} ; 
{\bf \bar q}_1 ,  {\bf  \bar p}_1;t) \times
{\cal K}({\bf q}, {\bf p} ; {\bf \bar q}_2 ,  {\bf  \bar p}_2;t) \nonumber
\\
= && 2^{4 d} \, \sum_{\stackrel{s_1,s_2}{ l_1, l_2} }
\; 
 \int {\rm d} {\bf x}_1{\rm d} {\bf x}'_1 {\rm d} {\bf x}_2  {\rm d} {\bf x}'_2 \,
\;\; e^{i (2{\bf p}  \cdot {\bf x}_1   -  2 {\bf \bar p}_1 \cdot {\bf  x}'_1) -{ {\it i } } (2{\bf p}  \cdot {\bf x}_2  -  2 {\bf \bar p}_2 \cdot {\bf  x}'_2)} \, \nonumber
 \\ \nonumber &&\;\;\times \,
K^{H_0}_{s_1}({\bf q}-{\bf x}_1,{\bar {\bf q}_1}-{\bf x}'_1;t)\;
 [K^{H_0}_{s_2}({\bf q}+{\bf x}_1,{\bar {\bf q}_1}+{\bf x}'_1;t)]^*
 \\  &&\;\;\times \, 
[K^{H}_{l_1}({\bf q}-{\bf x}_2,{\bar {\bf q}_2}-{\bf x}'_2;t)]^* \;
K^{H}_{l_2}({\bf q}+{\bf x}_2,{\bar {\bf q}_2}+{\bf x}'_2;t).
\eea
The four classical trajectories involved are illustrated in 
Fig.~\ref{Wigner Echo},  where, as before, a full (dashed) line correspond 
to $H_0$ ($H$).
\begin{figure}
\begin{center}
\resizebox{!}{6cm}{\includegraphics{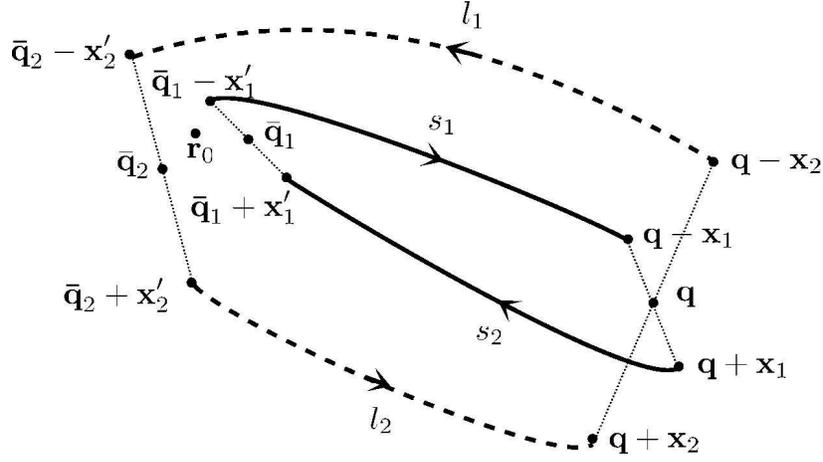}}
\end{center}
\vspace{-0.2cm}
\caption{Geometric illustration of the semiclassical  propagator  
for ${\cal M}_{\rm L}$ in the 
 Wigner function representation.  The full  lines correspond to  an 
unperturbed propagation and the dashed lines to a perturbed propagation.  } 
\label{Wigner Echo}
\end{figure}
We obtain the leading order quantum contributions by imposing a
stationary  phase approximation on the total phase
 \bea
 \Phi^{H_0} -  \Phi^{H}    &=& 
  2\{{\bf p}  \cdot  ({\bf x}_1 -  {\bf x}_2)  -   {\bf \bar p}_1 \cdot {\bf  x}'_1 -   {\bf \bar p}_2 \cdot {\bf  x}'_2\}
+ S^{H_0}_{s_1}({\bf q}-{\bf x}_1,{\bar {\bf q}_1}-{\bf x}'_1;t)
\\ & &
- S^{H_0}_{s_2}({\bf q}+{\bf x}_1,{\bar {\bf q}_1}+{\bf x}'_1;t)
- S^{H}_{l_1}({\bf q}-{\bf x}_2,{\bar {\bf q}_1}-{\bf x}'_2;t) +
S^{H}_{l_2}({\bf q}+{\bf x}_2,{\bar {\bf q}_1}+{\bf x}'_2;t)
\quad  \nonumber 
\eea
of each term in Eq.~(\ref{LE-Wigner}). These phases are minimized for
optimal matching of the two $H-$dependent symplectic areas  defined by the two evolved  
Wigner distribution and their respective  chords.    

We evaluate Eq.~(\ref{LE-Wigner}). The integral over
${\bf p}$ gives $\delta({\bf x}_1 -{\bf x}_2 )$, which restricts the choice
of pairs of trajectories $(s_i,l_i)$ to those with the same
final spatial point. We next make use of our choice of an initial 
Gaussian wavepacket and linearize all the actions around its center of mass. 
The starting point of all paths is then ${\bf r}_0$. 
\begin{figure}
\begin{center}
\resizebox{!}{6cm}{\includegraphics{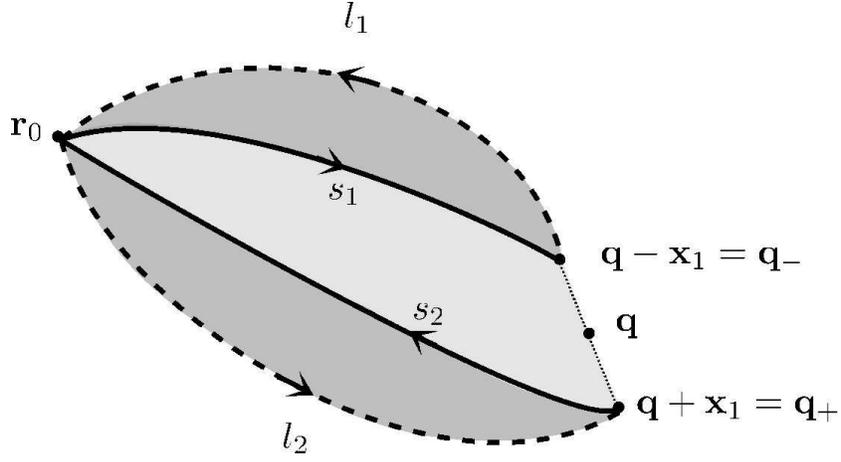}}
\end{center}
\vspace{-0.2cm}
\caption{Geometric illustration of the semiclassical  propagator  of the 
Loschmidt echo in the Wigner function representation at the level
of Eq.~(\ref{Wignerechoreduced}). 
The dark shaded phase space area gives the dominant contribution of the 
residual action $\Delta \Phi$ generated by $H_0$ on different
phase-space surfaces. Our stationary phase approximation
requires $s_i=l_i$, and thus cancels this contribution.
One then obtains Fig.~\ref{fig:wigner_echo_red2}, and
$\Delta \Phi$ is solely given by the contribution
which comes from the presence of the perturbation $\Sigma$
on the surface delimited by $l_1$, $l_2$ and
the chord joining $q_+$ and $q_-$.} 
\label{fig:wigner_echo_red}
\end{figure}
We next perform the integrations over the ${\bf \bar  q}$'s, the 
${\bf \bar  p}$'s and the ${\bf x}'$'s to obtain
 \begin{eqnarray}\label{Wignerechoreduced}
{\cal M}_{\rm L} (t )&=& 
\left(\frac{ 2 \nu^2}{  \pi  } \right)^{d}
 \int \! {\rm d} {\bf q} {\rm d} {\bf x}_1  \sum_{\stackrel{s_1,s_2}{l_1, l_2}}
 C^{1/2}_{s_1} C^{1/2}_{s_2} C^{1/2}_{l_1} C^{1/2}_{l_2}
e^{- \nu^2   ({\bf \delta p}_{s_1}^2 +{\bf \delta p}_{l_1}^2 + {\bf \delta p}_{s_2}^2 
+ {\bf \delta p}_{l_2}^2)/2}
e^{ i  \Delta \Phi}
 \end{eqnarray}
\noindent  where we wrote 
${\bf \delta p}_{s} = {\bf  p}_{s}- {\bf p}_0/2$ and 
\be\label{Wignerechophasereduced}
\Delta \Phi  =  S^{H_0}_{s_1}(  {\bf q} - {\bf x}_1,{\bf r}_0 ;t) 
- S^{H_0}_{s_2}( {\bf q} + {\bf x}_1, {\bf r}_0 ;t) 
-  S^{H}_{l_1}( {\bf q} - {\bf x}_1,{\bf r}_0  ;t) 
+ S^{H}_{l_2}( {\bf q} + {\bf x}_1,{\bf r}_0  ;t).
\ee
The situation at
this point in the calculation is sketched in Fig.~\ref{fig:wigner_echo_red}.   
There are two contributions to $\Delta \Phi$. 
If $s_1 \ne l_1$ and/or
$s_2 \ne l_2$ the dominant contribution comes from the action of $H_0$
on the difference in phase-space area covered by the two Wigner
functions (shaded area on Fig.~\ref{fig:wigner_echo_red}). This contribution
vanishes once we
enforce the stationary phase condition $s_i=l_i$, $i=1,2$. This is justified in the limit of
relevance for us, where the perturbation $\Sigma_1$ is so small that most of the action 
phase is provided by unperturbed dynamics.
Then, $\Delta \Phi$ is solely given by the contribution of
the perturbation $\Sigma = H_0-H$ on the exactly overlapping 
phase-space areas covered by the two Wigner functions, one of them
evolving with $H_0$, the second one with $H$.
Performing next a change of integration variables 
${\bf q}_{\pm} = {\bf q} \pm {\bf x}_1$, 
we reproduce Eq.~(\ref{mtot}), 
\begin{eqnarray}\label{mtot_repeat}
{\cal M}_{\rm L}(t) & = & \left(\frac{\nu^2}{\pi}\right)^d \int d{\bf r} \int d{\bf r}'
\sum_{s,l} C_s C_{l} \exp[i \delta S_s({\bf r},{\bf r}_0;t) - 
i \delta S_{l}({\bf r}',{\bf r}_0;t) ] \nonumber \\
& & \times \exp(-\nu^2 |{\bf p}_s-{\bf p}_0|^2-\nu^2 
|{\bf p}_{l}-{\bf p}_0|^2).
\end{eqnarray}
The decay of ${\cal M}_{\rm L}$, Eq.~(\ref{eq:final_decay}) derives
from Eq.~(\ref{mtot_repeat}) via separate calculation of the
correlated ($s=l$) and uncorrelated ($s \ne l$) contributions. 
Going back to the Wigner representation, it is seen that the 
two contributions correspond to 
\bea
{\cal M}_{\rm L}^{\rm (d)}(t)  &=& \int {\rm d}{\bf q} {\rm d}{\bf p} \,
\int {\rm d}{\bf \bar{q}}_1 {\rm d}{\bf \bar{p}}_1 \, 
\int {\rm d}{\bf \bar{q}}_2 {\rm d}{\bf \bar{p}}_2 \;
{\cal K}^{\rm cl}_{H_0} ({\bf q},{\bf p};
{\bar {\bf q}}_1,{\bar {\bf p}}_1;t)  \;  
W_\psi({\bar {\bf q}}_1,{\bar {\bf p}}_1;0) \nonumber \\
&& \;\;\;\;\;\;\;\;\;\;\;\;\;\;\;\;\;\;\;\;\;\;\;\;\;\;\;\;\;\;\;\;\;\;\;\;\;\;\;\;
\times \, {\cal K}^{\rm cl}_{H} ({\bf q},{\bf p};
{\bar {\bf q}}_2,{\bar {\bf p}}_2;t)
\; W_\psi({\bar {\bf q}}_2,{\bar {\bf p}}_2;0).\label{eq:lyap_wigner}\\
{\cal M}_{\rm L}^{\rm (nd)}(t)  &=& \int {\rm d}{\bf q} {\rm d}{\bf p} \,
\int {\rm d}{\bf \bar{q}}_1 {\rm d}{\bf \bar{p}}_1 \, 
\int {\rm d}{\bf \bar{q}}_2 {\rm d}{\bf \bar{p}}_2 \;
{\cal K}^{\rm qm}_{H_0} ({\bf q},{\bf p};
{\bar {\bf q}}_1,{\bar {\bf p}}_1;t)  \;  
W_\psi({\bar {\bf q}}_1,{\bar {\bf p}}_1;0) \nonumber \\
&& \;\;\;\;\;\;\;\;\;\;\;\;\;\;\;\;\;\;\;\;\;\;\;\;\;\;\;\;\;\;\;\;\;\;\;\;\;\;\;\;
\times \, {\cal K}^{\rm qm}_{H} ({\bf q},{\bf p};
{\bar {\bf q}}_2,{\bar {\bf p}}_2;t)
\; W_\psi({\bar {\bf q}}_2,{\bar {\bf p}}_2;0).\label{eq:grule_wigner}
\eea
\begin{figure}
\begin{center}
\resizebox{!}{5cm}{\includegraphics{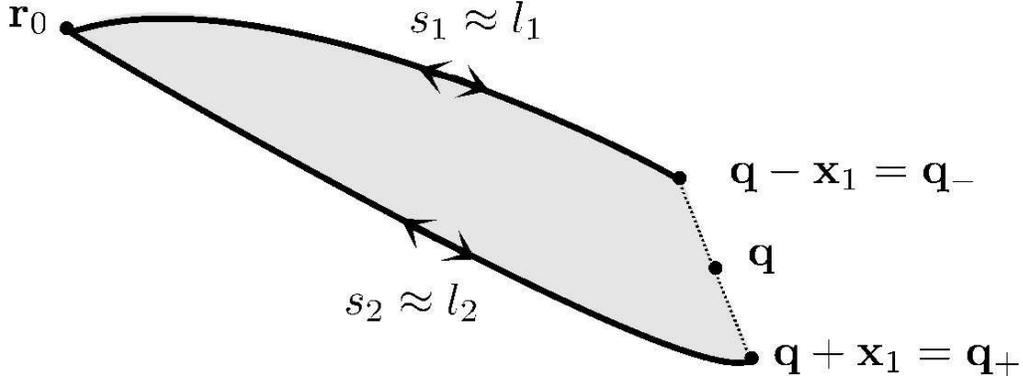}}
\end{center}
\vspace{-0.2cm}
\caption{Geometric illustration of the semiclassical  propagator  of the 
Loschmidt echo in the Wigner function representation after a stationary
phase condition has been imposed on $\Delta \Phi$. } 
\label{fig:wigner_echo_red2}
\end{figure}
The Lyapunov decay (power-law decay for
regular systems) arises from the classical, Liouville propagation of
the Wigner function, while the golden rule decay is generated by
the quantum corrections, and the associated perturbation-generated
phase-space action. There are no contributions coming from cross-terms
${\cal K}^{\rm cl} \cdot {\cal K}^{\rm qm}$. 
Eq.~(\ref{eq:grule_wigner}) gives the only contribution
sensitive to small phase-space structures, and while our derivation does
not contradict Zurek's argument, it is instructive to note that the route
we followed is somehow orthogonal to his. We identified the
(possibly fast) oscillating terms that remain in phase and that can thus still satisfy a stationary phase
condition, instead
of arguing about how easily they can be brought out of phase.

 As a final comment we note that in absence of perturbation the purity of
the density matrix must be identical to one, for all times. This is 
enforced by a sum rule
similar to Eq.~(\ref{sumrule1}). Similar to
Eqs.~(\ref{eq:lyap_wigner}) and (\ref{eq:grule_wigner}), ${\cal P}(t)$ is given
by the sum of a classical and a quantum term
\be\label{purity_split}
{\cal P} (t ) = \intps \; ({\cal K}^{\rm cl} W_\psi)^2
+({\cal K}^{\rm qm} W_\psi)^2.
\ee
The first term corresponds to the Liouville propagation of the 
Wigner function. Proceeding as for the Loschmidt echo [the final steps
leading to Eq.~(\ref{diagdecay})], one gets, for chaotic systems,
\be\label{class_wigner_decay}
 \intps \;  ({\cal K}^{\rm cl} W_\psi)^2 \propto 
 \exp [-\lambda t].
\ee
Although the Liouville propagation alone allows to satisfy 
the normalization condition, Eq.~(\ref{wigner_norm}), we see that it fails 
to fulfill the pure state criteria, Eq.~(\ref{wigner_pure}).  
Because Eq.~(\ref{purity_split}) gives one [this comes from a sum rule similar
to Eq.~(\ref{sumrule1})]~\cite{Alm03}, we conclude
that the quantum corrections are $\propto (1-\exp[-\lambda t])$. They
start to dominate the purity at the Ehrenfest time.
Our level of approximation is sufficient to ensure unitarity of the
time evolution at the level of the purity / product of two Wigner functions.

\subsection{What have we learned ?}

We have reproduced our
earlier qualitative argument that, in a chaotic system, 
the quantum contribution becomes important at the Ehrenfest time.
The decay of the Loschmidt echo does not start before $\tau_{\rm E}$,
a conclusion that was already drawn in Ref.~\cite{Sil03}. 
Does that influence decoherence by external degrees of freedom ?
Only the nonlocal propagation is sensitive to decoherence. Therefore,
if nondissipative decoherence mechanisms exist which annihilate the quantum terms
before they have a chance to appear, the resulting dynamics will be 
solely given by the classical Liouville time-evolution. In the next section, we
discuss this aspect in more details and present numerical and analytical
results which show how the coupling to external degrees of freedom render
the time-evolution of a quantum chaotic system identical to the 
Liouville evolution of its classical counterpart. 

With these considerations we close this discussion of irreversibility in quantum dynamical
systems with few degrees of freedom, and what determines it, and move on to a discussion
of entanglement generation in coupled dynamical systems. 


\section{Entanglement and Irreversibility in Bipartite Interacting Systems}\label{section:entanglement}

In this fourth section we extend our semiclassical and random matrix theories to systems of
few interacting sub-systems. Provided the interaction is weak enough -- explicit bounds are
given below -- shadowing can once again be invoked to justify the main operational
assumption in our semiclassical calculations, that noninteracting classical trajectories are
not affected by the presence of interactions. This assumption enables us to apply the
semiclassical machinery developed above to situations of interacting subsystems. Quite
surprisingly, nontrivial effects in entanglement generation 
can be captured by that approach, which
are partially reproduced by RMT -- a purely quantal approach. We carry on treating
irreversibility in partially controlled systems and consider two interacting subsystems, where only one
of them is controlled / time-reversed. We show that the degree of reconstruction of the initial
state of the controlled subsystem 
depends on the balance between the accuracy of the time-reversal operation and
the coupling between the two subsystems. 

\subsection{Dynamics of bipartite entanglement}

{\it When two systems, of which we know the states by their respective representatives, enter into temporary physical interaction due to known forces between them, and when after a time of mutual influence the systems separate again, then they can no longer be described in the same way as before, viz. by endowing each of them with a representative of its own. I would not call that one but rather the characteristic trait of quantum mechanics, the one that enforces its entire departure from classical lines of thought. By the interaction the two representatives [the quantum states] have become entangled.}
This is how entanglement was qualitatively characterized by  
Schr\"odinger some seventy years ago \cite{Sch35}. 
Entanglement is arguably the most puzzling property       
of multipartite interacting quantum systems, and often leads to
counterintuitive predictions due to, in Einstein's words, 
{\it spooky action at a distance}~\cite{Ein35}. Entanglement
has received a renewed, intense interest in recent years in the context 
of quantum cryptography and 
information theory~\cite{Sho94,Cir95,Los97,Sho97,Milb99,Nie00,Gis02,Ben07}.

In the spirit of Schr\"odinger's above formulation, one is naturally led to
wonder what determines the rate of entanglement 
production between coupled dynamical systems. Is this rate 
mostly determined by the interaction
between two, initially unentangled particles, or does it depend on the
underlying classical dynamics ? Or 
does it depend on the states initially occupied by
the particles ? These are some of the questions we
address in this chapter, where we consider an isolated bipartite system of 
two interacting, distinguishable particles. 

Beside our contributions, Refs.~\cite{Jac04a,Jac04b,Pet06b},
there have been several, mostly numerical works, that looked for 
connections between entanglement dynamics 
and the nature of the underlying classical 
dynamics~\cite{Fur98,Mil99,Lak01,Tan02,
Zni03a,Sco03,Gon03,Gho08}. Most of these works focused on the
purity ${\cal P}(t)$ [defined in Eq.~(\ref{eq:purity})]
or equivalently on the von Neumann entropy of the reduced one-particle density matrix. Claims have been made that entanglement is
favored by classical chaos, both in the rate it is 
generated \cite{Fur98,Mil99,Zni03a} and in the maximal amount it
can reach \cite{Lak01}. In particular, Miller and Sarkar gave strong numerical evidences 
for an entanglement production rate given by the 
system's Lyapunov exponents \cite{Mil99}. This is rather intriguing -- to
say the least --
how can the dynamics of entanglement, the quantity which Schr\"odinger himself
considered {\it not one, but rather the characteristic trait of quantum mechanics},
be governed by the Lyapunov exponent of the classical dynamics ?

The results of Miller and Sarkar of a
generation of entanglement governed by Lyapunov exponents
have however been challenged by Tanaka and collaborators~\cite{Tan02}, 
whose numerical investigations show no increase of the entanglement production rate upon 
increase
of the Lyapunov exponent in the strongly chaotic but weakly coupled regime.
The numerical investigations of Tanaka and co-authors
are remarkable in that, compared to earlier works, 
they are backed by analytical calculations
relating the rate of entanglement production to classical time correlators.
Ref.~\cite{Tan02} is seemingly 
in a paradoxical disagreement with the almost identical
analytical approach of Ref.~\cite{Zni03a}, where entanglement
production was found to be faster in chaotic systems than in
regular ones. This controversy was resolved in our two letters, 
Refs.~\cite{Jac04a,Pet06b}, where the semiclassical and RMT approaches that 
proved to
be so successful for the Loschmidt echo was extended to the
calculation of entanglement generation between
two interacting dynamical systems. The connection
can be made between the approach of Tanaka and RMT in the
golden rule regime -- they
both are valid when the Lyapunov exponent is very large, and accordingly
predict only an interaction-dependent decay of ${\cal P}(t)$.
The numerical results of Miller and Sarkar~\cite{Mil99}, on the other hand, 
were obtained for systems with moderate values of the Lyapunov exponent
and stronger interaction. In this case, the classical dynamics sets bounds
on entanglement generation -- its rate cannot exceed the classical
Lyapunov exponent.
As we now proceed to show, the decay of the purity for bipartite
systems behaves similarly as the Loschmidt echo, in that similar decay
regimes exist depending on the two-particle level spacing $\delta_2$, 
the interaction-induced broadening $\Gamma_2$ of 
noninteracting two-particle states and the two-particle bandwidth $B_2$. Most
notably, in the regime $\delta_2 \lesssim \Gamma_2 \ll B_2$,
${\cal P}(t)$ is determined by the same quantum-classical
competition between dephasing and decay of wavefunction overlap
that governs the behavior of ${\cal M}_{\rm L}$ -- this time,
both dephasing and the decay of 
wavefunction overlaps 
are generated by the coupling with the second
particle, the latter having a dynamics of its own.
Accordingly, one gets 
${\cal P}(t) \propto \exp[-{\rm min}(\lambda_1,\lambda_2,
2 \Gamma_2) t]$, with the Lyapunov exponents $\lambda_{1,2}$
of either subsystem. This
establishes the connection between purity and Loschmidt echo 
in the golden rule regime. 

The RMT calculation of ${\cal P}(t)$ proceeds as usual via a sequence
of contractions of wavefunctions, 
together with average expressions for the projection of
noninteracting two-particle states over the basis of 
interacting two-particle states~\cite{Wig57,Jac95,Fra95,Fyo95,Jac97a,Wei96}. 
The semiclassical approach we follow  relies on
the assumption that the interaction does not modify the classical
trajectories followed by each particle -- an assumption that can be 
formally justified by invoking structural stability theorems~\cite{Kat96}.
It turns out that once again, a direct one-to-one connection can be
made between the semiclassical pairing of trajectories and the RMT
contractions. 

Decoherence is nothing else but generation of entanglement 
with the environment, and it is very tempting to try and extrapolate
our approach towards a semiclassical theory
of decoherence. As a matter of fact, decoherence from the coupling
with few chaotic external degrees of freedom has attracted quite some 
attention~\cite{Coh02,Coh04,Bon06a,Bon06b,Ros06}, perhaps because the universality of 
quantum Brownian motion can be established in several, rather generic
situations, for instance
under the sole
assumption that the system-environment coupling
can be modeled by a random matrix~\cite{Lut99}, or for 
decoherence in the macroworld, where all time scales are slower 
than the process
of decoherence itself~\cite{Bra00,Bra03}. 
Strictly speaking, the theory of
decoherence usually invokes external baths with many degrees of freedom, and one might wonder,
again following Miller and Sarkar~\cite{Mil99} if large environments made of a collection
of many, rather simple sub-environments -- such as uncoupled harmonic oscillators --
can generally be traded for complex dynamical systems with few degrees of freedom. 
It is actually not uncommon that these findings -- that the 
generation of entanglement between two 
dynamical subsystems with few degrees of freedom depends on the Lyapunov
exponent -- are actually quoted as confirmation of the theory according to which decoherence by coupling to an external bath at large temperature proceeds at a rate
given by Lyapunov exponents~\cite{Zur03}. This point is briefly discussed below, where
we make the first step towards 
a generalization of our results for decoherence due to the coupling to
a complex environment. We show in particular 
how the partial Fourier transform of the one-particle reduced
density matrix -- the Wigner distribution -- becomes positive 
definite (and thus a true
probability distribution in phase-space)
and follows the uncoupled (chaotic) single-particle classical dynamics in 
the golden rule 
regime of interaction with a single second chaotic particle. 
These results pave the
way toward a semiclassical theory of decoherence in presence of many
chaotic and interacting degrees of freedom. This approach is currently under 
development~\cite{Fie03,Jac08}.

\subsection{Bipartite systems and the semiclassical approach to entanglement}

We first present a semiclassical calculation of the time-evolved density
matrix $\rho(t)$ for two interacting, distinguishable particles. 
Entanglement is investigated via the properties of the reduced
density matrix $\rho_1(t) \equiv {\rm Tr}_2 [\rho(t)]$, obtained 
from the two-particle density matrix by
tracing over the degrees of freedom of one (say, the second) particle.
We quantify entanglement with the purity ${\cal P}(t) 
\equiv {\rm Tr}[\rho_1^2(t)]$ 
of the reduced density matrix
and start our theoretical experiment with the two particles in a product state of two narrow
wavepackets -- this choice is motivated by our use of a trajectory-based semiclassical approach. 
In this way, ${\cal P}(t=0)=1$, and the average 
purity decays as  time goes by, while the two particles become more
and more entangled.
Because the global two-particle system is isolated, hence remains 
fully quantum mechanical at all times, the
two-particle density matrix is pure and ${\cal P}(t)$ 
is a good measure of entanglement. Compared to the von Neumann entropy 
or the concurrence, for instance, it moreover
presents the advantage of being analytically 
tractable. For the weak coupling situation we are interested in 
here, numerical works have moreover shown that von Neumann
and linear entropy ${\cal S}_{\rm lin}\equiv 1-{\cal P}(t)$  
behave very similarly \cite{Tan02}.
We thus expect the purity to give a faithful and generic
measure of entanglement.
We note that our semiclassical approach is straightforwardly extended 
to the case of undistinguishable 
particles, provided the nonfactorization of the reduced density  
matrix due to particle statistics
is properly taken care of.

Using semiclassics to investigate entanglement generation looks {\it a priori} like a shot in the
dark -- is there any hope to capture such a fundamentally quantal effect with an
expansion to lowest order in the effective Planck's constant ? Quite remarkably, the
{\it a posteriori} answer to this question is a firm ``YES'', as we now proceed to show. Still, we recall
that our approach is valid only in the short wavelength limit, and entanglement between 
particles occupying low-lying quantum levels is beyond the methodology we use here. 
Our approach is reminiscent of the semiclassical methods used
above for the Loschmidt echo, and relates the 
off-diagonal matrix elements of $\rho_1$ to classical action correlators. 
We find that, following an initial transient where $\rho_1$
relaxes but remains almost exactly pure, entanglement
production is exponential in chaotic systems, while it is algebraic
in regular systems. For not too strong interaction, the asymptotic rate of entanglement
production in chaotic systems depends on the strength of the interaction
between the two particles, and is explicitly given by a
classical time-correlator. As is the case for the Loschmidt echo,
this regime is also adequately captured by an approach based on 
RMT -- the time-correlator is then replaced by the
golden rule spreading of two-particle states due to the interaction. 
RMT for the entanglement generation in bipartite systems will be presented
in the next chapter.
For stronger coupling however,
the dominant stationary phase solution becomes interaction independent
and is determined only by the classical dynamics, the Lyapunov 
exponents giving an upper bound for the rate
of entanglement production. As for the Loschmidt echo,
the crossover between the two regimes
occurs once the golden rule width becomes comparable to the system's Lyapunov
exponent. Long-ranged interaction potentials can lead to 
significant modifications of this picture, especially at short
times, due to an anomalously slow decay
of off-diagonal matrix elements of $\rho_1({\bf x},{\bf y})$ within a 
bandwidth $|{\bf x}-{\bf y}| \lesssim \zeta$ set by the interaction correlator.

In Appendix~\ref{appendix:semiclassics_entangle} 
we reproduce in some more details the calculation 
we originally presented in Refs.~\cite{Jac04a,Jac04b,Pet06b}.
Here we only discuss some of the main steps. 
Our goal is to calculate the purity ${\cal P}(t) 
\equiv {\rm Tr}[\rho_1^2(t)]$  of the reduced density matrix for a bipartite 
systems of
two interacting systems with few degrees of freedom, 
with an initial two-particle product state where each particle
is prepared in a Gaussian wavepacket. As for the fidelity,
the two-particle semiclassical density matrix is written using semiclassical propagators,
which now include pairs of trajectories, one for each particle. One assumes that these
trajectories are unaffected by the interparticle interaction, and include the effect of the latter
solely in an additional interaction-dependent two-particle phase in the semiclassical propagator.
This propagator is given in Eq.~(\ref{twopart_propa}). The elements 
$\rho_1({\bf x},{\bf y};t) = \int d{\bf r} \langle {\bf x}, {\bf r}|
\rho(t) |{\bf y}, {\bf r} \rangle$ of the reduced density matrix are then calculated, and after the standard 
initial calculational steps (linearization around the initial position of the wavepackets and
integration of the resulting Gaussian integrals) one obtains
\begin{eqnarray}
\label{semicrhob}
\rho_1({\bf x},{\bf y};t) &=& \left(\frac{\nu^2}{\pi}\right)^{d_1/2} \sum_{s,l} \;
(C_s C_l )^{1/2} \; \exp[-\frac{\nu^2}{2} \{
({\bf p}_{s}-{\bf p}_1)^2 + ({\bf p}_{l}-{\bf p}_1)^2\}] 
 \\
&& \times  \; {\cal F}_{s,l}(t) \; 
\exp[i\{S_s({\bf x},{\bf r}_1;t)-S_{l}({\bf y},{\bf r}_1;t)\}] \nonumber \\[1mm]
\label{f-vernon-bi-pureb}
{\cal F}_{s,l}(t) & = & \left(\frac{\nu^2}{\pi}\right)^{d_2/2} \int d{\bf r}
\sum_{s',l'} (C_{s'} C_{l'})^{1/2} e^{-\frac{\nu^2}{2} \{
({\bf p}_{s'}-{\bf p}_2)^2 + ({\bf p}_{l'}-{\bf p}_2)^2\}} 
 \\
&& \times 
\exp[i\{S_{s'}({\bf r},{\bf r}_2;t)-S_{l'}({\bf r},{\bf r}_2;t)
+{\cal S}_{s,s'}({\bf x},{\bf r}_1;{\bf r},{\bf r}_2;t)
-{\cal S}_{l,l'}({\bf y},{\bf r}_1;{\bf r},{\bf r}_2;t)\}].\nonumber 
\end{eqnarray}
Eq.~(\ref{f-vernon-bi-pureb}) is nothing else but the
influence functional of Feynman and Vernon~\cite{Fey63}. 
M\"ohring and Smilansky derived a similar expression valid
when the second particle
(their environment / macrosystem) is classical~\cite{Moh80}.

One assumes that one-particle actions
vary faster than their two-particle counterpart, and accordingly
pair $s' \simeq l'$ -- this is motivated by a stationary phase conditions on Eq.~(\ref{semicrhob}).
To calculate the average $\langle \rho_1\rangle$ over the positions of the initial 
wavepackets, one enforces a second stationary phase condition, and pair
$s=l$, ${\bf x}={\bf y}$. One obtains 
$\langle \rho_1 ({\bf x},{\bf y};t) \rangle = 
\delta_{{\bf x},{\bf y}}/\Omega_1$, with the volume $\Omega_1$ occupied
by particle one.
Diagonal elements of the reduced density matrix acquire an ergodic 
value -- this is due to the average over initial
conditions -- and only they have a nonvanishing average.
Simultaneously this average conserves probabilities, ${\rm Tr} \rho_1 = 1$.
For each initial condition, $\rho_1(t)$
has however nonvanishing off-diagonal matrix elements, 
with a zero-centered distribution whose
variance is given by $\langle 
\rho_1({\bf x},{\bf y};t)  \rho_1({\bf y},{\bf x};t) \rangle$.
Beyond giving the variance of the distribution of off-diagonal matrix
elements, this quantity also appears in the average purity
 ${\cal P}(t) = \int {\rm d}{\bf x}
\int {\rm d}{\bf y} \langle \rho_1({\bf x},{\bf y};t)  
\rho_1({\bf y},{\bf x};t) 
\rangle$. To compute the latter, one 
thus has to go back one step, before this last stationary phase condition. 

The rest of the calculation
is detailed in Appendix~\ref{appendix:semiclassics_entangle}.
An important time scale, $\tau_{\cal U}$ is 
the time it takes for two initial classical points within a distance $\nu$ 
to move away a distance $\propto \zeta$ from each other.
In a chaotic system, this gives a logarithmic time, similar in physical
content to the Ehrenfest time, 
$\tau_{\cal U} = \lambda^{-1} \ln(\zeta/\nu)$, while in a regular
system, $\tau_{\cal U}$ is much longer, typically algebraic in 
$\zeta/\nu$.
The purity is straightforward to compute
from Eqs.~(\ref{2correl}) for $t>\tau_{\cal U}$ or (\ref{Fsmall})
for $t<\tau_{\cal U}$, using the correlators in Eq.~(\ref{correl1}) and (\ref{correl2}).
We get three
distinct regimes of decay: 

(a) an initial regime of classical
relaxation for $t<\tau_{\cal U}$, 

(b) a regime where quantum
coherence develops between the two particles so that
$\rho_1$ becomes a mixture, and 

(c) a saturation regime where the purity
reaches its minimal value. Let us look at these three regimes in more
details.

In the initial transient regime (a), $\rho_1$ evolves from a pure, but localized
$\rho_1(0)=|{\bf r}_1 \rangle \langle {\bf r}_1|$
to a less localized, but still almost pure $\rho_1(t)$, with an
algebraic purity
decay obtained from Eqs.~(\ref{Fsmall}) and (\ref{correl2}). One gets
\begin{equation}
{\cal P}(t < \tau_{\cal U}) \simeq
\frac{1}{\Omega_1 \Omega_2}
\left(\frac{1-\exp[-2 \gamma_2 L_1^2 t]}{2 \gamma_2 t}\right)^{d_1/2}
\times \left(\frac{1-\exp[-2 \gamma_2 L_2^2 t]}{2 \gamma_2 t}\right)^{d_2/2},
\end{equation}
which can easily be checked to 
go to unity for $t \rightarrow 0$ ($\Omega_i = L_i^{d_i}$). This gives a slow short-time
decay of the purity -- a slow entanglement generation -- and 
even in the case
of a correlator saturating at a finite, nonzero value
for $|t_1-t_2| \rightarrow \infty$, which may occur
in regular systems, this initial decay
will still be algebraic $\propto t^{-d_{1,2}}$. It is mostly this initial transient 
that differentiates the behavior of the purity from that of the fidelity. Such an
algebraic initial transient has also been calculated in Ref.~\cite{Gon03}.

In the asymptotic regime (b), the decay of ${\cal P}(t)$ is given by the correlator
$\langle {\cal S}_{s,s'}^2 \rangle$.
Because the four classical paths in that term come in two pairs,
the dependence on $|{\bf x}-{\bf y}|$ vanishes.
With Eqs.~(\ref{eq:red_threeterms}), (\ref{bounddecay}),
(\ref{2correl}) and (\ref{correl1}) one gets 
\begin{eqnarray}\label{eq:purity_decay}
{\cal P}(t) \propto \left\{
\begin{array}{cc}
\alpha_1 \, \Theta(t>\tau_{\lambda_1}) \,
e^{-\lambda_1 t} +\alpha_2 \, \Theta(t>\tau_{\lambda_2}) \,
e^{-\lambda_2 t} + \Theta(t>\tau_{\Gamma}) \, e^{-2 \Gamma_2 t}, 
& {\rm chaotic}, \\
\Theta(t>\tau_{\cal U}) \; [(t_1/t)^{d_1} + (t_2/t)^{d_2}], & {\rm regular}.
\end{array}
\right.
\end{eqnarray}
In regular systems, the algebraic decay sets in at $\tau_{\cal U}$ and
the time scales $t_{1,2}$ are system-dependent.
There are several onset times in chaotic systems. The golden rule
decay $\propto \exp[-2 \Gamma_2 t]$ sets in once enough action phase
has been generated by the interaction on a typical trajectory. The condition
for the corresponding onset time $\tau_\Gamma$ thus reads
\bea
\Big|\int_0^{\tau_\Gamma} {\rm d}t {\cal U}({\bf q}_s(t),{\bf q}_{s'}(t)) \Big| 
\approx
\left(\int_0^{\tau_\Gamma} {\rm d}t \ {\rm d}t' 
\langle {\cal U}({\bf q}_s(t),{\bf q}_{s'}(t)) 
{\cal U}({\bf q}_s(t'),{\bf q}_{s'}(t')) \rangle \right)^{1/2} = 1,
\eea
from which one estimates $\tau_\Gamma \approx \Gamma_2^{-1}$. The onset
time $\tau_{\lambda_i}$ for the Lyapunov decay is similar to the Ehrenfest 
time. At shorter times, there is no Lyapunov
decay, as two nearby trajectories stay together, within a resolution scale
determined by ${\cal U}$~\cite{Pet06b}.
In the numerics
to be presented below, $\tau_\Gamma, \tau_{\lambda_i} > \tau_{\cal U}$, and a proper
rescaling of the data for different sets of parameters first requires
shifts $t \rightarrow t-t_\Gamma,\, t-\tau_{\lambda_i}$ of the time axis.

Finally the saturation value in regime (c) can also be estimated 
semiclassically, starting before the stationary phase approximation leading
to Eq.~(\ref{Sigma2}). Two pairings, one for the trajectories of the first particle,
one for those of the second particle lead to exact cancellation of the action phase,
but simultaneously restrict the endpoints of those trajectories. 
Assuming ergodicity,
and once again using the sum rule (\ref{sumrule1}), one obtains
\begin{equation}\label{eq:purity_sat}
{\cal P}(\infty) = 2 \Theta(t>\tau_{\rm E}^{(1)}) (\nu^{d_1}/\Omega_1)+
2 \Theta(t>\tau_{\rm E}^{(2)}) 
(\nu^{d_2}/\Omega_2) + O(\nu^{2d_{1,2}}/\Omega_{1,2}^{2}), \;\;\;\;\;\; {\rm chaotic}.
\end{equation}
Each saturation term sets in at the corresponding Ehrenfest time. The fact
that the fastest possible, Lyapunov decay brings the purity down to 
its saturation level at precisely that time is of
course not a coincidence. As is the case for the fidelity, the saturation level occurs
at the inverse size $N_i^{-1} = \nu^{d_i}/\Omega_i$ of Hilbert space.
There is no reason to expect a
universal saturation value in regular systems where ergodicity is not granted. 

Analyzing these results, we note that
Eqs.~(\ref{2correl}) and (\ref{Fsmall}) are reminiscent of
the results obtained for ${\cal P}(t)$ by
perturbative treatments in Refs.~\cite{Tan02,Zni03a}, but they
apply well beyond the linear response regime. Our weak coupling
condition
that the one-particle actions $S$ vary faster than the two-particle
actions ${\cal S}$ roughly gives an upper bound $\Gamma_2 \le B_2$ for the
interaction strength. 
The linear response regime
is however restricted by a much more stringent condition $\Gamma_2 \le 
\delta_2 \ll B_2$. The decay regime
(II) of ${\cal P}(t)$ reconciles the {\it a priori} contradicting claims
of Refs.~\cite{Fur98,Mil99,Zni03a} 
and Ref.~\cite{Tan02}. For weak coupling,
the decay of ${\cal P}(t)$ is given by classical correlators, and thus
depends on the interaction strength, in agreement with Ref.~\cite{Tan02}. 
However, ${\cal P}(t)$ cannot decay faster than the bound
given in Eq.~(\ref{bounddecay}), so that at stronger coupling, and in
the chaotic regime, one recovers the results of Ref.~\cite{Mil99}.
Simultaneously, regime (II) also explains the data in Fig.~2 and 4 of
Ref.~\cite{Zni03a}, showing an
exponential decay of ${\cal P}(t)$ in the chaotic regime, and
a power-law decay with an exponent close to 2
in the regular regime (this power-law decay was left unexplained by 
the authors of Ref.~\cite{Zni03a}).

Our semiclassical treatment thus presents a unified picture for the
role of the classical dynamics in entanglement generation, and
we summarize it now.
To leading order in the semiclassical small parameter
$N_{1,2}^{-1} = \nu^{d_{1,2}}/\Omega_{1,2}$, and neglecting the onset times (i.e.
considering $t > \tau_{\cal U}$, $\tau_{\lambda_i}$ and $\tau_{\rm E}^{(i)}$)
the purity of the reduced one-particle density matrix in a quantum chaotic dynamical 
system of two interacting particles evolves as 
\be\label{purity}
{\cal P}(t) \simeq \exp[-2 \Gamma_2 t] 
+  \sum_{i=1,2} \alpha_i \exp[-\lambda_i t]  
+ N_1^{-1}
+ N_2^{-2}.
\ee
The first term is the standard, interaction-dependent
quantum term giving the golden rule decay of the purity.
Being given by a classical correlator
evaluated along classical trajectories,
$\Gamma_2$ does not depend on $\hbar_{\rm eff}$. 
The second,
classical term decays with the Lyapunov
exponents $\lambda_{1,2}$ and has weakly time-dependent
prefactors $\alpha_i = {\cal O}(1)$.
Finally, the two saturation terms set in at the relevant Ehrenfest time
$\tau_{\rm E}^{(i)}$, $i=1,2$ indexing the particle number.
For classically regular systems, Eq.~(\ref{purity}) is replaced by
\be\label{purity_reg}
{\cal P}(t) \simeq (t_1/t)^{d_1} + (t_2/t)^{d_2}.
\ee
This equation corrects a mistake made in Ref.~\cite{Jac04a}. Accordingly, the 
results presented here 
are now compatible with those of \v{Z}nidari\v{c} and Prosen, Ref.~\cite{Zni05a}.

The validity of Eq.~(\ref{purity})  is determined by
$\delta_2 \le \Gamma_2 \le B_2$, where $\delta_2=B_2 \nu_1 \nu_2/
(\Omega_1 \Omega_2)$ and $B_2$
are the two-particle bandwidth and level spacing respectively
\cite{Jac01}. This range of validity is parametrically large in the 
semiclassical limit $\nu_i/\Omega_i \rightarrow 0$. In this range, 
${\cal U}$ is quantum-mechanically strong
as individual levels are broadened beyond their average spacing, 
but classically weak, as $B_2$ is unaffected by ${\cal U}$. 
We note that our semiclassical approach preserves all required symmetries,
in particular the properties of the 
reduced density matrix ${\rm Tr}_1[\rho_1(t)]=1$, $\rho_1=\rho_1^\dagger$, 
as well as the symmetry ${\rm Tr}_1[\rho_1^2(t)]={\rm Tr}_2[\rho_2^2(t)]$.

Eq.~(\ref{purity}) expresses the decay of ${\cal P}(t)$
as a sum over dynamical, purely classical contributions, and quantal ones, 
depending on the 
interaction strength. Because the decaying terms are exponential,
with prefactors of order unity, the purity
can be rewritten
\begin{equation}\label{puritysum}
{\cal P}(t) \simeq
\exp[-{\rm min} (\lambda_1,\lambda_2,2 \Gamma_2 ) t] + 
N_1^{-1} + N_2^{-1},
\end{equation}
a form which expresses more explicitly how
Eq.~(\ref{puritysum}) reconciles the results of 
Refs.~\cite{Mil99} and \cite{Tan02}. 
The regime of validity of Eq.~(\ref{puritysum})
is parametrically large in the semiclassical limit $N_{1,2} 
\rightarrow \infty$. 

Four more remarks are in order here. 
First, the power-law decay
of ${\cal P}(t)$ predicted above for regular systems,
is to be taken as an average over initial conditions ${\bf r}_{1,2}$
(in that respect see Refs.~\cite{Jac03} and \cite{Pro03c}), but
may also hold for individual initial conditions, as e.g.
in \cite{Zni03a}. Second, there are cases when the
correlators (\ref{correl1}) and (\ref{correl2}) decay exponentially in time 
with a rate related to the spectrum of Lyapunov exponents.
This also may induce a dependence of ${\cal P}(t)$
on the Lyapunov exponents, which can be captured by the linear response
approach of Ref.~\cite{Tan02}. We note however that this is not necessarily a
generic situation, as many fully chaotic, but nonuniformly hyperbolic
systems have power-law decaying correlations. 
Third, we mention that
because of the second line in Eq.(\ref{2correl}),
the connection 
between decoherence and Loschmidt Echo breaks down at short times where
the decay of ${\cal P}(t)$ is significantly slower than the decay of ${\cal M}_{\rm L}$.
The calculations presented in some details in this chapter and 
in particular our main result, Eq.~(\ref{puritysum}), complement
Refs.~\cite{Pet06b,Jac04a}

Outside the semiclassical regime of validity of Eq.~(\ref{purity}), the purity has a Gaussian decay, either given by first-order perturbation theory, or by the system's bandwidth. These two decays cannot
be captured by semiclassics. Instead we follow our standard procedure 
and present a detailed RMT calculation of the purity decay in these two regimes.

\subsection{RMT approach to entanglement in bipartite interacting systems}
\label{section:P_RMT}

The semiclassical results just derived suggest that the purity of the
reduced one-particle density matrix in a two-particle problem behaves 
just like the fidelity in a Loschmidt echo experiment. 
This similarity is complete 
in the golden rule regime -- up to short-time corrections -- and
only necessitates to replace one-particle energy scales
by their two-particle counterpart -- the level spacing $\delta_2$, 
the golden rule broadening $\Gamma_2 $ and the energy bandwidth $B_2$.
The RMT calculation we are about to present
is very enlightening in that it clearly indicates the origin of this
similarity, and extends it beyond the golden rule regime. 

Two-particle RMT for ${\cal P}(t)$ is not very different from 
one-particle RMT for ${\cal M}_{\rm L}$. The interaction between particles,
together with the tracing over the degrees of freedom of the second particle
effectively results in a perturbation operator acting on the degrees 
of freedom of the first particle. Without restriction on generality other than
considering chaotic dynamics, the 
statistical properties of that operator are the same as those of the
perturbation $\Sigma$ for ${\cal M}_{\rm L}$. This is so 
because a two-body interaction
operator acting on two chaotic particles generically gives a full
matrix, when expressed in the basis of 
noninteracting states~\cite{Wei96}. 
This is no longer the case for larger number $M$ of particles,
unless one considers $M$-body interactions~\cite{Bro81,Abe90,Geo97,Jac97a}.

We start by rewriting the purity as
\begin{eqnarray}\label{eq:purity_start_rmt}
{\cal P}(t) &=& \int {\rm d} {\bf x} \, {\rm d}{\bf y} \,  {\rm d}{\bf r} \,  {\rm d}{\bf r}' \,
\langle {\bf x} \otimes {\bf r}| \,e^{-i {\cal H} t} \rho_0 e^{i {\cal H} t} \, | {\bf y} \otimes {\bf r} \rangle
\langle {\bf y} \otimes {\bf r}'| \, e^{-i {\cal H} t} \rho_0 e^{i {\cal H} t} \, | {\bf x} \otimes {\bf r}' \rangle ,
\end{eqnarray}
where as before, $\rho_0 = |\psi_1 , \psi_2 \rangle \langle \psi_1 , \psi_2|$.
As for the Loschmidt echo, our RMT strategy consists in inserting 
resolutions of the identity into Eq.~(\ref{eq:purity_start_rmt})
and then use generalization of the RMT averages of Eq.~(\ref{contractions}). 
We write
\begin{eqnarray}\label{resol_id1}
I & = & \sum_{\alpha_1,\alpha_2} |\alpha_1 \rangle \langle \alpha_1 | \otimes |\alpha_2 \rangle \langle \alpha_2 | \\\label{resol_id2}
I & = &
\sum_{\Lambda} |\Lambda \rangle \langle \Lambda |,
\end{eqnarray}
where $|\alpha_{1,2}\rangle $ are single-particle eigenstates of $H_{1,2}$,
and $|\Lambda \rangle$ is a two-particle eigenstate of ${\cal H}$. We recall
that the particles are assumed distinguishable. We need RMT averages. 
Restricting ourselves to 
the leading-order contribution in $N_1^{-1}$ and $N_2^{-1}$ and neglecting 
in particular
weak localization corrections,
Eqs.(\ref{contractions}) translates into
\begin{subequations}\label{rmt_contract_tip}
\begin{eqnarray}
\overline{\langle  \alpha_1 , \alpha_2 | \phi_1  ,  \phi_2 \rangle} &=& 0, \qquad \\
\overline{\langle  \alpha_1 , \alpha_2 | \phi_1 , \phi_2  \rangle \langle \phi_1 , \phi_2
  |  \beta_1 ,  \beta_2 \rangle}  &=& N_1^{-1} \, N_2^{-2}  \; \delta_{\alpha_1,\beta_1} \, \delta_{\alpha_2,\beta_2}, \qquad \\
\overline{\langle  \alpha_1 ,  \alpha_2 | \phi_1 , \phi_2 \rangle \langle \phi_1 , \phi_2 |  \beta_1 ,  \beta_2 \rangle
\langle  \gamma_1 , \gamma_2 | \phi_1 , \phi_2 \rangle \langle \phi_1 , \phi_2 | 
\delta_1 , \delta_2 \rangle}  &=& N_2^{-2} N_1^{-2}  \qquad \\
\;\;\; \times (\delta_{\alpha_1,\beta_1} \delta_{\gamma_1,\delta_1} + \delta_{\alpha_1,\delta_1} \delta_{\beta_1,\gamma_1}) 
 (\delta_{\alpha_2,\beta_2} \delta_{\gamma_2,\delta_2} &+& \delta_{\alpha_2,\delta_2} \delta_{\beta_2,\gamma_2}). \nonumber
\end{eqnarray}
\end{subequations}
In these expression, $|\alpha_i \rangle$ and $|\beta_i \rangle$ 
denote eigenfunctions of $H_0$ while
$|\phi_{1,2}\rangle$ can be either $|\psi_{1,2}\rangle $, 
$|{\bf x} \rangle$, $|{\bf y}\rangle$ or $|{\bf r}'\rangle$,
$|{\bf r}'\rangle$ in Eq.~(\ref{eq:purity_start_rmt}). 
Within RMT, ${\cal P}(t)$ is given by the sum of three 
terms
\bea\label{calP_sum}
{\cal P}(t) & = & {\cal P}_1(t) + {\cal P}_2(t) + {\cal P}_3(t), \\
{\cal P}_1(t) & = & N_1^{-1} + N_2^{-1} , \\
{\cal P}_2(t) & = & N_1^{-1} N_2^{-1}.
\eea
The time-dependent decay of the purity is dominantly determined by
${\cal P}_3(t)$, which we now proceed to calculate. 
The calculation of the saturation contributions ${\cal P}_{1,2}(t)$
proceeds along the same lines, and we therefore only write 
the final results here.

We sandwich each of the two initial density matrices $\rho_0$ in 
Eq.~(\ref{eq:purity_start_rmt}) between resolutions of identity as in
Eq.~(\ref{resol_id1}). We next perform the RMT averages 
(\ref{rmt_contract_tip}) with all terms involving $\psi_{1,2}$. This gives
the three terms in Eq.(\ref{calP_sum}), and in particular, 
\bea
{\cal P}_3(t) & = & N_1^{-2} N_2^{-2} \sum_{\alpha_1,\beta_1,\gamma_1,\delta_1}
\sum_{\alpha_2,\beta_2,\gamma_2,\delta_2} 
\langle \alpha_1,\alpha_2 |
e^{-i {\cal H}t} | \beta_1,\beta_2 \rangle
\langle \gamma_1,\beta_2 | e^{i {\cal H}t} | \delta_1,\alpha_2 \rangle
\nonumber \\
&& \;\;\;\;\;\;\;\; \;\;\;\;\;\;\;\; \;\;\;\;\;\;\;\;\;\;\;\;\;\; \;\;\;\;\;\;\;\; \times
\langle \delta_1,\gamma_2 | e^{-i {\cal H}t} | \gamma_1,\delta_2 \rangle
\langle \beta_1,\delta_2 | e^{i {\cal H}t} | \alpha_1,\gamma_2 \rangle.
\eea
It is easily checked that ${\cal P}_3(t=0)=1$, which confirms that it
is the dominant term at not too long times. 
We next insert four resolutions of identity as 
in Eq.~(\ref{resol_id2}) around the time-evolution operators
$\exp[\pm i {\cal H}t]$. There are three different regimes of interaction
and, as for the Loschmidt echo, they are
differentiated by the three energy scales, $\delta_2$, 
$\Gamma_2 \simeq 2 \pi 
\overline{|\langle \alpha_1,\alpha_2|{\cal U}|\beta_1,\beta_2\rangle|^2}$,
and $B_2$.
The projection of interacting states over noninteracting ones is 
regime-dependent and given by~\cite{Boh69,Wig57,Jac95,Fra95,Fyo95,Fla00,Abe90,Geo97}
\bea\label{eq:2part_spread}
\overline{|\langle \alpha_1, \alpha_2 | \Lambda \rangle|^2}
& = & \left\{
\begin{array}{cc}
\delta_{(\alpha_1,\beta_1),\Lambda}, & \Gamma_2 < \delta_2,  \\
(\Gamma_2 \delta_2 / 2 \pi) \big/[(E_\Lambda-\epsilon_{\alpha_1}-\epsilon_{\alpha_2})^2+\Gamma_2^2/4], & \;\; \delta_2 \lesssim \Gamma_2 \ll B_2, \\
N_1^{-1} N_2^{-1}, & \Gamma_2 \gtrsim B_2,
\end{array}
\right.
\eea
whereas 
$\overline{\langle \alpha_1, \alpha_2 | \Lambda \rangle 
\langle \Lambda | \beta_1, \beta_2 \rangle} = 0$ if 
$\alpha_1 \ne \beta_1 $ or $\alpha_2 \ne \beta_2$.
The corresponding three asymptotic decays of the purity read, to leading order,
\begin{eqnarray}
{\cal P}(t) = \left\{
\begin{array}{cc}
\exp[-\sigma_2 t^2] &  \Gamma_2 < \delta_2,  \\
\exp[-2 \Gamma_2 t] & \delta_2 \lesssim \Gamma_2 \ll B_2, \\
\exp[-B_2^2 t^2] &  \Gamma_2 \gg B_2,
\end{array}
\right.
\end{eqnarray}
with the RMT result $\sigma_2^2 \equiv {\rm Tr} \; {\cal U}^2/(N_1 N_2)$. 
Comparison with Eq.~(\ref{eq:echo_leading}) establishes the similarity
between the two-particle purity and the Loschmidt echo. Moreover,
the equivalence between semiclassics and RMT in the golden rule regime
that was already observed at the level of ${\cal M}_{\rm L}$ also
prevails for ${\cal P}(t)$.

\begin{figure}
\includegraphics[width=12cm]{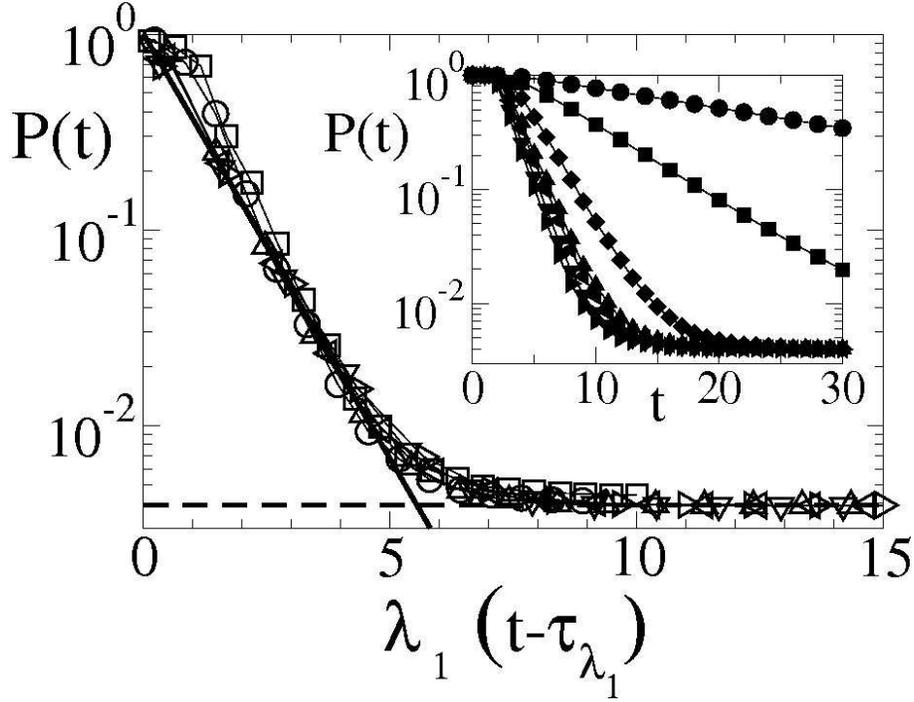}
\caption{\label{fig1_qmclcorr}
${\cal P}(t)$ for the coupled kicked rotator model of Eq.~(\ref{2krot}) with
$N=512$, $K_1=K_2 \in[4,12]$, and
$\epsilon=4/N^2$ giving $2 \Gamma_2 = 13.6 \gg \lambda_1 = \lambda_2$. 
Data are averages over
twenty different initial states. The time axis has been shifted by the onset
time $\tau_{\lambda_1}$ of Eqs.~(\ref{eq:purity_decay}) 
and (\ref{puritysum_num}),
and rescaled with $\lambda_1 \in [0.5,1.35]$.
The full line indicates $ \propto \exp[-\lambda_1 t]$, and
the dashed line gives the asymptotic saturation ${\cal P}(\infty) = 2N^{-1}$, in agreement
with the theoretical predictions.
Inset: Purity for $K_1=K_2=5.09$ for
$ \epsilon = 0.2$ (circles), $0.4$ (squares), $0.8$ (diamonds), 
 $1.6, 2,3$ and $4$ (triangles). (Figure taken from Ref.~\cite{Pet06b}. Copyright (2006) by the American Physical Society.)}
\end{figure}

\subsection{Numerical experiments on entanglement generation}
\label{section:nums_entangle}

To numerically check our results, we consider the Hamiltonian of 
Eq.~(\ref{2hamiltonian}) for the specific case of two coupled kicked rotators~\cite{Izr90}. Some details
of the model are given in Appendix~\ref{appendix:2krot}. Here we only mention the three relevant
energy scales. The two-particle bandwidth and level spacing are given by
$B_2 = 2 \pi$, $\delta_2 = 2 \pi/(N_1 N_2)$, and we consider
two equally large Hilbert spaces with $N\equiv N_1=N_2$. Also, 
the interaction term with strength $\epsilon$ between the two systems gives rise to a broadening
$\Gamma_2 \simeq 0.43\epsilon^2 N^2$ of the two-particle energy levels.

We check the validity of our prediction
\bea\label{puritysum_num}
{\cal P}(t) &\simeq& \alpha_1 \Theta(t>\tau_{\lambda_1})
\exp[-\lambda_1 t] +\alpha_2 \Theta(t>\tau_{\lambda_2})
\exp[-\lambda_2 t] + \Theta(t>\tau_{\Gamma}) \exp[-2 \Gamma_2 t] \nonumber \\
&& + \Theta(t > \tau^{(1)}_{\rm E}) N_1^{-1} + \Theta(t > \tau^{(2)}_{\rm E}) N_2^{-1},
\eea
for the decay of the purity in chaotic systems.
The behavior of ${\cal P}(t)$ is shown in Figs.~\ref{fig1_qmclcorr}, ~\ref{fig6_qmclcorr} and~\ref{fig7_qmclcorr}. We first focus on symmetric two-particle systems where both particles have the
same size of Hilbert space and the same Lyapunov exponent.
Fig.~\ref{fig1_qmclcorr} illustrates perhaps the most spectacular finding
of the analytical approach presented above, 
that under proper conditions, the generation of
entanglement is given by a classical Lyapunov exponent. The inset shows that,
as the interaction strength $\epsilon$ increases, so does the rate of entanglement generation,
up to some value $\epsilon_c$ after which it saturates. The main part of 
Fig.~\ref{fig1_qmclcorr}
furthermore shows that in the saturated regime, 
the decay rate of the purity is given by the
classical Lyapunov exponent, ${\cal P}(t) \propto \exp[-\lambda_{1,2} t]$.
The rescaling of the time axis $t \rightarrow
\lambda_1 t$ allows to bring together six curves with 
$\lambda_1 \in [0.5,1.35]$, varying by almost a factor three. Third, Fig.~\ref{fig1_qmclcorr}
shows that in the chaotic regime considered here, 
${\cal P}(t \rightarrow \infty) = 2 N^{-1}$. 

\begin{figure}
\includegraphics[width=12cm]{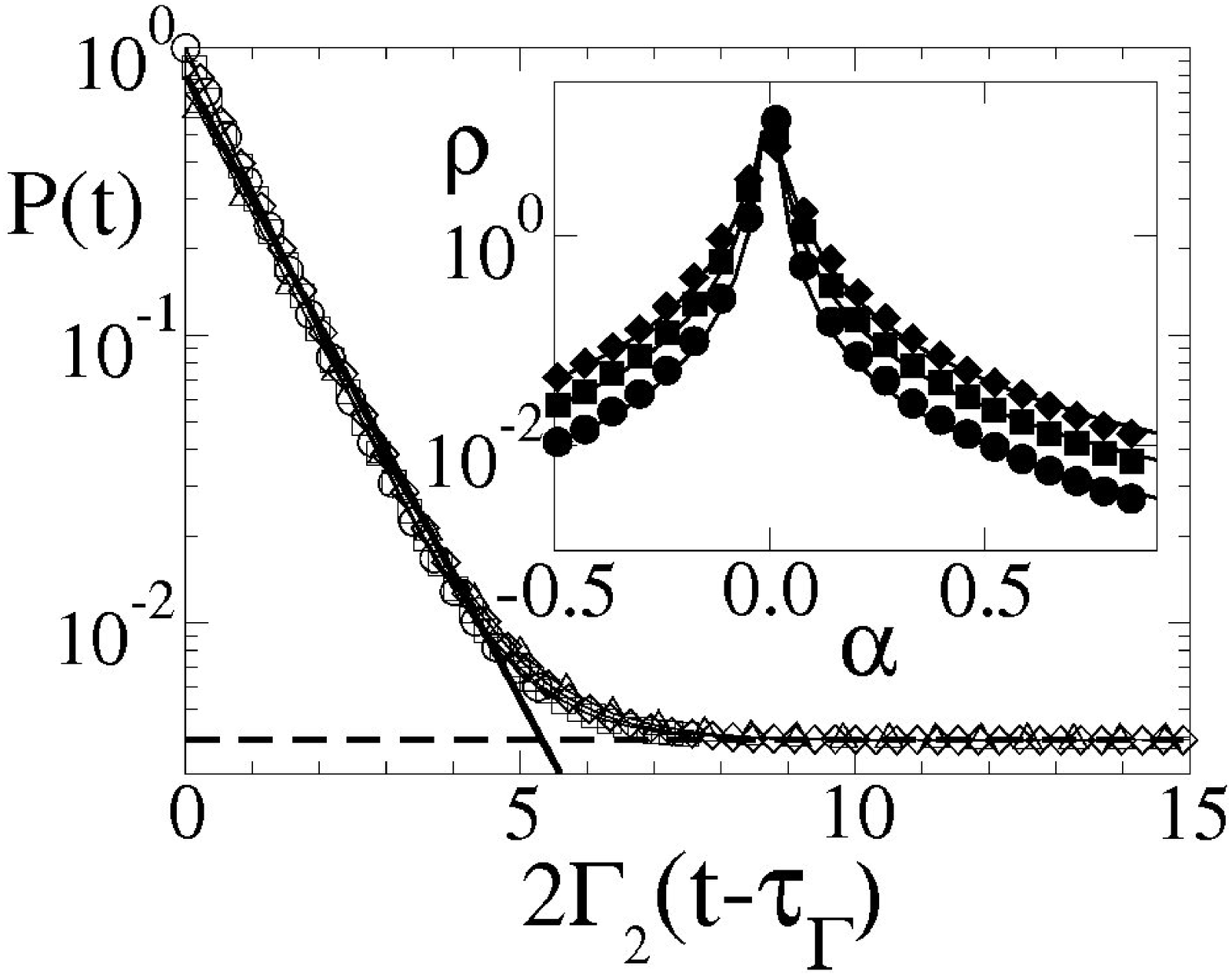}
\caption{\label{fig6_qmclcorr}
${\cal P}(t)$ for the coupled kicked rotator model of Eq.~(\ref{2krot}) with
  $N= 512$, in the golden rule regime with
$\Gamma_2 < \lambda_i$, for
 $K_0=K_1 =50.09$, and different 
$\Gamma_2 \approx 0.43 (\epsilon N)^2$,   with $(\epsilon N)^2=0.06 $ (circles),  
 $0.3 $ (squares),   $0.6 $ (diamonds) and  $0.9 $ (triangles). 
Data points are averages over twenty different initial Gaussian wavepackets.
The time axis has been shifted by the onset  time $\tau_\Gamma$ 
of Eqs.~(\ref{eq:purity_decay}) and (\ref{puritysum_num}), and rescaled with  $2 \Gamma_2  \in [5.\,  10^{-2},   \,  8.\,  10^{-1}]$. The full line indicates the decay $\propto  \exp[-2 \Gamma_2 t]$ without any free parameter.  
The dashed line gives the saturation  ${\cal P}(\infty) = 2N^{-1}$. 
Inset: local spectral density of
states $\rho(\alpha)$ 
of eigenstates of a noninteracting double kicked rotator
over the eigenstates of an interacting  double 
kicked rotator, both with $K_1= K_2= 50.09$ . Both system
sizes are $N=64$,   with $(\epsilon N)^2=0.037 $ (circles),   $0.1 $ (squares),    $0.163 $ 
(diamonds). The solid lines are Lorentzian
with widths $\Gamma_2 \approx 0.016$,   $0.042$ and $0.07$. 
From these and other data at different $N$  
we extract $\Gamma_2 = 0.43 \; (\epsilon N)^2$. 
}
\end{figure}

We next focus on the golden rule decay. We have 
found that (i) prior to saturation, ${\cal P}(t)$ decays exponentially with a rate close to
twice the golden rule
rate, $\propto \exp[-0.85 \, \epsilon^2 N^2 t]$, provided $\Gamma_2 = 0.43 \epsilon^2 N^2 > 
\delta_2 = 2 \pi/(N^2)$ is satisfied, and that (ii)
$\epsilon_c$ behaves consistently with Eq.~(\ref{puritysum}). This is illustrated in 
Fig.~\ref{fig6_qmclcorr}. The inset shows the behavior of the  local spectral density of noninteracting
eigenstates over interacting eigenstates. The curves are well fitted with Lorentzians of width
$\Gamma_2 \approx 0.43 \epsilon^2 N^2$. With this extracted value of $\Gamma_2$ in mind,
we next plot the purity ${\cal P}(t)$ in the regime $\delta_2 < \Gamma_2 \ll B_2$ with 
$\Gamma_2 < \lambda_{1,2}$ in the main panel of Fig.~\ref{fig6_qmclcorr}. Once the
horizontal axis is rescaled as $t \rightarrow
2\Gamma_2 t$ four curves corresponding to 
$ 2\Gamma_2 \in [5.\,  10^{-2},   \,  8.\,  10^{-1}] $ are brought together, confirming the
golden rule decay ${\cal P}(t) \propto \exp[-2 \Gamma_2 t]$ with the broadening of
two-particle levels due to the interaction. 

In our third figure, Fig.~\ref{fig7_qmclcorr} we investigate the 
independence of ${\cal P}(t)$ on $\lambda_2$ in the regime 
$\lambda_2 \gg \lambda_1$.
The main plot shows ${\cal P}(t)$ for $\lambda_1\simeq 0.97$,   
and four values of $\lambda_1 \in [0.97,  \,  3.2]$. Varying $\lambda_2$
by more than a factor of three has no effect on the asymptotic decay of 
${\cal P}(t)$. We conclude that its decay is given by $\exp[-{\rm min}(\lambda_1,\lambda_2)t]$, 
in agreement with Eqs.~(\ref{puritysum}) and (\ref{puritysum_num}).
In the inset, data moreover confirm the behavior given in Eq.~(\ref{eq:purity_sat})
of the long time saturation of the purity, 
${\cal P}(\infty)=  N_1^{-1} +N_2^{-1} $.

\begin{figure}
\includegraphics[width=12cm]{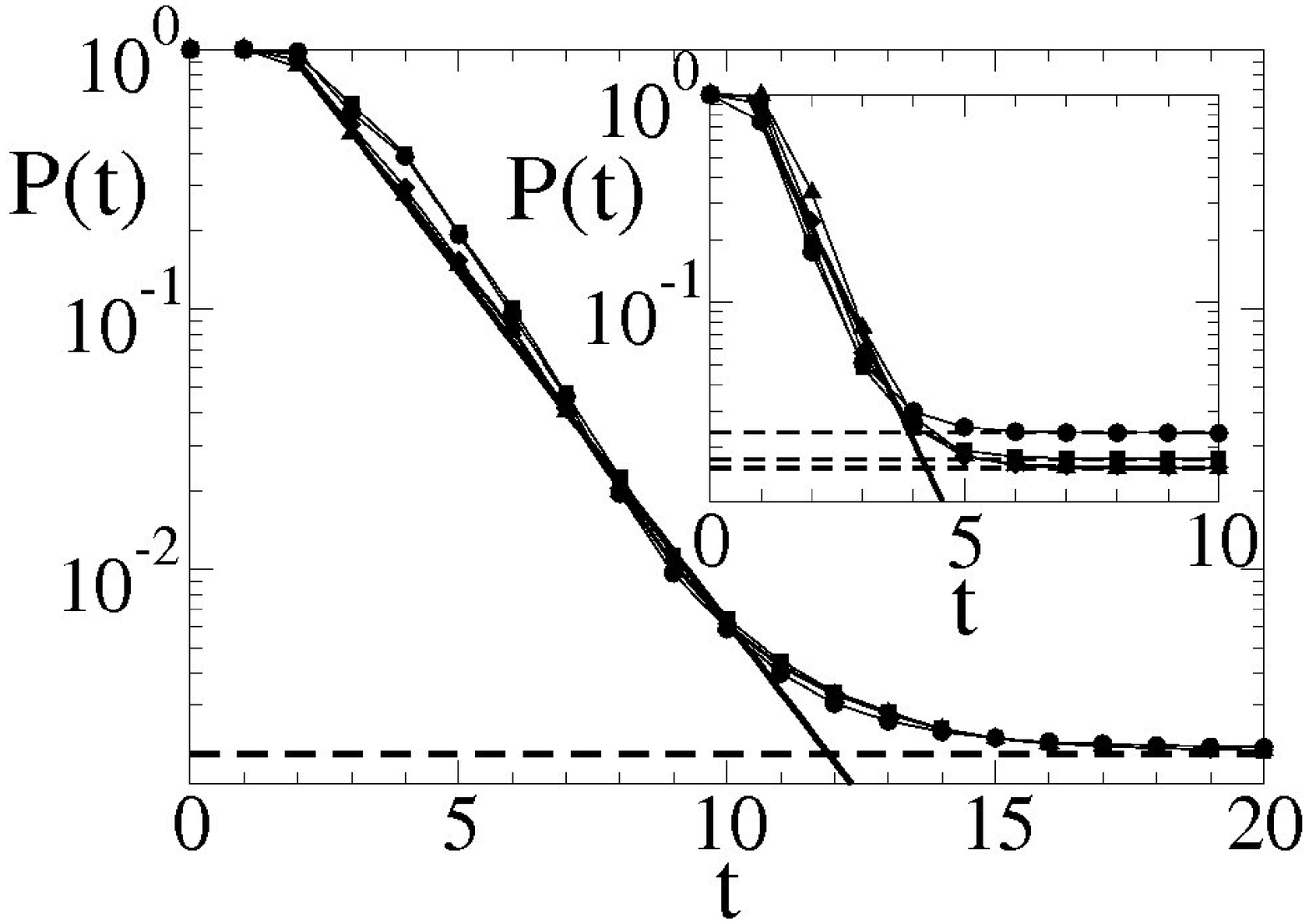}
\caption{\label{fig7_qmclcorr}
${\cal P}(t)$ for the coupled kicked rotator model of Eq.~(\ref{2krot}) with
$N=1024$, in the golden rule regime with 
$2 \Gamma_2 \gg \lambda_1$,
 $K_1= 5.09$,  $\epsilon^2 N^2 = 4$ and $K_2= 5.09$,   (circles),  
 $10.09  $ (squares),    $20.09 $ (diamonds),   $50.09 $ (triangles). 
The full line indicates the decay $\propto  \exp[-\lambda_1 t]$.
The dashed line gives the saturation ${\cal P}(\infty) =2N^{-1} $. 
Inset : Purity  ${\cal P}(t)$ in the regime 
$2 \Gamma_2 \gg \lambda_1$  for
 $K_1= 10.09$,   $K_2=50.09$, $\epsilon^2 N^2 =4 $ and 
$N_1=64$, $N_2 = 128$ (circles),  
 $512$ (squares),    $2048 $ (diamonds)  $8192 $ (triangles).  
 The full line indicates the 
decay $\propto  \exp[-\lambda_1 t]$. The dashed lines give the 
long-time saturation, 
${\cal P}(\infty) =N_1^{-1}+N_2^{-1}$.
All data points are averages over 
$20$ different initial Gaussian wavepackets. }
\end{figure}

These numerical data fully confirm our semiclassical and RMT
analytical theories, specifically our final result,
Eq.~(\ref{puritysum_num}).

\subsection{Towards decoherence : classical phase-space behavior}
\label{towards_decoherence}

Decoherence is nothing else but entanglement with a large, complex,
uncontrolled environment. It is thus very tempting to extrapolate the
analytical results obtained earlier in this section to the 
problem of decoherence -- a semiclassical theory of decoherence
would certainly be very helpful in investigating the conditions
under which quantum mechanics delivers classical mechanics (as we believe
it should). One central question in that respect 
is whether the observed classical entanglement rate 
translates into a Lyapunov decoherence rate for systems
coupled to a true environment -- much more complex and bigger than
a single-particle dynamical system. The times scales in such an environment 
are much shorter, it has moreover a much bigger Hilbert space, and
it cannot be initially prepared in a pure Gaussian wavepacket, or any
other specific state. As a minimal, analytically tractable 
first-step approach, we can
take these conditions into account in our semiclassics
by considering (i) $\lambda_2 \gg \lambda_1$, 
(ii) $N_2 \rightarrow \infty$
and (iii) an initial mixed environment density matrix 
$\rho_{\rm env}=\sum_\alpha |C_a|^2 |\phi_\alpha \rangle 
\langle \phi_\alpha |$, with a set 
$\{\phi_\alpha\}$ of $M \gg 1$ nonoverlapping 
Gaussian wavepackets. 
The semiclassical calculation gives that
Eq.~(\ref{puritysum_num}) is replaced by
\begin{equation}\label{decoherence}
{\cal P}(t) \simeq \alpha_1 \, \Theta(t>\tau_1) \,
\exp[-\lambda_1 t] + \frac{\alpha_2} {M} \, \Theta(t>\tau_1)
\, \exp[-\lambda_2 t] 
+ \exp[-2 \Gamma_2 t] 
+ N_1^{-1} \; \Theta(t > \tau^{(1)}_{\rm E}) .
\end{equation}
The Lyapunov decay of the purity thus seems to survive in the case
of a particle coupled to an environment, but even if $\lambda_2$  
remains finite, there is no decay with the Lyapunov exponent of the
environment in the limit $M \rightarrow \infty$ of a very complex environment, 
because the initial state is no longer meaningful
classically -- the initial state of the environment cannot be prepared!
This is similar to the behavior of the Loschmidt echo
for superpositions (see Eqs.~(\ref{eq:decay_pure}) and 
(\ref{eq:decay_mixed}) and below). 
The same disappearance of the $\lambda_2$-term occurs for an incoherent
superposition of $M$ Gaussians, but this term does not exist to start with if the initial state of the
second particle is a random pure state, a random mixture, or a thermal state.
Ref.~\cite{Lee05a}
investigated decoherence of a two-level system coupled to an external 
dynamical system, and found that in some circumstances, it occurs
at a rate given by the Lyapunov exponent of the external system.
This finding might be valid when the external dynamical
system is a detector over which one has some control, and whose initial
state can accordingly be prepared.
It does not apply to general cases of
decoherence by a complex environment.

There is another, perhaps more quantitative argument suggesting that
the behavior of the purity in bipartite quantum dynamical systems 
is reflected in the decoherence of dynamical systems coupled to 
complex environments.
The standard approach to decoherence starts from
a master equation valid in the regime of 
weak system-environment coupling~\cite{Joo03,Zur03}. 
The master equation is a generalization of Eq.~(\ref{Moyalbracexp}),
which takes into account the coupling to an external environment.
In the case when the potential in the system's Hamiltonian only depends
on the spatial degrees of freedom, 
the time-evolution of the system's Wigner
function is determined by
\begin{equation}\label{Wigner_o}
\frac{\partial W_\psi}{\partial t}  = \Big\{ H,W_\psi \Big\} + \sum_{n \ge 1}
\frac{(-1)^{n}}{2^{2n} (2n+1)!} \frac{\partial^{2n+1}}{\partial {\bf q}^{2n+1}} V \frac{\partial^{2n+1}}{\partial {\bf p}^{2n+1}} W_\psi
+ 2 \gamma \frac{\partial}{\partial {\bf p}} (pW_\psi) + D \frac{\partial^2}{\partial {\bf p}^2} W_\psi.
\end{equation}
The first term on the right-hand side of Eq.~(\ref{Wigner_o}) is the classical 
Poisson
bracket. As discussed in Chapter~\ref{chapter:Wigner},
the second term
exists already in closed systems and generates quantum corrections
to the dynamical evolution of $W_\psi$. 
This term starts to become comparable to the Poisson
bracket at the Ehrenfest
time~\cite{Zur03}. Up to there, the equation describes the
time-evolution of the Wigner function in an isolated system,
Eq.~(\ref{Moyalbracexp}). 
The last two terms on the right-hand side of Eq.~(\ref{Wigner_o})
are induced by the coupling to the environment. The third term is a friction
term, inducing dissipation and deviations from the unperturbed dynamics
generated by $H$, and the fourth term induces diffusion in momentum. For details
on how Eq.~(\ref{Wigner_o}) is derived, we refer the reader
to Refs.~\cite{Zur03,Zur93,Joo03}.

Starting from Eq.(\ref{Wigner_o}), the following scenario has been proposed 
for the emergence of classical mechanics out of quantum mechanics~\cite{Zur03,Joo03}.
In the limit of weak system-environment coupling, 
$\gamma \rightarrow 0$,
but finite diffusion constant, $D \propto \gamma T = {\rm Cst}$ --
this implicitly assumes high temperatures -- the friction 
term vanishes, leaving the classical dynamics unaffected. 
Simultaneously, for large enough $D$, 
the momentum diffusion term 
induces enough noise so as to kill the quantum corrections before they become 
important. One then hopes that this can occur without relaxing the dynamics generated
by the Hamiltonian. If and when this is the case, the result is a true quantum-classical correspondence. 
But is this possible at all ? Our answer is yes, at least as long as external couplings can be sent to
zero in the semiclassical limit. Then, shadowing and structural stability theorems can rigorously
be invoked, which replace each unperturbed (without coupling to external degrees of freedom) 
classical trajectories with a shadowing perturbed ones, in the sense that the two orbits remain
very close to one another for arbitrarily long times. Still the coupling is strong enough that it
generates enough dephasing. Perhaps the main knowledge we have gained
in our semiclassical investigations 
is that mathematically rigorous theorems explain how decoherence can beat relaxation -- how
external sources of noise, external baths or environments to which a quantal system is weakly 
coupled can decohere that system without changing the dynamics it follows -- with the original
$H$ in the first term on the right-hand side of Eq.~(\ref{Wigner_o}), and not some new $H'=H'(D,\gamma,T)$.

The time-evolution of $W_\psi$ is then solely governed by the classical Poisson bracket,
that is to say, classical dynamics emerges out of quantum mechanics.
Refs.~\cite{Hab98,Tos05} provided for some numerical illustration of this 
scenario.
Accordingly, claims have been made of an environment-induced entropy 
production governed by the system's
Lyapunov exponent $\lambda$ 
\cite{Zur03,Pat99,Mon00}, without rigorous analytical 
derivation, nor strong numerical evidence (Refs.~\cite{Pat99,Mon00} show 
entropy production at a single, fixed value of the Lyapunov exponent).  A trajectory-based 
semiclassical treatment has been applied
to a stochastic Schr\"odinger equation in Ref.~\cite{Kol96a,Kol96b},
concluding that decoherence can occur at a Lyapunov rate. In this chapter,
we verify the validity of this scenario, and investigate if it is at all related
to the extrapolation~(\ref{decoherence}) of our results on 
entanglement generation presented above. To this end, we
consider a minimal toy model, where the environment is modeled 
by a second dynamical
system. We establish the connection between our main result in this section, 
Eq.~(\ref{puritysum}), and its extrapolations, Eqs.~(\ref{Wigner_o}) and (\ref{decoherence}),
can be argued in the following way.
The purity measures the weight of off-diagonal
elements of $\rho_1(t)$, and hence of the importance of coherent
effects, tuned by the second term
on the right-hand side of Eq.~(\ref{Wigner_o}).
According to Eq.~(\ref{decoherence}), in the 
regime $2 \Gamma_2 \gg \lambda_1$, 
${\cal P}(t)$ reaches its minimal value at the Ehrenfest time, i.e. before
quantum effects have a chance to appear. The latter are dephased by the interparticle
coupling and their contribution to the purity of the reduced density matrix
decays exponentially with $2 \Gamma_2$ -- they essentially are killed before they
have a chance to appear if $2 \Gamma_2 \gg \lambda_1$. In that regime, one therefore 
expects the
quantum-classical correspondence to become complete in the semiclassical
limit $N_{1,2} \rightarrow \infty$. 
Let us see in some more details
if a toy model of two interacting particles can still lead to a true
quantum-classical crossover. 

\begin{figure}
\includegraphics[width=7cm]{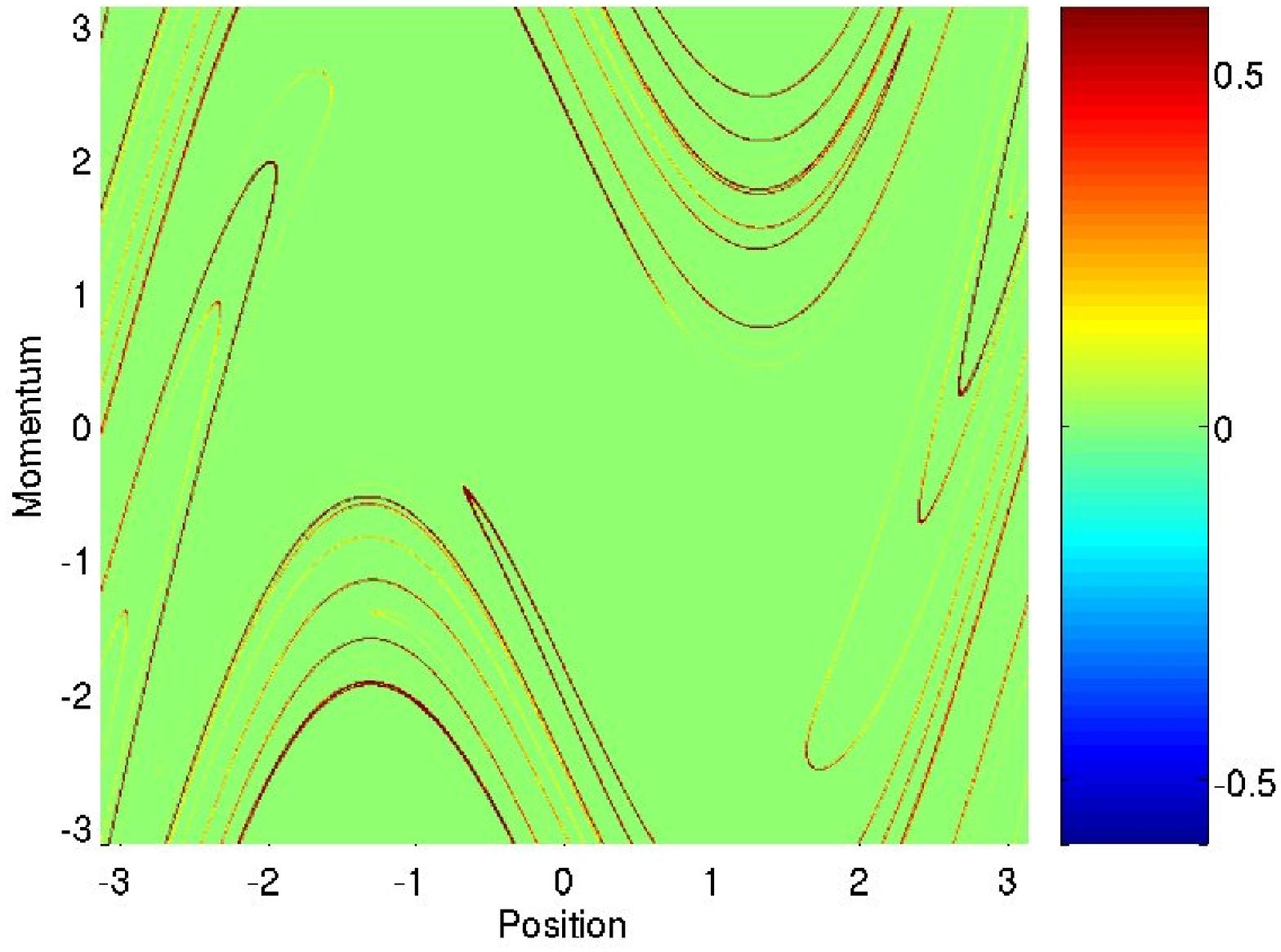}
\includegraphics[width=7cm]{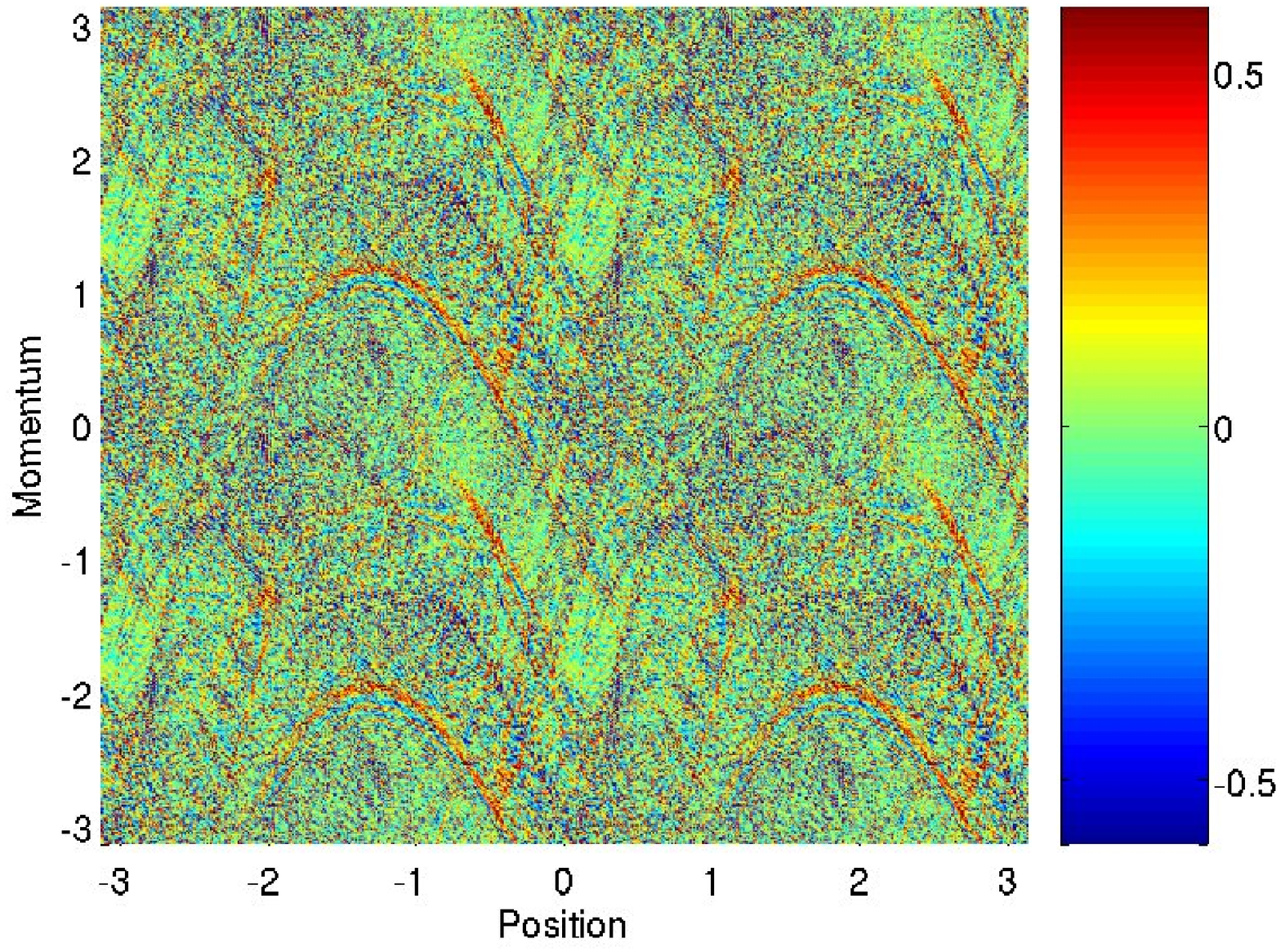} \\
\includegraphics[width=7cm]{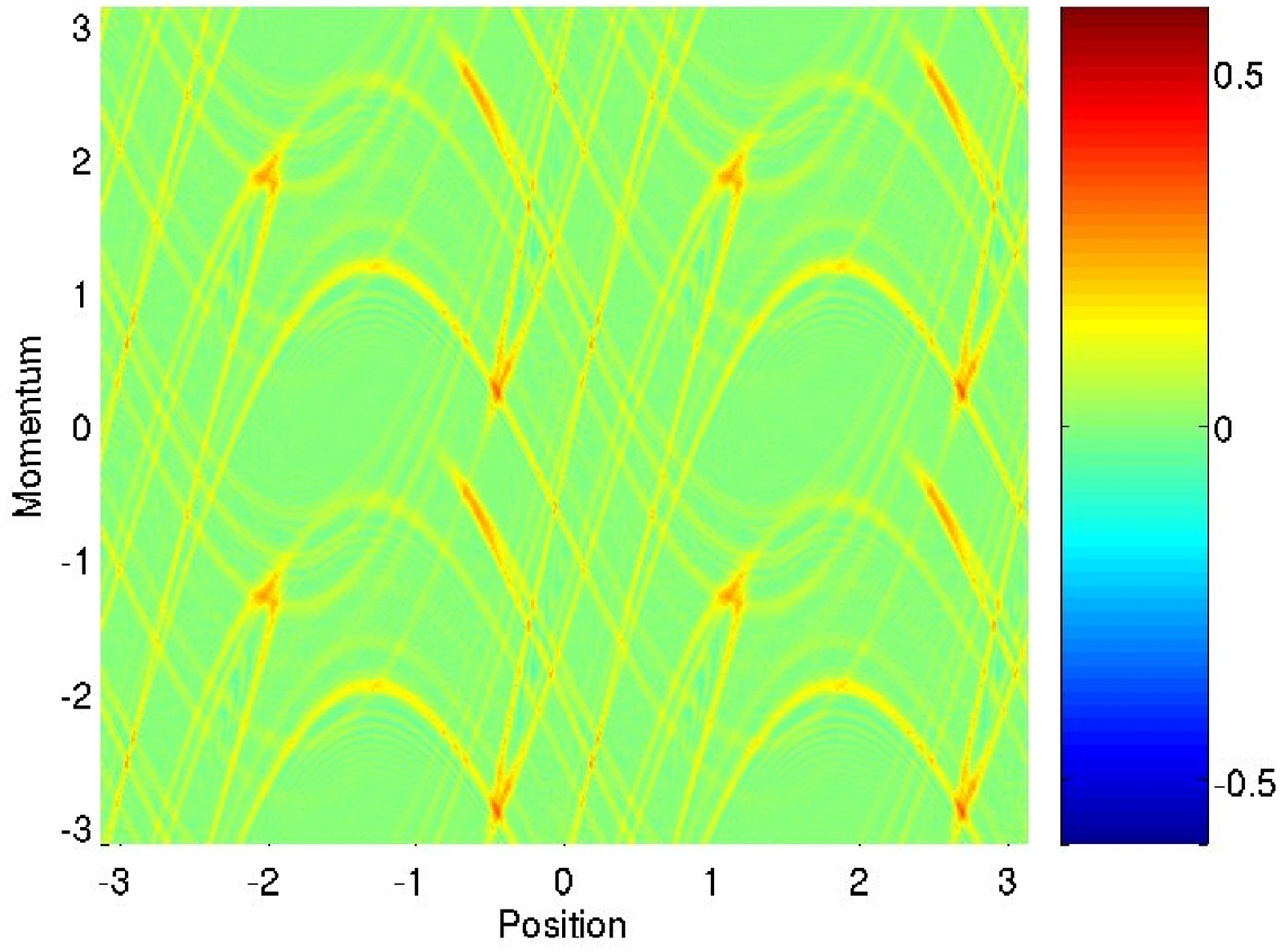}
\includegraphics[width=7cm]{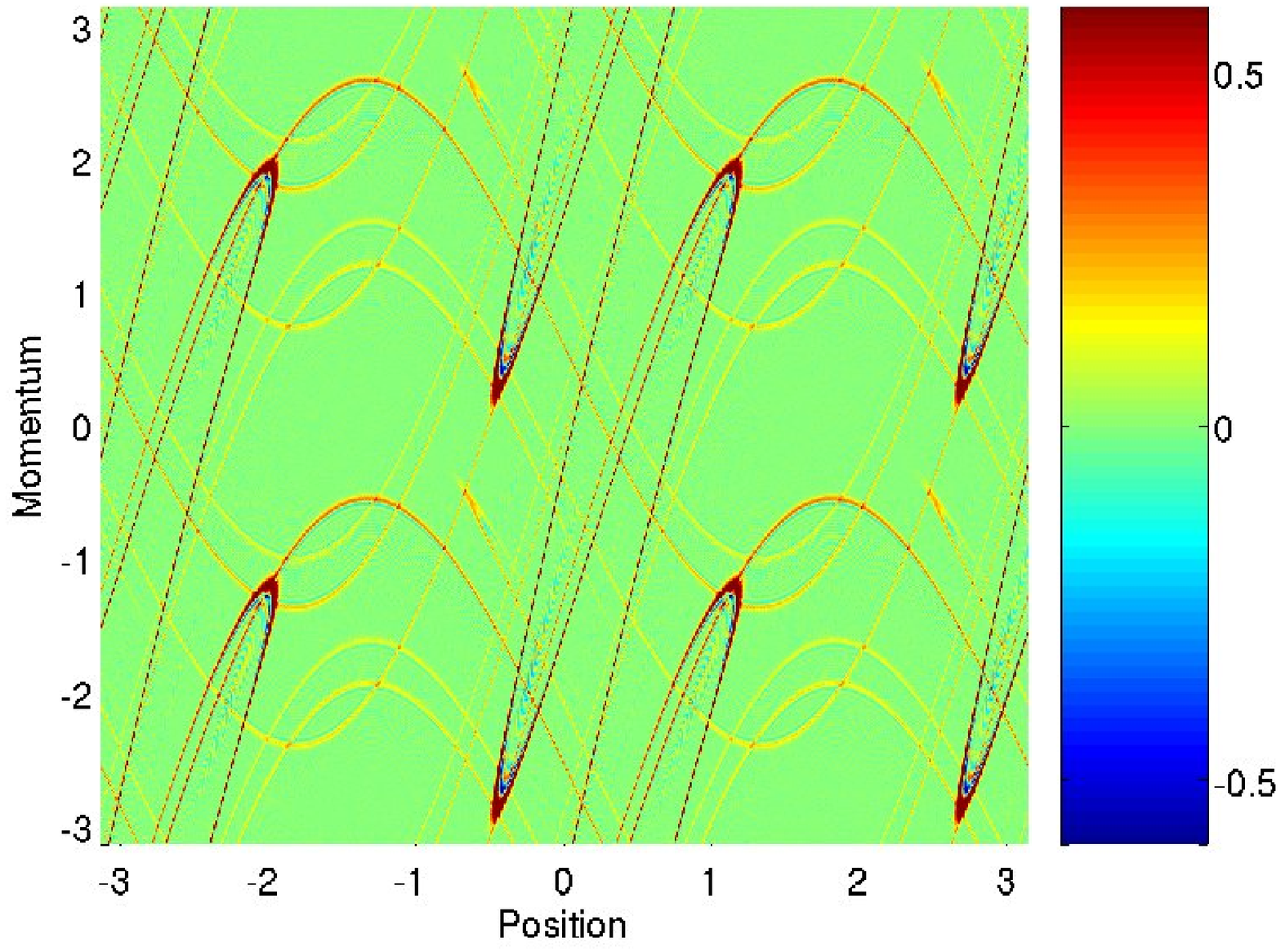}
\caption{\label{fig2_qmclcorr} Phase-space plots for a classical distribution (top left),
uncoupled (top right) and coupled (bottom left and right, $\epsilon=4$) quantum
Wigner distributions, after five iterations of the kicked 
rotator map of Eqs.~(\ref{2hamiltonian}) and (\ref{2krot}). In all cases, the system has $K_1=3.09$, and
the initial distributions are Gaussian centered in the chaotic sea 
at $(x,p)=(1,2)$. Bottom panels: Wigner functions for the 
quantum system coupled to a second kicked rotator
with $K_2 = 100$. One has $2 \Gamma_2=13.6 > \lambda_2 \gg \lambda_1$, 
so that the purity behaves as ${\cal P}(t) \simeq \exp[-\lambda_1 t]$.
The left panel has $N_1=N_2=512$ and the right panel has
$N_1=N_2=2048$. The presence of ghost images in the
Wigner function -- giving replicas of the true structures at
$x \rightarrow x+\pi$ and $p \rightarrow p+\pi$ (see the two bottom panels) 
-- is an artifact of the periodic boundary conditions and our torus 
quantization. This point has been discussed in Ref.~\cite{Arg05}.
(Figures taken from Ref.~\cite{Pet06b}. Copyright (2006) by the American Physical Society.)}
\end{figure}

We follow the lines of Ref.~\cite{Pet06b} to present numerical evidences
supporting this reasoning. We turn our attention to the 
quantum-classical correspondence in
phase space. We compare in
Fig.~\ref{fig2_qmclcorr} the Liouville evolution 
of a classical distribution in an uncoupled dynamical system
with that of the Wigner function $W_{\rho_1}({\bf q},{\bf p},;t) = 
\pi^{-d} \int {\rm d} {\bf x} \exp[2 i {\bf p} {\bf x}]
\rho_1({\bf q}-{\bf x},{\bf q}+{\bf x};t)$ corresponding to the reduced
density matrix of the corresponding quantum system coupled to a second
dynamical system. The Wigner function
is quantum-mechanically evolved 
from a localized wavepacket with the same initial location and extension as
the classical distribution. 
The quantum time-evolution is given by the coupled kicked rotator model of 
Eq.~(\ref{2krot}), while the classical evolution is governed
by a single, uncoupled standard map --
the classical counterpart of the kicked rotator.
Three quantum phase-space plots are
shown: (i) (top right) for an uncoupled system, $\epsilon=0$; (ii) and (iii) 
(bottom left and right) for a coupled system
$\epsilon = 4$, in the regime ${\cal P}(t) \simeq \exp[-\lambda_1 t]$ where the
argument we just presented predicts quantum-classical correspondence. 
The bottom left panel
has a system size $N_1=N_2=512$ while the bottom right panel has $N_1=N_2=2048$. All plots
show phase-space distributions after 5 kicks, a duration comparable 
to $\tau_{\rm E}$.
Two things are clear from these figures. First, a coupling is necessary and sufficient
to achieve phase-space quantum-classical correspondence. Second, the
correspondence becomes better as we move deeper in the semiclassical
regime $N_1,N_2 \rightarrow \infty$. 
Because that limit, to be consistent, requires to keep $\Gamma_2$ constant,
this quantum classical correspondence emerges 
{\it  even though the interaction Hamiltonian vanishes in that limit} !

It seems thus that the coupling to a single dynamical particle is sufficient
to drive a full quantum-classical transition in a parametrically large
range of parameters $\delta_2=B_2/(N_1 N_2) \lesssim
\Gamma_2 \ll B_2$, where the coupling is classically weak. 
Care should be taken in interpreting this result, however, as our
approach explicitly excludes dissipation 
effects~\cite{Cal81,Cal83} and moreover 
neglects possible non-universal, low-temperature contributions
to the coupling correlator~\cite{Hak85}. It is highly desirable to 
extend our analytical approaches to the case of more complex, multipartite
environments. 

\subsection{Irreversibility in partially controlled interacting systems:
the Boltzmann echo}\label{section:boltzmannecho}

We have argued that any quantum reversibility experiment 
is unavoidably polluted 
by its coupling to external degrees of freedom, over which one 
has no control and whose dynamics
cannot be time-reversed. Therefore, if taken naively,
Asher Peres' line of reasoning leading to the introduction of
the Loschmidt echo, Eq.~(\ref{eq:def_LE}), as measure of reversibility neglects the fact that any 
time-reversal operation correctly operates at best 
only on part of the system -- Asher Peres knew of course much better than 
that.
This is so, for instance because the system is composed of 
so many degrees of freedom that the time arrow can be inverted only for
a fraction of them. To capture the physics of echo experiments one thus
has to take into account 
that 

(a) the system decomposes into two
interacting subsystems 1 and 2,

(b) the initial state of the controlled
subsystem 1 is prepared, i.e. well defined, and its final state is measured 
and compared to the initial one,

(c) both the initial and final states of 
the uncontrolled subsystem 2 are unknown, and 

(d) the
Hamiltonian of system 1 is time-reversed with some tunable accuracy, however

(e) both the Hamiltonian of system 2 and the interaction between the two
subsystems are uncontrolled.

\noindent These considerations lead us to introduce the
{\it Boltzmann echo} of Eq.~(\ref{irrevtest}) as 
measure of quantum reversibility, 
instead of the Loschmidt echo of Eq.~(\ref{eq:def_LE}).
In this chapter we follow our letter~\cite{Pet06a} and
present both a semiclassical and a RMT calculation of the 
partial fidelity [we rewrite Eq.~(\ref{irrevtest}) here for convenience]
\begin{eqnarray}\label{irrevtest_repeat}
{\cal M}_{\rm B} (t) = \Big \langle \big\langle \psi_0  \big|    
{\rm Tr}_2 \left[
\exp[-{\it i }{\cal H}_{\rm b} t]  \exp[-{\it i}{\cal H}_{\rm f} t ] \rho_0 
\exp[ {\it i }{\cal H}_{\rm f} t ] \exp[ {\it i}{\cal H}_{\rm b} t ]
\right]\big| \psi_0 \big\rangle \Big \rangle,
\end{eqnarray}
where the forward and backward (partially time-reversed) Hamiltonians 
read 
\begin{subequations}\label{hamiltonians}
\begin{eqnarray}
{\cal H}_{\rm f} & = & H_1 \otimes I_2 + I_1 \otimes H_2 + {\cal U}_{\rm f}, \\
{\cal H}_{\rm b} & = & -[H_1+\Sigma_1] \otimes I_2 
+ I_1 \otimes [H_2+\Sigma_2] + {\cal U}_{\rm b}.
\end{eqnarray}
\end{subequations}
The experiment starts with an initial product density matrix 
$\rho_0  =  |\psi_0\rangle\langle \psi_0| \otimes \rho_2$,
which is propagated forward in time with ${\cal H}_{\rm f}$. 
After a time  $t$, 
we invert the dynamics of system 1, with $\Sigma_1$ modelling the imperfection in that time-reversal
operation. This operation might or might not affect the dynamics of system 2,
which is allowed by the presence of
$\Sigma_2$. We will see below, however, that tracing
over the degrees of freedom of 
system 2 makes ${\cal M}_{\rm B}$ independent of either $H_2$ or
$\Sigma_2$. We leave open the possibility 
that the interaction between the two systems is affected
by the time-reversal operation, i.e. ${\cal U}_{\rm f}$ may or may not
be equal to ${\cal U}_{\rm b}$. Because one has no control over system 2,
the corresponding degrees of freedom are traced out. For the same reason,
the outermost brackets in Eq.~(\ref{irrevtest_repeat}) indicate an average over the initial
density matrix $\rho_2$ for system 2. 
We dubbed ${\cal M}_{\rm B}$ the {\it Boltzmann echo} in Ref.~\cite{Pet06a} to stress its 
connection to Boltzmann's
counterargument to Loschmidt that time cannot be  
inverted for all components of a system with many degrees of freedom. 
We note that, for some specific choices of parameters, ${\cal M}_{\rm B}$ is identical
to the reduced fidelity introduced in Ref.~\cite{Zni03a}.

Clearly, the analytical approaches that worked for the purity ${\cal P}(t)$ in the previous
section also apply here. We therefore start with a presentation of the 
semiclassical calculation of the Boltzmann echo for two classically chaotic 
subsystems. Following a well established routine,
we next compare our results with those obtained using RMT. We finally present numerical
checks of our theories.

Our main result is that, in the regime of classically weak
but quantum mechanically strong
imperfection $\Sigma_1$ and couplings ${\cal U}_{\rm f,b}$, 
${\cal M}_{\rm B}(t)$ is parametrically given by the sum of two exponentials
and a long-time saturation term,
\begin{eqnarray}\label{eq:BEmod} 
{\cal M}_{\rm B}(t) \simeq \exp\left[-  \left(  \Gamma_{\Sigma_1}+ 
\Gamma_{\rm f} +
\Gamma_{\rm b} \right) t \right]+  
\alpha_1  \exp\left[-\lambda_1 t\right] + N_1^{-1},
\end{eqnarray}
with a weakly time-dependent prefactor $\alpha_1={\cal O}(1)$, the Lyapunov exponent
 $\lambda_1$ of system 1, and two perturbation/interaction-dependent rates
 $\Gamma_{\Sigma_1}$ and $\Gamma_{\rm f,b}$
given by classical correlators for $\Sigma_1$ and ${\cal U}_{\rm f,b}$ 
respectively -- we make this quantitative below. These rates can be regarded as the golden 
rule width of the Lorentzian
broadening of the levels of $H_1$ induced by $\Sigma_1$ and
${\cal U}_{\rm f,b}$ respectively. Together with the one-  and two-particle
level spacings $\delta_{1,2}$ and bandwidths $B_{1,2}$, they define
the range of validity of the semiclassical approach as
$\delta_1 \lesssim \Gamma_{\Sigma_1} \ll B_1$, $\delta_2 \lesssim 
\Gamma_{\rm f,b} \ll B_2$. 
The second term on the right-hand side of Eq.~(\ref{eq:BEmod}) 
exists exclusively for a
classically meaningful initial state $\psi_1$ such as a Gaussian wavepacket
or a position state, but the first term is much more generic. It emerges
from both semiclassics and RMT and does not depend on the
initial preparation $\psi_1$ of system 1. Other regimes of decay exist
in different regimes of perturbation and coupling.
For quantum mechanically weak $\Gamma_{\Sigma_1} \ll \delta_1$ and
$\Gamma_{\rm f,b} \ll \delta_2$,
one has a Gaussian decay,
\begin{equation}\label{gaussiand}
{\cal M}_{\rm B}(t)= \exp\left[-\left( \overline{\Sigma_1^2}/4+
\overline{ {\cal U}_{\rm f}^2} /2 + \overline{ {\cal U}_{\rm b}^2 }/2  \right) t^2\right] + N_1^{-1} ,
\end{equation}
in terms of
the typical squared matrix elements 
of $\Sigma_1$ and ${\cal U}_{\rm f,b}$. The perturbation 
$\Sigma_1$ and the coupling  ${\cal U}$ can be tuned independently of one
another. Accordingly, the Gaussian decays individually turn into 
exponential decays as $\Gamma_{\Sigma_1} \ll \delta_1$ or
$\Gamma_{\rm f,b} \ll \delta_2$ are no longer satisfied. 
For instance in the regime $\Gamma_{\rm f,b} \ll \delta_2$ 
and $\delta_1 \lesssim \Gamma_{\Sigma_1} \ll B_1$, one has
\begin{eqnarray}\label{eq:BEmod_mixed} 
{\cal M}_{\rm B}(t) \simeq \exp\left[- \Gamma_{\Sigma_1} t -\left( \overline{ {\cal U}_{\rm f}^2} /2 + \overline{ {\cal U}_{\rm b}^2 }/2\right) t^2 \right]+  
\alpha_1  \exp\left[-\lambda_1 t\right]  + N_1^{-1}.
\end{eqnarray}
The presence of the Gaussians is however irrelevant most of the time,
except perhaps in crossover regimes. The two conditions
$\Gamma_{\rm f,b} \ll \delta_2$ 
and $\delta_1 \lesssim \Gamma_{\Sigma_1} \ll B_1$ imply that the Gaussians
are turned on long after the exponential terms have led to the 
saturation of ${\cal M}_{\rm B}$. 
Also, at short times a parabolic decay
of ${\cal M}_{\rm B}$ prevails for any coupling strength.
Finally, if system 1 is integrable, the decay of ${\cal M}_{\rm B}$
is power-law in time. The dynamics of system 2, both in the forward and backward propagations,
is irrelevant because of the trace one takes over the corresponding degrees of freedom.
System 2 matters only in that it is coupled to system 1 with ${\cal U}_{\rm f,b}$.

The equivalence between Boltzmann and Loschmidt echoes is broken by
$\Gamma_{\rm f,b}$, the decoherence rate of system 1 induced by the coupling
to system 2 (or by $\overline{ {\cal U}_{\rm f,b}^2}$ at weak interaction). 
Skillful experimentalists can thus investigate decoherence
in echo experiments with weak time-reversal imperfection $\Sigma_1$ for
which $\Gamma_{\Sigma_1} \ll \Gamma_{\rm f,b}$, and thus
${\cal M}_{\rm B}(t) \simeq
\exp[-(\Gamma_{\rm f}+\Gamma_{\rm b}) t]$ (or 
${\cal M}_{\rm B}(t) \simeq \exp[-(
\overline{ {\cal U}_{\rm f}^2}+ \overline{ {\cal U}_{\rm b}^2 }
) \; t^2/2] $ at weak interaction)
as $\Sigma_1$ is reduced. 
The NMR experiments of Ref.~\cite{Pas00} reported a 
$\Sigma_1$-independent decay of polarization 
echoes as the time-reversal operation is performed with better
and better accuracy, corresponding to a reduction of $\Sigma_1$.
This might well indicate that other, uncontrolled sources of irreversibility
are at work, whose degrees of freedom are out of reach of the experimental
apparatus, and whose effect is to give an lower bound for the decay
rate of ${\cal M}_{\rm B}$. This point is discussed below and details of the semiclassical
calculation, are discussed
in Appendix~\ref{appendix:semiclassics_boltzmann}. The RMT approach to the Boltzmann echo
reproduces Eqs.~(\ref{eq:BEmod}), (\ref{gaussiand}) and (\ref{eq:BEmod_mixed}) 
in the appropriate limit $\lambda_1 \rightarrow \infty$. It
is presented in Appendix~\ref{appendix:RMT}.

\subsection{The Boltzmann echo and its relevance to NMR experiments}

Analyzing Eqs.~(\ref{eq:BEmod}) and (\ref{eq:BEmod_mixed}), 
we first note that ${\cal M}_{\rm B}(t)$ 
depends neither on $H_2$ nor on $\Sigma_2$. 
This is so because one traces over the 
uncontrolled degrees of freedom, and
this holds independently of the dynamics generated by $H_2$, and the strength of $\Sigma_2$ --
the result is still valid, even for classically strong $\Sigma_2$. Most importantly,
besides strong similarities
with the Loschmidt echo, such as competing golden rule and Lyapunov decays, 
the Boltzmann echo can exhibit a $\Sigma_1$-independent 
decay given by the decoherence rates $\Gamma_{\rm f,b}$ in the limit
$\Gamma_{\Sigma_1} \ll \Gamma_{\rm f,b}$. Extending our analysis to the regime 
$\Gamma_{\Sigma_1} \ll \delta_1$, $\Gamma_{\rm f,b} \ll \delta_2$ by means of
quantum perturbation theory, we find a Gaussian decay of
${\cal M}_{\rm B}(t)$, Eq.~(\ref{gaussiand}).
It is thus possible to reach either a Gaussian or an exponential, 
$\Sigma_1$-independent decay, depending on the balance between the accuracy $\Sigma_1$
with which the time-reversal operation is performed and the coupling between controlled
and uncontrolled degrees of freedom. This might explain the experimentally
observed saturation of the polarization echo as $\Sigma_1$ is reduced
\cite{Pas00}. A more precise analysis of these experiments
in the light of the results presented here is necessary, however this behavior is appealing in that
it is the only one on the market which predicts a saturation of the echo decay rate upon
reduction of $\Sigma_1$ -- the experimentally observed phenomenon. The idea that the Boltzmann
echo might be the solution to the puzzle posed by this experimental observation has been
further developed by Zurek, Cucchietti and Paz~\cite{Zur07}.

\begin{figure}
\includegraphics[width=12cm,angle=0]{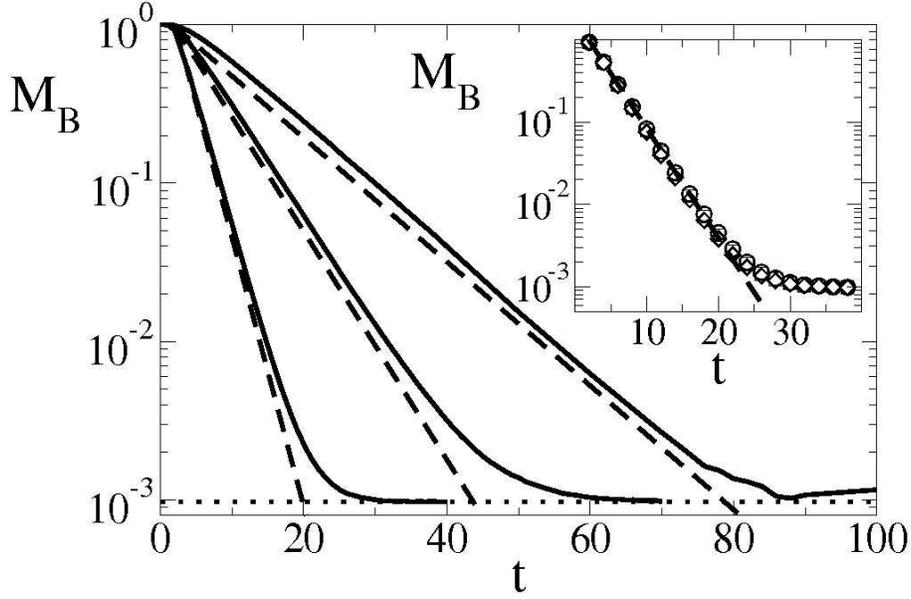}
\caption{\label{fig:fig1_Becho}
Main plot: Boltzmann echo for the quantized double kicked rotator model of Eq.~(\ref{2krot_repeat}) 
with 
$N=1024$, $K_1=K_2 = 10.$, and
$\Sigma_1 = 0.0018$ ($ \Gamma_{\Sigma_1} \simeq 0.09$). 
Data have been calculated from 50 different initial states.
The full lines correspond to $ \epsilon= 0 $, 0.0018 and 0.0037 
(from right to left)
and the dashed lines give the predicted exponential decay of
Eq.~(\ref{eq:BEmod}), with $\Gamma_{\cal U} = 1.2\, 10^{4} \epsilon^2,
 \Gamma_{\Sigma_1}= 2.6 \, 10^{4} \delta K_1^2$, $\lambda_0 = 1.6 \gg  \Gamma_{\cal U}, \Gamma_{\Sigma_1}$ (dashed lines have been
slightly shifted for clarity).
The dotted line gives the saturation $N^{-1}.\,$
Inset :  ${\cal M}_{\rm B}$ for $\epsilon =0.0037$,  and 
$ \delta K_1= 0.0003$ (circles; $ \Gamma_{\Sigma_1} \simeq 2.\, 10^{-3} $), $ \delta K_1= 0.0006$ (squares; $ \Gamma_{\Sigma_1} \simeq 9 .\,10^{-3}$), and $0.0009$ (diamonds; $ \Gamma_{\Sigma_1} \simeq 0.02 $). 
The dashed line indicates the theoretical prediction 
${\cal M}_{\rm B}(t) = \exp[-0.3 t]$.
(Figure taken from Ref.~\cite{Pet06a}. Copyright (2006) by the American Physical Society.)}
\end{figure}

\subsection{Numerical experiments on the Boltzmann echo} 

We numerically check our results, and consider a 
Hamiltonian for two interacting kicked rotators. Here we consider that
the first particle is the system, which is time-reversed with some finite accuracy, and the second
particle mimics the external degrees of freedom over which one has no control.
We thus consider the same model of two coupled kicked rotators as in our investigations
of entanglement dynamics in Chapter~\ref{section:nums_entangle}, 
\begin{subequations}
\label{2krot_repeat}
\begin{eqnarray}
H_i & = & p_i^2 / 2 + K_i \cos(x_i) \; \sum_n \delta(t-n),\\
{\cal U} & = & \epsilon \; \sin(x_1-x_2-0.33) \; \sum_n \delta(t-n),
\end{eqnarray}
\end{subequations}
where, however we have to define time-reversed one- and two-particle Hamiltonians.

\begin{figure}
\includegraphics[width=12cm,angle=0]{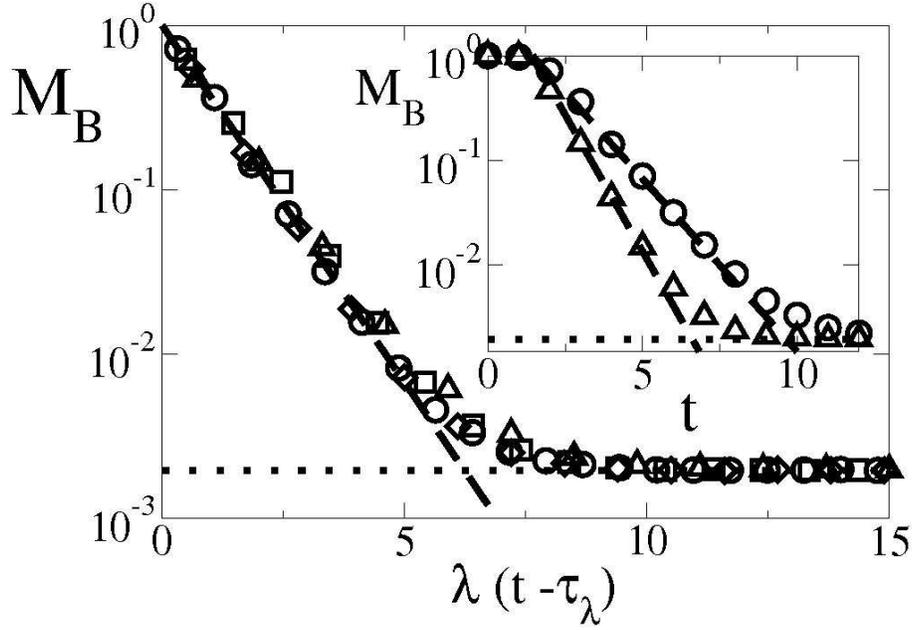}
\caption{\label{fig:fig2_Becho}
Main plot: Boltzmann echo for the quantized double kicked rotator 
model of Eq.~(\ref{2krot_repeat}) with 
$N=512$, $K_1=K_2= \in[6,12]$,
$\delta K _1=\delta K _2=0$  and $ \epsilon =   0.0245  $ 
(giving $ \Gamma_{\cal U} \geq  \lambda_0$). 
Data have been calculated from 50 different initial states.
The time axis has been shifted by the onset time $\tau_{\lambda} $  
and rescaled with $\lambda_0 \in [0.76,1.3]$.
The dashed line indicates the exponential decay with the effective
Lyapunov exponent $\lambda_0$ and
the dotted line gives the long-time saturation $M_{\rm B}(\infty) = N^{-1}.\,$
Inset : 
Same data as used in the main plot for $K_1=K_2=6$ and $12$ 
but without rescaling nor shift of the time axis. 
The dashed lines indicate the respective Lyapunov decays with $\lambda_0=0.76$
and $1.3$.
}
\end{figure}

The time-reversed
one-particle Hamiltonian is obtained through $K_1 \rightarrow 
K_1 +\delta K_1$ [from the definition of the Boltzmann echo, Eq.~(\ref{irrevtest_repeat}),
the Hamiltonian of particle 2 is not
time-reversed, but one can also add a change in its dynamics due to that operation,
$K_2 \rightarrow 
K_2 +\delta K_2$], and we restrict our investigations to the case 
${\cal U}={\cal U}_{\rm f}={\cal U}_{\rm b}$ and write 
$\Gamma_{\cal U} = \Gamma_{\rm f,b}$. 
Except for the partial time-reversal
operation working on $H_1$ only, we follow the same numerical 
procedure as in our
investigations of entanglement in Section~\ref{section:entanglement}.
Here, we only recall that our quantization procedure amounts to consider 
discrete values
$p_{i,l}=2 \pi l/N_i$ and $x_{i,l}=2 \pi l/N_i$, $l=1,...N_i$, 
for the canonically conjugated
momentum and position of particle $i=1,2$. We take $N=N_1=N_2$ and the total
Hilbert space size is $N^2$.

\begin{figure}
\includegraphics[width=12cm,angle=0]{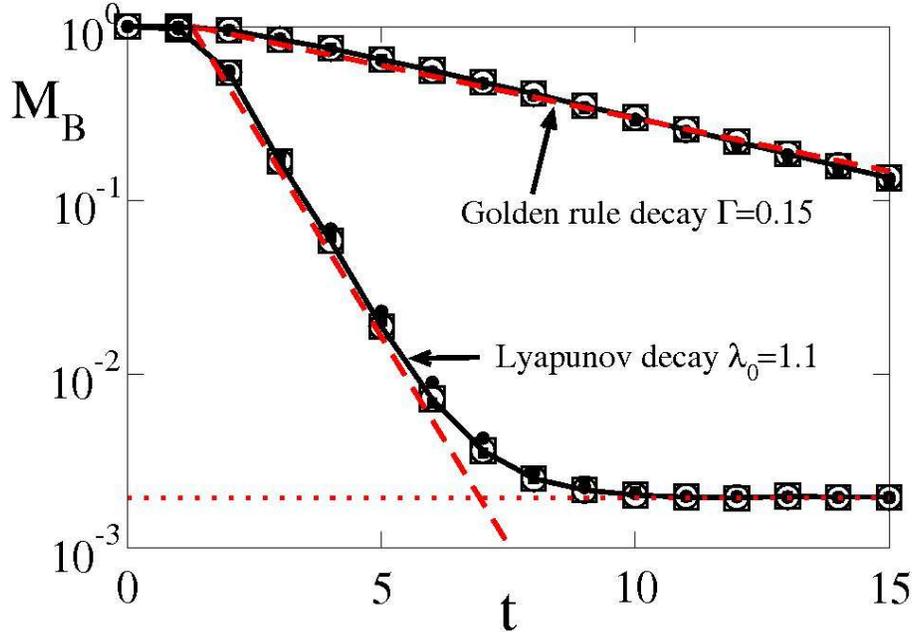}
\caption{\label{fig:fig3_Becho}
Main plot: Boltzmann echo for the quantized double 
kicked rotator of Eq.~(\ref{2krot_repeat}) 
with $N=512$. Two sets of data are shown, corresponding to a golden rule
decay with $\Gamma=0.15$ and a Lyapunov decay with $\lambda_0=1.1$.
All data have $K_1=10.$
Both full black lines have $K_2 = 10$, $\delta K _2= 0$, with  
$\delta K _1= \epsilon  = 0.0036$ (upper, golden rule curve) 
and $\delta K _1=0$, $ \epsilon =   0.0245$ 
(lower, Lyapunov curve). The empty symbols correspond to variations of
$\delta K _2 =  0.0036 $ (squares) and $ = 0.0122$ (circles).  
The full symbols correspond to variations of $ K _2 =5$ (circles)  
and $ K _2 =20$ (squares). This shows that ${\cal M}_{\rm B}$
is insensitive to both $H_2$ and $\Sigma_2$ in all regimes.}
\end{figure}

We first set
$K_2=K_1 \gtrsim 9$ in the chaotic regime, and restrict ourselves to
$\delta K_2=0$. From our earlier 
investigations of the local density of states 
(see Chapter~\ref{section:k_rot} and Fig.~\ref{fig6_qmclcorr}), 
we already know that $\Gamma_{\Sigma_1} = 0.024 \delta K_1^2 N^2$ and 
$\Gamma_{\cal U} = 0.43 \epsilon^2 N^2$.
The main panel in Fig.~\ref{fig:fig1_Becho} 
shows that for 
$B_1 \gg \Gamma_{\Sigma_1}  \gtrsim \delta_1$,  
$B_2 \gg \Gamma_{\cal U}  \gtrsim \delta_2$,
Eq.~(\ref{eq:BEmod}) is satisfied. Additionally, the inset of
Fig.~\ref{fig:fig1_Becho} illustrates that when 
$\Gamma_{\Sigma_1} \ll 2 \Gamma_{\cal U}$,
the observed decay is only sensitive to ${\cal U}$ , 
and one effectively obtains a $\Sigma_1$-independent decay. 

In Fig.~\ref{fig:fig2_Becho}, we next 
confirm the existence of the Lyapunov decay 
[second term in Eq.~(\ref{eq:BEmod})]. For a modest, but still
finite variation of the effective Lyapunov $\lambda_0 \in [0.76,1.3]$,
we can rescale three different set of data so that they all fall
on the same exponentially decaying curve. The inset in
Fig.~\ref{fig:fig2_Becho} shows that the raw data significantly 
differ from one another.

In Fig.~\ref{fig:fig3_Becho} we finally show that ${\cal M}_{\rm B}$ is 
independent of $H_2$ and $\Sigma_2$ in the golden rule regime, for
both Lyapunov and golden rule decay. As long as either
$\Gamma_{\Sigma_1} + 2 \Gamma_{\cal U}$ (golden rule decay) or
$\lambda_0$ (Lyapunov decay) are fixed, varying $K_2$ or $\delta K_2$
has no influence on the decay of ${\cal M}_{\rm B}$.
All our 
numerical results confirm the validity of Eq.~(\ref{eq:BEmod}).
We also investigated numerically other regimes of interaction and perturbation
which agreed well with Eq.~(\ref{eq:BEmod_mixed}). We can therefore
conclude that our analytical investigations successfully passed
the numerical test with the best possible grade.

\section{Conclusions, and where to go from here}\label{section:conclusion}

Taken literally, the correspondence principle requests that there be direct manifestations
of chaos in quantized dynamical systems. There are  
two distinct avenues one might chose to follow in the search for such manifestations
-- in the energy or in the time domain. 
Investigating the quantum-classical correspondence and searching for traces of chaos in
spectral and eigenfunction 
properties of quantum mechanical systems has been the method of choice in the field
of quantum chaos for quite some time, certainly for reasons that are easy to understand. 
As a matter of fact, there is a mathematical
one-to-one relation between quantal and classical integrability in that
classical systems with a complete set of integrals of motion acquire a complete set of good quantum 
numbers once quantized. Classical Poisson brackets going into quantum commutators,
good quantum numbers are equally well defined as integrals of motion. However, nonintegrable
systems are not always chaotic, moreover, integrability is proven once enough integrals
of motion/good quantum numbers have been found -- but determining the latter is often a pretty 
hard task, and showing that integrals of motion are missing in a given system is often
not trivial. Finding a criterion for chaotic behavior in the dynamics of quantum
systems would therefore be much more practical. 
There is however, at least
{\it a priori}, no such relation when looking at the dynamics of quantum systems and their classical
version: the Schr\"odinger time-evolution, in either spatial or momentum space generates a
dynamics that is fundamentally different
from the classical Hamiltonian/Liouville time-evolution in phase-space. 
Asymptotic local exponential instability does not exist in quantum mechanics, where
classical concepts of locality break down and quantum coherence set in. Yet, we have just presented
a wealth of rather clear manifestations of classical behavior, including the appearance of
classical Lyapunov exponents at short pre-Ehrenfest times, in quantum dynamical quantities in
the short-wavelength limit. Let us review and interpret these results and put them in the perspective
of the common wisdom of quantum chaos. 

Our first result is that
the fidelity of Eq.~(\ref{eq:def_LE2}) for a classically meaningful initial quantum state 
typically exhibits a power-law decay 
if $H_0$ is regular, but an exponential decay if $H_0$ is chaotic. 
Moreover, the latter exponential decay is
often -- but not always -- governed by the Lyapunov exponent of the underlying classical dynamics.
This very contrasted change in behavior makes a lot of sense, given that Liouville distribution
as well as wavepackets spread
algebraically in regular systems, but exponentially in chaotic systems -- at least for short enough times.
Accordingly, the 
exponential Lyapunov decay of the fidelity is classical in nature -- though to observe it, one needs
enough perturbation-induced 
dephasing -- as is the power-law decay 
in the regular regime. From a mathematical point of view, 
the origin of both the Lyapunov decay in chaotic systems and the 
algebraic decay in regular systems can be traced back in our calculation to the
asymptotic expression for the determinant of the stability matrix of classical orbits, the 
$C_s= |{\rm det} D_s|$
in the starting point of our semiclassical theory, Eq.~(\ref{propwp}). This determinant measures
how fast one moves away from a given nearby orbit, when one is very close to it but not quite
on it. The entries in the stability matrix are given by second derivatives 
$[D_s(t)]_{ij} = \partial^2 S_s({\bf r},{\bf r}_0;t)/\partial r_{i} \partial r_{0j}$ of the classical
action on that orbit with respect to its classical initial and final points. There is no $\hbar$ in $C_s$, 
it is a purely classical quantity. For regular systems, the stability is algebraic and one
gets $C_s \propto t^{-d}$. For chaotic systems, on the other hand, local exponential
instability gives $C_s \propto \exp[-\lambda t]$ with the Lyapunov exponent of the
classical dynamics. This is not the full story, however, and while the fidelity decay 
 is only very weakly affected by quantum coherence, random dephasing due to the
finite accuracy $\Sigma = H-H_0$  in the time-reversal operation modifies the exponent
of the power-law decay. Compared to the typical decay $\propto t^{-d}$ of the classical
fidelity in regular systems~\cite{Pro02,Ben02,Eck03,Ben03c,Ben03b}, 
one gets anomalous exponents for the decay of the quantum 
fidelity in regular systems. Yet, the physics is
that the fidelity decays algebraically in regular systems and exponentially, at a rate given by
the Lyapunov exponent, in chaotic systems because of the associated decay of
the overlap of the envelope of the wavefunction with itself, when it is propagated with
two slightly different Hamiltonians. This conclusion fully agrees with the common belief that
quantum dynamics follows classical dynamics for times shorter than the 
Ehrenfest time~\cite{Lar68,Ber78,Ber79,Chi81,Chi88} -- this is
discussed above in 
Paragraph~\ref{afterthoughts}. For regular systems, the Ehrenfest time becomes
infinite and the power-law decay of the average fidelity extends up to the Heisenberg time, i.e.
the breaktime for semiclassics. 

While there is a rather straightforward interpretation for 
these behaviors of the fidelity
at short times, perhaps the biggest surprise we tried to convey in this review is that
a priori purely quantal phenomena are successfully captured by 
semiclassical approaches, and the most striking example is provided by the rate of
entanglement generation between two interacting particles. In this review, entanglement
was quantified by the purity ${\cal P}(t)$ of the reduced density matrix. This is appropriate
since we considered bipartite systems with a pure global density matrix. We showed how
our semiclassical and RMT approaches that proved so successful for the fidelity can also
be applied to calculate ${\cal P}(t)$, under the assumption that the interaction between the
two subsystems is so weak that classical trajectories are not affected by it. Even in that
regime, we showed that highly nontrivial effects occur, and that they are qualitatively and
quantitatively captured by our analytical approaches. We showed how, under
certain circumstances, the classical dynamics determines how
entanglement between two interacting dynamical systems is
generated, and here again we showed that the occurrence of Lyapunov exponents giving
the decay rate of ${\cal P}(t)$ in certain regimes is of purely classical origin -- again its
origin is to find in the stability of classical orbits. Why this stability affects entanglement generation
is possible to interpret as arising from the increase of the number of wavefunction components
in the basis determined by the interaction -- in our investigations, the real-space basis. At
short times, our initially narrow wavepackets have only few nonzero
components in that basis, but as time
goes by, their number increases. The entangling action of the interaction
acts between any pair of these components. The number of such pairs increases
exponentially in a chaotic system, where accordingly entanglement is generated exponentially fast. 
The same argument explains the algebraic generation of entanglement in regular
systems -- in that classical regime  the spreading of the wavepackets is what determines
the dynamics of entanglement. 
We also mention that, once ergodicity and the lack of interaction-generated
classical relaxation are assumed, the Lyapunov decay of ${\cal P}(t)$ mathematically
emerges in the very same way as it does for the fidelity. 

Once our semiclassical approach 
was extended to interacting systems in Ref.~\cite{Jac04a,Pet06b}, 
it only made sense
to introduce the Boltzmann echo of Eq.~(\ref{irrevtest}). Our analysis of that quantity 
has been perhaps a bit underestimated until now and we stress here that it not only
allowed us to understand better
in what regime the fidelity gives an appropriate description of echo experiments, it
has moreover
provided the only theory on the market that explains how a saturation of the
decay of the echo signal can occur as the time-reversal operation is made more and more
accurate, i.e. as the perturbation is made weaker and weaker. 
This is the main experimental result of Ref.~\cite{Pas00} which 
provided Jalabert and Pastawski's original motivation
for their search for a perturbation independent decay~\cite{Jal01}, and, one thing leading to another,
eventually gave birth to the field of {\it echology}, which we attempted to summarize
in Sections~\ref{section:fidelity} and \ref{section:wigner}. 
Going from the Loschmidt to the Boltzmann echo, showing how in many 
instances they are equivalent and how semiclassics can be extended to treat interacting
systems, we therefore 
feel that a full cycle of investigations has been successfully completed.

Depending on the balance between the Lyapunov exponent and the strength of the
perturbation (of the imperfection in the time-reversal operation or of the interaction between
subsystems), the exponential Lyapunov decay can
turn into a perturbation-dependent, dephasing generated decay. The very existence of this
{\it golden rule decay}, as it was first dubbed in Ref.~\cite{Jac01}, is quite surprising. It presupposes
that a regime of parameters exist, where external perturbations lead to the accumulation
of pseudo-random relative phases in wavefunction components without relaxing the dynamics,
i.e. with no noticeable change in classical trajectories. How is this possible ? The answer is provided
by rigorously proven theorems on hyperbolic dynamical systems and the shadowing of 
their orbits once these systems are perturbed. It is quite remarkable that such theorems have applications
in quantum mechanics, even in the semiclassical limit. The fact that a classically weak
perturbation must have a vanishing strength in the limit $\hbar_{\rm eff} \rightarrow 0$ justifies
rigorously (in a physicist's sense) to invoke shadowing in our treatment of the fidelity in that 
regime. Perhaps equally surprising, even this weak a perturbation generates nontrivial behaviors
such as quantum irreversibility and entanglement generation,
and this opens up doors to future analytical investigations of interacting systems. We will come
back to this momentarily. For the time being, we stress that shadowing does not exist in 
slightly perturbed regular systems, so that our semiclassical approach is a bit harder to
justify there. One might still expect that perturbed trajectories stay close together for some
time in regular systems -- after all they have a much stronger, algebraically decaying stability --
which might well save the day. One should nevertheless always bear in mind that regular systems
exhibit much larger fluctuations than their chaotic counterparts so that our statistical approach 
might well fail for regular systems, where the average behavior may well be dominated
by exceptional events not captured by our approach.
In any event, one of the key aspects of the investigations presented here is that 
dephasing without relaxation can be rigorously justified in hyperbolic systems. 

For all quantities discussed in this review, and this has already been 
mentioned several times above, 
there is a quantum-classical competition between dynamically driven
effects and dephasing effects. In the asymptotic regime, our semiclassical approach expresses 
${\cal M}_{\rm L}(t)$, ${\cal M}_{\rm B}(t)$, ${\cal M}_{\rm D}(t)$ and ${\cal P}(t)$ as
sums over two dominant terms, one of dynamical origin, one of dephasing origin, 
and in chaotic systems we found that both of them decay exponentially with time. 
Who wins the fight is then straightforwardly determined by a direct
comparison between the two rates of these exponentials,
i.e. the measures $\Gamma$ of the strength of the perturbation and
$\lambda$ of the rate of dynamical stretching. To observe a Lyapunov exponent
in any of these quantities, one must have $\Gamma > \lambda$ (where $\Gamma = 2 \Gamma_2$
for interacting systems). Once the initial surprise is passed that different regimes are
determined by a direct comparison between
a purely quantum mechanical and a classical quantity, this relation is rather transparent and 
appealing. It states that classical effects, decaying on a time scale $\lambda^{-1}$, 
are observable only once/if quantum mechanical dephasing effects, with a typical time
scale $\Gamma^{-1}$, set in so fast that quantum coherence is 
practically lost before $\lambda^{-1}$. Only then does one have a chance
to observe the classical Lyapunov exponents in the quantities we discussed in this review. 
Is there anything useful for decoherence one can learn from that ? The answer is yes. 
Let us go back to the standard formulation of decoherence starting from the
master equation of Eq.~(\ref{Wigner_o}) valid in the regime of 
weak system-environment coupling \cite{Joo03,Zur03}. 
We rewrite here this master equation which determines the time-evolution of the system's Wigner
function $W({\bf p},{\bf q};t)$ as
\begin{eqnarray}\label{Wigner_oo}
\partial_t W &=& \Big\{ H,W \Big\} + \sum_{n \ge 1}
\frac{(-\hbar)^{2n}}{2^{2n} (2n+1)!} \partial_{\bf q}^{2n+1} V \partial_{\bf p}^{2n+1} W
+ 2 \gamma \partial_{\bf p} (pW) + D \partial^2_{\bf p} W.
\end{eqnarray}
The right-hand side is the classical Poisson
bracket. Alone, it would give the classical equation of motions. 
The second term, written here for the case of a momentum-independent
potential $V({\bf q})$, 
exists already in closed systems, without environment, and generates quantum corrections
to the dynamical evolution of $W$ -- in Eq.~(\ref{Wigner_oo}),
we gave the $\hbar$-dependence of that term
to make this more explicit. This term starts to become comparable to the Poisson
bracket at the Ehrenfest time $\tau_{\rm E}$ -- in absence of
environment, $\gamma \propto D T=0$, this equation establishes that the 
quantum mechanical dynamics follows the classical one for short times. A true quantum-classical
correspondence requires that, one way or another, there is a certain regime of parameter where
the last three terms on the right-hand side of Eq.~(\ref{Wigner_oo}) 
cancel out.
The last two terms 
are induced by the coupling to the environment, and it has been argued that 
in the limit of weak coupling, $\gamma \rightarrow 0$,
but finite diffusion constant, $D \propto \gamma T = {\rm Cst}$, the friction term $\propto \gamma$
vanishes, but the momentum diffusion term $\propto D$
induces enough noise so as to kill the quantum corrections (the second term on the
right-hand side) before they become important. Is this all ? If the answer is yes, then
this leaves the classical dynamics unaffected and
the time-evolution of $W$ is then solely governed by the classical Poisson bracket,
that is to say, classical dynamics emerges out of quantum mechanics. As appealing as it
is, one might wonder whether this argument is generically applicable -- when is it
possible to generate enough dephasing without relaxation ? In other words, is it possible
to kill the second term on the right-hand side of Eq.~\ref{Wigner_oo} with the last two terms,
without having to substitute $H \rightarrow H'(H,\gamma,D)$ in the Poisson bracket ?
The answer is yes, and 
it turns out that the validity of this standard argument for the quantum-classical correspondence is
directly related to our investigations of the purity, in that shadowing
might possibly be invoked to legitimate the standard view on decoherence. 
The purity measures the weight of off-diagonal
elements of $\rho_1(t)$, and hence of the importance of coherent
effects. In the regime $2 \Gamma_2 \gg \lambda$, when the interaction generates
fast enough dephasing (but $\Gamma_2 < B$ and
the interaction is still semiclassically small enough
that  it is classically negligible for $\hbar_{\rm eff} \rightarrow 0$), 
${\cal P}(t)$ furthermore reaches its minimal value at the Ehrenfest time.
Thus, quantum effects are killed before they have a chance to appear. In that regime, 
one therefore expects the
quantum-classical correspondence to become complete in the semiclassical
limit. Numerical evidences supporting this reasoning have been presented above in 
Chapter~\ref{towards_decoherence}.

The agreement between our predictions and
exact quantum mechanical calculations is quantitative. This is not 
trivial at all, given that semiclassical approaches take into account 
leading-order (in $\hbar_{\rm eff}$) corrections to classical dynamics only. 
Given the transparent physical content of semiclassics, it is
certainly advantageous to try and apply the methods developed above
and the gained knowledge in decoherence, entanglement and 
quantum reversibility, to other problems in complex quantum systems.
Now that we have outlined how RMT and semiclassical methods can be 
successfully 
applied to quantum dynamical problems, one might wonder what is next. It 
seems pretty clear that the current flow of the interdisciplinary 
field of quantum chaos goes toward many-body physics, and we believe that
the topics outlined here are no exception to that trend.  
Recent works indeed abound on the
dynamics of multipartite entanglement and 
decoherence~\cite{Car04,Min05c,Min05a},
entanglement and decoherence in many-body lattice 
systems~\cite{Chan07,San03,San04,San05,San05a,Brow07,Vio07a,Ami08,Zha07},
reversibility in many-body cold atomic gases~\cite{Liu05,Cuc06b,Bod07,Man08}
and close to many-body quantum phase transitions~\cite{Som04,Cuc06b}.
Most of these works considered discrete lattice models which often exhibit
quantum chaotic -- i.e. RMT-like -- spectral and wavefunction 
properties~\cite{Flam96a,Flam96b,Geo97,Jac97b,Fla00,Flam01,Abe90}, even in
absence of disorder or randomness~\cite{Mon93,Poi93}. 
This should certainly motivate the 
extension of the RMT approach developed in this review to many-body 
systems. 
The same approach might be useful in analyzing 
the fidelity in many-body spin models of quantum 
computers~\cite{Flam99,Geo00,Sil01,Fra04},
or decoherence due to many-body baths made out of spins~\cite{Lag05}.
We foresee in that context that RMT might allow to perform controlled 
analytical calculations beyond models of noninteracting
many-spin bath~\cite{Zur92,Cuc05,Zur07}. 

In parallel to this extension of RMT to discrete many-body systems,
continuous systems might be treated semiclassically. It has already been
noted that numerical investigations have shown that many-body chaotic
systems also exhibit 
 stability over quite long 
times~\cite{Hay03a,Hay03b}. This properties of theirs might be put
to use for a semiclassical treatment of not too strongly interacting
many-body dynamical systems. In the spirit of this review, RMT and
semiclassical approaches may be applied in parallel, for instance, to
treat decoherence due to complex interacting environments, going beyond the
bath of noninteracting harmonic oscillators of Caldeira and 
Leggett~\cite{Cal81,Cal83}. One might finally wonder how
the assumption we made above that two-body interactions
do not alter classical trajectories can be lifted in order to extend our semiclassical
approach to treat dissipation in strongly interacting quantum dynamical systems. The analytical
approaches we presented in this review seem very promising, however much is left to be done.

\newpage 

\section*{Acknowledgments}
While working on the topics surveyed in this review, we had the pleasure and privilege to
collaborate with \.{I}nanc Adagideli, Carlo Beenakker, Diego Bevilaqua,
Rick Heller and Peter Silvestrov. 
We would like to express our gratitude to
each of them. We also greatly benefited
from and enjoyed sometimes lively and controversial but always interesting and 
fruitful discussions on these and
related topics with S. {\AA}berg, G. Casati, N. Cerrutti,
D. Cohen, F. Cucchietti, D. Dalvit, J. Emerson, S. Fishman, A. Goussev, 
M. Guti\'errez, F. Haake,
R. Jalabert, C. Jarzynski, 
P. Levstein, C. Lewenkopf, E. Mucciolo, H. Pastawski, J.P. Paz, K. Richter,
H.-J. St\"ockmann, A. Tanaka,
S. Tomsovic, J. Vanicek, D. Waltner,
R. Whitney, D. Wojcik, S. Wu and W. Zurek, among others.
Our work on these projects was at one point or another 
funded by the Dutch Science Foundation NWO/FOM,
the U.S. Army Research Office, the Swiss
National Science Foundation, the Alexander von Humboldt foundation and the National
Science Foundation under grant No. DMR--0706319. 

\newpage 
\appendix

\section{Semiclassical theory}\label{appendix:semiclassics}

\subsection{General considerations}

In the search for semiclassical approximations to quantum mechanical wavefunction amplitudes
one expresses the latter as
\begin{equation}\label{eikonal}
\psi({\bf r}) = \sum_n A_n({\bf r}) \, \exp[i \phi_n({\bf r})].
\end{equation}
The sums runs over all possible ways to reach ${\bf r}$ from a given initial condition. 
The total number of ways depends on the
underlying classical dynamics. In regular systems, this number remains finite, with
a finite number of different momenta ${\bf p}_n({\bf r}) 
= \nabla \phi_n({\bf r})$ restricted by integrals of motion.
With the Bornian interpretation of the wavefunction, $|A_n({\bf r})|^2$ gives the classical
probability to reach ${\bf r}$ following the $n^{\rm th}$ way, and the phase is given by
a well defined action integral along the corresponding classical path,
$\phi_n({\bf r}) = \int_n {\bf p} {\rm d}{\bf r}'$. Such an approximation is standardly called
WKB or eikonal approximation. 

The situation is altogether different in chaotic systems, where
the number of terms in the sum in Eq.~(\ref{eikonal}) blows up to infinity with time. Still, useful
semiclassical expression can be derived that look very similar to Eq.~(\ref{eikonal}). 
The time-evolution kernel
$K^{H_0}({\bf r},{\bf r}_0;t) = \langle {\bf r}| \exp[-i H_0 t] | {\bf r}_0 \rangle $ propagates the quantum
amplitude from ${\bf r}_0$ to ${\bf r}$. Its semiclassical approximation starts from the path integral
formulation for $K$ -- the Feynman-Kac formula -- and enforces a stationary phase condition on it. 
The result is that $K$ becomes a sum over all possible classical trajectories generated by
$H_0$ connecting ${\bf r}$ and ${\bf r}_0$ in the time $t$~\cite{Gut90,Haa01} 
\begin{subequations}\label{propwpa}
\begin{eqnarray}
\psi({\bf r},t) = 
\langle {\bf r}|
\exp(-i H_0 t) |\psi_0\rangle  =  \int d{\bf r}_0'
\sum_s K_s^{H_0}({\bf r},{\bf r}_0';t) \psi_0({\bf r}_0'), \\
K_s^{H_0}({\bf r},{\bf r}_0';t)  =  \frac{C_s^{1/2}}{(2 \pi i)^{d/2}} 
\exp[i S_s^{H_0}({\bf r},{\bf r}_0';t)-i \pi \mu_s/2].
\end{eqnarray}
\end{subequations}
The classical trajectories are labeled $s$ and correspond to all possible initial momenta 
compatible with $\psi_0$, pointing
out of ${\bf r}_0'$ and going to ${\bf r}$ after a time-evolution of duration $t$.
For each $s$, the partial semiclassical propagator $K_s$ contains
the action integral $S_s^{H_0}({\bf r},{\bf r}_0';t)$ along $s$,
the determinant $C_s=|{\rm det} D_s|$ of the stability matrix 
$D_s = -\partial {\bf p}_s/\partial{\bf r}_0'$
(${\bf p}_s$ is the final momentum on $s$), and 
a topological Morse index $\mu_s$. The latter
counts the number of conjugate points encountered
by $s$ between ${\bf r}_0'$ and ${\bf r}$, where 
$D_s$
has a diverging eigenvalue. Going through a conjugate point, one eigenvalue of $D_s$
goes to infinity and back to a finite value with a sign change. Each such sign change leads to a unit 
increment of the Morse index. The value of this index turns out to be irrelevant in all the calculations
presented in this review, however its presence is conceptually of utmost importance. It was indeed
one of the main difficulties of constructing a semiclassical theory for nonintegrable systems to
extend the propagator beyond the first passage at a conjugate point, and for a long time, it was believed
that semiclassics would break down at the Ehrenfest time $\tau_{\rm E}$, 
since the latter gives an estimate for
the average time to reach a first conjugate point. Semiclassics can be extended beyond $\tau_{\rm E}$, 
however, provided one takes into account the increment in the topological Morse index at each
conjugate point.
Numerous numerical experiments have confirmed that semiclassics allows to calculate the time
evolution of smooth, initially localized wavepackets 
up to algebraically long times in the effective Planck's constant
$\propto {\cal O}(\hbar_{\rm eff}^{-a})$ (with $a>0$)~\cite{Tom91,Hel93}.
Conjugate points do not pose much of a problem in that they are isolated -- the semiclassical
propagation is singular only for discrete positions, and these singularities are anyway blurred
by the finite spatial resolution brought about by quantum mechanics.

Early semiclassics focused on the density of states of dynamical systems, with the crowning 
achievement provided by Gutzwiller's celebrated trace formula, giving the oscillatory part of the
density of states -- more precisely the leading order correction to the Thomas-Fermi density of
states -- given by a sum over all periodic orbits in the system. Even more interesting physical
quantities are based on products of even numbers of propagators or Green's function.
Examples include the conductance/conductivity in solid-state systems, the density-density
correlation function, or its Fourier transform, the form factor, in dynamical systems, or
the fidelity ${\cal M}_{\rm L}(t)$ and all its offsprings discussed in this review. Generally speaking, 
the semiclassical calculation of such quantities is a two-stage process. First, one identifies
contributions satisfying a stationary phase condition, i.e. sets of trajectories $(s_1,s_2,...,s_{2n})$
such that the action difference $\delta S_{s_1,s_2,...,s_{2n}} = S_{s_1}-S_{s_2}+...-S_{s_{2n}}$
is stationary under the variation of an external parameter, in most instances, the energy. 
The formal justification for that step is, for instance, that one performs 
an average over a finite energy interval, which cancels out contributions that are not
stationary. Whereas recent semiclassical investigations have brought the identification
of stationary phase conditions to a sophisticated, almost artistic level 
~\cite{Heu07,Mul04,Heu06,Jac06,Whi06,Pet08,Rah05,Rah06,Ric02,Sie01,Sie02}, 
most, if not all interesting observable effects in the Loschmidt echo are captured
within diagonal approximations, where classical trajectories are paired. 
Why is that so ? To incorporate interesting
interference effects into semiclassics, formulas must be derived that still contain finite
phase differences. When considering only a single unperturbed system, this requires to 
search for nontrivial trajectory pairing. This is the situation encountered, e.g., in the 
semiclassical theory of quantum 
transport or when calculating the form factor for closed chaotic systems
~\cite{Heu07,Mul04,Heu06,Jac06,Whi06,Pet08,Rah05,Rah06,Ric02,Sie01,Sie02}.
Life is a bit easier when one is 
interested in quantities depending on two different Hamiltonians. In this case, diagonal pairing
still retains coherent interferences arising from phase differences due to the perturbation
acting on only one of the trajectories.

The second step in semiclassical calculations 
is to construct a sum rule which relates sums over classical
trajectories by integrals. In the case of the Loschmidt echo, the relevant sum rule are
similar to
\begin{equation}\label{sumrule1}
\left(\frac{\nu^2}{\pi} \right)^d \left| \int {\rm d}{\bf r} \sum_s C_s \exp[-\nu^2 ({\bf p}_s -{\bf p}_0)^2] \right|^2 = 1.
\end{equation}
There are two ways to justify this sum rule. Ref.~\cite{Jal01} noted that $C_s$ gives the Jacobian
of the transformation from position to momentum integration variable. 
The resulting Gaussian integral over ${\bf p}_s$ 
is then properly normalized and this gives Eq.~(\ref{sumrule1}).
To be strictly valid, however, 
this argument further requires that the times considered are long enough that almost all
trajectories in the sum in Eq.~(\ref{sumrule1}) are ergodic. This is in the same spirit as the
Hanay-Ozorio de Almeida sum rule invoked to calculate the form factor~\cite{Ozo88,Han84}. 
Eq.~(\ref{sumrule1})
is however equally valid at short, pre-Ehrenfest times, when the diagonal approximation is exact. 
This is easily seen by semiclassically calculating the left hand-side of the equation
\begin{equation}\label{sumrule2}
\left| \int {\rm d}{\bf r} \langle {\bf r} | \exp[-i H t] | \psi_0 \rangle \right|^2 = 1,
\end{equation}
and enforcing a diagonal pairing on the pair of classical trajectories in the squared amplitude.

\subsection{Average fidelity}
\label{appendix:semiclassics_avg}

Inserting Eq.~(\ref{propwpa}) into Eq.~(\ref{eq:def_LE}) we rewrite the fidelity as
\begin{eqnarray}\label{mofta}
{\cal M}_{\rm L} (t) & = & \Bigg|\int d{\bf r} \int d{\bf r}_0' \int d{\bf r}_0'' \;
\psi_0({\bf r}_0')\psi_0^*({\bf r}_0'') \;
\sum_{s_1,s_2} K_{s_1}^{H_0}({\bf r},{\bf r}_0 ';t) \;
 [K_{s_2}^H({\bf r},{\bf r}_0'';t)]^* \Bigg|^2 \, .
\end{eqnarray}
Up to now, the only approximation made is the stationary phase condition extracting the semiclassical
propagator from the path integral formulation of the quantum propagator. The above expression is
expected to be accurate for (i) times longer than $\tau_{\rm min} = 1/E$, and (ii) in the semiclassical
limit of large quantum numbers. The second condition readily imposes that $\tau_{\rm min}$ is short,
that enforcing a stationary phase condition is justified, and that the Heisenberg time, 
$\tau_{\rm H}$, is long.

Noting that $\psi_0$ is a narrow Gaussian wavepacket centered on
${\bf r}_0$ and that it thus restricts the range of ${\bf r}_0'$ and
${\bf r}_0''$, we linearize the action around ${\bf r}_0$ 
as $S_s({\bf r},{\bf r}_0';t) \simeq 
S_s({\bf r},{\bf r}_0;t) - ({\bf r}_0'-{\bf r}_0) \cdot {\bf p}_s$,
with initial momentum
${\bf p}_s=-\partial S_s/\partial {\bf r}_0$.
We then calculate the integrals over ${\bf r}_0'$ and ${\bf r}_0''$. 
We are going to see
momentarily that semiclassically motivated stationary phase approximations reduce the
four-fold sum over classical paths to three dominant terms, two involving a two-fold sum, one
involving a single sum over classical paths. These three contributions are sketched on the right-hand
side of Fig.~\ref{fig:fig_LEavg}.

We next enforce a stationary phase approximation on
the action phase difference $S_{s_1}({\bf r},{\bf r}_0;t)-
S_{s_2}({\bf r},{\bf r}_0;t)$ appearing in Eq.~(\ref{mofta}). The reason for 
this is that we calculate the fidelity averaged over an ensemble of
initial Gaussian wavepackets $\psi_0$. As the center of mass
${\bf r}_0$ of these initial states is moved, the difference
$S_{s_1}({\bf r},{\bf r}_0;t)-S_{s_2}({\bf r},{\bf r}_0;t)$ fluctuates, so that
the only contributions that survive the average are those which minimize
these fluctuations. The dominant such contribution is obtained from 
the diagonal approximation $s_1=s_2\equiv s$, 
from which one gets the leading-order
semiclassical fidelity
\begin{eqnarray}\label{semicl}
{\cal M}_{\rm L}(t) & = & (4 \pi \nu^2)^d \left|\int d{\bf r} 
\sum_s [K_s^H({\bf r},{\bf r}_0;t)]^* K_s^{H_0}({\bf r},{\bf r}_0;t)  
\exp(-\nu^2 |{\bf p}_s-{\bf p}_0|^2) \right|^2.
\end{eqnarray}
 
Eqs. (\ref{propwpa}--\ref{semicl}) are equally valid
for regular and chaotic Hamiltonians, as long as semiclassics
applies. Squaring the amplitude in 
Eq.\ (\ref{semicl}) leads to a double sum over classical paths $s$ and $s'$ 
and a double integration over coordinates ${\bf r}$ and
${\bf r}'$,
\begin{eqnarray}\label{mtota}
{\cal M}_{\rm L}(t) & = & (\nu^2/\pi)^d \int d{\bf r} \int d{\bf r}'
\sum_{s,s'} C_s C_{s'} \exp[i \delta S_s({\bf r},{\bf r}_0;t) - 
i \delta S_{s'}({\bf r}',{\bf r}_0;t) ] \nonumber \\
& & \times \exp(-\nu^2 |{\bf p}_s-{\bf p}_0|^2-\nu^2 
|{\bf p}_{s'}-{\bf p}_0|^2),
\end{eqnarray}
\noindent with $ \delta S_s({\bf r},{\bf r}_0;t) =
S_s^{H_0}({\bf r},{\bf r}_0;t)-S_s^{H}({\bf r},{\bf r}_0;t)$.
Accordingly, ${\cal M}_{\rm L}(t)={\cal M}_{\rm L}^{\rm (d)}(t)+{\cal M}_{\rm L}^{\rm (nd)}(t)$
 splits into two contributions, depending on whether the trajectories
$s$ and $s'$ are correlated ($s\simeq s'$, within a spatial resolution $\nu$) 
or not ($s\ne s'$). We call the
correlated contribution the {\it diagonal contribution}, and
the uncorrelated one the {\it nondiagonal contribution} by
some abuse of language, even though both contributions already emerge from
the diagonal approximation $s_1 \simeq s_2$ we made to go from Eq.~(\ref{mofta}) to 
Eq.~(\ref{semicl}). The decay of the diagonal contribution is
governed by the decay of overlap of 
$|\psi_{\rm F} \rangle = \exp[-i H_0 t] | \psi_0 \rangle$ and 
$|\psi_{\rm R} \rangle = \exp[-i H t] | \psi_0 \rangle$, while the
behavior of the nondiagonal contribution is determined by the
$\Sigma$-induced dephasing between the wavepacket propagating along 
$s$ and the one propagating along $s'$.
 Below we show that the diagonal contribution sensitively depends
on whether $H_0$ is regular or chaotic, while the nondiagonal contribution
is generically insensitive to the nature
of the classical dynamics set by $H_0$, provided that
the perturbation Hamiltonian $\Sigma$ induces enough mixing of eigenstates 
of $H_0$, and in particular that it has no common integral of motion
with $H_0$.

We first consider the diagonal contribution ${\cal M}_{\rm L}^{\rm (d)}(t)$.
With $s \simeq s'$, and hence ${\bf r} \simeq {\bf r}'$, both conditions
to be satisfied with a spatial resolution $\nu$, 
we expand the phase difference in Eq.~(\ref{mtota}) as
\begin{equation}\label{eq:expansion_phase}
\delta \Phi_s \equiv \delta S_s({\bf r},{\bf r}_0;t) - 
\delta S_{s'\simeq s}({\bf r}',{\bf r}_0;t) = \int_0^t 
d{\tilde t} \ \nabla \Sigma [{\bf q}({\tilde t})] \cdot
\bigl({\bf q}(\tilde{t}\;) - {\bf q}'(\tilde{t}\;)\bigr).
\end{equation}
The points ${\bf q}$ and ${\bf q}'$ lie on $s \simeq s'$
with ${\bf q}(t)={\bf r}$, ${\bf q}'(t)={\bf r}'$, and
${\bf q}(0)={\bf q}'(0)={\bf r}_0$. Regular 
systems having a linear increase 
of the distance between two nearby initial
conditions have to be differentiated from chaotic ones which exhibit
local exponential sensitivity to initial conditions. Asymptotically, one writes
\begin{subequations}\label{eq:divergence}
\begin{eqnarray}
|{\bf q}(\tilde{t}) - {\bf q}'(\tilde{t})| &\simeq& (\tilde{t}/t)
|{\bf r}-{\bf r}'|, \;\;\;\;\;\;\;\;\;\;\;\;\;\;\;\;\;\;\; {\rm regular \; systems}, \\
|{\bf q}(\tilde{t}) - {\bf q}'(\tilde{t})| &\simeq& 
\exp[\lambda(\tilde{t}-t)] \; |{\bf r}-{\bf r}'|, \;\;\;\;\;\; {\rm chaotic \; systems}.
\end{eqnarray}
\end{subequations}
In both instances, the spatial integrations 
and the sums over classical paths
in Eq. (\ref{mtota}) lead to the phase averaging
\begin{equation}\label{clt}
\exp(i \delta \Phi_s ) \rightarrow 
\langle \exp(i \delta \Phi_s ) \rangle \simeq \exp[-{\textstyle \frac{1}{2}}
\langle \delta \Phi_s^2 \rangle ],
\end{equation}
\noindent which is justified by our 
assumption that $\Sigma$ varies rapidly along a
classical trajectory. Because of the further assumption
that $\Sigma$ and $H_0$ have no common integral
of motion, we expect a typically fast decay of correlations,
both for regular and chaotic systems,
\begin{equation}\label{correl}
\langle \partial_i \Sigma[{\bf q}(\tilde{t})] \partial_j 
\Sigma[{\bf q}(\tilde{t}')] \rangle = U \delta_{ij} 
\delta({\tilde t}-{\tilde t}'). 
\end{equation}
\noindent Two remarks are in order here. First, it is obvious that 
this latter assumption is easily violated by specific choices of
perturbation on regular or integrable systems. Second, the fast
decay (\ref{correl}) of correlations is generic in chaotic systems
(see e.g. Ref.~\cite{Col04}). This allows us to generalize the results
of Ref.~\cite{Jal01}, which were derived with a specific perturbation
in the form of a distribution of smooth impurities. 
The perturbation considered from here on 
is instead not specified, except for 
the decay (\ref{correl}) of its correlations.

With Eqs.~(\ref{eq:expansion_phase}), (\ref{eq:divergence}),
(\ref{clt}), and (\ref{correl}), 
Eq.~(\ref{mtota}) gives for the diagonal contribution to 
the Loschmidt echo
\begin{eqnarray}\label{echo_diaga}
{\cal M}_{\rm L}^{\rm (d)}(t) & = & (\nu^2/\pi)^d \int d{\bf r}_+ \int d{\bf r}_- \sum_s C_s^2
\exp\bigl(-\frac{1}{2} U \tau \; {\bf r}_-^2\bigr) 
\exp(-2 \nu^2 |{\bf p}_s-{\bf p}_0|^2),
\end{eqnarray}
with $\tau=t/6$ for regular systems and 
$\tau=\lambda^{-1}(1-\exp[-\lambda t]) \simeq \lambda^{-1}$ for chaotic
systems. The rest of the calculation is straightforward. 
The Gaussian integration over ${\bf r}_- \equiv 
{\bf r}-{\bf r}'$ ensures that ${\bf r} \approx {\bf r}'$, 
and hence ${\bf r}_+ \equiv ({\bf r}+{\bf r}')/2 \approx {\bf r}$.
The change of variables from ${\bf r}_+$
to ${\bf p}_s$ delivers a second Gaussian integral by means of
\begin{eqnarray}\label{eq:chg_var}
\int d{\bf r} \sum_s C_s^2 = \left\{
\begin{array}{cc}
& \big[ t_0 \big/ ( t+t_0 ) \big]^d \; \int d{\bf p}_s,  
\;\;\;\;\;\;\;\; {\rm regular \; systems},\\
& \exp[-\lambda t] \int d{\bf p}_s, \;\;\;\;\;\;\;\;\;\;\;\;\;\;\;\; {\rm chaotic \; systems}.
\end{array}
\right.
\end{eqnarray}
In this latter expression we took into account the algebraic stability 
of regular systems with $C_s \propto t^{-d}$ (regularized at short times with $t_0$) 
to be contrasted with the
exponential instability of chaotic systems with $C_s \propto \exp[-\lambda 
t]$. One finally arrives at
\begin{eqnarray}\label{diagdecaya}
{\cal M}_{\rm L}^{\rm (d)}(t) & \propto & \left \{
\begin{array}{cc}
t^{-d}, & {\rm regular \; systems \; with \;} U \tau < \nu^{-2}, \\
t^{-3d/2}, & {\rm regular \; systems \; with \;} U \tau > \nu^{-2}, \\
\exp[-\lambda t], &{\rm chaotic \; systems},
\end{array} \right.
\end{eqnarray} 
where, because the integral over ${\bf r}_-$ in Eq.~(\ref{echo_diaga})
is restricted to ${\bf r}_- \le \nu$, 
$\exp[-U\tau {\bf r}_-^2/2]$ matters only if $U \tau > \nu^{-2}$.
In this case there is an additional contribution $\propto t^{d/2}$
to the decay of ${\cal M}_{\rm L}$, otherwise, the decay is only 
given by $C_s \propto t^{-d}$. In the semiclassical limit $\nu \rightarrow 0$, there is a crossover
from a $t^{-d}$ behavior at short times to a $t^{-3d/2}$ behavior at longer times.
These decays are rather insensitive to the choice (\ref{correl}) 
of a $\delta$-function force correlator. 
Even a power-law decaying correlator 
$\propto |{\tilde t}-{\tilde t}'|^{-a}$ 
reproduces Eqs.\ (\ref{diagdecaya}) at large enough times, provided $a \ge 1$.
Still our assumption of short-ranged correlations, Eq.~(\ref{correl}), is not always
satisfied in regular systems, where it is actually the rule, rather than the exception,
that correlators such as the one in Eq.~(\ref{correl}) decay more slowly than $t^{-1}$.
Assuming a constant correlator
\begin{equation}\label{correl_long}
\langle \partial_i \Sigma[{\bf q}(\tilde{t})] \partial_j 
\Sigma[{\bf q}(\tilde{t}')] \rangle = U' \delta_{ij} 
\end{equation}
results in $\tau = t^2/8$ in Eq.~(\ref{echo_diaga}), which can lead, for $U' \tau > \nu^{-2}$
to an accelerated, but still power-law decay of the diagonal 
contribution to the fidelity, ${\cal M}_{\rm L}(t) \propto t^{-2d}$, in regular systems. 
We believe that the decay of the average fidelity in regular systems is generically algebraic,
however, the exponent with which ${\cal M}_{\rm L}(t)$ decays can vary from case to case.

We next calculate 
the nondiagonal contribution ${\cal M}_{\rm L}^{\rm (nd)}(t)$ to Eq.~(\ref{mtota}).
One argues that the action phases accumulated on $s \ne s'$
are uncorrelated to perform the phase averaging separately for $s$ and $s'$
with 
\begin{equation}\label{eq:phasea}
\langle \exp[i \delta S_s] \rangle = 
\exp(-{\textstyle \frac{1}{2}} \langle \delta S_s^2 \rangle ) = \mbox{}
\exp\left(-{\textstyle \frac{1}{2}} \int_0^t d{\tilde t} \int_0^t d{\tilde t}' 
\langle \Sigma[{\bf q}(\tilde{t})]
\Sigma[{\bf q}(\tilde{t}')] \rangle \right).
\end{equation}
Here ${\bf q}(\tilde{t})$ lies on path $s$ with 
${\bf q}(0)={\bf r}_0$ and ${\bf q}(t)={\bf r}$. 
We next note that, for chaotic systems, one generically observes fast, exponential decays of
correlations.
Assuming additionally
that $\Sigma$ and $H_0$ have no common integral of motion, so that
$\delta S_s$ fluctuates fast and randomly enough, 
the correlator of $\Sigma$ 
gives the golden rule decay 
\begin{equation}\label{eq:nondiagdecaya}
{\cal M}_{\rm L}^{({\rm nd})}(t) \propto \exp(-\Gamma t), \;\;\;\; \;{\rm with} \;\;\;\;
\Gamma t \equiv  \frac{1}{2} \int_0^t d{\tilde t} \int_0^t d{\tilde t}' 
\langle \Sigma[{\bf q}(\tilde{t})]\,
\Sigma[{\bf q}(\tilde{t}')] \rangle,
\end{equation}
regardless of whether $H_0$ is chaotic or regular. 

Our semiclassical approach thus predicts that the
Loschmidt echo is given by the sum of the diagonal and nondiagonal terms,
\begin{eqnarray}\label{eq:final_decaya}
{\cal M}_{\rm L} (t) = {\cal M}_{\rm L}^{\rm (d)}(t)
+ {\cal M}_{\rm L}^{\rm (nd)}(t) \propto \left\{
\begin{array}{cc}
t^{-d}, & {\rm regular \; systems,} \, U \tau < \nu^{-2}, \\
t^{-3d/2}, & {\rm regular \; systems,} \, U \tau > \nu^{-2}, \\
\alpha e^{-\lambda t} + e^{-\Gamma t}, & {\rm chaotic \; systems}.
\end{array} \right.
\end{eqnarray} 

It has apparently never been noticed that the semiclassical approach
also gives the long-time saturation of the Loschmidt echo.
To see this we go back one step before the diagonal approximation
leading to Eq.~(\ref{semicl}). We have
\begin{eqnarray}\label{full_semicla}
{\cal M}_{\rm L}(t) & = & (\nu^2/\pi)^d \int d{\bf r} \; \int d{\bf r}' 
\sum_{s_1,s_2,s_3,s_4} K_{s_1}^{H_0}({\bf r},{\bf r}_0;t)
[K_{s_2}^{H}({\bf r},{\bf r}_0;t)]^*  [K_{s_3}^{H_0}({\bf r}',{\bf r}_0;t)]^*
K_{s_4}^{H}({\bf r}',{\bf r}_0;t) \nonumber \\
&& \times \exp\big(-\nu^2 \big[|{\bf p}_{s_1}-{\bf p}_0|^2
+|{\bf p}_{s_2}-{\bf p}_0|^2 + |{\bf p}_{s_3}-{\bf p}_0|^2 
+|{\bf p}_{s_4}-{\bf p}_0|^2\big]/2 \big).
\end{eqnarray}
Pairing the trajectories as $s_1=s_3$ and $s_2=s_4$ exactly cancels all action phases,
and simultaneously requires ${\bf r} \simeq {\bf r}'$ within the
wavelength resolution $\nu$. Assuming ergodicity, one substitutes
\begin{equation}\label{semicl_sat}
\int {\rm d}{\bf r}' \Theta(\nu-|{\bf r} - {\bf r}'|) \rightarrow 
(\nu^d/\Omega) \int {\rm d}{\bf r}' ,
\end{equation}
with the system's spatial volume $\Omega$.

The rest of the calculation is straightforward, and follows steps already 
described above. The $C_s$'s are used to transform from spatial integration
variables to momentum integration variables. One is then left with
two normalized Gaussian integrals, multiplied by a prefactor
$(\nu^d/\Omega) = \hbar_{\rm eff}$. Hence one gets a time-independent 
contribution
\begin{equation}\label{longtimesaturationLEa}
{\cal M}_{\rm L}(\infty) = \hbar_{\rm eff} \; \Theta(t > \tau_{\rm E}),
\end{equation}
corresponding to the long-time saturation of ${\cal M}_{\rm L}$. This term 
requires that different paths exist between ${\bf r}_0$ and ${\bf r}
\simeq {\bf r}'$ (see the rightmost contribution sketched in 
Fig.~\ref{fig:fig_LEavg}) and therefore does not exist for times
shorter than the Ehrenfest time
$\tau_{\rm E}$ ~\cite{Lar68,Ber78,Ber79,Chi81,Chi88}. 
It is given by the time it takes 
the classical dynamics 
to increase the distance between two trajectories
from $\nu$ to $L$.
The trajectory pairings that lead to these results, 
Eqs.~(\ref{eq:final_decaya}) and (\ref{longtimesaturationLEa})  
are summarized in Fig.~\ref{fig:fig_LEavg}.

\subsection{Mesoscopic fluctuations of the Loschmidt echo}
\label{appendix:semiclassics_flucs}

We want to calculate ${\cal M}_{\rm L}^2$. Squaring Eq.~(\ref{moft}), we see
that it is given by
eight sums over classical paths and twelve spatial integrations. Eight of these integrals can
be calculated once we
note, as before, that $\psi_0$ is
a narrow Gaussian wavepacket, and accordingly linearize
all eight action integrals around ${\bf r}_0$, $
S_{s}({\bf r},{\bf r}_0';t) \simeq S_{s}({\bf r},{\bf r}_0;t)
-({\bf r}_0'-{\bf r}_0) \cdot {\bf p}_{s}.$
We can then perform the Gaussian integrations over the eight initial positions
${\bf r}_0'$, ${\bf r}_0''$... and so forth. In this way 
${\cal M}_{\rm L}^2(t)$ is expressed as a sum over eight trajectories connecting
${\bf r}_0$ to four
independent final points ${\bf r}_j$ over which one integrates,
\begin{equation}\label{eq:m2a}
{\cal M}_{\rm L}^2(t) = \int \prod_{j=1}^4 d{\bf r}_j
\sum_{s_i;i=1}^8 \;  \exp[i (\Phi^{H_0}-\Phi^{H}-\pi \Xi /2)] 
\; \prod_i C_{s_i}^{1/2} \left(\frac{\nu^2}{\pi}\right)^{d/4}
\exp(-\nu^2\delta{\bf p}_{s_i}^2/2).
\end{equation}
Here we introduced the sum $\Xi=\sum_{i=0}^{3} (-1)^i (\mu_{s_{2i+1}} -
\mu_{s_{2i+2}} )$ of Maslov indices and the momentum
difference $\delta{\bf p}_{s_i}={\bf p}_{s_i}-{\bf p}_0$.
The right-hand side of Eq.~(\ref{eq:m2a}) 
is schematically described in Fig.~\ref{fig1_echoflucs}. 
Eq.~(\ref{eq:m2a}) is dominated by terms 
where the variation of the difference of the two action phases
\begin{subequations}
\begin{eqnarray}
\Phi^{H_0}&=&S^{H_0}_{s_1}({\bf r}_1,{\bf r}_0;t)-
S^{H_0}_{s_3}({\bf r}_2,{\bf r}_0;t) +
S^{H_0}_{s_5}({\bf r}_4,{\bf r}_0;t)-
S^{H_0}_{s_7}({\bf r}_3,{\bf r}_0;t), \\
\Phi^{H}&=&
S^{H}_{s_2}({\bf r}_1,{\bf r}_0;t)\;-
S^{H}_{s_4}({\bf r}_2,{\bf r}_0;t)\; +
S^{H}_{s_6}({\bf r}_4,{\bf r}_0;t)\;-
S^{H}_{s_8}({\bf r}_3,{\bf r}_0;t),
\end{eqnarray}
\end{subequations}
is minimal. The four dominant contributions to 
the fidelity variance are depicted on the right-hand side of 
Fig.~\ref{fig2_echoflucs}. We now proceed to calculate them one by one.

The first one corresponds to $s_1=s_2 \simeq
s_7=s_8$ and $s_3=s_4 \simeq s_5=s_6$, which
requires ${\bf r}_1 \simeq {\bf r}_3$, ${\bf r}_2 \simeq {\bf r}_4$, and gives
a contribution
\begin{eqnarray}\label{sigma1}
\sigma_1^2 & = & \left(\frac{\nu^2}{\pi}\right)^{2d} \Bigg\langle 
\int d{\bf r}_1 d{\bf r}_3 \sum
C^2_{s_1} \;
\exp[-2 \nu^2 \delta{\bf p}_{s_1}^2 + i \delta \Phi_{s_1}]
\Theta(\nu-|{\bf r}_1-{\bf r}_3|)
\Bigg\rangle^2.
\end{eqnarray}
Here $\delta \Phi_{s_1}= \int_0^t dt' \nabla \Sigma [{\bf q}(t')] [
{\bf q}_{s_1}(t')-{\bf q}_{s_7}(t')]$ originates from the same linearization of $\Sigma$
on $s=s_{1,2} \simeq s'=s_{7,8}$ that was used earlier 
in the calculation of the average
fidelity, and
${\bf q}_{s_1}(\tilde{t})$ lies on $s_1$ with ${\bf q}(0)={\bf r}_0$
and ${\bf q}(t)={\bf r}_1$. In Eq.~(\ref{sigma1})
the integrations are restricted by $|{\bf r}_1-{\bf r}_3|\le \nu$ because
of the finite resolution with which two paths can be equated (this
is also enforced by the presence of $\delta \Phi_s$ as we will see
momentarily).
For long enough times, $t \gg t^*$ with $t^*$ defined by the first root of
$\big| \int_0^{t^*} \Sigma({\bf q}_s(t),t) \big| = 1$ on a typical
trajectory $s$,
the phases $\delta \Phi_s$ fluctuate randomly and exhibit no correlation 
between different trajectories. This justifies to apply
the Central Limit Theorem
(CLT) $\langle \exp[i \delta \Phi_s] \rangle = \exp[-\langle \delta \Phi_s^2 
\rangle/2] \simeq \exp[- \int d \tilde{t} \langle \nabla \Sigma(0) \cdot 
\nabla \Sigma(\tilde{t}) \rangle |{\bf r}_1-
{\bf  r}_3|^2/2 \lambda]$. Using Eq.~(\ref{correl}), one then 
obtains a similar Gaussian damping of relative coordinates as in 
Eq.~(\ref{echo_diaga}). We perform 
the change of integration variable given in the 
second line of Eq.~(\ref{eq:chg_var})
to get the first contribution to $\sigma^2({\cal M}_{\rm L})$,
\begin{subequations}\label{sigma1lambdaa}
\begin{eqnarray}
\sigma_1^2&=& \alpha^2 \exp[-2\lambda t], 
\end{eqnarray}
\end{subequations}
where $\alpha$ is the same as in Eq.~(\ref{eq:final_decay}).

The second dominant term is obtained from  
$s_1=s_2 \simeq s_7=s_8$, $s_3=s_4$ and $s_5=s_6$, with
${\bf r}_1 \simeq {\bf r}_3$, or equivalently $s_1=s_2$, $s_7=s_8$ and  
$s_3=s_4 \simeq s_5=s_6$ with ${\bf r}_2 \simeq {\bf r}_4$. 
It comes with a multiplicity of two, and reads
\begin{eqnarray}\label{sigma2}
\sigma^2_2 &=&  2 \left(\frac{\nu^2}{\pi}\right)^{2d} 
\left\langle\int d{\bf r}_2 \sum C_{s_3} 
\exp[-\nu^2 \delta{\bf p}_{s_3}^2+i \delta S_{s_3}] 
\right\rangle^2\nonumber \\
&& \times \Bigg \langle
\int d{\bf r}_1 d{\bf r}_3 \sum C^2_{s_1} \;
\exp[-2 \nu^2 \delta{\bf p}_{s_1}^2
+i\delta \Phi_{s_1}] \Theta(\nu -|{\bf r}_1-{\bf r}_3|)
\Bigg\rangle ,
\end{eqnarray}
again with the restriction $|{\bf r}_1-{\bf r}_3|\le \nu$.
To calculate the first bracket on the right-hand side of
Eq.~(\ref{sigma2}), we first average the complex exponential, assuming 
again that enough time has elapsed so that actions are randomized. 
The CLT gives $\langle \exp[i \delta S_{s_3}] \rangle = 
\exp(-{\textstyle \frac{1}{2}} \langle \delta S_{s_3}^2 \rangle )$ with
\begin{equation}\label{phase}
\langle \delta S_{s_3}^2 \rangle  = \mbox{}
\int_0^t d{\tilde t} \int_0^t d{\tilde t}' 
\langle \Sigma[{\bf q}(\tilde{t})]
\Sigma [{\bf q}(\tilde{t}')] \rangle.
\end{equation}
Here ${\bf q}(\tilde{t})$ lies on $s_3$ with ${\bf q}(0)={\bf r}_0$
and ${\bf q}(t)={\bf r}_2$. We already observed above that in hyperbolic systems,
correlators typically decay exponentially fast~\cite{Col04}, which justifies 
the assumption made in Eq.~(\ref{correl}) of $\delta$-correlated perturbations
\begin{equation}
\langle \Sigma[{\bf q}(\tilde{t})]
\Sigma[{\bf q}(\tilde{t}')] \rangle \propto \delta(\tilde{t}-\tilde{t}').
\end{equation} 
Here we depart slightly from Ref.~\cite{Pet05} which instead considered
an exponentially decaying correlator, with a decay rate bounded from above
by the smallest positive Lyapunov
exponent. These two choices differ only by 
exponentially small corrections in the limit of large enough times, $t \gtrsim \lambda^{-1}$,
for which even algebraic decaying correlations deliver the same answer [see the discussion
below Eq.~(\ref{diagdecay})]. One obtains
$\langle \delta S_{s_3}^2 \rangle = \Gamma t$. In the RMT approach, $\Gamma$ is
identified with the golden rule spreading of eigenstates of $H$ over those of $H_0$ \cite{Jac01}.
It is dominated by the short-time behavior of
$\langle \Sigma[{\bf q}(\tilde{t})] \Sigma[{\bf q}(0)] \rangle$.
Expressions similar to Eq.~(\ref{phase})
relating the decay of ${\cal M}_{\rm L}$ to
perturbation correlators have been derived in 
Refs.~\cite{Pro03a,Gor04a} using a more restricted, linear response 
approach. We next use the sum rule of Eq.~(\ref{sumrule1}) 
to finally obtain
\begin{equation}\label{sigma2lga}
\sigma_2^2 \simeq 2 \alpha \exp[-\lambda t] \exp[-\Gamma t].
\end{equation}

The third and last dominant time-dependent term arises
from either $s_1=s_7$, $s_2=s_8$, $s_3=s_4$, $s_5=s_6$ and 
${\bf r}_1\simeq{\bf r}_3$, or $s_1=s_2$, $s_3=s_5$, $s_4=s_6$, $s_7=s_8$ and 
${\bf r}_2\simeq{\bf r}_4$. It thus also has a multiplicity of two and reads
\begin{eqnarray}\label{sigma3}
\sigma^2_3 &=& 2 \left(\frac{\nu^2}{\pi}\right)^{2d}
\Bigg\langle
\int d{\bf r}_1 d{\bf r}_2 d{\bf r}_3 d{\bf r}_4
\sum C_{s_1}C_{s_2}C_{s_3}C_{s_5} \; \exp[-\nu^2 (\delta{\bf p}_{s_1}^2+
\delta{\bf p}_{s_2}^2+\delta{\bf p}_{s_3}^2+\delta{\bf p}_{s_5}^2)] 
\nonumber \\
&& \;\;\;\;\;\;\;\;\;\;\;\;\;\;\;\;\;\;\;\;\;
\times \exp[i (\delta S_{s_3}-\delta S_{s_5})] \;
\Theta(\nu - |{\bf r}_1-{\bf r}_3|) \Bigg\rangle.
\end{eqnarray}
To take the restriction into account
that the integrations have to be performed with
$|{\bf r}_1-{\bf r}_3|\le \nu$,
we assume ergodicity and set
\begin{equation}\label{saturation}
\Big\langle \int d{\bf r}_1 d{\bf r}_2 d{\bf r}_3 d{\bf r}_4 \ldots
\Theta(\nu - |{\bf r}_1-{\bf r}_3|) \Big\rangle
= \hbar_{\rm eff} \; \Big\langle \int d{\bf r}_1 d{\bf r}_2 d{\bf r}_3 
d{\bf r}_4 \ldots \Big\rangle \; \Theta(t-\tau_{\rm E}), 
\end{equation}
which is valid for
times larger than the Ehrenfest time. For shorter times,
$t<\tau_{\rm E}$, the third diagram on the right-hand side 
of Fig.~\ref{fig2_echoflucs} 
goes into the second one. Once again we use the CLT to
average the phases. One gets Eq.~(\ref{sigma3g}).

\subsection{Displacement echo}
\label{appendix:semiclassics_displ}

We semiclassically evaluate
${\cal M}_{\rm D}$ defined in Eq.~(\ref{decho_repeat}). As for the Loschmidt echo, we consider
an initial Gaussian wavepacket, $
\psi_0 ({\bf r}) =(\pi \nu^2)^{-d/4}
\exp[i{\bf p}_0 \cdot ({\bf r}-{\bf r}_0)-|{\bf r}-{\bf r}_0|^2/2 \nu^2].$
We semiclassically propagate $|\psi_0 \rangle$ with the help of
the Gutzwiller--van Vleck propagator~\cite{Tom91,Cvi05,Gut90,Haa01}, 
and expand linearly around $\bf{r}_0$,
\begin{eqnarray}
\langle {\bf r'} | \exp[-i H t] | \psi_0 \rangle \simeq 
\left( -\frac{i \nu}{\sqrt{\pi}} \right)^{d/2} 
\sum_s \sqrt{C_s} \exp[i S_s-i \pi \mu_s/2-\nu^2 ({\bf p}_s-{\bf p}_0)/2].
\end{eqnarray}
Here, the sum runs over all possible classical trajectories $s$ connecting
$\mathbf{r}_0$ and $\mathbf{r'}$ in the time $t$, 
$\mathbf{p}_s=\left. -\partial S_s / \partial \mathbf{r}\right|_{\mathbf{r}=\mathbf{r}_0}$
is the initial momentum on $s$, $S_s$ is the classical action accumulated
on $s$, $\nu_s$ is the Maslov index and 
$C_s=-\partial^2 S_s(\mathbf{r}',\mathbf{r};t) \big/ \partial r_i\partial r^{\prime}_j 
\big|_{\mathbf{r}=\mathbf{r}_0}$. The kernel of 
${\cal M}_{\rm D}(t)$ involves a double sum over classical trajectories $s_1$ 
and $s_2$, 
which can be interpreted as the overlap between a wavepacket that is boosted 
and subsequently propagated with a wavepacket that is first propagated and 
subsequently boosted. 
Enforcing a stationary phase condition kills all but the contributions 
with the smallest actions. As for the standard Loschmidt echo 
[see above Eq.~(\ref{semicl})], one therefore enforces
a stationary phase condition which, to leading order, requires
$s_1=s_2$ . 
Taking the squared amplitude of the kernel,
one obtains the semiclassical expression for the
displacement echo (corresponding to Eq.~(\ref{mtot}) for the Loschmidt echo)
\begin{eqnarray}\label{mtot_displ}
{\cal M}_{\rm D} (t ) &=& 
\left(\frac{\nu^2} {\pi}\right)^{d} 
\int {\rm d}{\bf r} \, {\rm d}{\bf r}^{\prime} 
 \, \sum_{s,s'}  \,
 C_{s}\,  C_{s^{\prime}}\,\exp[i {\bf P} \cdot ({\bf r} -{\bf r}^{\prime})]\\
&\times & \exp\left\{-\frac{\nu^2}{2}\Big[(\mathbf{p}_s-\mathbf{p}_0)^2 
+(\mathbf{p}_s-\mathbf{p}_0-\mathbf{P})^2 
+(\mathbf{p}_{s'}-\mathbf{p}_0)^2+
(\mathbf{p}_{s'}-\mathbf{p}_0-\mathbf{P})^2 \Big] \right\}.  \nonumber 
\end{eqnarray}
We calculate the ensemble-averaged displacement echo 
over a set of initial 
Gaussian wavepackets with varying center of mass ${\bf r}_0$ for which,
as for the Loschmidt echo,
there are two qualitatively different contributions. The first 
contribution, ${\cal M}_{\rm D}^{\rm (d)}$,
comes from pairs $s \simeq s'$ of correlated 
trajectories that remain within a distance $\lesssim \nu$ of each other 
for the whole duration of the experiment, while the second contribution,
${\cal M}_{\rm D}^{\rm (nd)}$, arises from pairs of uncorrelated 
trajectories $s \ne s'$. For the first contribution, we write
$\exp[i {\bf P} ({\bf r}-{\bf r}')] \approx 1$, which is true in the 
semiclassical limit where $\nu \rightarrow 0$, and set $s=s'$.
One then has
\begin{eqnarray}\label{corr1}
{\cal M}_{\rm D}^{\rm (d)}(t)&=& 
\left(\frac{\nu^2} {\pi}\right)^{d} 
\int {\rm d}{\bf r} {\rm d}{\bf r}^{\prime} \, \Theta(\nu-|{\bf r}-{\bf r}'|) 
\Big\langle \sum_{s} \,
 C_{s}^2\, e^{-\nu^2\left[ (\mathbf{p}_s-\mathbf{p}_0)^2+(\mathbf{p}_s-\mathbf{p}_0-\mathbf{P})^2 \right]}\Big\rangle,
\end{eqnarray}
where the Heaviside function 
$\Theta(\nu-|{\bf r}-{\bf r}'|)$ restricts the integrals to
$|{\bf r} -{\bf r}^{\prime}| \le \nu$. The calculation of (\ref{corr1})
is straightforward. The integral over ${\bf r}'$ gives a factor $\nu^d$.
One then changes integration variable as in Eq.~(\ref{eq:chg_var}).
A Gaussian integration finally delivers the correlated
contribution to ${\cal M}_{\rm D}(t)$,
\begin{eqnarray}\label{corr2a}
{\cal M}_{\rm D}^{\rm (d)}(t)&=& 
\alpha \, \exp[-({\bf P} \nu)^2/2] \, \exp[-\lambda t].
\end{eqnarray}
Here, $\alpha={\cal O}(1)$ is only weakly time-dependent~\cite{Jal01,Pet07a}.

For the uncorrelated part, an ergodicity assumption is justified at 
sufficiently large times, under which one gets
\begin{subequations}\label{uncorrelated}
\begin{eqnarray}
{\cal M}_{\rm D}^{\rm (nd)} (t ) 
&=& f({\bf P}) \; \tilde{\cal M}_{\rm D}^{\rm (nd)} (t) , \\
f({\bf P}) &=& \Omega^{-2}\int {\rm d}{\bf r} {\rm d}{\bf r}^{\prime} 
\exp[i {\bf P} \cdot ({\bf r} -{\bf r}^{\prime})] , \\
\tilde{\cal M}_{\rm D}^{\rm (nd)}(t) &=& \left( \frac{\nu^2} {\pi}\right)^{d} 
\Big( \int {\rm d}{\bf x} 
 \, \sum_{s} \,
 C_{s}\,  \exp -\frac{\nu^2}{2}\Big[ (\mathbf{p}_s-\mathbf{p}_0)^2+(\mathbf{p}_s-\mathbf{p}_0-\mathbf{P})^2 \Big] \Big)^2,
\end{eqnarray}
\end{subequations}
where as usual $\Omega \propto L^d$ is the system's volume. 
It is straightforwardly seen that
$\tilde{\cal M}_{\rm D}^{\rm (nd)}(t)= \exp[-({\bf P} \nu)^2/2]$,
and $f({\bf P})=g(|{\bf P}|L)/(|{\bf P}|L)^2$, in terms of
an oscillatory function $g(|{\bf P}|L)=4 \sin^2(|{\bf P}|L/2)$ for $d=1$
and $g(|{\bf P}|L)=4 J_1^2(|{\bf P}|L)$ for $d=2$. For $d=3$, 
$g$ is given by Bessel and Struve functions. Finally, the uncorrelated contribution
reads
\begin{eqnarray}
{\cal M}_{\rm D}^{\rm (nd)}(t) &=& 
\exp[-({\bf P} \nu)^2/2] \; g(|{\bf P}|L)\Big/(|{\bf P}| L)^2.
\end{eqnarray}
Together with Eq.~(\ref{corr2a}) this gives the total displacement echo 
\begin{eqnarray}\label{corrtota}
{\cal M}_{\rm D}(t) =
\exp[-({\bf P} \nu)^2/2] \, \left [\alpha \, \exp[-\lambda t] \, +
\frac{g(|{\bf P}|L)}{(|{\bf P}| L)^2} \right].
\end{eqnarray}
As is the case for the Loschmidt echo, 
the semiclassical approach also delivers the long-time
saturation ${\cal M}_{\rm D}(\infty ) = \hbar_{\rm eff} = 
N^{-1}$, valid for displacements
such that $g(|{\bf P}|L) \big/ (|{\bf P}| L)^2 \ll N^{-1}$.

\subsection{Bipartite entanglement}
\label{appendix:semiclassics_entangle}

Our goal is to calculate the purity ${\cal P}(t) 
\equiv {\rm Tr}[\rho_1^2(t)]$  of the reduced density matrix for a bipartite systems of
two interacting few-degrees of freedom dynamical system. 
We start with an initial two-particle product state
$|\psi_1 \rangle \otimes |\psi_2 \rangle \equiv
|\psi_1, \psi_2 \rangle$. The state of each particle is a
Gaussian wavepacket $\psi_{1,2}({\bf y}) = 
(\pi \nu^2)^{-d_{1,2}/4} \exp[i {\bf p}_{1,2} \cdot ({\bf y}-{\bf r}_{1,2})-
|{\bf y}-{\bf r}_{1,2}|^2/2 \sigma^2]$. 
We write the two-particle Hamiltonian as 
\begin{equation}\label{2hamiltonian}
{\cal H} = H_1 \otimes I_2 + I_1 \otimes H_2 + {\cal U},
\end{equation}
where the two particles are subjected to possibly different
Hamiltonians $H_{1,2}$. The interaction potential ${\cal U}$ appears
in the semiclassical 
calculation only via its correlator along classical trajectories.
Therefore there is no need to specify it, beyond saying that it
depends only on the
distance between the particles, and that it is characterized by 
a typical length scale $\zeta > \nu$. This can be its range, or the scale
over which it fluctuates. The two-particle density matrix evolves according to
\begin{subequations}
\begin{eqnarray}
\rho(t) &=& \exp[-i {\cal H} t] \rho_0 \exp[i {\cal H} t] \, , \\
\rho_0 & = & |\psi_1,\psi_2\rangle\langle \psi_1,\psi_2|.
\end{eqnarray}
\end{subequations}
The elements 
$\rho_1({\bf x},{\bf y};t) = \int d{\bf r} \langle {\bf x}, {\bf r}|
\rho(t) |{\bf y}, {\bf r} \rangle$ of the reduced density matrix read 
\begin{eqnarray}
\rho_1({\bf x},{\bf y};t) &=& (\pi \nu^2)^{-(d_1+d_2)}
\int d{\bf r} \int \prod_{i=1}^4 d{\bf y}_i \;\;
e^{-\{({\bf y}_1-{\bf r}_1)^2 
+({\bf y}_2-{\bf r}_2)^2 + ({\bf y}_3-{\bf r}_1)^2 + ({\bf y}_4-{\bf r}_2)^2 \}
/2 \nu^2} \nonumber \\
&& \times \;e^{i {\bf p}_1 \cdot ({\bf y}_1-{\bf y}_3)}
\, e^{i {\bf p}_2
\cdot ({\bf y}_2-{\bf y}_4)} \; \;
\langle {\bf x}, {\bf r}| e^{-i{\cal H}t} |{\bf y}_1, {\bf y}_2
\rangle
\langle {\bf y}_3, {\bf y}_4 |
e^{i {\cal H}t} |{\bf y}, {\bf r} \rangle.
\end{eqnarray}
We next introduce the semiclassical two-particle propagator
\begin{eqnarray}\label{twopart_propa}
 \langle {\bf x}, {\bf r}| e^{-i{\cal H}t} |{\bf y}_1,
{\bf y}_2 \rangle &=& 
(2 \pi i)^{-(d_1+d_2)/2} \; \sum_{s,s'} {\cal C}_{s,s'}^{1/2} \; e^{i \{S_s({\bf x},{\bf y}_1;t) +
S_{s'}({\bf r},{\bf y}_2;t) +
{\cal S}_{s,s'}({\bf x},{\bf y}_1;{\bf r},{\bf y}_2;t)\}}, 
\end{eqnarray}
which is expressed as a sum over pairs of classical trajectories, labeled $s$
and $s'$, respectively 
connecting ${\bf y}_1$ to ${\bf x}$ and ${\bf y}_2$ to ${\bf r}$
in the time $t$. Each such pair of paths gives a 
contribution containing one-particle actions $S_s$ and $S_{s'}$
(they include the Maslov indices)
and two-particle action integrals
\begin{equation}
{\cal S}_{s,s'}=\int_0^t dt_1 {\cal U}({\bf q}_s
(t_1), {\bf q}_{s'}(t_1)),
\end{equation}
accumulated along $s$ and $s'$,  
and the determinant ${\cal C}_{s,s'} = C_s \, C_{s'}$ of the stability
matrix corresponding to the two-particle 
dynamics in the $(d_1+d_2)-$dimensional space. 
Eq.~(\ref{twopart_propa}) relies on the assumption that individual particle
trajectories can be identified and are not modified by the 
interaction between the two particles. The only effect of the interaction
is to contribute a two-particle term in 
the action accumulated on those trajectories. As for the Loschmidt echo, this
approximation is justified by the structural stability of chaotic systems,
where perturbed (with interaction)
trajectories are shadowed by unperturbed (noninteracting) trajectories.
Numerical investigations have shown that structural stability also exists in chaotic 
many-body systems~\cite{Hay03a,Hay03b}.

With the above definition, ${\cal C}_{s,s'}$ is real 
and positive. Because we consider sufficiently
smooth interaction potentials, varying over a distance much 
larger than the de Broglie wavelength, $\zeta \gg \nu$,
we set ${\cal S}_{s,s'}({\bf x},{\bf y}_1;
{\bf r},{\bf y}_2;t) \simeq {\cal S}_{s,s'}({\bf x},{\bf r}_1;
{\bf r},{\bf r}_2;t)$. Still we must keep in mind that ${\bf r}_{1}$ and
${\bf r}_{2}$, taken as arguments of
the two-particle action integrals have a quantum-mechanical 
uncertainty $O(\nu)$. 
We next use 
the narrowness of the initial wavepackets to linearize
the one-particle actions in ${\bf y}_i-{\bf r}_j$ ($i=1,\ldots 4$; $j=1,2$).
This gives us four Gaussian integrals over the ${\bf y}_i$'s which
we perform to obtain
\begin{eqnarray}
\label{semicrho}
\rho_1({\bf x},{\bf y};t) &=& \left(\frac{\nu^2}{\pi}\right)^{d_1/2} \sum_{s,l} \;
(C_s C_l )^{1/2} \; \exp[-\frac{\nu^2}{2} \{
({\bf p}_{s}-{\bf p}_1)^2 + ({\bf p}_{l}-{\bf p}_1)^2\}] 
 \\
&& \times  \; {\cal F}_{s,l}(t) \; 
\exp[i\{S_s({\bf x},{\bf r}_1;t)-S_{l}({\bf y},{\bf r}_1;t)\}] \nonumber \\
\label{f-vernon-bi-pure}
{\cal F}_{s,l}(t) & = & \left(\frac{\nu^2}{\pi}\right)^{d_2/2} \int d{\bf r}
\sum_{s',l'} (C_{s'} C_{l'})^{1/2} e^{-\frac{\nu^2}{2} \{
({\bf p}_{s'}-{\bf p}_2)^2 + ({\bf p}_{l'}-{\bf p}_2)^2\}} 
 \\
&& \times 
\exp[i\{S_{s'}({\bf r},{\bf r}_2;t)-S_{l'}({\bf r},{\bf r}_2;t)
+{\cal S}_{s,s'}({\bf x},{\bf r}_1;{\bf r},{\bf r}_2;t)
-{\cal S}_{l,l'}({\bf y},{\bf r}_1;{\bf r},{\bf r}_2;t)\}].\nonumber 
\end{eqnarray}
Eq.~(\ref{f-vernon-bi-pure}) expresses the 
influence functional of Feynman and Vernon~\cite{Fey63} as a sum over classical
trajectories. 

We consider the weak coupling regime, where the one-particle actions
vary faster than their two-particle counterpart.
We thus perform a 
stationary phase approximation on the one-particle actions of the
environment and accordingly pair the trajectories $s' \simeq l'$, since they 
have the same endpoints.
We get the
semiclassical Feynman-Vernon influence functional
\begin{eqnarray}
\label{f-vernon-bi-pure-semicl}
{\cal F}_{s,l}(t) & = & \left(\frac{\nu^2}{\pi}\right)^{d_2/2} \ \int d{\bf r}
\sum_{s'} C_{s'} e^{-\nu^2 
({\bf p}_{s'}-{\bf p}_2)^2 }
e^{i\{{\cal S}_{s,s'}({\bf x},{\bf r}_1;{\bf r},{\bf r}_2;t)
-{\cal S}_{l,s'}({\bf y},{\bf r}_1;{\bf r},{\bf r}_2;t)\}}.
\end{eqnarray}
It is straightforward to see that our procedure is probability-conserving,
${\rm Tr}[\rho_1(t)]=1$, and that it preserves the Hermiticity of the
reduced density matrix $\rho_1({\bf x},{\bf y};t)=[\rho_1({\bf y},
{\bf x};t)]^*$, as required. 

Enforcing a further stationary phase condition on Eq.~(\ref{semicrho})
amounts to performing an average 
over different initial conditions ${\bf r}_{1,2}$. 
It results in $s=l$, ${\bf x}={\bf y}$, and thus
$\langle \rho_1 ({\bf x},{\bf y};t) \rangle = 
\delta_{{\bf x},{\bf y}}/\Omega_1$, with the volume $\Omega_1$ occupied
by particle one.
Diagonal elements of the reduced density matrix acquire an ergodic 
value -- this is due to the average over initial
conditions -- and only they have a nonvanishing average.
For each initial condition, $\rho_1(t)$
has however nonvanishing off-diagonal matrix elements, 
with a zero-centered distribution whose
variance is given by $\langle 
\rho_1({\bf x},{\bf y};t)  \rho_1({\bf y},{\bf x};t) \rangle$.
Beyond giving the variance of the distribution of off-diagonal matrix
elements, this quantity also appears in the purity
 ${\cal P}(t) = \int {\rm d}{\bf x}
\int {\rm d}{\bf y} \rho_1({\bf x},{\bf y};t)  \rho_1({\bf y},{\bf x};t) 
\rangle$, and we therefore proceed to calculate it.

Squaring Eq.~(\ref{semicrho}), averaging over ${\bf r}_{1,2}$ and
enforcing a stationary phase approximation on the $S$'s, one gets
\begin{eqnarray}\label{Sigma2}
\langle 
\rho_1({\bf x},{\bf y};t)  \rho_1({\bf y},{\bf x};t) \rangle&=&
\left(\frac{\nu^2}{\pi}\right)^{d_1+d_2}  \int d{\bf r} d{\bf r}'
\sum_{s,s'} \; \sum_{l,l'} \; C_s \, C_l \, C_{s'} \, C_{l'} \; 
\langle {\cal G}_{s,s';l,l'} \rangle
\\
&& \times \; \exp[-\nu^2 
({\bf p}_{s}-{\bf p}_1)^2 +
({\bf p}_{l}-{\bf p}_1)^2 +
({\bf p}_{s'}-{\bf p}_2)^2 +
({\bf p}_{l'}-{\bf p}_2)^2] , \nonumber \\[3mm]
\label{Faction}
\langle {\cal G}_{s,s';l,l'} \rangle &=& \;\;\;\;
\Big\langle \exp[i \{ {\cal S}_{s,s'}({\bf x},{\bf r}_1;{\bf r},{\bf r}_2;t) 
- {\cal S}_{l,s'}({\bf y},{\bf r}_1;{\bf r},{\bf r}_2;t) \}] \\
&& \;\;\; \times \exp[i \{ 
{\cal S}_{l,l'}({\bf y},{\bf r}_1;{\bf r}',{\bf r}_2;t)
- {\cal S}_{s,l'}({\bf x},{\bf r}_1;{\bf r}',{\bf r}_2;t) \}]\Big\rangle .\nonumber 
\end{eqnarray}
In our analysis of Eqs.~(\ref{Sigma2}) and (\ref{Faction}) we note that the time-dependence
of $\langle |\rho_1|^2\rangle$ is given by the sum of three positive
contributions,
\begin{equation}\label{eq:red_threeterms}
\langle 
\rho_1({\bf x},{\bf y};t)  \rho_1({\bf y},{\bf x};t) \rangle
= \Sigma_1({\bf x},{\bf y};t) + \Sigma_2({\bf x},{\bf y};t) + \Sigma_3 ({\bf x},{\bf y};t) \, .
\end{equation}
First, those particular paths for which ${\bf r}={\bf r}'$ and $s'=l'$, 
accumulate no phase (${\cal G}_{s,s';l,s'}=1$) and thus have to be considered
separately. On average, their contribution does not depend on 
${\bf x}$ nor ${\bf y}$, and decays in time only because of their decreasing 
measure with respect to all the paths with ${\bf r} \ne {\bf r}'$. 
By analogy with the calculation of ${\cal M}_{\rm L}$ 
we readily anticipate that this contribution is governed by the decay
of overlap of two initially identical wavepackets interacting with 
a second particle in different states -- giving a Lyapunov, exponential 
decay in the chaotic case, a power-law decay in the regular case.
Second, similar contributions with $s=l$ also exist, which however affect only
the variance of the diagonal matrix elements
and do not depend on ${\bf x} \simeq {\bf y}$.
We find that, on average, these two diagonal contributions give
\begin{eqnarray}\label{bounddecay}
\Sigma_1({\bf x},{\bf y};t) & \simeq & \left\{ 
\begin{array}{cc}
\Omega_1^{-2} \exp[-\lambda_2 t]  \;\;\; ; \;\;\; {\rm chaotic}, \\
\Omega_1^{-2} \left(t_0\big/t\right)^{d_2} 
\;\;\;\;\;\;\;\;  ; \;\;\; {\rm regular}.
\end{array} \right. \\
\Sigma_2({\bf x},{\bf y};t) & \simeq & \left\{ 
\begin{array}{cc}
\Omega_1^{-1} \delta_\nu({\bf x}-{\bf y}) \exp[-\lambda_1 t]  \;\;\; ; \;\;\; {\rm chaotic}, \\
\Omega_1^{-1} \delta_\nu({\bf x}-{\bf y}) \left(t_0\big/t\right)^{d_1} 
\;\;\;\;\;\;\;\; ; \;\;\; {\rm regular},
\end{array} 
\right. ,
\end{eqnarray}
with the spatial volume $\Omega_1$ occupied by particle one.
Despite the local nature of $\Sigma_1$, 
both terms give a contribution
of the same order to the average 
purity. Three facts are worth noting. First, these contributions do not
depend on the interaction strength, second they 
give a lower bound for the decay of $\langle |\rho_1|^2 \rangle$.
Third, in the regular regime, 
both $\Sigma_1$ and $\Sigma_2$ give a power-law decay
with the classical exponent $d_{1,2}$ and not the anomalous
exponent $3 d_{1,2}/2$ one would expect from the semiclassical
analysis of the Loschmidt echo. This is so because we assumed
that the interaction potential is smooth on a distance much larger
than the particle's de Broglie wavelength. Accordingly 
we approximate ${\cal S}_{s,s'}({\bf x},{\bf y}_1;
{\bf r},{\bf y}_2;t) \simeq {\cal S}_{s,s'}({\bf x},{\bf r}_1;
{\bf r},{\bf r}_2;t)+({\bf y}_1-{\bf r}_1) \cdot \nabla_{{\bf y}_1}
{\cal S}_{s,s'}({\bf x},{\bf y}_1;
 {\bf r},{\bf y}_2;t)+({\bf y}_2-{\bf r}_2) \cdot \nabla_{{\bf y}_2}
{\cal S}_{s,s'}({\bf x},{\bf y}_1; {\bf r},{\bf y}_2;t)
\approx {\cal S}_{s,s'}({\bf x},{\bf r}_1;
{\bf r},{\bf r}_2;t)$, since the envelope of the initial Gaussian wavepackets
$\psi_{1,2}$ requires $({\bf y}_i-{\bf r}_i) \lesssim \nu$.

The third contribution to $\langle |\rho_1|^2 \rangle$ 
is uncorrelated in the sense that it does not require further
pairing of trajectories. Its
decay with time is thus governed by the dephasing due to the
particle-particle interaction contained 
$\langle {\cal G} \rangle$.
From Eq.~(\ref{Faction}), it is natural to expect that 
$\langle {\cal G} \rangle$ 
is a decreasing function of $|{\bf x}-{\bf y}|$ and $t$ only, and that the
CLT applies in the form
\begin{equation}
\langle {\cal G}_{s,s';l,l'} \rangle =
\exp[-\big\langle ({\cal S}_{s,s'} 
- {\cal S}_{l,s'} +
{\cal S}_{l,l'}
- {\cal S}_{s,l'} )^2/2\big\rangle].
\end{equation}
Sums and integrals in Eq.~(\ref{Sigma2}) can then be
performed separately to give
\begin{eqnarray}\label{2correl}
\Sigma_3 ({\bf x},{\bf y};t) & = & \Omega_1^{-2} \, 
\exp[-2 \, (\langle {\cal S}_{s,s'}^2 \rangle
-\langle {\cal S}_{s,s'} {\cal S}_{l,s'} \rangle
+\langle {\cal S}_{s,s'} {\cal S}_{l,l'} \rangle
-\langle {\cal S}_{l,s'} {\cal S}_{l,l'} \rangle)], \\
\langle {\cal S}_{s,s'} {\cal S}_{l,l'} \rangle &=& 
\int_0^t dt_1 dt_2 \; \langle 
{\cal U}({\bf q}_s(t_1), {\bf q}_{s'}(t_1))\;
{\cal U}({\bf q}_l(t_2),{\bf q}_{l'}(t_2)).
\end{eqnarray}
The four correlators are different in the number of trajectories appearing
twice for each particle. It is easily seen, however, that unpaired
trajectories lead to a fast decay of the corresponding correlator.
This decay occurs on a time scale $\tau_{\cal U}$ which we estimate as
the time it takes for two initial classical points within a distance $\nu$ 
to move away a distance $\propto \zeta$ from each other.
In a chaotic system, this gives a logarithmic time, similar in physical
content to the Ehrenfest time, 
$\tau_{\cal U} = \lambda^{-1} \ln(\zeta/\nu)$, while in a regular
system, $\tau_{\cal U}$ is much longer, typically algebraic in 
$\zeta/\nu$. For $t > \tau_{\cal U}$, 
the last three correlators in Eq.(\ref{2correl}) disappear and
only $\langle {\cal S}_{s,s'}^2 \rangle$ survives. Because the four 
classical paths in that term come in two pairs,
they have no dependence on $|{\bf x}-{\bf y}|$. This is due
to the average we take over initial conditions together with the
dynamical spread of the wavepacket. 

At short times $t< \tau_{\cal U}$, on
the other hand,
the four correlators almost cancel one another, and Eq.~(\ref{2correl}),
which was obtained with
$\langle {\cal S}_{s,s'} {\cal S}_{l,s'}\rangle = \langle
{\cal S}_{l,l'} {\cal S}_{s,l'} \rangle$ and similar equalities,
does not hold anymore.
A Taylor expansion of the differences of the 
two-particle action integrals in Eq.(\ref{Faction}) gives
\begin{eqnarray}\label{Fsmall}
\Sigma_3(|{\bf x}-{\bf y}| \le \zeta;t)
&=&  \left(\frac{\nu^2}{\pi}\right)^{d_1+d_2}  \int d{\bf r} d{\bf r}'
\sum_{s,s'} \; \sum_{l,l'} \; C_s \, C_l \, C_{s'} \, C_{l'} \nonumber \\
&& \;\;\;\;\;\;\;\;\;\;\; \times \, 
\exp \big[-\nu^2 ({\bf p}_{s}-{\bf p}_1)^2 + ({\bf p}_{l}-{\bf p}_1)^2 +
({\bf p}_{s'}-{\bf p}_2)^2 + ({\bf p}_{l'}-{\bf p}_2)^2\big] \nonumber \\
&& \;\;\;\;\;\;\;\;\;\;\; \times \, \exp\Big[-2 \sum_{\alpha,\beta=1}^{d_1}
({\bf x}-{\bf y})_{\alpha} ({\bf x}-{\bf y})_{\beta}
D^{(1)}_{\alpha,\beta}({\bf x},{\bf y},{\bf r},{\bf r}';t)\Big] \nonumber \\
&& \;\;\;\;\;\;\;\;\;\;\; \times \, \exp\Big[-2 \sum_{\alpha,\beta=1}^{d_2}
({\bf r}-{\bf r}')_{\alpha} ({\bf r}-{\bf r}')_{\beta}
D^{(2)}_{\alpha,\beta}({\bf x},{\bf y},{\bf r},{\bf r}';t)\Big],
\end{eqnarray}
where
\begin{eqnarray}
D^{(1)}({\bf x},{\bf y},{\bf r},{\bf r}';t) &=& 
\int_0^t dt_1 \; dt_2 \langle \partial_{\alpha}^{(s)} {\cal U}
({\bf q}_s(t_1), {\bf q}_{s'}(t_1)) \;
\partial_{\beta}^{(s)} {\cal U}({\bf q}_s(t_2),{\bf q}_{s'}(t_2))
\rangle , \\
D^{(2)}({\bf x},{\bf y},{\bf r},{\bf r}';t) &=& 
\int_0^t dt_1 \; dt_2 \langle \partial_{\alpha}^{(s')} {\cal U}
({\bf q}_s(t_1), {\bf q}_{s'}(t_1)) \;
\partial_{\beta}^{(s')} {\cal U}({\bf q}_s(t_2),{\bf q}_{s'}(t_2))
\rangle,
\end{eqnarray}
depend on the endpoints 
${\bf x}$, ${\bf y}$, ${\bf r}$ and ${\bf r}'$ of $s$ and $s'$. 

So far we have learned that the variance of
off-diagonal matrix elements 
of $\rho_1$ is determined by classical correlators, with the
important caveat that they are bound downward by
the expressions given in Eq.~(\ref{bounddecay}). 
The rest of the discussion requires to specify the time-dependence
of these correlators as in Appendix~\ref{appendix:semiclassics_avg}. 
We make the same observation as above [see the discussions on 
Eqs.~(\ref{correl}) and (\ref{eq:nondiagdecay})] that,
provided these correlators 
decay faster than $\propto |t_1-t_2|^{-1}$, 
the off-diagonal matrix elements exhibit a dominant
exponential decay in time. This condition is rather nonrestrictive
and is surely satisfied in a 
chaotic system~\cite{Col04}. We therefore assume from now on a 
fast decay of correlations,
\begin{eqnarray}\label{correl1}
\langle 
{\cal U}({\bf q}_s(t_1), {\bf q}_{s'}(t_1))\;
{\cal U}({\bf q}_s(t_2),{\bf q}_{s'}(t_2)) \rangle & = & \Gamma_2 \;
\delta(t_1-t_2), \\
\label{correl2}
\langle \partial_{\alpha}^{(s,s')} {\cal U}
({\bf q}_s(t_1), {\bf q}_{s'}(t_1)) \;
\partial_{\beta}^{(s,s')} {\cal U}({\bf q}_s(t_2),{\bf q}_{s'}(t_2))\rangle
& = & \gamma_2 \; \delta_{\alpha,\beta} \; \delta(t_1-t_2).
\end{eqnarray}
The purity is straightforward to compute
from Eqs.~(\ref{2correl}) for $t>\tau_{\cal U}$ or (\ref{Fsmall})
for $t<\tau_{\cal U}$, using the correlators in Eq.~(\ref{correl1}) and (\ref{correl2})
and is discussed in the body of the text.

\subsection{The Boltzmann echo}
\label{appendix:semiclassics_boltzmann}

As starting point of our semiclassical calculation, we take 
chaotic one-particle Hamiltonians $H_{1,2}$, and an
interaction potential ${\cal U}$ that is smooth over a semiclassically
large distance, in the sense that 
it is characterized by a typical classical length scale, much larger than the de 
Broglie wavelength $\sigma$ of particle 1.
We furthermore assume that it
depends only on the distance between the particle 1 and 2.
For pedagogical reasons, the initial states are
narrow Gaussian wavepackets for both particles,
$\psi_i({\bf q})= 
(\pi \nu^2)^{-d_i/4} \exp[{\it i}{\bf p}_i \cdot 
({\bf q}-{\bf r}_{i})-|{\bf q}-{\bf r}_{i}|^2/2 \nu^2]$, though 
within our semiclassical approach,
more general states can be taken for
the uncontrolled system 2, such as random pure states 
$\rho_2 =\sum_{\alpha\beta} a_{\alpha} 
a_{\beta}^{\ast}  \vert \phi_{\alpha} \rangle \langle \phi_\beta \vert$,
random mixtures  
$\rho_2 =\sum_{\alpha} |a_{\alpha}|^2 
\vert \phi_{\alpha} \rangle \langle \phi_\alpha \vert$ or 
thermal mixtures $\rho_2 =\sum_n 
\exp\left[-\beta E_n\right]\vert n\rangle \langle n \vert$, without
affecting our result.
Also, arbitrary initial states for both subsystems 
can be considered within the RMT approach presented in the next appendix.

We first write ${\cal M}_{\rm B}(t)$ as 
\begin{widetext}
  \begin{eqnarray}\label{MB_start}
{\cal M}_{\rm B}(t) &=& \int {\rm d}{\bf z}_2 \;\;\;
\Bigg |
\int \prod_{i=1}^2 {\rm d} {\bf x}_i  \prod_{j=1}^{3} {\rm d} {\bf q}_j \;
\psi_1({\bf q}_1)\psi_2({\bf q}_2)\psi^{\dagger}_1({\bf q}_3) \nonumber \\
&& \;\;\;\;\;\;\;\;\;\;\;\;\;\;\; \times
 \left\langle {\bf q}_3, {\bf z}_2 \left\vert e^{-{\it i }{\cal H}_{\rm b} t }
\right\vert {\bf x}_1, {\bf x}_2  \right\rangle
\left\langle{\bf x}_1, {\bf x}_2  \left\vert e^{-{\it i }{\cal H}_{\rm f}  t} 
\right\vert {\bf q}_1, {\bf q}_2 \right\rangle
\Bigg |^2 .
\end{eqnarray}
We next generalize the two-particle semiclassical propagator of
Eq.~(\ref{twopart_propa}) to treat partial time-reversal. 
The propagator is given by
\begin{eqnarray}
 \left\langle  {\bf x}_1, {\bf x}_2 \left\vert e^{-{\it i}{\cal H}_{a} 
t} \right \vert {\bf q}_1, {\bf q}_2 \right\rangle \; &=& 
(2 \pi i)^{-(d_1+d_2)/2} 
\sum_{s_1,\,s_2} {\cal C}_{s_1,s_2}^{1/2} 
\exp[i \left\{  \epsilon^{(a)}  
S^{(a)}_{s_1}({\bf x}_1,{\bf q}_1;t) + 
S^{(a)}_{s_2}({\bf x}_2,{\bf q}_2;t) \right \}] \nonumber \\
&& \;\;\;\;\;\;\;\;\;\;\;\;\;\;\;\;\;\;\;\;\;\;\;\;\;\;\;\;\;\;\;\;
\times \exp[i \left\{ {\cal S}^{(a)}_{s_1,s_{2}}({\bf x}_{1},
{\bf q}_1;{\bf x}_{2}, {\bf q}_{2};t)\right \}] ,
 \label{SCproF} 
\end{eqnarray}
where $a={\rm f,b}$ labels forward or backward 
evolution and $ \epsilon^{(f)} = - \epsilon^{(b)} = 1$. This propagator
is expressed as sums over pairs of classical trajectories, 
labeled $s_i$ for particle $i$ connecting ${\bf q}_i$ to 
${\bf x}_i$ in the time $t$ with dynamics determined by 
$H_i$ or $H_i+\Sigma_i$. Under our assumption of a classically weak
coupling, classical trajectories are only determined by the 
one-particle Hamiltonians, and at this point, the reader certainly anticipates that
our justification for this approximation relies on structural
stability. Each pair of paths gives a 
contribution containing one-particle action integrals denoted by $S_{s_i}$
(where we included the Maslov indices) and two-particle action 
integrals  ${\cal S}^{({\rm f,b})}_{s_1,s_2}=\int_0^{t} {\rm d}\tau \
{\cal U}_{{\rm f,b}}[{\bf q}_{s_1}(\tau), 
{\bf q}_{s_2}(\tau)]$ accumulated along $s_1$ and $s_2$ and the 
 determinant ${\cal C}_{s_1,s_2}=C_{s_1} C_{s_2}$ of the stability 
matrix corresponding to the two-particle dynamics in the 
$(d_1+d_2)-$dimensional space.

We insert the semiclassical expression (\ref{SCproF}) into
Eq.~(\ref{MB_start}). There are four propagators in total, and one thus
faces a sum over eight classical trajectories $s_i$, and $l_i$,
$i=1,2,3,4$.
Our choice of initial Gaussian wave packets justifies
to linearize the one-particle action integrals 
in ${\bf q}_{j}-{\bf r}_{i}$. We furthermore 
set ${\cal S}^{(a)}_{s_1,s_2}({\bf x}_1,{\bf q}_1;{\bf x}_2,
{\bf q}_2;t) \simeq {\cal S}^{(a)}_{s_1,s_2 }
({\bf x}_1,{\bf r}_1; 
{\bf x}_2,{\bf r}_2;t)$, keeping in mind that ${\bf r}_{1}$ and
${\bf r}_{2}$, taken as arguments of the two-particle action integrals, 
have an uncertainty ${\cal O}(\nu)$. We then perform 
six Gaussian integrations to get
\begin{eqnarray}\label{eq:irrevtesbeforeSPA}
{\cal M}_{\rm B}(t)  &=&
   (\nu^2/\pi)^{(2 d_1+d_2)/2 }\int
 \prod_{i=1}^2 {\rm d}{\bf x}_i {\rm d}{\bf y}_i {\rm d}{\bf z}_2  \sum_{
\rm paths}
  {\cal A}_{s_1} {\cal A}_{s_2} {\cal A}_{s_3}^{\dagger} 
{\cal A}_{s_4}^{\dagger} {\cal A}_{l_1}^{\dagger} {\cal A}_{l_3}
 C_{l_2}^{{1\over 2}} C_{l_4}^{ {1\over 2} \dagger} \nonumber \\
&& \;\;\;\;\;\;\;\;\;\;\;\;\;\;\;\;\;\;\;\;\;\;\;\;\;\;\;\;\;\;\;\;\;\;\;\;\;\;\;\;\;\;\;\;\;\;\;\;\;\;\;\;\;\;
\times \exp\left[ i \left( \Phi_{1}+   \Phi_{2}  +     
\Phi_{12}  \right)\right].
\end{eqnarray}
In this expression, paths with odd (even) indices correspond to system 1 (2),
and paths denoted $s$ ($l$) correspond to the forward (backward) 
time-evolution. We furthermore defined
${\cal A}_{s_i} \equiv C_{s_i}^{\frac{1}{2}} \exp[- \nu^2
({\bf p}_{s_{i}} - {\bf p}_{i})^2 / 2 ]$. 
The semiclassical expression to ${\cal M}_{\rm B}$ is obtained by enforcing
a stationary phase condition on Eq.~(\ref{eq:irrevtesbeforeSPA}), i.e. keeping
only terms which minimize the variation of the three action phases
\end{widetext}
 \begin{subequations}
\begin{eqnarray}
 \Phi_{1}&=& S^{(\rm f)}_{s_1}({\bf x}_1,{\bf r}_1;t) -  
S^{(\rm b)}_{l_1} ({\bf x}_1,{\bf r}_1 ; t)                  
- S^{(\rm f)}_{s_3}({\bf y}_1,{\bf r}_1;t) +  S^{(\rm b)}_{l_3} ({\bf y}_1,
{\bf r}_1;t),  
\label{phi1}\\
  \Phi_{2} &=& S^{(\rm f)}_{s_2}({\bf x}_2,{\bf r}_2;t)  + 
S^{(\rm b)}_{l_2}({\bf z}_2,{\bf x}_2;t) 
                                - S^{(\rm f)}_{s_4}({\bf y}_2,{\bf r}_2;t)   
- S^{(\rm b)}_{l_4}({\bf z}_2,{\bf y}_2;t),  \label{phi2} \\
   \Phi_{12}&=& {\cal S}^{(\rm f)}_{s_1,s_{2} } + {\cal S}^{(\rm b)}_{l_1,l_{2} } 
                                    -  {\cal S}^{(\rm f)}_{s_3,s_{4} } - 
{\cal S}^{(\rm b)}_{l_3,l_4 }.  \label{phi12}
\end{eqnarray}
\end{subequations}
The semiclassically dominant terms are identified by
path contractions required by stationary phase conditions. We consider the weak interaction limit
where larger phases are due to the uncoupled dynamics, and accordingly first enforce a stationary
phase condition on $\Phi_1$ and $\Phi_2$.
The first stationary phase approximation 
over $\Phi_{1}$ corresponds to contracting unperturbed paths with perturbed 
ones, $s_1 \simeq l_1$ and  $s_3 \simeq l_3$. This pairing 
is allowed by our assumption of a classically
weak $\Sigma_1$, and is justified by structural stability,
rigorously for hyperbolic systems~\cite{Kat96,Cer02,Van03a} and numerically
for more generic chaotic systems Ref.~\cite{Gre90}.
The phase $\Phi_{1}$ is then given by the difference of
action integrals of the perturbation $\Sigma_1$ on paths $s_1$ and $s_3$,
$\Phi_{1} = \delta S_{s_1}({\bf x}_1,{\bf r}_1;t)  - 
\delta S_{s_3}({\bf y}_1,{\bf r}_1;t) $, with
  $\delta S_{s_i} =\int_{0}^{t} {\rm d}\tau \ \Sigma_1[{\bf q}_{s_i}(\tau)]$.
Here, ${\bf q}_{s_i}(\tau)$ lies on $s_i$ with 
${\bf q}_{s_i}(0) ={\bf r}_1$ 
and ${\bf q}_{s_1}(t)= {\bf x}_1$, ${\bf q}_{s_3}(t)= {\bf y}_1$.
A similar procedure for  
$\Phi_{2}$ requires $s_2 \simeq s_4$ and $l_2 \simeq l_4$, 
and thus ${\bf x}_2 \simeq {\bf y}_2$. These contractions lead to an exact
cancellation of the one-particle phase 
$\Phi_{2} = 0$ accumulated by system 2, and one gets a sum over 
four trajectories
\begin{eqnarray}\label{eq:irrevtesafterSPA}
 {\cal M}_{\rm B}(t)  &=&
   (\nu^2/\pi)^{\frac{2 d_1+d_2}{2} }\int
 \prod_{i=1}^2 {\rm d}{\bf x}_i {\rm d}{\bf y}_j {\rm d}{\bf z}_2 \; 
\Theta(\nu-|{\bf x}_2-{\bf y}_2|) \nonumber \\
&&\times  \sum 
 \vert {\cal A}_{s_1} \vert^2  \vert  {\cal A}_{s_2} \vert^2   
\vert  {\cal A}_{s_3}\vert ^2
  \vert C_{l_2}\vert
 \exp[ i \left( \delta S_{s_1}   -   \delta S_{s_3} +\delta \Phi_{12}.
\right)] . \mbox{} \quad
\end{eqnarray}
The Heaviside function $\Theta(\nu-|{\bf x}_2-{\bf y}_2|) $ restricts
the spatial integrations to
$|{\bf x}_2-{\bf y}_2|\leq \nu$ because of the finite resolution with 
which two paths can be equated. 

The semiclassical Boltzmann echo
(\ref{eq:irrevtesafterSPA}) is dominated by two contributions.
The one is non diagonal in that all paths are 
uncorrelated. Applying the CLT one has
\begin{subequations}
\begin{eqnarray}
\left\langle \exp[i \left\{ \delta S_{s_1} - \delta S_{s_3} + \delta
\Phi_{12}  \right\} ]\right\rangle &=& 
\exp\big[-\left\langle  \delta S^2_{s_1}\right\rangle
- \big\langle ({\cal S}^{(\rm f)}_{s_1,s_{2} })^2  \big\rangle
- \big\langle ({\cal S}^{(\rm b)}_{s_1,s_{2} })^2   \big\rangle\big], \\
 \langle \delta S_{s_1}^2 \rangle ]  &=&
\int_0^{t} {\rm d}\tau {\rm d}\tau' 
\langle \Sigma_1 [{\bf q}_{s_1}(\tau)] \;
\Sigma_1[{\bf q}_{s_1}(\tau')] \rangle, \\
\big\langle ({\cal S}^{(\rm f,b)}_{s_1,s_{2} })^2  \big\rangle 
&=& \int_0^{t } {\rm d}\tau \;  {\rm d}\tau' 
\langle {\cal U}_{\rm f,b}[{\bf q}_{s_1}(\tau), {\bf q}_{s_2}(\tau)] \;
{\cal U}_{\rm f,b}[{\bf q}_{s_1}(\tau'), {\bf q}_{s_2}(\tau')]  \rangle. \qquad
\end{eqnarray}
\end{subequations}
Once again we use the property that 
correlators typically decay exponentially fast in chaotic systems to write
 $\left\langle  \delta S^2_{s_1}  \right\rangle \simeq  \Gamma_{\Sigma_1} \; t $  
and 
$\big\langle ({\cal S}^{(\rm f,b)}_{s_1,s_{2} })^2  \big\rangle
 \simeq \Gamma_{\rm f,b} \; t $. 
Using next the two sum rules [similar to Eq.~(\ref{sumrule1})]
\begin{subequations}\label{eq:sumrule}
\begin{eqnarray}
(\nu^2/\pi)^{{{\rm d}_i \over 2}} \int {\rm d}{\bf x}_i  \sum_{s_i} \vert {\cal A}_{s_i} \vert^2 =1, \;\;\;\;\;\;\;\;\;\;
  \int {\rm d}{\bf x}_i  \int {\rm d}{\bf y}_i \; \Theta(\nu -| {\bf y}_i 
-{ \bf x}_i|) \sum_{l_i} \vert C_{l_i} \vert =1,
\end{eqnarray}
\end{subequations}
one obtains the nondiagonal contribution to the Boltzmann echo,
\begin{eqnarray}
\label{eq:irrevtes1}
 {\cal M}^{(\rm nd)}_{\rm B}(t)  & \simeq & \exp\left[-\left( \Gamma_{\Sigma_1} +
\Gamma_{\rm f}+\Gamma_{\rm b} \ \right) t  \right].
\end{eqnarray}

The second contribution is diagonal in the classical paths 
followed by the first particle, 
with $s_1 \simeq  s_3$ and
${\bf x}_1 \simeq {\bf y}_1$. It is thus given by a sum over three
trajectories. From
Eq.~ (\ref{eq:irrevtesafterSPA}) it reads
\begin{eqnarray}\label{eq:irrevtesdiag}
 {\cal M}^{(\rm d)}_{\rm B}(t)  =&&\!\!\! \!\! 
   (\nu^2/\pi)^{\frac{2 d_1+d_2}{2} }\int
 \prod_{i=1}^2 {\rm d}{\bf x}_i {\rm d}{\bf y}_i d{\bf z}_2  
\; \Theta(\nu-|{\bf x}_i-{\bf y}_i|) 
 \nonumber \\ &\times& \; \sum_{s_1,s_2,l_2} \;
 \vert {\cal A}_{s_1} \vert^4  
\vert  {\cal A}_{s_2}\vert ^2
 \vert C_{l_2}\vert
  e^{ {\it i} \left[ \Delta S_{s_1}+ \Delta{\cal S}_{s_1,s_2}^{({\rm f}) } + \Delta{\cal S}_{s_1,l_2} ^{({\rm b}) }
\right]},\qquad
\end{eqnarray}
where 
$\Delta S_{s_1} =  \int_{0}^{t} {\rm d}\tau \nabla_1 
\Sigma_1[{\bf q}_{s_1}(\tau)] 
\cdot [{\bf q}_{s_3}(\tau)-{\bf q}_{s_1}(\tau)]$ and 
$\Delta  {\cal S}^{({\rm f,b})}_{s_1,s_2}=  \int_{0}^{t} {\rm d}\tau
\nabla_1 {\cal U}_{{\rm f,b}}[{\bf q}_{s_1}(\tau),{\bf q}_{s_2}(\tau)] \cdot [
{\bf q}_{s_3}(\tau)-{\bf q}_{s_1}(\tau)]
$. 
We perform a change of coordinates $\int d{\bf x}_1 \sum  |{C}_{s_1}| 
 = \int d{\bf p_1}$, and use both the asymptotics
$|C_{s_1}| \propto \exp\left[-\lambda_1 t \right]$ valid for chaotic 
systems 
and the sum rules of Eqs.~(\ref{eq:sumrule}) to get
 \begin{eqnarray}\label{eq:rrevtes2}
{\cal M}^{(\rm d)}_{\rm B}(t) & \simeq & 
 \alpha_1  \exp\left[-\lambda_1 t \right].
  \end{eqnarray}
Here, $\alpha_1$ is only algebraically time-dependent with
$\alpha_1(t=0) = {\cal O} (1)$. 
We finally note that the long-time saturation at the inverse
Hilbert space size of system 1,
${\cal M}_{\rm B}(\infty) = N_1^{-1}$,
is obtained from Eq.~(\ref{eq:irrevtesbeforeSPA}) with the contractions
$s_1\simeq s_3$, $s_2\simeq s_4$, $l_1\simeq l_3$ and $l_2\simeq l_4$.
Summing the saturation contribution with
the diagonal (\ref{eq:rrevtes2})
and nondiagonal (\ref{eq:irrevtes1})
contributions, one obtains our main result, Eq.~(\ref{eq:BEmod}).

\section{Random matrix theory of the Boltzmann echo}\label{appendix:RMT}

RMT treatments for the Loschmidt echo and for entanglement generation in bipartite
interacting systems have been presented in Chapter~\ref{section:ML_RMT} and
Chapter~\ref{section:P_RMT}
respectively. They are based on eigenfunction correlators, Eqs.~(\ref{contractions}) 
and (\ref{rmt_contract_tip}). 
The RMT calculation of ${\cal M}_{\rm B}$ we present here is based on similar relations.
It does not represent any additional technical difficulty, and we thus confine it
to this appendix. Our main task here is to show how the RMT result for ${\cal M}_{\rm B}(t)$ 
is compatible
with the semiclassical result, Eq.~(\ref{eq:BEmod}) in the limit $\lambda \rightarrow \infty$. 
The approach follows the same lines as 
the calculation presented 
in Chapters~\ref{section:ML_RMT} and \ref{section:P_RMT}.
Our starting point is
\begin{equation}\label{eq:mb_start_rmt}
{\cal M}_{\rm B}(t) = N_2^{-1} \sum_{{\bf \phi}_2, {\bf \psi}_2} \, 
\langle \psi_1, \phi_2 | \, \exp[-{\it i }{\cal H}_{\rm b} t]  \exp[-{\it i}{\cal H}_{\rm f} t ] \, \rho_0 \, \exp[ {\it i }{\cal H}_{\rm f} t ] \exp[ {\it i}{\cal H}_{\rm b} t ] \, | \psi_1, \phi_2 \rangle ,
\end{equation}
where we take an initial product state
$\rho_0 = |\psi_1 , \psi_2 \rangle \langle \psi_1 , \psi_2|$.
Our RMT strategy consists in inserting 
resolutions of the identity into Eq.~(\ref{eq:mb_start_rmt})
and then use averages similar to those we already encountered in Eq.~(\ref{rmt_contract_tip}). 
Compared to the purity, the Boltzmann echo requires to consider
four different complete 
sets of eigenvectors $\{\alpha_i^{(f,b)}\}$, for the
uncoupled forward $(f)$ and backward $(b)$ dynamics of particle $i=1,2$ 
and two two-particle eigenstates basis 
$\{ \Lambda^{(f,b)}\}$. This
renders the calculation somehow longer and more tedious, 
but does not add any additional technical difficulty.
We first insert four resolutions of the identity
\bea
I & = & \sum_{\alpha_1,\alpha_2} |\alpha_1^{(f,b)},\alpha_2^{(f,b)} \rangle
\langle \alpha_1^{(f,b)},\alpha_2^{(f,b)} | \,
\eea
into Eq.~(\ref{eq:mb_start_rmt}) to obtain
\bea
{\cal M}_{\rm B}(t) & = & N_2^{-1} \sum_{{\bf \phi}_2, {\bf \psi}_2} \,
\sum_{\alpha'{\rm s},\beta'{\rm s}}
\overline{ \langle \psi_1, \phi_2 | \alpha_1^{(b)},\alpha_2^{(b)} \rangle \,
\langle \alpha_1^{(f)},\alpha_2^{(f)}| \psi_1, \psi_2\rangle \,
\langle  \psi_1, \psi_2| \beta_1^{(f)},\beta_2^{(f)}\rangle \, 
\langle \beta_1^{(b)},\beta_2^{(b)} |  \psi_1, \phi_2\rangle} \, \nonumber \\
&& \;\;\;\;\;\;\;\;\;\;\;\;\;\;\;\;\;\;\;\;  \times \langle \alpha_1^{(b)},\alpha_2^{(b)} | \exp[-{\it i }{\cal H}_{\rm b} t]  \exp[-{\it i}{\cal H}_{\rm f} t ] | \alpha_1^{(f)},\alpha_2^{(f)} \rangle 
 \, \nonumber \\
&& \;\;\;\;\;\;\;\;\;\;\;\;\;\;\;\;\;\;\;\;  \times \langle \beta_1^{(f)},\beta_2^{(f)} | \exp[-{\it i }{\cal H}_{\rm b} t]  \exp[-{\it i}{\cal H}_{\rm f} t ] | \beta_1^{(b)},\beta_2^{(b)} \rangle .
\eea
We next use the leading-order RMT averages (we neglect subdominant
weak localization corrections)
\begin{subequations}
\begin{eqnarray}
\overline{\langle \phi_2 | \alpha_2^{(b)} \rangle
\langle \beta_2^{(b)} | \phi_2\rangle} =
\overline{\langle \psi_2 | \beta_2^{(f)} \rangle
\langle \alpha_2^{(f)} | \psi_2\rangle}
& = & \delta_{\alpha_2,\beta_2}
\, N_2^{-1}, \\ \label{secondeq}
\overline{\langle \psi_1 | \alpha_1^{(b)} \rangle \,
\langle \alpha_1^{(f)} | \psi_1 \rangle \,
\langle  \psi_1 | \beta_1^{(f)} \rangle \, 
\langle \beta_1^{(b)} |  \psi_1 \rangle }
&=& 
\langle \alpha_1^{(f)} | \alpha_1^{(b)} \rangle 
\, \langle \beta_1^{(f)} | \beta_1^{(b)} \rangle  \, N_1^{-2} +
\delta_{\alpha_1,\beta_1}   N_1^{-2}, 
\end{eqnarray}
\end{subequations}
where we eased the notation a bit by dropping the 
subindices $^{(f,b)}$ in the Kronecker delta's. The second term on the 
right-hand of Eq.(\ref{secondeq}) leads to the long-time saturation
${\cal M}_{\rm B}(\infty) = N_1^{-1}$.
The dominant contribution to ${\cal M}_{\rm B}$ thus reads
\bea\label{thirdeq}
{\cal M}_{\rm B}(t) &=& N_1^{-2} N_2^{-1} \sum_{\alpha'{\rm s}}
\sum_{\beta'{\rm s}} 
\, \langle \alpha_1^{(f)} | \alpha_1^{(b)} \rangle \,
\langle \beta_1^{(b)}| \beta_1^{(f)} \rangle \,\\
&& \;\;\;\;\;\;\;\;\;\;\;\;\;\; \times
\langle \alpha_1^{(b)},\alpha_2^{(b)}|  e^{-{\it i }{\cal H}_{\rm b} t}  
e^{-{\it i}{\cal H}_{\rm f} t } \,
|\alpha_1^{(f)}, \beta_2^{(f)} \rangle 
\langle \beta_1^{(f)},\beta_2^{(f)}|\, e^{i {\cal H}_{\rm f} t}
e^{ i{\cal H}_{\rm b} t } \,
|\beta_1^{(b)}, \alpha_2^{(b)} \rangle. \nonumber
\eea
We next insert 
\be
I = \sum_{\Lambda^{(f,b)}} |\Lambda^{(f,b)} \rangle \langle \Lambda^{(f,b)}|
\ee
left and right of all the time-evolution operators in Eq.~(\ref{thirdeq}),
and finally use
\bea
\langle \Lambda_i^{(f)} | \Lambda_j^{(b)}\rangle = 
\sum_{\alpha_1^{(f)},\alpha_2^{(f)}}
\sum_{{\beta_1^{(b)},\beta_2^{(b)}}}
\langle \Lambda_i^{(f)} | \alpha_1^{(f)},\alpha_2^{(f)} \rangle
\langle \alpha_1^{(f)},\alpha_2^{(f)}|\beta_1^{(b)},\beta_2^{(b)}\rangle\langle \beta_1^{(b)},\beta_2^{(b)} | \Lambda_j^{(b)} \rangle,
\eea
as well as  a similar expression with $b \leftrightarrow f$. After some algebra --
invoking further RMT averages as in Eq.~(\ref{secondeq}) among others -- 
one finally obtains
\bea\label{fourtheq}
{\cal M}_{\rm B}(t) & = & N_1^{-2} \sum_{\alpha_1'{\rm s}}
\big|\langle \alpha_1^{(f)} |
\alpha_1^{(b)} \rangle\big|^2 e^{-i (\alpha_1^{(b)}
-\alpha_1^{(f)})t} \; \sum_{\beta_1'{\rm s}}
\big|\langle 
\beta_1^{(b)}| \beta_1^{(f)} \rangle\big|^2 e^{i ({\beta_2}^{(b)}
-{\beta_1}^{(f)})t}  \\
&\times& \sum_{\Lambda_1^{(b)}} \big|\langle 
\alpha_1^{(b)}, \alpha_2^{(b)} | \Lambda_1^{(b)} \rangle\big|^2 
e^{-i(\Lambda_1^{(b)} -\alpha_1^{(b)}
-\alpha_2^{(b)})t} \, \sum_{\Lambda_1^{(f)}} \big|\langle 
\alpha_1^{(f)}, \alpha_2^{(f)}| \Lambda_1^{(f)} \rangle\big|^2 
e^{-i(\Lambda_1^{(f)} -\alpha_1^{(f)}
-\alpha_2^{(f)})t}\nonumber \\
&\times& \sum_{\Lambda_2^{(b)}} \big|\langle \Lambda_2^{(b)} |
\beta_1^{(b)}, \alpha_2^{(b)} \rangle\big|^2 
e^{i(\Lambda_2^{(b)} -\beta_1^{(b)}
-\alpha_2^{(b)})t} \;\; \sum_{\Lambda_2^{(f)}} \big|\langle \Lambda_2^{(f)} |
\beta_1^{(f)}, \alpha_2^{(f)} \rangle\big|^2 
e^{i(\Lambda_2^{(f)} -\beta_1^{(f)}
-\alpha_2^{(f)})t}, \nonumber 
\eea
where eigenenergies are denoted by $\Lambda_i^{(f,b)}$ and $\alpha_i^{(f,b)}$ and $\beta_i^{(f,b)}$. 
We are almost done. Each of the six terms in the above
expression gives the Fourier transform of the projection of 
one- or two-particle eigenfunctions of a perturbed Hamiltonian over
the eigenfunctions of the corresponding unperturbed Hamiltonian.
For the two terms in the first line of (\ref{fourtheq}), the perturbation 
is $\Sigma_1$, while for the last four terms, the perturbation is
${\cal U}_{\rm f,b}$. In both cases, the three usual first-order 
perturbative, golden rule and strongly perturbed regimes have to be considered
separately, with the corresponding delta-peaked, Lorentzian and 
ergodic eigenfunction projections [see Eqs.~(\ref{eq:1part_spread}) 
and (\ref{eq:2part_spread})]. Replacing the sums by integral over
energies the first line of (\ref{fourtheq}) gives a factor
\bea
\sim \left\{ 
\begin{array}{cc}
\exp[-\overline{\Sigma_1^2} t^2] & {\rm first \, order }, \, \Gamma_{\Sigma_1} < \delta,  \\
\exp[-\Gamma_{\Sigma_1} t]  & {\rm golden \, rule }, \, \delta \lesssim \Gamma_{\Sigma_1} \ll B,  \\
\exp[-B_1^2 t^2]  & \;\; {\rm strong \, perturbation}, \, \Gamma_{\Sigma_1} > B. 
\end{array} \right.
\eea
while the second and third line combine to give
\bea
\sim \left\{ 
\begin{array}{cc}
\exp[-(\overline{{\cal U}_{\rm f}^2}+\overline{{\cal U}_{\rm b}^2}) t^2] 
& {\rm first \, order }, \, \Gamma_{\rm f,b} < \delta_2, \\
\exp[-(\Gamma_{\rm f} + \Gamma_{\rm b}) t]  & {\rm golden \, rule }, \, \delta_2 \lesssim \Gamma_{\rm f,b} \ll B_2,\\
\exp[-B_2^2 t^2]  &  \;\;  {\rm strong \, perturbation}, \, \Gamma_{\rm f,b} > B_2 . 
\end{array} \right.
\eea
Taking the saturation term into account, we finally recover our 
results Eqs.~(\ref{eq:BEmod}), (\ref{gaussiand})
and (\ref{eq:BEmod_mixed}), for the RMT-compatible case of infinite
Lyapunov exponent.

\section{Numerical models}\label{appendix:nums}

\subsection{The kicked top}
\label{appendix:ktop}

The kicked top \cite{Haa87,Haa01} has a time-dependent Hamiltonian
\begin{equation}
H_{0}=(\pi/2)S_{y}+(K/2S)S_{z}^{2}\sum_{n}\delta(t-n). \label{H0def}
\end{equation}
The model describes a vector spin of conserved 
integer or half-integer magnitude $S$ that undergoes a free precession
around the $y$-axis perturbed periodically 
by sudden rotations of period $\tau \equiv 1$ around the $z$-axis over an angle proportional to $S_{z}$. Classically, this dynamics is captured by the
map
\begin{eqnarray}
\left\{
\begin{array}{cc}
x_{n+1} = &z_n \cos(K x_n) + y_n \sin(K x_n) \\
y_{n+1} = &-z_n \sin(K x_n) + y_n \cos(K x_n) \\
z_{n+1} = &-x_n, 
\end{array}
\right.
\end{eqnarray}
Quantum-mechanically, the unitary time evolution after $n$ periods is 
given by the $n$-th power of the {\it Floquet operator}
\begin{equation}
F_{0}=\exp[-i(K/2S)S_{z}^{2}]\exp[-i(\pi/2)S_{y}]. \label{Fdef}
\end{equation}
Depending on the kicking strength $K$, the classical dynamics is regular,
partially chaotic, or fully chaotic at large $K$. From the data shown on 
the top left panel of Fig.~\ref{fig:fig1_echo1}, we see that the kicked top
is chaotic, with vanishingly small islands of stability for $K \gtrsim 9$.

The Floquet operator $F_0$ gives the forward time-evolution and for the fidelity, we need
to define a perturbed reversed time-evolution. Therefore, for the reversed time evolution we introduce a
perturbation in the form of a
periodic rotation of constant angle around the $x$-axis, slightly
delayed with respect to the kicks in $H_0$,
\begin{equation}
H_{1}=\phi S_{x}\sum_{n}\delta(t-n -\epsilon). \label{H1def}
\end{equation}
The corresponding Floquet operator is $F=\exp(-i\phi S_{x})F_{0}$. The parameter $\phi$ gives the
strength of the perturbation.

Both $H$ and $H_{0}$ conserve the spin magnitude $S$. However, because the Hamiltonian is
time-dependent, the energy is not conserved and a transition to chaos occurs as $K$ is increased.
We choose the initial wave
packets as coherent states of the spin SU(2) group~\cite{Per86}, i.e.\ states
which minimize the Heisenberg uncertainty in phase space. In our case the latter is the
sphere of radius $S$, on which the Heisenberg resolution is determined by the
effective Planck constant
$\hbar_{\rm eff} \sim S^{-1}$. 

\subsection{The one-particle kicked rotator}
\label{appendix:1krot}

The second dynamical system we use in our numerics is the kicked rotator 
model. Its Hamiltonian reads
\cite{Izr90}
\begin{equation}\label{kickrot}
H_0 = \frac{\hat{p}^2}{2} + K_0 \cos \hat{x} \sum_n \delta(t-n).
\end{equation}
Eq.~(\ref{kickrot}) gives the time-dependent Hamiltonian formulation of
the celebrated standard map~\cite{Chi08}. The latter gives a local description of 
nonlinear resonances which correctly describes a large variety of dynamical systems --
hence its name. For $K=0$, the system is trivially integrable. 
Nonlinear resonances arise as $K$ is increased, and for $K \approx 1$, the last invariant
torus globally bounding the dynamics in momentum is destroyed.
We concentrate on the regime $K > 7$, for which the dynamics is fully
chaotic with a Lyapunov exponent $\lambda \simeq \ln[K/2]$. 
We quantize this Hamiltonian on a torus, which 
requires to consider discrete values
$p_l=2 \pi l/N$ and $x_l=2 \pi l/N$, $l=1,...N$, for the canonically conjugated
momentum and position. Here, $N$ is an integer proportional to the inverse effective Planck's constant, 
$\hbar_{\rm eff}=N^{-1}$, i.e. the semiclassical limit correspond to taking the large $N$ limit. It increases
the system size and accordingly the computation time. 

The fidelity is computed for discrete times $t=n$, as
\begin{eqnarray}\label{krot_fid}
{\cal M}_{\rm L}(n) & = & |\langle \psi_0 | \left(F^\dagger\right)^n 
\left(F_{0}\right)^n  | \psi_0 \rangle|^2
\end{eqnarray}
using the unitary Floquet operators 
\begin{eqnarray}
F_{0}=\exp[-i \hat{p}^2/2 \hbar_{\rm eff}] \exp[-iK_0 \cos \hat{x}/ 
\hbar_{\rm eff}] \, , \\
F_{\delta K}=\exp[-i \hat{p}^2/2 \hbar_{\rm eff}] \exp[-i(K_0+\delta K) \cos \hat{x}/ 
\hbar_{\rm eff}] .
\end{eqnarray}
The quantization procedure results in a matrix form of the Floquet operators,
whose matrix elements in $x-$representation are given by
\begin{eqnarray}
\left(F_{0}\right)_{l,l'} & = & \frac{1}{\sqrt{N}} \exp[i 
\frac{\pi (l-l')^2}{N}] \exp[-i \frac{N K_0}{2 \pi} \cos \frac{2 \pi l'}{N}] \, , \\
\left(F_{\delta K}\right)_{l,l'} & = & \frac{1}{\sqrt{N}} \exp[i 
\frac{\pi (l-l')^2}{N}] \exp[-i \frac{N (K_0+\delta K)}{2 \pi} \cos \frac{2 \pi l'}{N}] \, .
\end{eqnarray}
Numerically, the time-evolution of $\psi_0$ in
the fidelity, Eq.~(\ref{krot_fid}),
is calculated by recursive calls to a fast-Fourier transform routine.
Thanks to this algorithm, the matrix-vector multiplication
$F_{0,\delta K} \, \psi_0$ requires $O(N \ln N)$ operations instead
of $O(N^2)$, and thus 
allows to deal with much larger system sizes with the kicked rotator than with the kicked top. 
The data presented in Chapter~\ref{section:k_rot} 
correspond to system sizes of up to
$N \le  262144 = 2^{18}$ which still allowed to collect enough statistics
for the calculation of the variance $\sigma^2({\cal M}_{\rm L})$ of the
Loschmidt echo. 
Because our algorithm
relies on fast-Fourier transforms, our system sizes in this review are powers of 2 whenever
we use the kicked rotator.

\subsection{The two- and $N$-particle kicked rotator}
\label{appendix:2krot}

In our investigations of entanglement generation and of the Boltzmann echo, we rely on a
model of two interacting kicked rotators. The model still keeps most of the 
algorithmic advantages of the
single-particle kicked rotator, in particular, one can still reach semiclassically large system sizes
that allow to search and find Lyapunov decays over several decades, even for two interacting
particles and the associated squaring of the system size. The model is defined by
\begin{subequations}
\label{2krot}
\begin{eqnarray}
H_i & = & p_i^2 / 2 + K_i \cos(x_i) \; \sum_n \delta(t-n),\\
{\cal U} & = & \epsilon \; \sin(x_1-x_2-0.33) \; \sum_n \delta(t-n).
\end{eqnarray}
\end{subequations}
The interaction potential ${\cal U}$ is long-ranged, with a 
strength $\epsilon$ and acts at the same time as the kicks. It has already been
mentioned above that the chaoticity of the dynamics can be tuned from
fully integrable ($K_i=0$) to fully chaotic [$K_i\agt 7$, with Lyapunov exponent
$\lambda_i\approx\ln (K_i/2)$]. For $1<K_i<7$ the dynamics is mixed, and one may consider
all possibilites of regular, mixed or chaotic dynamics individually for particle one and two.
In this work, however we restrict ourselves to the case of two chaotic particles, and vary 
$K_{1,2} \in [3,12]$ to get a maximal variation of $\lambda_i$, while
making sure that both initial Gaussian wavepackets 
$\psi_1$ and $\psi_2$ lie in the chaotic sea. 
We follow the usual quantization procedure on the torus $x,p\in(-\pi,\pi)$ for each kicked rotator. 
There is no procedure of quantum symmetrization involved as we consider distinguishable
particles.
The two-particle bandwidth and level spacing are given by
$B_2 = 2 \pi$, $\delta_2 = 2 \pi/(N_1 N_2)$, and we numerically extracted the level
broadening of interacting two-particle levels
$\Gamma_2 \simeq 0.43\epsilon^2 N_1N_2$ from exact diagonalization calculations
of the local spectral density of eigenstates of the ${\cal U}=0$ Hamiltonian 
over the eigenstates
of the full, interacting two-particle Hamiltonian (this local spectral density of states 
is shown in the inset to 
Fig.~\ref{fig6_qmclcorr}). The time evolved density matrix 
is computed by means of a two-dimensional fast Fourier transforms.
The algorithm requires only ${\cal O}(N_1 N_2 \ln N_1 N_2)$ operations, which
allowed us to reach system sizes up to $N_{1,2}=2048$, more than one order
of magnitude larger than any previously investigated case for entanglement generation
between two interacting dynamical systems. The data we
present are restricted to $N_1=N_2\equiv N$, except in the inset to 
Fig.~\ref{fig7_qmclcorr}.

The model is easily generalized to $N$ interacting particles,
\begin{subequations}
\label{Nkrot}
\begin{eqnarray}
H_i & = & p_i^2 / 2 + K_i \cos(x_i) \; \sum_n \delta(t-n),\\
{\cal U}_{ij} & = & \epsilon_{ij} \; \sin(x_i-x_j-0.33) \; \sum_n \delta(t-n).
\end{eqnarray}
\end{subequations}
for $i,j=1,2,..., N$. It is not clear to us how the (anti)symmetrization of the $N$-body wavefunction 
required by quantum mechanics for distinguishable particles
can
be achieved in this model, without negatively affecting the performance of our algorithm.



\end{document}